\documentclass[12pt,draftcls, onecolumn]{IEEEtran}	
\usepackage[utf8]{inputenc}
\usepackage[english]{babel}
\usepackage{float}
\usepackage{bookmark}
\usepackage{acronym}
\usepackage[toc]{glossaries}
\usepackage{array}
\usepackage{dsfont}
\usepackage{amsmath}
\usepackage{dsfont}
\usepackage{amssymb}
\usepackage{amsfonts}
\usepackage{graphics}
\usepackage{graphicx}
\usepackage{grffile}
\usepackage{mathtools}
\usepackage{nccmath}
\usepackage{caption}
\usepackage{relsize}
\usepackage{comment}
\usepackage{multirow}
\usepackage{multicol}
\usepackage{subfiles}
\usepackage{wrapfig}
\usepackage[a4paper, total={7in, 10in}]{geometry}
\usepackage{collectbox}
\usepackage{setspace}
\usepackage{amsthm}
\usepackage{epsfig,verbatim}
\usepackage{blindtext}
\usepackage{textcomp}
\usepackage[justification=centering,font=footnotesize,labelfont=bf]{caption}
\usepackage{romannum}
\usepackage{subcaption}
\usepackage{chngcntr}
\usepackage[shortlabels]{enumitem}

\def\BibTeX{{\rm B\kern-.05em{\sc i\kern-.025em b}\kern-.08em
    T\kern-.1667em\lower.7ex\hbox{E}\kern-.125emX}} 
\theoremstyle{plain} 

\newtheorem{MyRemark}{Remark}

\newtheorem{definition}{Definition}

\newtheorem{lemma}{Lemma}

\DeclareMathOperator{\Tr}{Tr}

\DeclareMathOperator*{\argmin}{argmin}
\makeatletter
\newcommand{\mybox}{%
    \collectbox{%
        \setlength{\fboxsep}{1pt}%
        \fbox{\BOXCONTENT}%
    }%
}
\makeatother
\newcommand{\RNum}[1]{\uppercase\expandafter{\romannumeral #1\relax}}
\begin{document}
\title{A New Frame Synchronization Algorithm
for Linear Periodic Channels with
Memory}
\author{Oren Kolaman, \IEEEmembership{Student Member, IEEE},
Ron Dabora,
\IEEEmembership{Senior Member, IEEE}}
\maketitle
\begin{abstract}
Identifying the start time of a sequence of symbols received at the receiver, commonly referred to as \emph{frame synchronization}, is a critical task for achieving good performance in digital communications systems employing time-multiplexed transmission. In this work we focus on \emph{frame synchronization} for linear channels with memory in which the channel impulse response is periodic and the additive Gaussian noise is correlated and cyclostationary. Such channels appear in many communications scenarios, including narrowband power line communications and interference-limited wireless communications. We derive frame synchronization algorithms based on simplifications of the optimal likelihood-ratio test, assuming the channel impulse response is unknown at the receiver, which is applicable to many  practical scenarios. The computational complexity of each of the derived algorithms is characterized, and a procedure for selecting nearly optimal synchronization sequences is proposed. The algorithms derived in this work achieve better performance than the noncoherent correlation detector, and, in fact, facilitate a controlled tradeoff between complexity and performance.
\end{abstract}



\pagenumbering{arabic}
\section{Introduction}
\label{chap:intro}


In digital communications,  blocks of information bits are mapped into  sequences of symbols taken from a finite constellation set. The transmitter partitions the stream of  symbols into frames of finite duration, which are then sent to a receiver through a channel. In order for the receiver to successfully recover the information bits from the received distorted and noisy version of the transmitted signal, the receiver must first identify the physical parameters of the transmitted signal; In particular it has to first identify the starting time of a received frame before any subsequent processing task is applied. The process of identifying the beginning of a received information frame within a measured incoming stream of samples at the receiver is called frame synchronization (FS) \cite{scholtz1980frame}, \cite{robertson1995optimal}. As an example, in time division multiple access (TDMA) systems, FS identifies the boundaries between the multiplexed users \cite{mengali2013synchronization}. 
In the current work, the received signal consists of a  convolution between the transmitted signal and a periodic channel impulse response (CIR), which generalizes upon the common linear time-invariant (LTI) channel model. Then, an additive cyclostationary Gaussian noise (ACGN) process with a finite correlation length is summed with the output of the convolution to form the overall received signal. The ACGN model generalizes upon the common additive stationary Gaussian noise model. The channel model described above represents many communications scenarios, including narrowband (NB) power line communications (PLC) and interference-limited wireless communications \cite[Sec. 5]{gardner2006cyclostationarity}, \cite{nassar2012local}, \cite{shaked2017joint}. Instances of these scenarios include non-orthogonal multiple access (NOMA)  \cite{Andrews:2014}, \cite{Dai:2015}, full duplex communications, cognitive communications \cite{Hong:2009}, digital subscriber line (DSL) communications \cite{Campbell:1983} as well as additional communications mediums and paradigms, including  the ``Internet of Things" \cite{McHenry:2015}. In \cite{shaked2017joint}, we considered the joint estimation of the carrier frequency offset and the channel impulse response for this channel model, and proposed a compressed-sensing based approximate joint maximum likelihood estimator for these quantities.  In this paper we study FS for the above received signal model, using the common pilot-assisted approach, in which the transmitted signal contains a specifically designed synchronization sequence which is embedded into the random data \cite{scholtz1980frame}, \cite{massey1972optimum}. Accordingly, the objective of the FS algorithm developed in this work is to identify the presence of the synchronization sequence within an observation window. The unique aspects of our work are the general channel model for which the algorithm is derived, as well as the rigorous approach used in the derivation, in contrast to the ad-hoc solutions of previous works. Novelty follows as previous works did not explicitly account for the channel memory in the derivation of the estimator, due to its associated computational complexity. Moreover, our derived algorithms do not assume knowledge of the CIR coefficients.
To the best of our knowledge, an FS algorithm for an unknown linear channel with memory and ACGN, derived based on the likelihood ratio test (LRT) has not been previously presented. 

\begin{figure}[!ht]
\includegraphics[width=155mm,height=35mm]{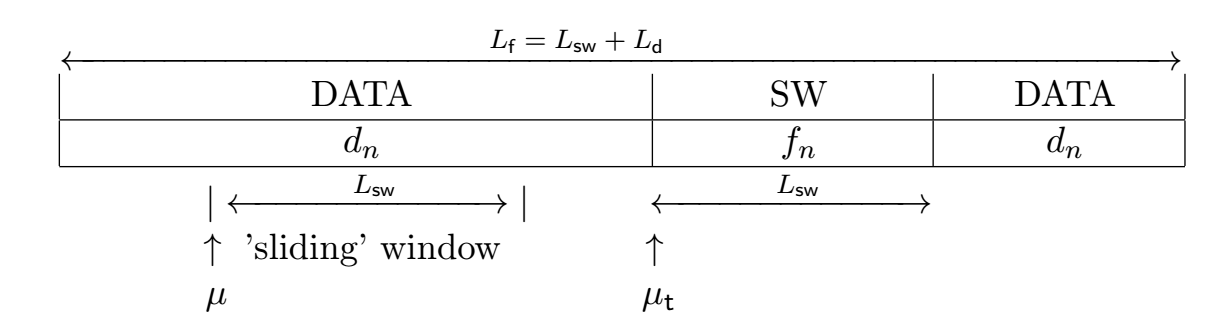}
\centering
\captionsetup{justification=raggedright}
\caption[Frame synchronization process]{Frame synchronization process: $\mu_{\textsf{t}}$ is the true position of the synchronization sequence within the observation window and $\mu$ is the start position of the sliding search window within the received observation window. \label{Structure of Transmited Frame Detection}}
\end{figure}

The common approach for frame synchronization is  the \emph{marker concept}, in which a known synchronization sequence, also referred as the synchronization word (SW), is incorporated into each transmitted data frame, 
as conceptually
\begin{figure}[!ht]
\includegraphics[width=155mm,height=30mm]{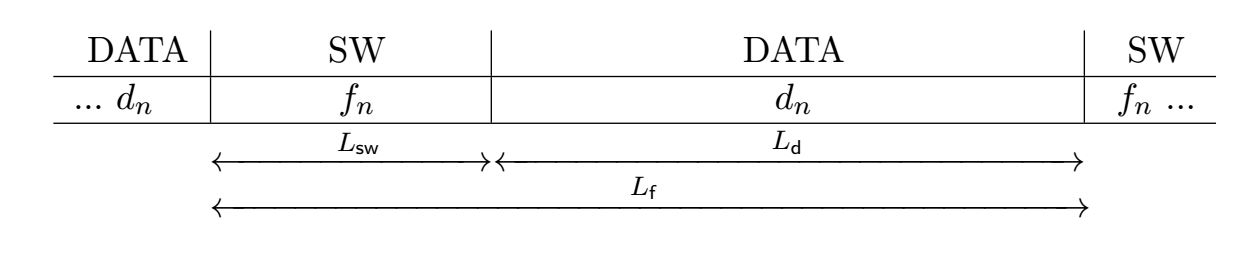}
\centering
\captionsetup{justification=raggedright}
\caption[The structure of the transmitted frame]{The structure of transmitted frame. Overall, the frame contains $L_{\textsf{f}}\in \mathcal{N}$ complex symbols and is divided into two parts: A synchronization word whose length is $L_{\textsf{sw}}\in \mathcal{N}$ and $L_{\textsf{d}}=(L_{\textsf{f}}-L_{\textsf{sw}})$ randomly selected data symbols denoted $\{d_n\}$.  \label{Structure of Transmited Frame}}
\end{figure}
depicted in Fig \ref{Structure of Transmited Frame}. Assuming there is only one SW in each data frame,
the synchronization algorithm observes the received signal and evaluates a mathematical function over a sliding search window whose duration is equal to the duration of the SW, as conceptually depicted in Fig. \ref{Structure of Transmited Frame Detection}.
There are two main detection approaches which utilize the marker concept:
The first approach evaluates a test metric for all possible  positions within the observation window and chooses the start position of the SW as the value of $\mu$ which maximizes the test metric. This approach is referred to as one-shot point estimation (OSPE) FS, \cite{ramakrishnan2010frame,liang2015sequential}. The second approach is comparing the test metric computed for a specific position $\mu$ within the observation window to a pre-determined threshold, such that once the threshold value is exceeded, the start position of the synchronization sequence is declared. Otherwise, the test metric is evaluated at the next position. This approach is commonly referred to as sequential FS  \cite{ramakrishnan2010frame,liang2015sequential}. A study of the theoretical and implementation tradeoffs between OSPE FS and sequential FS for the general problem of FS in the presence of a large residual carrier
frequency offset, is reported in \cite{ramakrishnan2010frame}.
OSPE FS algorithms for memoryless additive white Gaussian noise (AWGN) channels have been proposed in \cite{massey1972optimum,nielsen1973some,lui1987frame,choi2002frame}, and for AWGN channels with memory in \cite{moon1991ml} and \cite{wang2004continuous}. In \cite{massey1972optimum}, Massey showed that the optimal rule, in the sense of maximizing the probability of correctly locating the SW, for OSPE FS over memoryless AWGN channels with coherent binary phase-shift keying (BPSK) transmission is a correlator between the synchronization sequence and the received signal, modified by an additive correction term which is a function of the received signal. In \cite{nielsen1973some},  the performance of the optimal detector presented in \cite{massey1972optimum} is analyzed, and an explicit expression for the probability of  synchronization error, leading to incorrect SW start position estimation within the data frame, is derived. It is observed in \cite{nielsen1973some} that the performance of the high SNR approximate optimal detector derived in \cite{massey1972optimum}, is practically indistinguishable from the performance of the optimal detector. The work  \cite{lui1987frame} extended \cite{massey1972optimum} to a general $M$-ary phase-coherent and phase-noncoherent signaling over AWGN channels. The work in \cite{choi2002frame} derived an OSPE FS algorithm for transmission of $M$-ary symbols over memoryless AWGN channels with unknown phase and frequency offsets. In \cite{moon1991ml}, the work of \cite{massey1972optimum} was extended to AWGN intersymbol interference (ISI) channels with BPSK transmission assuming the CIR is known at the receiver. In \cite{wang2004continuous}, an algorithm for OSPE FS over AWGN channels with ISI unknown at the receiver, and with an arbitrary symbol constellation, was presented. The work of \cite{wang2004continuous} first estimates the unknown CIR via a maximum likelihood estimation assuming an arbitrary location for the SW. Then, the estimated CIR is plugged back into the likelihood function to obtain the likelihood value for the assumed location, thus facilitating a search for the maximizing value of the location. This scheme was next extended to handle also unknown CFO.

Sequential FS algorithms over memoryless AWGN channels have been proposed in \cite{gansman1997optimum,chiani2004optimum,chiani2005practical,chiani2010noncoherent,liang2015sequential,hasselmann1981techniques}.  The work in \cite{liang2015sequential} developed a sequential FS algorithm for BPSK transmission over memoryless AWGN channels with an unknown noise variance and an unknown channel gain. The work in \cite{gansman1997optimum} presented a sequential FS algorithm based on hypothesis testing, for pilot-symbol assisted modulation (PSAM) transmission \cite{cavers1991analysis}, in which the pilot and the data symbols are selected from disjoint sets, and the signal is received over a memoryless AWGN channel such that the channel gain has an unknown random phase and frequency offsets.  In \cite{chiani2004optimum}, an  hypothesis testing-based optimal FS algorithm for BPSK transmission over AWGN channels was proposed using a SW. The work \cite{chiani2004optimum} assumed that the lengths of the frames are varying and are unknown at the receiver. The work in \cite{chiani2005practical} extended the work of \cite{chiani2004optimum} to the case in which the probability distribution of the BPSK data symbols is unknown, which led to the derivation of a generalized likelihood ratio test (GLRT). In \cite{hasselmann1981techniques}, a sequential FS algorithm  was presented for BPSK transmission over memoryless AWGN channels, received with unknown random carrier phase offset. To the best of our knowledge, a sequential FS algorithm for channels with memory in which the CIR is unknown has not yet been derived, and this is the contribution of the current work.

\label{pg:Main_Contribution-start}
\textbf{Main Contributions}: In this work we derive a sequential FS algorithm for channels with linear, periodically time-varying CIR and with ACGN, based on the likelihood ratio test (LRT) metric.  The unique aspects of our work are the general channel model for which the algorithm is derived, as well as the rigorous approach used in the derivation, in contrast to the ad-hoc methods used in previous works. Novelty follows as previous works did not explicitly account for the channel memory in the derivation of the detector, due to its associated computational complexity. Moreover, our derived algorithm does not assume knowledge of the CIR coefficients. To the best of our knowledge, an FS algorithm for an unknown linear channel with memory and ACGN, derived based on the LRT has not been previously presented. Furthermore, in the proposed algorithm we apply approximations which facilitate a tradeoff between complexity and performance. The computational complexity of the proposed scheme is then explicitly derived. Lastly, we propose a statistics-based approach for constructing good SWs. We demonstrate the superiority of the proposed algorithm over the very common correlator detector in an extensive simulation study. 
It should be noted that in this work we do not consider designing new low-correlation sequences, since an important insight which arises from the simulations is that once the CIR has memory and the noise has correlation, then the low-correlation property becomes less relevant for frame synchronization performance. This point is  clearly demonstrated by the numerical evaluations, which  include  an  exhaustive  search  over  all  SWs  and  show  that  the good SWs  for  the  considered channels have a relatively large correlation.
\label{pg:Main_Contribution-end}

The rest of this paper is organized as follows: Section \ref{chap:Preliminaries} presents the notations and a brief background on cyclostationary random processes. Section \ref{chap:signal_Model_and_Problem_Formulation} details the received signal model and the statement of the synchronization problem in the context of the model. Next, Section \ref{chap:LRT_Derivation} derives the exact LRT assuming known CIR and subsequently, Section \ref{Sec:SALRT} presents complexity reduction of the exact LRT assuming channel knowledge, and 
Section \ref{chap:SALRT_Algorithm_Derivation} presents our main result, which is a derivation of a low complexity approximate synchronization algorithm for unknown CIR, along with its computational complexity. Lastly, Section \ref{chap:Simulation} presents the simulation results and discussion and Section \ref{chap:Conclusions} presents some concluding remarks.


\section{Preliminaries and Background}
\label{chap:Preliminaries}
\subsection{Notations}
We use upper-case letters, e.g., $X$, to denote random variables (RVs), lower-case letters, e.g., $x$, to denote deterministic values (including realizations of random variables), and calligraphic letters, e.g., $\mathcal{X}$, to denote sets.
An exception to this rule are the deterministic constants $K, M, N, P_{\textsf{h}}, P_{\textsf{z}}, L_{\textsf{sw}}, L_{\textsf{f}}, L_{\textsf{ch}}, L_{\textsf{z}}, L_{\textsf{h}},$ $ L_{\textsf{tot}}, L_{\textsf{est}}, L_{\textsf{EQ}}, L_{\textsf{d}}$, and $N_{\textsf{s}}$, which are defined explicitly as appropriate. Column vectors are denoted with boldface letters, e.g., $\mathbf{x}$ denotes a deterministic vector and $\mathbf{X}$ denotes a random vector. 
We use upper-case double-stroke letters to denote matrices, e.g., $\mathds{A}$; the element at the $i$-th row and the $l$-th column of $\mathds{A}$ is denoted with $[\mathds{A}]_{i,l}$. $\mathds{O}_{N \times M}$ and $\mathds{I}_{N \times M}$ denote an $N \times M$ matrix of zeros, and an  $N \times M$ matrix whose diagonal elements are all $1$ and all off-diagonal elements are zero, respectively. $\mathcal{C},~\mathcal{R},~\mathcal{N},~\mbox{and}~\mathcal{Z}$ denote the sets of complex numbers, real numbers, natural numbers, and integers, respectively. Complex conjugate, transpose, Hermitian transpose, Euclidean norm, and stochastic expectation, are denoted by $(\cdot)^*,~(\cdot)^T,~(\cdot)^H, ~\| \cdot\|,~\mathds{E}\{ \cdot \}$. We use $|\cdot|$ to denote magnitude when applied to scalars, and the determinant when applied to matrices, and $\delta[\cdot]$ is used for denoting the Kronecker impulse function. We use $j$ to denote the imaginary unit, that is $j^2=-1$. 
\subsection{Cyclostationary Stochastic Processes}
In the following we briefly overview some aspects of wide-sense cyclostationary (WSCS) random processes.
\begin{definition} \label{WSCS_definition}
\cite{gardner2006cyclostationarity}, \cite{giannakis1998cyclostationary} A discrete-time proper-complex stochastic process $R[n]$ is called a WSCS stochastic process, if its mean and its autocorrelation functions are periodic with period  $N_0 \in \mathcal{N}$, that is:
\begin{ceqn}
\begin{align*}
\mathds{E}\big\{R[m]\big\} &=\mathds{E}\Big\{R[m+N_0]\Big\},~m \in \mathcal{Z},\\
c_R[m,l]&\triangleq \mathds{E}\Big\{R[m+l]\cdot\big(R[m]\big)^*\Big\}\\
&=c_R[m+N_0,l],~m,l \in \mathcal{Z}~.
\end{align*} 
\end{ceqn}
\end{definition}
The corresponding definition for continuous-time (CT) processes simply replaces the time variable $m$, $l \in\mathcal{Z}$ with $t,\tau \in \mathcal{R}$ and the period $N_0\in\mathcal{N}$ with $T_0 \in \mathcal{R}^{++}$.
WSCS stochastic processes are encountered in many outcomes of man-made actions as well as in natural phenomena with periodic characteristics. Instances of cyclostationarity, include communications \cite[Clause 7.1]{gardner2006cyclostationarity}, econometric modeling \cite{parzen1979approach}, atmospheric science \cite{hasselmann1981techniques}, and biological systems \cite{newton1982using}.  A comprehensive  overview of DT cyclostationary processes which are at the focus of this work is provided  in \cite{giannakis1998cyclostationary}. 

The decimated components decomposition (DCD) is a mapping which transforms a complex scalar DT WSCS process $R[m]$ with period $N_0$ into an equivalent $N_0 \times 1$ multivariate wide-sense stationary (WSS) process $\mathbf{R}[n] \in \mathbb{C}^{~N_0~\times~1}$  \cite[Clause 3.10]{gardner2006cyclostationarity}, \cite{giannakis1998cyclostationary}, defined as:
\begin{figure}[!ht]
\includegraphics[width=175mm,height=50mm]{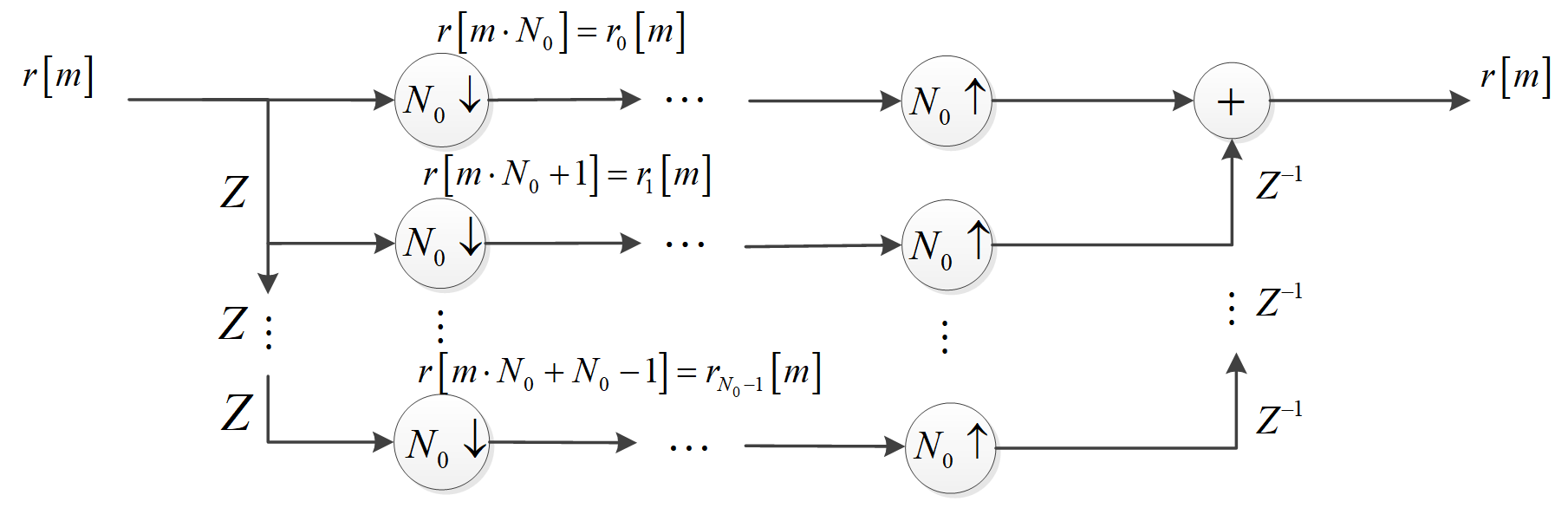}
\centering
\captionsetup{justification=centering}
\caption[Decimated components decomposition system mode]{Schematic description of the decimated components decomposition  \label{Decimated}}
\end{figure}
\begin{ceqn}
\begin{align*}
&\mathbf{R}[n]=
\left( \begin{array}{l}
R_0[n]\\
R_1[n]\\
\vdots \\
R_{N_0-1}[n]
\end{array}\right),\big[\mathbf{R}[n]\big]_q=R_{q}[n]\triangleq R[nN_0-q],\\
&n\in \mathbb{Z} ,~ N_0 \in \mathcal{N},~q=0,1,...,N_0-1~.
\end{align*} 
\end{ceqn}
This transformation is schematically described in Fig. \ref{Decimated}.
Letting $0\leq a_1,a_2\leq N_0-1$ the auto-correlation matrix of $\mathbf{R}[n]$, defined as $\mathds{R}_{\mathbf{R}}[n,l]\triangleq\mathds{E}\big\{\mathbf{R}[n+l](\mathbf{R}[n])^H\big\}$ is obtained as:
\begin{ceqn}
\begin{align*}
\Big[\mathds{R}_{\mathbf{R}}[n,l]\Big]_{a_1+1,a_2+1}
&=\mathds{E}\Big\{R\big[(n+l)N_0-a_1\big]\cdot \big(R[nN_0-a_2]\big)^*\Big\}\\
&= c_R[n N_0 -a_2,lN_0+a_2-a_1]\\
&\stackrel{(a)}{=}c_R[-a_2,lN_0+a_2-a_1]
\end{align*}
\end{ceqn}
where (a) follows from the fact that $R\big[n]$ is a WSCS stochastic process with period $N_0$. Observe that $\mathds{R}_{\mathbf{R}}[n,l]=\mathds{R}_{\mathbf{R}}[l]\in \mathcal{C}^{N_0~\times~N_0}$ which is independent of $n$. It is also straightforward to verify that $\mathds{E}\big\{\mathbf{R}[n]\big\}$ is a constant vector, implying that $\mathbf{R}[n]$ is WSS.
The cyclostationary process $R[m]$ can be recovered from the vector process $\mathbf{R}[n]$ via: 
\begin{ceqn}
\begin{align*}
R[m]=\sum\limits_{i=0}^{N_0-1} \sum\limits_{l=-\infty}^{\infty}\big[\mathbf{R}[l]\big]_i \cdot \delta[m-i-lN_0],
\end{align*} 
\end{ceqn} 
 as depicted in Figure \ref{Decimated}.



\section{Signal Model and Problem Formulation}
\label{chap:signal_Model_and_Problem_Formulation}

In this section we first introduce the models for the DT transmitted signal and for the corresponding received signal, and then formulate the frame synchronization problem as a hypothesis testing problem.
\subsection{Received Signal Model}
\label{Received Signal Model}
We consider communications over an additive noise channel with a linear periodically time-varying (LPTV) CIR, denoted with  $h[m,l]$, such that when the channel input is $\delta[m-l_0] ,~m,l_0\in \mathcal{Z}$, i.e., an impulse which is introduced at the input of the channel at time $l_0$, the output is $h[m,m-l_0]$. We let $P_{\textsf{h}}\in \mathcal{N}$ denote the periodicity of the CIR, i.e.,
\begin{ceqn}
\begin{align*}
&h[m+P_{\textsf{h}},l]=h[m,l],~m,l\in \mathcal{Z}. 
\end{align*}
\end{ceqn}
We observe that when $P_{\textsf{h}}=1$, then the periodic CIR $h[m,l]$ specializes to an LTI CIR, as for any $m\in\mathcal{Z}$ it follows that  $h[m,l]=h[0,l]$, and we may thus denote $h[0,l]\equiv h[l]$.
The communications channel is assumed to be casual with a finite memory, and a maximal length denoted by $L_{\textsf{h}}\in \mathcal{N}$:
\begin{ceqn}
\begin{align*}
&\exists m_1 \in \mathcal{Z}~\mbox{s.t.} ~ h[m_1,0]\ne 0,~\exists m_2 \in \mathcal{Z}~\mbox{s.t.} ~h[m_2,L_{\textsf{h}}]\ne 0 , \nonumber\\ &\mbox{and}~h[m,l]=0 ,~\forall~ l<0~\mbox{and}~\forall~ l>L_{\textsf{h}},~\forall~m \in \mathcal{Z}.
\end{align*}
\end{ceqn}
The received signal is contaminated by an ACGN\footnote{In Appendix \ref{app:Noise_Model} it is shown that under rather general conditions it follows that the baseband representation of a real bandlimited passband WSCS Gaussian channel is a proper complex WSCS Gaussian process.} process $Z[m]$, which is a zero mean, baseband proper complex process. The process $Z[m]$ has a period of
$P_{\textsf{z}}\in \mathcal{N}$ and a finite correlation memory of $L_{\textsf{z}} \in \mathcal{N}$:
%
\begin{ceqn}
\begin{subequations}
\begin{align}
\mathds{E}\big\{Z[m]\big\}&=0 \label{Zero_Noise_Expectation}\\
c_{\textsf{z}}\big[m,l\big]&\triangleq \mathds{E}\Big\{Z[m+l]\cdot\big(Z[m]\big)^{*}\Big\}=c_{\textsf{z}}\big[m+P_{\textsf{z}},l\big]\label{Noise_covariance} \\	\label{Noise_Memory} 			
c_{\textsf{z}}\big[m,l\big]&=0 ~~\mbox{for}~~ |l|> L_{\textsf{z}},~~\forall~ m\in \mathcal{Z}~,
\end{align}
\end{subequations}
\end{ceqn}
where $c_{\textsf{z}}\big[m,l\big]$ {\em is assumed known at the receiver. }
We observe that with $P_{\textsf{z}}=1$ the periodic autocorrlation function $c_{\textsf{z}}\big[m,l\big]$ specializes to a time-invariant autocorrelation
function $c_{\textsf{z}}\big[l\big]$, corresponding to a WSS random noise process.
The symbol sequence $S[m]$ is an independent and identically distributed (i.i.d) sequence of complex symbols, selected from a finite constellation set $\mathcal{S}=\{s_0,s_1,...,s_{N_{\textsf{s}}-1}\}, \mbox{ such that}~s_i \in \mathcal{C},~i=0,1,...,N_{\textsf{s}}-1$, according to a uniform distribution, with the first and second moments given by:
\begin{subequations}
\begin{ceqn}
\begin{align}
\mathds{E}\big\{S[m]\big\}&=0,~\forall~m \in \mathcal{Z}. \label{Zero_Symbols_Expectation} \\
\mathds{E}\Big\{S[m_1]\cdot\big(S[m_2]\big)^*\Big\}&=\sigma_{\textsf{s}}^2 \cdot \delta[m_1-m_2],~\sigma_{\textsf{s}}^2\in \mathcal{R}^{++},~\forall~m_1,m_2 \in \mathcal{Z}.\label{Symbol_varaince}\\
\mathds{E}\Big\{S[m_1]\cdot S[m_2]\Big\}&=\tilde{\sigma}_{\textsf{s}}^2 \cdot \delta[m_1-m_2],~\tilde{\sigma}_{\textsf{s}}^2\in \mathcal{C},~\forall~m_1,m_2 \in \mathcal{Z}. \label{Psaudo_Symbol_varaince}
\end{align}
\end{ceqn}
\end{subequations}
Eq. (\ref{Psaudo_Symbol_varaince}) represents the pseudo-covariance of the symbol sequence $S[m]$. The symbol and noise processes, $S[m]$ and $Z[m]$,  are mutually independent. 

Letting, let $L_{\textsf{ch}}\triangleq\mbox{max}\{L_{\textsf{z}},L_{\textsf{h}}\}$ denote the memory of the channel, we express the relationship between the realizations of the DT received signal $R[m]$, the transmitted symbol process $S[m]$ and the noise process $Z[m]$ as:
\begin{ceqn}
\begin{align} \label{Rreceived_signal_model}
&~~~~~~~~~~~~r[m]=\sum\limits_{l=0}^{L_{\textsf{ch}}} h[m,l]s[m-l]+z[m],~~~m\in \mathcal{Z}.
\end{align}
\end{ceqn}															
This model is a generalization of the LTI channel with additive WSS noise. It is assumed that $h[m,l]$ is deterministic and unknown except for its period $P_{\textsf{h}}$ and memory length $L_{\textsf{ch}}$, and that the values of the coefficients $h[m,l]$ remain the same throughout the frame synchronization process.
A signal frame consists of  $L_{\textsf{f}}\in \mathcal{N}$ complex symbols and is conceptually divided into two parts as depicted in Figure \ref{Structure of Transmited Frame}:
One part is an SW, which consists of $L_{\textsf{sw}}\in \mathcal{N}$ complex symbols $\mathbf{f}^T_{\textsf{sw}}=[f_{L_{\textsf{sw}-1}},f_{L_{\textsf{sw}-2}},...,f_0]$, where $L_{\textsf{sw}}\ll L_{\textsf{f}}$. 
The second part is $L_{\textsf{d}}=(L_{\textsf{f}}-L_{\textsf{sw}})$ randomly selected data symbols denoted $\{d_n\}$. Letting $N$ be a multiple of the least common multiple (LCM) of the discrete periods $P_{\textsf{h}}$ and $P_{\textsf{z}}$ that satisfies $N>L_{\textsf{ch}}$, it follows that there exist $k_1,k_2\in \mathcal{N}$ s.t. $k_1P_{\textsf{h}}=k_2P_{\textsf{z}}=N$.
The length of the synchronization sequence, $L_{\textsf{sw}}$, is set to satisfy $L_{\textsf{sw}}>L_{\textsf{ch}}+1.$
Upon transmission, the transmitter applies the following preprocessing: The SW of $L_{\textsf{sw}}$ complex symbols is divided into $M \in \mathcal{N}$ blocks, each containing $K\triangleq (N-L_{\textsf{ch}})$ complex symbols, such that $L_{\textsf{sw}}=KM$. With this partition, we can write  $\mathbf{f}^T_{\textsf{sw}}=[\mathbf{t}^{T}_{M-1},\mathbf{t}^{T}_{M-2},...,\mathbf{t}^{T}_0]$, where$~\mathbf{t}_i=[f_{iK+K-1},f_{iK+K-2},...,f_{iK}]^{T},~i=0,1,...,M-1$.
\begin{figure}[!ht]
\includegraphics[width=175mm,height=32mm]{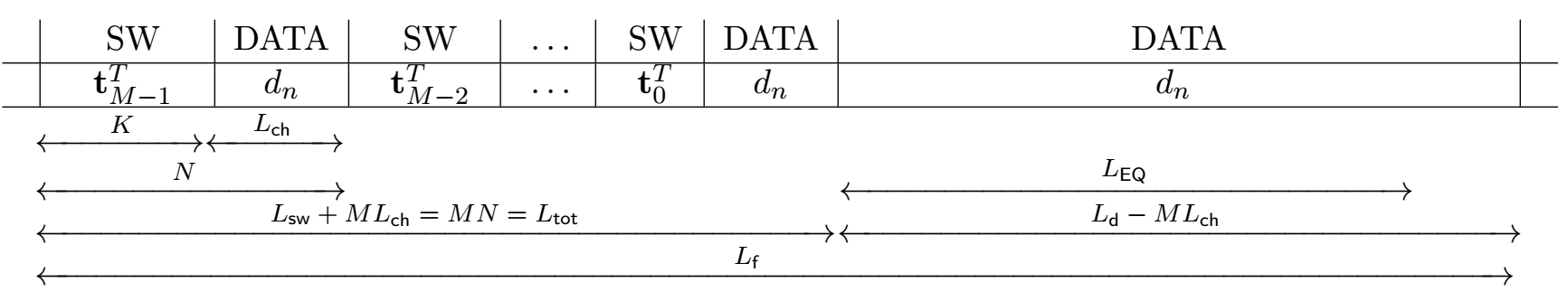}
\centering
\captionsetup{justification=centering}
\caption{Structure of a transmitted frame containing SW and random data\label{Structure of Transmited Sw}}
\end{figure}
At the end of each $\mathbf{t}_i$ block of length $K$, the transmitter inserts $L_{\textsf{ch}}$ data symbols\footnote{It is possible to add a guard interval instead of data symbols if so desired, yet using data symbols will result in a higher spectral efficiency.}
as depicted in Figure \ref{Structure of Transmited Sw}, such that the total length of the sequence transmitted for synchronization is  $L_{\textsf{tot}}\triangleq L_{\textsf{sw}}+ML_{\textsf{ch}}=KM+ML_{\textsf{ch}}=(N-L_{\textsf{ch}})\cdot M+ML_{\textsf{ch}}=NM$, where it is assumed that $L_{\textsf{tot}}\ll L_{\textsf{f}}$.
The objective of the synchronization detector is to identify the beginning of a frame by observing $L_{\textsf{tot}}$ subsequent samples collected from the received signal, stated as:
\begin{ceqn}
\begin{align*} 
&r\big[m-i\big]=\sum\limits_{l=0}^{L_{\textsf{ch}}} h\big[m-i,l\big]s\big[m-i-l\big]+z\big[m-i\big],~i=0,1,...,L_{\textsf{tot}}-1
\end{align*}
\end{ceqn}
\subsection{Statement of the Hypothesis Testing Problem}
\label{Statement_of_the_Hypothesis Testing_Problem}
We begin by defining the quantities to be used in the derivations of the algorithm. We use the column vector notation $\mathbf{x}_{T_{\textsf{x}},d_{\textsf{x}}}[n]\in \mathcal{C}^{L_{\textsf{x}}~\times~1}$ with constants $T_{\textsf{x}},~L_{\textsf{x}}\in \mathcal{N},~d_{\textsf{x}} \in \{0,1,...T_{\textsf{x}}-1\} \triangleq \mathcal{T}_{\textsf{x}}$, defined as:
\begin{ceqn}
\begin{align*}
&\mathbf{x}_{T_{\textsf{x}},d_{\textsf{x}}}[n]\triangleq
\left( \begin{array}{l}
x\big[n \cdot T_{\textsf{x}}-d_{\textsf{x}}\big]\\
x\big[n \cdot T_{\textsf{x}}-d_{{\textsf{x}}}-1\big]\\
\vdots \\
x\big[n \cdot T_{\textsf{x}}-d_{\textsf{x}}-(L_{\textsf{x}}-1)\big]
\end{array}\right),~0\leq d_x \leq T_{\textsf{x}}-1 .~~~~~
\end{align*}
\end{ceqn}
With this notation, we express the received vector $\mathbf{r}_{L_{\textsf{tot}},d_{\textsf{r}}}[n]\in \mathcal{C}^{L_{\textsf{tot}}~\times~1}$ of $L_{\textsf{tot}}$ subsequent samples as:
\begin{ceqn}
\begin{align} \label{Received_Vector}
&\mathbf{r}_{L_{\textsf{tot}},d_{\textsf{r}}}[n]\triangleq
\left( \begin{array}{l}
r\big[n \cdot L_{\textsf{tot}}-d_{\textsf{r}}\big]\\
r\big[n \cdot L_{\textsf{tot}}-d_{\textsf{r}}-1\big]\\
\vdots \\
r\big[n \cdot L_{\textsf{tot}}-d_{\textsf{r}}-(L_{\textsf{tot}}-1)\big]
\end{array}\right),~0\leq d_r \leq L_{\textsf{tot}}-1~,
\end{align} 
\end{ceqn}
choosing proper values for $d_{\textsf{r}}$ and $n$, allows us to collect different $L_{\textsf{tot}}$ subsequent samples from the received signal frame. 
Next, we define the notation $h_{n,k,l}^{(d_{\textsf{r}})}\triangleq h\big[n \cdot L_{\textsf{tot}}-d_{\textsf{r}}-k,l\big],~k=0,1,...,L_{\textsf{tot}}-1$ and express $\mathbf{r}_{L_{\textsf{tot}},d_{\textsf{r}}}[n]$ in a matrix form using the following matrices and vectors :
\begin{ceqn}
\begin{align*}
\mathbf{s}^{(m)}_{L_{\textsf{tot}},d_{\textsf{r}}}[n]&\triangleq
\left(\begin{array}{l}
s\big[n \cdot L_{\textsf{tot}}-m\cdot N-d_{\textsf{r}}\big]\\
s\big[n \cdot L_{\textsf{tot}}-m\cdot N-d_{\textsf{r}}-1\big]\\
\vdots \\
s\big[n \cdot L_{\textsf{tot}}-m\cdot N-d_{\textsf{r}}-(K-1)\big]
\end{array}\right)\in \mathcal{C}^{K~\times~1},~m=0,1,...,M-1~,\\
%
\mathbf{d}^{(m)}_{L_{\textsf{tot}},d_{\textsf{r}}}[n]&\triangleq
\left(\begin{array}{l}
s\big[n \cdot L_{\textsf{tot}}-m\cdot N-d_{\textsf{r}}-K\big]\\
s\big[n \cdot L_{\textsf{tot}}-m\cdot N-d_{\textsf{r}}-(K+1)\big]\\
\vdots \\
s\big[n \cdot L_{\textsf{tot}}-m\cdot N-d_{\textsf{r}}-(N-1)\big]
\end{array}\right)\in \mathcal{C}^{(N-K)~\times~1},~m=0,1,...,M-1~,
\end{align*}
\end{ceqn}
and for transmission block $m$ we write
%
\begin{ceqn}
\begin{align}
\mathbf{g}^{(m)}_{L_{\textsf{tot}},d_{\textsf{r}}}[n]&\triangleq
\left(\begin{array}{l}
\mathbf{s}^{(m)}_{L_{\textsf{tot}},d_{\textsf{r}}}[n]\\
\mathbf{d}^{(m)}_{L_{\textsf{tot}},d_{\textsf{r}}}[n]
\end{array}\right)\in \mathcal{C}^{N~\times~1},~m=0,1,...,M-1. \label{Data_block_m}
\end{align}
\end{ceqn}
%
Lastly, define:
\begin{ceqn}
\begin{align*}
&~~~~~~~~\mathbf{l}_{L_{\textsf{tot}},d_{\textsf{r}}}[n]\triangleq
\left(\begin{array}{l}
s\big[n \cdot L_{\textsf{tot}}-M\cdot N-d_{\textsf{r}}\big]\\
s\big[n \cdot L_{\textsf{tot}}-M\cdot N-d_{\textsf{r}}-1\big]\\
\vdots \\
s\big[n \cdot L_{\textsf{tot}}-M\cdot N-d_{\textsf{r}}-(L_{\textsf{ch}}-1)\big]
\end{array}\right)\in \mathcal{C}^{L_{\textsf{ch}}~\times~1}.    
\end{align*}
\end{ceqn}
Each vector $\mathbf{s}^{(m)}_{L_{\textsf{tot}},d_{\textsf{r}}}[n]$ consists of only random data symbols when the synchronization word is outside the observation window and of the SW symbols when the synchronization word is within the observation window. The vectors $\mathbf{d}^{(m)}_{L_{\textsf{tot}},d_{\textsf{r}}}[n]$ and $\mathbf{l}_{L_{\textsf{tot}},d_{\textsf{r}}}[n]$ consist of only random data symbols. With these definitions we can express the overall signal used for generating the observed samples as:
\begin{ceqn}
\begin{align}
%
\mathbf{s}_{L_{\textsf{tot}},d_{\textsf{r}}}[n]&\triangleq
\left( \begin{array}{l}
\mathbf{g}^{(0)}_{L_{\textsf{tot}},d_{\textsf{r}}}[n]\\
\mathbf{g}^{(1)}_{L_{\textsf{tot}},d_{\textsf{r}}}[n]\\
\vdots\\
\mathbf{g}^{(M-1)}_{L_{\textsf{tot}},d_{\textsf{r}}}[n]\\
\mathbf{l}_{L_{\textsf{tot}},d_{\textsf{r}}}[n]
\end{array}\right)=
\left( \begin{array}{l}
\mathbf{s}^{(0)}_{L_{\textsf{tot}},d_{\textsf{r}}}[n]\\
\mathbf{d}^{(0)}_{L_{\textsf{tot}},d_{\textsf{r}}}[n]\\
\mathbf{s}^{(1)}_{L_{\textsf{tot}},d_{\textsf{r}}}[n]\\
\mathbf{d}^{(1)}_{L_{\textsf{tot}},d_{\textsf{r}}}[n]\\
\vdots\\
\mathbf{s}^{(M-1)}_{L_{\textsf{tot}},d_{\textsf{r}}}[n]\\
\mathbf{d}^{(M-1)}_{L_{\textsf{tot}},d_{\textsf{r}}}[n]\\
\mathbf{l}_{L_{\textsf{tot}},d_{\textsf{r}}}[n]
\end{array}\right)\in \mathcal{C}^{(L_{\textsf{tot}}+L_{\textsf{ch}})~\times~1}~,\nonumber
\end{align}
and the respective noise component as 
\begin{align}
\mathbf{z}_{L_{\textsf{tot}},d_{\textsf{r}}}[n]&\triangleq
\left( \begin{array}{l}
z\big[n \cdot L_{\textsf{tot}}-d_{\textsf{r}}\big]\\
z\big[n \cdot L_{\textsf{tot}}-d_{\textsf{r}}-1\big]\\
\vdots \\
z\big[n \cdot L_{\textsf{tot}}-d_{\textsf{r}}-(L_{\textsf{tot}}-1)\big]
\end{array}\right)\in \mathcal{C}^{L_{\textsf{tot}}~\times~1}. \label{Z_S_Pre_Process}
\end{align} 
\end{ceqn}
Lastly, we define the matrix $\mathds{A}_{{L_{\textsf{tot}}},d_{\textsf{r}}}[n] \in \mathcal{C}^{L_{\textsf{tot}}~\times~(L_{\textsf{tot}}+L_{\textsf{ch}})}$: 
\arraycolsep=1.0pt\def\arraystretch{0.7}
\begin{ceqn}
\begin{align}
&\mathds{A}_{{L_{\textsf{tot}}},d_{\textsf{r}}}[n]\triangleq
\left( \begin{array}{lllllllllll}
h_{n,0,0}^{(d_{\textsf{r}})} &h_{n,0,1}^{(d_{\textsf{r}})} &h_{n,0,2}^{(d_{\textsf{r}})}&\ldots &h_{n,0,L_{\textsf{ch}}}^{(d_{\textsf{r}})}&0&0&\ldots&0&\ldots&0 \\
0& h_{n,1,0}^{(d_{\textsf{r}})} & h_{n,1,1}^{(d_{\textsf{r}})} &\ldots &h_{n,1,L_{\textsf{ch}}-1}^{(d_{\textsf{r}})}&h_{n,1,L_{\textsf{ch}}}^{(d_{\textsf{r}})}&0&\ldots&0&\ldots&0  \\
0&0 & h_{n,2,0}^{(d_{\textsf{r}})} & \ldots & h_{n,2,L_{\textsf{ch}}-2}^{(d_{\textsf{r}})}&h_{n,2,L_{\textsf{ch}}-1}&h_{n,2,L_{\textsf{ch}}}^{(d_{\textsf{r}})}&\ldots&0&\ldots&0 \\
\vdots & \vdots & \vdots & \ddots &\vdots&\vdots&\vdots &\ddots &\vdots &\ddots &\vdots  \\
0& 0 &0 &\ldots & 0 &0 &0 &\ldots&h_{n,L_{\textsf{tot}}-1,0}^{(d_{\textsf{r}})}&\ldots&h_{n,L_{\textsf{tot}}-1,L_{\textsf{ch}}}^{(d_{\textsf{r}})}
\end{array} \right) . \label{A_Pre_Matrix}
\end{align}
\end{ceqn}
%
%
%
Using the definitions above, the relationship between the  signal received within the observation window and the transmitted signal can be expressed in a matrix form as :
\begin{ceqn}
\begin{equation}  \label{Pre_Process}
\mathbf{r}_{L_{\textsf{tot}},d_{\textsf{r}}}[n]= \mathds{A}_{{L_{\textsf{tot}}},d_{\textsf{r}}}[n] \cdot \mathbf{s}_{L_{\textsf{tot}},d_{\textsf{r}}}[n]+ \mathbf{z}_{L_{\textsf{tot}},d_{\textsf{r}}}[n],\quad n \in \mathcal{Z}.
\end{equation}
\end{ceqn}
At the receiver, the received vector signal $\mathbf{r}_{L_{\textsf{tot}},d_{\textsf{r}}}[n]$ is applied post-processing, denoted $\mathsf{P}\{\mathbf{r}_{L_{\textsf{tot}},d_{\textsf{r}}}[n]\}$, designed to extract only the relevant information for synchronization as follows: The received vector $\mathbf{r}_{L_{\textsf{tot}},d_{\textsf{r}}}[n]$ is first divided into $M$ component vectors of length $N$. Then, for each component vector, the last $L_{\textsf{ch}}$ samples are discarded resulting in the following processed vector:
%
\begin{ceqn}
\begin{align}
\mathbf{r}^{(\mathsf{P})}_{L_{\textsf{tot}},d_{\textsf{r}}}[n]&\equiv  \mathsf{P}\{\mathbf{r}_{L_{\textsf{tot}},d_{\textsf{r}}}[n]\}\nonumber\\
&\triangleq
\left( \begin{array}{l}
\tilde{\mathbf{r}}_{N,d_{\textsf{r}}}[n\cdot M]\\
\tilde{\mathbf{r}}_{N,d_{\textsf{r}}}[n\cdot M-1]\\
\vdots\\
\tilde{\mathbf{r}}_{N,d_{\textsf{r}}}[n\cdot M-(M-1)]\\
\end{array}\right)\in \mathcal{C}^{L_{\textsf{sw}}~\times~1}, \nonumber\\
\tilde{\mathbf{r}}_{N,d_{\textsf{r}}}[n]&\triangleq\left( \begin{array}{l}
r[n\cdot N-d_{\textsf{r}}] \\
r[n\cdot N-d_{\textsf{r}}-1] \\
\vdots\\
r[n\cdot N-d_{\textsf{r}}-(K-1)]\\
\end{array} \right)\in \mathcal{C}^{K~\times~1} \label{Post_Processing}.
\end{align}
\end{ceqn}
Next, for $l=0,1,...,K-1$,~ $m=0,1,...,M-1$ and $k  = 0,1,...,L_{\textsf{tot}}-1$, we define the following matrices and vectors:
\arraycolsep=1.0pt\def\arraystretch{0.25}
\begin{ceqn}
\begin{align}
\mathbf{h}^{(d_{\textsf{r}},l)}[n,k]&\triangleq
\left( \begin{array}{l}
\mathds{O}_{l \times 1}\nonumber\\
~\\
\tilde{\mathbf{h}}^{(d_{\textsf{r}})}[n,k]\\
~\\
\mathds{O}_{(N-l-L_{\textsf{ch}}-1) \times 1}
\end{array}\right)\in \mathcal{C}^{N~\times~1}\nonumber\\
\tilde{\mathbf{h}}^{(d_{\textsf{r}})}[n,k]&\triangleq
\left(\begin{array}{l}
h_{n,k,0}^{(d_{\textsf{r}})}\\
h_{n,k,1}^{(d_{\textsf{r}})}\\
\vdots \\
h_{n,k,L_{\textsf{ch}}}^{(d_{\textsf{r}})}
\end{array}\right)\in \mathcal{C}^{(L_{\textsf{ch}}+1)~\times~1}\nonumber\\
\mathds{B}^{(d_{\textsf{r}},m)}[n]&\triangleq
\left( \begin{array}{l}
\big(\mathbf{h}^{(d_{\textsf{r}},0)}[n,m\cdot N] \big)^{T}\\
~\\
~\\
\big(\mathbf{h}^{(d_{\textsf{r}},1)}[n,m\cdot N+1]\big)^{T}\\
~\\
~\\
\vdots
~\\
~\\
\big(\mathbf{h}^{(d_{\textsf{r}},K-1)}[n,m\cdot N+K-1] \big)^{T}\\
\end{array}\right)\in \mathcal{C}^{K~\times~N}.
\label{eqn:Construct_B}
\end{align}
\end{ceqn}
We also define
\begin{ceqn}
\begin{align}
\mathds{A}^{(\mathsf{P})}_{{L_{\textsf{tot}}},d_{\textsf{r}}}[n] &\triangleq
\left( \begin{array}{lllll}
\mathds{B}^{(d_{\textsf{r}},0)}[n]&\mathds{O}_{K \times N}&\ldots&\mathds{O}_{K \times N}\\
\mathds{O}_{K \times N}&\mathds{B}^{(d_{\textsf{r}},1)}[n]&\ldots&\mathds{O}_{K \times N}\\
\mathds{O}_{K \times N}&\mathds{O}_{K \times N}&\ldots&\mathds{O}_{K \times N}\\
\vdots&\vdots&\ddots&\vdots\\
\mathds{O}_{K \times N}&\mathds{O}_{K \times N}&\ldots&\mathds{B}^{(d_{\textsf{r}},M-1)}[n]\\
\end{array}\right)\in \mathcal{C}^{L_{\textsf{sw}}~\times~L_{\textsf{tot}}} \nonumber \\
\tilde{\mathbf{s}}_{L_{\textsf{tot}},d_{\textsf{r}}}[n] & \triangleq 
\left( \begin{array}{l}
\mathbf{s}^{(0)}_{L_{\textsf{tot}},d_{\textsf{r}}}[n]\\
\mathbf{d}^{(0)}_{L_{\textsf{tot}},d_{\textsf{r}}}[n]\\
\mathbf{s}^{(1)}_{L_{\textsf{tot}},d_{\textsf{r}}}[n]\\
\vdots\\
\mathbf{s}^{(M-1)}_{L_{\textsf{tot}},d_{\textsf{r}}}[n]\\
\mathbf{d}^{(M-1)}_{L_{\textsf{tot}},d_{\textsf{r}}}[n]\\
\end{array}\right) \in \mathcal{C}^{L_{\textsf{tot}}~\times~1} \nonumber\\
\tilde{\mathbf{z}}_{N,d_{\textsf{r}}}[n]&\triangleq \left( \begin{array}{l}
z[nN-d_{\textsf{r}}] \\
z[nN-d_{\textsf{r}}-1] \\
\vdots\\
z[nN-d_{\textsf{r}}-(K-1)]\\
\end{array} \right) \in \mathcal{C}^{K~\times~1}\nonumber\\ 
\mathbf{z}^{(\mathsf{P})}_{L_{\textsf{tot}},d_{\textsf{r}}}[n] & \triangleq 
\left( \begin{array}{l}
\tilde{\mathbf{z}}_{N,d_{\textsf{r}}}[n\cdot M]\\
\tilde{\mathbf{z}}_{N,d_{\textsf{r}}}[n\cdot M-1]\\
\vdots\\
\tilde{\mathbf{z}}_{N,d_{\textsf{r}}}[n\cdot M-(M-1)]\\
\end{array}\right) \in \mathcal{C}^{L_{\textsf{sw}}~\times~1}.\label{Vec_Def}
\end{align} 
\end{ceqn}
Using the relationship in (\ref{Pre_Process}) and the above definitions, we obtain the following relationship for the post-processed received vector:
\begin{ceqn}
\begin{align*}
\mathbf{r}^{(\mathsf{P})}_{L_{\textsf{tot}},d_{\textsf{r}}}[n]&=
\mathds{A}^{(\mathsf{P})}_{{L_{\textsf{tot}}},d_{\textsf{r}}}[n] \cdot \tilde{\mathbf{s}}_{L_{\textsf{tot}},d_{\textsf{r}}}[n]+ \mathbf{z}^{(\mathsf{P})}_{L_{\textsf{tot}},d_{\textsf{r}}}[n].
\end{align*}
\end{ceqn}
Recalling that the channel period is $N$, it follows that $h_{n,k,l}^{(d_{\textsf{r}})}=h_{n,k+N,l}^{(d_{\textsf{r}})}$, hence we conclude that $\mathds{B}^{(d_{\textsf{r}},0)}[n]=\mathds{B}^{(d_{\textsf{r}},1)}[n]=...=\mathds{B}^{(d_{\textsf{r}},M-1)}[n]$, and use $\mathds{B}^{(d_{\textsf{r}})}[n]$ to denote $\mathds{B}^{(d_{\textsf{r}},m)}[n]=\mathds{B}^{(d_{\textsf{r}})}[n]~, m=0,1,...,M-1$. With these definitions we can represent the vectors  $\tilde{\mathbf{r}}_{N,d_{\textsf{r}}}[n\cdot M-m] ,m=0,1,...,M-1$, which corresponds to the segments of $\mathbf{r}^{(\mathsf{P})}_{L_{\textsf{tot}},d_{\textsf{r}}}[n]$, as:
\begin{ceqn}
\begin{align} 
\tilde{\mathbf{r}}_{N,d_{\textsf{r}}}[n\cdot M-m]&=\mathds{B}^{(d_{\textsf{r}})}[n] \cdot \mathbf{g}^{(m)}_{L_{\textsf{tot}},d_{\textsf{r}}}[n]+\tilde{\mathbf{z}}_{N,d_{\textsf{r}}}[n\cdot M-m],~m=0,1,...,M-1~.\label{R_Sub_Vec_Equation}
\end{align}
\end{ceqn}
It is clarified that the samples of the received samples which are discarded by the preprocessing are ignored only for the purpose of the frame synchronization algorithm. Naturally, these samples can be stored for processing by the decoder in the data recovery step once the frame synchronizer has detected the beginning of a frame.
Next, recall the partition of the SW, into $M$ blocks as defined at Section \ref{Received Signal Model} :
\begin{ceqn}
\begin{align*}
\mathbf{f}_{\textsf{sw}}\triangleq
\left( \begin{array}{l}
f_{L_{\textsf{sw}}-1}\\
f_{L_{\textsf{sw}}-2}\\
\vdots\\
f_0\\
\end{array}\right)=
\left( \begin{array}{l}
\mathbf{t}_{M-1}\\
\mathbf{t}_{M-2}\\
\vdots\\
\mathbf{t}_0\\
\end{array}\right) \in \mathcal{C}^{L_{\textsf{sw}}~\times~1}.    
\end{align*}
\end{ceqn}
In the following, when $\tilde{\mathbf{s}}_{L_{\textsf{tot}},d_{\textsf{r}}}[n]$  contains the synchronization sequence we denote it as $\tilde{\mathbf{s}}^{(\textsf{sw})}_{L_{\textsf{tot}},d_{\textsf{r}}}[n]$, and it is defined as:
\begin{ceqn}
\begin{align}
%
\tilde{\mathbf{s}}^{(\textsf{sw})}_{L_{\textsf{tot}},d_{\textsf{r}}}[n]&\triangleq
\left( \begin{array}{l}
\tilde{\mathbf{g}}^{(0)}_{L_{\textsf{tot}},d_{\textsf{r}}}[n]\\
\tilde{\mathbf{g}}^{(1)}_{L_{\textsf{tot}},d_{\textsf{r}}}[n]\\
\vdots\\
\tilde{\mathbf{g}}^{(M-1)}_{L_{\textsf{tot}},d_{\textsf{r}}}[n]
\end{array}\right)\triangleq
\left( \begin{array}{l}
\mathbf{t}_{M-1}\\
~\\
\mathbf{d}^{(0)}_{L_{\textsf{tot}},d_{\textsf{r}}}[n]\\
~\\
\mathbf{t}_{M-2}\\
~\\
\mathbf{d}^{(1)}_{L_{\textsf{tot}},d_{\textsf{r}}}[n]\\
~\\
\vdots\\
~\\
\mathbf{d}^{(M-2)}_{L_{\textsf{tot}},d_{\textsf{r}}}[n]\\
~\\
\mathbf{t}_0\\
~\\
\mathbf{d}^{(M-1)}_{L_{\textsf{tot}},d_{\textsf{r}}}[n]
\end{array}\right) \in \mathcal{C}^{L_{\textsf{tot}}~\times~1},\nonumber \\
\tilde{\mathbf{g}}^{(m)}_{L_{\textsf{tot}},d_{\textsf{r}}}[n]&\triangleq
\left( \begin{array}{l}
\mathbf{t}_{M-1-m}\\
~\\
\mathbf{d}^{(m)}_{L_{\textsf{tot}},d_{\textsf{r}}}[n]\\
\end{array}\right) \in \mathcal{C}^{~N~\times~1} ,~m=0, 1, ..., M-1. \label{Sync_block_m}
\end{align}
\end{ceqn}
When $ \tilde{\mathbf{s}}_{L_{\textsf{tot}},d_{\textsf{r}}}[n]$ does not contain the synchronization sequence (namely, it contains only data), we denote it as $\tilde{\mathbf{s}}^{(\textsf{data})}_{L_{\textsf{tot}},d_{\textsf{r}}}[n]\in \mathcal{C}^{L_{\textsf{tot}}~\times~1}$. In the following, to reduce notational clutter, we set $d_{\textsf{r}}=0$. The hypothesis testing problem with $d_{\textsf{r}}=0$ (it is emphasized that this only simplifies notations in the subsequent derivations) after the post-processing $\mathsf{P}\{\cdot\}$ can be written as:
\begin{ceqn}
\begin{subequations} \label{Hypothesis_Problem}
\begin{align}
&H_0 : \mathbf{R}^{(\mathsf{P})}_{L_{\textsf{tot}},0}[n]= \mathds{A}^{(\mathsf{P})}_{{L_{\textsf{tot}}},0}[n] \cdot \tilde{\mathbf{S}}^{(\textsf{data})}_{L_{\textsf{tot}},0}[n]+ \mathbf{Z}^{(\mathsf{P})}_{L_{\textsf{tot}},0}[n]\\
&H_1 : \mathbf{R}^{(\mathsf{P})}_{L_{\textsf{tot}},0}[n]= \mathds{A}^{(\mathsf{P})}_{{L_{\textsf{tot}}},0}[n] \cdot \tilde{\mathbf{S}}^{(\textsf{sw})}_{L_{\textsf{tot}},0}[n]+ \mathbf{Z}^{(\mathsf{P})}_{L_{\textsf{tot}},0}[n]
\end{align}
\end{subequations}
\end{ceqn}
where, $\mathds{B}^{(0)}[n]\equiv\mathds{B}[n]$, $\mathbf{h}^{(0,m)}[n,k]\equiv\mathbf{h}^{(m)}[n,k]$, $\tilde{\mathbf{h}}^{(0)}[n,k]\equiv\tilde{\mathbf{h}}[n,k]$ and $h_{n,k,l}^{(0)}\equiv h_{n,k,l}$.
Our objective is to derive an algorithm to decide between $H_0$ and $H_1$ such that for any given probability of false alarm the probability of detection is maximized. In the derivations and subsequent analysis which follows, we shall not consider in the hypothesis testing setup the mixed scenario in which the observation window contains data symbols and part of the SW symbols, since in such cases a properly designed synchronization sequence mimics random data  \cite{chiani2006sequential}.


\section{Derivation of the LRT for the Frame Synchronization Problem}
\label{chap:LRT_Derivation}

In this section we develop the LRT \cite{van2004detection} for testing the hypothesis $H_0$ vs. hypothesis $H_1$, stated in (\ref{Hypothesis_Problem}).
\subsection{Statistics of the Received Signal for Hypothesis \texorpdfstring{$H_0$}~}
\label{sec:stat_H0}
In the following we prove that $\tilde{\mathbf{R}}_{N,0}[n]$ defined in (\ref{Post_Processing}), subject to the $H_0$ hypothesis is a WSS multivariate random process.
It is straight forward to show that $\mathds{E}\big\{\tilde{\mathbf{R}}_{N,0}[n]\big|H_0\big\}=\mathbf{0}$, see Appendix \ref{app:Derivations_Chapter_LRT} for the details.
Next, we show that the autocorrelation matrix for $\tilde{\mathbf{R}}_{N,0}[n]$ subject to the $H_0$ hypothesis, denoted
$\mathds{R}_{\tilde{\mathbf{R}}_{N,0}|H_0}[n,l]\triangleq \mathds{E}\Big\{\tilde{\mathbf{R}}_{N,0}[n+l]\cdot\big(\tilde{\mathbf{R}}_{N,0}[n]\big)^{H} |H_0\Big\}$ is a function only of $l$, and is independent of $n$.
Applying the statistics of the source sequence together with the periodicity of the noise and the CIR we obtain that the $\big((a_1+1),(a_2+1)\big)$ element of the correlation matrix is given by::
\begin{align} 
&\mathds{E}\bigg\{R[n\cdot N-a_1]\cdot\Big(R\big[(n+l)\cdot N-a_2\big]\Big)^{*}\bigg|H_0\bigg\} \nonumber\\
&\qquad =\sum\limits_{l_1=0}^{L_{\textsf{ch}}}\sum\limits_{l_2=0}^{L_{\textsf{ch}}} h\big[(n+l)\cdot N-a_1,l_1\big]\cdot h\big[n\cdot N-a_2,l_2\big] \nonumber\\
&\qquad \qquad ~\cdot \mathds{E}\bigg\{S\big[(n+l)\cdot N-a_1-l_1\big]\cdot \Big(S\big[n\cdot N-a_2-l_2\big]\Big)^*\bigg|H_0\bigg\}  \nonumber \\
&\qquad \qquad \qquad ~~~~~~~~~~~~~~~~+\mathds{E}\bigg\{Z\big[(n+l)\cdot N-a_1\big]\cdot \Big(Z\big[n\cdot N-a_2\big]\Big)^* \bigg|H_0\bigg\}  \nonumber \\
&\qquad \stackrel{(a)}{=}\sum\limits_{l_1=0}^{L_{\textsf{ch}}}\sum\limits_{l_2=0}^{L_{\textsf{ch}}} h[-a_1,l_1]\cdot\big(h[-a_2,l_2]\big)^* \cdot \sigma_s^2\cdot\delta[l\cdot N-a_1-l_1+a_2+l_2]+c_{\textsf{z}}[-a_2,l\cdot N+a_2-a_1]~,~~~~~~~~~~~~~~~~~~~~~~~~~~~~~~~~~~~~~~~~~~~~~~~~~~~~~~~~~~~~~~~~~~~~~~~~~~~~~~~~~~~~~~~~~ \nonumber\\
&\qquad \qquad \qquad \qquad \qquad \qquad \qquad \qquad ~a_1,a_2\in \{0,1,...,K-1\}, \label{Fisrt_wss_2}
\end{align}
where $(a)$ follows since $N=k_1P_{\textsf{h}}$, thus $h[n,l_1]=h[n+k_1P_{\textsf{h}},l_1]=h[n+N,l_1]$, and from the symbol and noise statistics defined in Eqs. (\ref{Symbol_varaince}) and (\ref{Noise_covariance}). 
From (\ref{Fisrt_wss_2}) it directly follows that the elements of the correlation matrix $\mathds{R}_{\tilde{\mathbf{R}}_{N,0}|H_0}[n,l]$ are independent of $n$, hence, we can write $\mathds{E}\big\{\tilde{\mathbf{R}}_{N,0}[n+l](\tilde{\mathbf{R}}_{N,0}[n])^{H}\big|H_0\big\}=\mathds{R}_{\tilde{\mathbf{R}}_{N,0}|H_0}[l],~ l\in \mathcal{Z}$. To complete the proof, we must show that the pseudo-autocorrelation matrix for $\tilde{\mathbf{R}}_{N,0}[n]$ subject to the $H_0$ hypothesis, denoted 
$\tilde{\mathds{R}}_{\tilde{\mathbf{R}}_{N,0}|H_0}[n,l]\triangleq \mathds{E}\big\{\tilde{\mathbf{R}}_{N,0}[n+l]\cdot(\tilde{\mathbf{R}}_{N,0}[n])^{T} |H_0\big\}$ is also a function only of $l$, and is independent of $n$. Following parallel arguments of those leading to (\ref{Fisrt_wss_2}), the $\big((a_1+1),(a_2+1)\big)$ element in the pseudo-autocorrelation matrix $\tilde{\mathds{R}}_{\tilde{\mathbf{R}}_{N,0}|H_0}[n,l],~a_1,a_2\in \{0,1,...,K-1\}$, can be stated as:
\begin{align} 
&\mathds{E}\bigg\{R\big[(n+l)\cdot N-a_1]\cdot R\big[(n+l)N-a_2\big]\bigg|H_0\bigg\}\nonumber\\
&=\sum\limits_{l_1=0}^{L_{\textsf{ch}}}\sum\limits_{l_2=0}^{L_{\textsf{ch}}} h\big[(n+l)\cdot N-a_1,l_1\big]\cdot h\big[n\cdot N-a_2,l_2\big] \cdot \mathds{E}\bigg\{S\big[(n+l)\cdot N-a_1-l_1\big]\cdot S\big[n\cdot N-a_2-l_2\big]\bigg|H_0\bigg\} \nonumber\\
&~~~~~~~~~~~~~~~~~~~~~~~~~~~~~~~~~~~~~~~~~~~~~~~~~~~~~~~~~~~~~~~~~~+\mathds{E}\bigg\{Z\big[(n+l)\cdot N-a_1\big]\cdot Z\big[n\cdot N-a_2\big] \bigg|H_0\bigg\}  \nonumber\\ 
&\stackrel{(a)}{=} \sum\limits_{l_1=0}^{L_{\textsf{ch}}}\sum\limits_{l_2=0}^{L_{\textsf{ch}}} h[-a_1,l_1]\cdot h[-a_2,l_2]\cdot \mathds{E} \bigg\{S \big[(n+l)\cdot N-a_1-l_1\big]\cdot S\big[n\cdot N-a_2-l_2\big] \bigg|H_0\bigg\} \nonumber \\
&~~~~~~~~~~~~~~~~~~~~~~~~~~~~~~~~~~~~~~~~~~~~~~~~~~~~~~~~~~~~~~~~~~~~~~~~~~~~~~~~~+\tilde{c}_{\textsf{z}}\big[n\cdot N-a_2,l\cdot N+a_2-a_1\big] \nonumber \\
&\stackrel{(b)}{=}\sum\limits_{l_1=0}^{L_{\textsf{ch}}}\sum\limits_{l_2=0}^{L_{\textsf{ch}}} h[-a_1,l_1]\cdot h[-a_2,l_2] \cdot \mathds{E} \bigg\{\Big(S\big[n\cdot N-a_2-l_2\big]\Big)^2\bigg\}\cdot\delta[l\cdot N-a_1-l_1+a_2+l_2] +0\nonumber\\
 \label{Pseudo_Auto_corroleation_Proof}
&\stackrel{(c)}{=}\sum\limits_{l_1=0}^{L_{\textsf{ch}}}\sum\limits_{l_2=0}^{L_{\textsf{ch}}} h[-a_1,l_1]\cdot h[-a_2,l_2]\cdot  \tilde{\sigma}_{\textsf{s}}^2   \cdot\delta[l\cdot N-a_1-l_1+a_2+l_2]~,~~~~~~~~~~~~~~~~~~~~~~~~~~~~~~~~~~
\end{align}
where $(a)$ follows since $N=k_1P_{\textsf{h}}$, thus $h[n,l_1]=h[n+k_1P_{\textsf{h}},l_1]=h[n+N,l_1]$, and $\tilde{c}_{\textsf{z}}\big[n\cdot N-a_2,l\cdot N+a_2-a_1\big]\triangleq \mathds{E}\bigg\{Z\big[(n+l)\cdot N-a_1\big]\cdot\Big(Z\big[n\cdot N-a_2\big]\Big)\bigg|H_0\bigg\}$ is the pseudo-autocorrelation function as defined in Appendix \ref{app:Noise_Model}, $(b)$ follows since 
$\mathds{E}\bigg\{S\big[(n+l)\cdot N-a_1-l_1\big]\cdot\Big(S\big[n\cdot N-a_2-l_2\big]\Big)\bigg|H_0\bigg\}=0,~-a_2-l_2\neq l\cdot N-a_1-l_1$, and since $Z[n]$ is proper complex, hence its pseudo-autocorrelation function $\tilde{c}_{\textsf{z}}\big[m,l\big]=0,~\forall m,l \in \mathcal{Z}$, as defined in Appendix \ref{app:Noise_Model}, (c) follows from the pseudo covariance of $S[n]$ defined in Eq. (\ref{Psaudo_Symbol_varaince}).
From (\ref{Pseudo_Auto_corroleation_Proof}) it directly follows that the elements of the pseudo-autocorrelation matrix  $\tilde{\mathds{R}}_{\tilde{\mathbf{R}}_{N,0}|H_0}[n,l]$ are independent of $n$, and combined with $\mathds{E}\big\{\tilde{\mathbf{R}}_{N,0}[n]\big|H_0\big\}=\mathbf{0}$, and with (\ref{Fisrt_wss_2}) it follows that when $H_0$ is valid then $\tilde{\mathbf{R}}_{N,0}[n]$ is a WSS random process. Next, we define the mathematical notation $c_{a_1,a_2}\triangleq c_{\textsf{z}}[-a_2,a_2-a_1]$. We recall that since $N$ is the period of the correlation function then for $l=0,~c_{\textsf{z}}[-a_2,l\cdot N+ a_2-a_1]=c_{\textsf{z}}[-a_2,a_2-a_1]=c_{\textsf{z}}[N-a_2,a_2-a_1]=c_{a_1-N,a_2-N},~a_1,a_2 \in \mathcal{Z}$.  Accordingly, from (\ref{Noise_Memory}) it follows that $c_{a_1,a_2}=0$ for $|a_1-a_2|>L_{\textsf{z}}$.
It is also noted that after the post-processing $\mathsf{P}\{\cdot\}$, $~\tilde{\mathbf{R}}_{N,0}[n\cdot M-m]$ subject to the $H_0$ hypothesis is a function of mutually independent and identically distributed random vectors,$~\mathbf{G}^{(m)}_{L_{\textsf{tot}},0}[n]~\mbox{and}~\tilde{\mathbf{Z}}_{N,0}[n\cdot M-m]$ as expressed in Eq. (\ref{R_Sub_Vec_Equation}), hence$~\tilde{\mathbf{R}}_{N,0}[n\cdot M-m]$ with $m \in \mathcal{M}\triangleq\{0,1,...,M-1\}$ are mutually independent and identically distributed random vectors w.r.t $n$ and $m$.
Next, we define the noise correlation matrix $\mathds{C}_{{\textsf{z}}}\in \mathcal{R}^{K~\times~K}$ as: 
\begin{align}
&[\mathds{C}_{{\textsf{z}}}]_{a_1+1,a_2+1}=c_{a_1,a_2}  \label{Noise_Matrix}
\end{align}
and prove that the covariance
of $\tilde{\mathbf{R}}_{N,0}[n\cdot M-m],~m=0,1,...,M-1$, subject to the $H_0$ hypothesis and given $\mathbf{g}^{(m)}_{L_{\textsf{tot}},0}[n]$, is equal to $\mathds{C}_{{\textsf{z}}}$, that is:
\begin{align*}
&\mathds{C}_{\tilde{\mathbf{R}}_{N,0}[n\cdot M-m]|\mathbf{g}^{(m)}_{L_{\textsf{tot}},0}[n],H_0}\\
&\triangleq \mathds{E}\Big\{\tilde{\mathbf{R}}_{N,0}[n\cdot M-m]\cdot\big(\tilde{\mathbf{R}}_{N,0}[n\cdot M-m]\big)^{H}\Big|\mathbf{g}^{(m)}_{L_{\textsf{tot}},0}[n],H_0\Big\}\\
&~~~~~~~~~~~~~~~~~~~~~~~~~-\mathds{E}\Big\{\tilde{\mathbf{R}}_{N,0}[n\cdot M-m]\Big|\mathbf{g}^{(m)}_{L_{\textsf{tot}},0}[n],H_0\Big\}\cdot\mathds{E}\Big\{\big(\tilde{\mathbf{R}}_{N,0}[n\cdot M-m]\big)^{H}\Big|\mathbf{g}^{(m)}_{L_{\textsf{tot}},0}[n],H_0\Big\}\\
&=\mathds{C}_{{\textsf{z}}}.~~~~~~~~~~~~~~~~~~~~~~~~~~~~~~~~~~~~~~~~~~~~~~~~~~~~~~~~~~~~~~~~~~~~~~~~~~~~~~~~~~~~~~~~~~~~~~~~~~~~~~~~~~~~~~~~~~~~~~~~~~~~~~~~~~~~~~~~~~~~~~~~~~~~~~~~~
\end{align*}
To that aim, first examine $\mathds{E}\Big\{\tilde{\mathbf{R}}_{N,0}[n\cdot M-m]\Big|\mathbf{g}^{(m)}_{L_{\textsf{tot}},0}[n],H_0\Big\}$:
From periodicity of the channel, $h[n,l]=h[n+k_1P_{\textsf{h}},l]=h[n+N,l]$, the relationship $NM=L_{\textsf{tot}}$, and the fact that $\mathds{E}\big\{Z[n]|\mathbf{g}^{(m)}_{L_{\textsf{tot}},0}[n],H_0\big\}=0,~\forall n \in \mathcal{Z}$, it directly follows that:
\begin{ceqn}
\begin{align} \label{conditional_1}
&\mathds{E}\Big\{\tilde{\mathbf{R}}_{N,0}[n\cdot M-m]\Big|\mathbf{g}^{(m)}_{L_{\textsf{tot}},0}[n],H_0\Big\}=\mathds{B}[n] \cdot \mathbf{g}^{(m)}_{L_{\textsf{tot}},0}[n]~~~
\end{align}
\end{ceqn}
Using (\ref{conditional_1}) we can express
\begin{align*}
&\mathds{E}\Big\{\tilde{\mathbf{R}}_{N,0}[n\cdot M-m]\Big|\mathbf{g}^{(m)}_{L_{\textsf{tot}},0}[n],H_0\Big\}\cdot\mathds{E}\Big\{\big(\tilde{\mathbf{R}}_{N,0}[n\cdot M-m]\big)^{H}\Big|\mathbf{g}^{(m)}_{L_{\textsf{tot}},0}[n],H_0\Big\}\\
&~~~~=\mathds{B}[n] \cdot \mathbf{g}^{(m)}_{L_{\textsf{tot}},0}[n]\cdot \big(\mathbf{g}^{(m)}_{L_{\textsf{tot}},0}[n]\big)^H\cdot\big(\mathds{B}[n]\big)^H\\
&~~~~\triangleq\mathds{V}[n,m]~,    
\end{align*}
where the elements of matrix $\mathds{V}[n,m]$ at the $(a_1+1)$'th row and $(a_2+1)$'th column, where $a_1,a_2\in \{0,1,...,K-1\}$ are given by:
\begin{align} \label{Matrix_element_1}
&\big[\mathds{V}[n,m]\big]_{a_1+1,a_2+1}\nonumber\\
&=\bigg(\sum\limits_{l_1=0}^{L_{\textsf{ch}}} h[-a_1,l_1]s[nL_{\textsf{tot}}-mN-a_1-l_1]\bigg)\cdot \bigg(\sum\limits_{l_2=0}^{L_{\textsf{ch}}} h[-a_2,l_2]s[nL_{\textsf{tot}}-mN-a_2-l_2]\bigg)^{*} \nonumber~~~~~~~~~~~~~~~~~~~~~~~~\\ 
&=\sum\limits_{l_1=0}^{L_{\textsf{ch}}}\sum\limits_{l_2=0}^{L_{\textsf{ch}}} h[-a_1,l_1]\cdot \big(h[-a_2,l_2]\big)^* \cdot s[nL_{\textsf{tot}}-mN-a_1-l_1]\cdot \Big(s[nL_{\textsf{tot}}-mN-a_2-l_2]\Big)^{*}. 
\end{align}
From the definition of $\tilde{\mathbf{R}}_{N,0}[n]$ it immediately follows that the element of the matrix $\mathds{U}[n,m]\triangleq \mathds{E}\Big\{\tilde{\mathbf{R}}_{N,0}[n\cdot M-m]\cdot\Big(\tilde{\mathbf{R}}_{N,0}[n\cdot M-m]\Big)^{H}\Big|\mathbf{g}^{(m)}_{L_{\textsf{tot}},0}[n],H_0\Big\},$ at the $(a_1+1)$'th row, and $(a_2+1)$'th column, $a_1,a_2\in \{0,1,...,K-1\}$, is:
\begin{align}
&\big[\mathds{U}[n,m]\big]_{a_1+1,a_2+1}\nonumber\\
&=\sum\limits_{l_1=0}^{L_{\textsf{ch}}}\sum\limits_{l_2=0}^{L_{\textsf{ch}}} h[-a_1,l_1] \cdot \big(h[-a_2,l_2]\big)^* \cdot s[nL_{\textsf{tot}}-mN-a_1-l_1]\cdot \Big(s[nL_{\textsf{tot}}-mN-a_2-l_2]\Big)^*\nonumber\\
&~~~~~~~~~~~~~~~~~~~~~~~~~~~~~~~~~~~~~~~~~~~~~~~~~~~~~~~~~~~~~~~~~~~~~~~~~~~~~~~~~~~~~~~~~+c_{\textsf{z}}[-a_2,a_2-a_1].
 \label{Matrix_element_2}
\end{align}
Thus, we obtain $\mathds{C}_{\tilde{\mathbf{R}}_{N,0}[n\cdot M-m]|\mathbf{g}^{(m)}_{L_{\textsf{tot}},0}[n],H_0}=\mathds{U}[n,m]- \mathds{V}[n,m]$, hence, the element at the $(a_1+1)\mbox{'th}$ row and the $(a_2+1)\mbox{'th}$ column of $\mathds{C}_{\tilde{\mathbf{R}}_{N,0}[n\cdot M-m]|\mathbf{g}^{(m)}_{L_{\textsf{tot}},0}[n],H_0}$ where $~a_1,a_2\in \{0,1,...,K-1\}$ is given by:
\begin{ceqn}
\begin{align}
&\Big[\mathds{C}_{\tilde{\mathbf{R}}_{N,0}[n\cdot M-m]|\mathbf{g}^{(m)}_{L_{\textsf{tot}},0}[n],H_0}\Big]_{a_1+1,a_2+1}\nonumber\\
&=\big[\mathds{U}[n,m]\big]_{a_1+1,a_2+1} - \big[\mathds{V}[n,m]\big]_{a_1+1,a_2+1}\nonumber \\
&\stackrel{(a)}{=} \sum\limits_{l_1=0}^{L_{\textsf{ch}}}\sum\limits_{l_2=0}^{L_{\textsf{ch}}} h[-a_1,l_1] \cdot \big(h[-a_2,l_2]\big)^*\cdot s[nL_{\textsf{tot}}-mN-a_1-l_1]\cdot \Big(s\big[nL_{\textsf{tot}}-mN-a_2-l_2\big]\Big)^*\nonumber \\
&~~~~~~~~~~~~~~~~~~~~~~~~~~~~~~~~~~~~~~~~~~~~~~~~~~~~~~~~~~~~~~~~~~~~~~~~~~~~~~~~~~~~~~~~~+c_{\textsf{z}}[-a_2,a_2-a_1] \nonumber\\
&~~~~~~~~~~~ -\sum\limits_{l_1=0}^{L_{\textsf{ch}}}\sum\limits_{l_2=0}^{L_{\textsf{ch}}} h[-a_1,l_1] \cdot \big(h[-a_2,l_2]\big)^*\cdot s[nL_{\textsf{tot}}-mN-a_1-l_1]\cdot \Big(s\big[nL_{\textsf{tot}}-mN-a_2-l_2\big]\Big)^*  \nonumber \\
&= c_{\textsf{z}}[-a_2,a_2-a_1], \label{Matrix_Proof}
\end{align}
\end{ceqn}
where $(a)$ follows from (\ref{Matrix_element_1}) and (\ref{Matrix_element_2}). It follows from (\ref{Matrix_Proof}) that:
\begin{ceqn}
\begin{align*}
&\mathds{C}_{\tilde{\mathbf{R}}_{N,0}[n\cdot M-m]|\mathbf{g}^{(m)}_{L_{\textsf{tot}},0}[n],H_0}=\mathds{C}_{{\textsf{z}}}~~.
\end{align*}
\end{ceqn}
Lastly, for representing the PDF of $\mathbf{R}^{(\mathsf{P})}_{L_{\textsf{tot}},0}[n]$ under the $H_0$ hypothesis we define the following vectors:
\begin{align*}
&\mathbf{a}_{l} \triangleq
\left( \begin{array}{l}
s_{(m_{0,l}\bmod N_{\textsf{s}})}\\
s_{(m_{1,l}\bmod N_{\textsf{s}})}\\
\vdots \\
s_{(m_{N-1,l}\bmod N_{\textsf{s}})}\\ 
\end{array} \right) \in \mathcal{C}^{N~\times~1},
\end{align*}
where $m_{d,l}=\left\lfloor \frac{l}{(N_{\textsf{s}})^d} \right\rfloor,~l=0,1,...,(N_{\textsf{s}})^{N}-1,~d=0,1,...,N-1$, and $s_i, ~i=0,1,...,N_{\textsf{s}}-1$ are the elements of the constellation set $\mathcal{S},~|\mathcal{S}|= N_{\textsf{s}}$. The set 
$\big\{\mathbf{a}_{l}\big\}_{l=0}^{(N_{\textsf{s}})^{N}-1}$ represents $(N_{\textsf{s}})^{N}$ vectors of length $N$, each containing a different combination of $N$ symbols from $\mathcal{S}$.
The PDF of $\mathbf{R}^{(\mathsf{P})}_{L_{\textsf{tot}},0}[n]$ under the $H_0$ hypothesis is expressed as:
\begin{align*}
&f\Big(\mathbf{R}^{(\mathsf{P})}_{L_{\textsf{tot}},0}[n]\Big|H_0\Big)\\
&\stackrel{(a)}{=}\prod_{m=0}^{M-1} f\Big(\tilde{\mathbf{R}}_{N,0}[n\cdot M-m]\Big|H_0\Big)\\
&\stackrel{(b)}{=} \sum\limits_{l=0}^{(N_{\textsf{s}})^{MN}-1} \frac{1}{(N_{\textsf{s}})^{MN}}\cdot\prod_{m=0}^{M-1} f\Big(\tilde{\mathbf{R}}_{N,0}[n\cdot M-m]\Big|\mathbf{g}^{(m)}_{L_{\textsf{tot}},0}[n]=\mathbf{a}_{q_{(l,m)}},~H_0\Big)\\
&\stackrel{(c)}{=} \sum\limits_{l=0}^{(N_{\textsf{s}})^{L_{\textsf{tot}}}-1} \frac{1}{(N_{\textsf{s}})^{L_{\textsf{tot}}}} \prod_{m=0}^{M-1}\frac{1}{\pi^K \cdot \big|\mathds{C}_{{\textsf{z}}}\big|} \\
&~~~~~~~~~~~~\cdot \mbox{exp}\bigg(-\Big(\big(\tilde{\mathbf{r}}_{N,0}[n\cdot M-m]-\mathds{B}[n] \cdot \mathbf{a}_{q_{(l,m)}} \big)^H\cdot \mathds{C}^{-1}_{{\textsf{z}}}\cdot \big(\tilde{\mathbf{r}}_{N,0}[n\cdot M-m]-\mathds{B}[n] \cdot \mathbf{a}_{q_{(l,m)}}\big)\Big)\bigg)\\
&= \frac{1}{\big(N_{\textsf{s}}\big)^{L_{\textsf{tot}}} \cdot \Big(\pi^{K}  \cdot \big|\mathds{C}_{{\textsf{z}}}\big|\Big)^{M}} \\
&~~~\cdot\!\!\! \sum\limits_{l=0}^{(N_{\textsf{s}})^{L_{\textsf{tot}}}-1}\!\!\!\! \mbox{exp}\bigg(\! -\sum\limits_{m=0}^{M-1}\Big(\big(\tilde{\mathbf{r}}_{N,0}[n\cdot M-m]-\mathds{B}[n] \cdot \mathbf{a}_{q_{(l,m)}} \big)^H \cdot \mathds{C}^{-1}_{{\textsf{z}}}\cdot \big(\tilde{\mathbf{r}}_{N,0}[n\cdot M-m]-\mathds{B}[n] \cdot \mathbf{a}_{q_{(l,m)}}\big)\Big)\bigg)
\end{align*}
where $(a)$ follows from the discussion after Eq. (\ref{Pseudo_Auto_corroleation_Proof}), which establishes that subject to the $H_0$ hypothesis, the post-processing $\mathsf{P}\{\cdot\}$ results in mutually independent random vectors $\tilde{\mathbf{R}}_{N,0}[n\cdot M-m]$, $m \in \mathcal{M}$, $(b)$ follows from the law of total probability \cite[Ch. 4.7]{mendenhall2012introduction}, where 
$q_{(l,m)}=\left( \left\lfloor \frac{l}{(N_{\textsf{s}})^{mN}}\right\rfloor \mod (N_{\textsf{s}})^N \right)$, $(c)$ follows form (\ref{R_Sub_Vec_Equation}) which implies that $f\Big(\tilde{\mathbf{R}}_{N,0}[n\cdot M-m]\Big|\mathbf{g}^{(m)}_{L_{\textsf{tot}},0}[n]=\mathbf{a}_{l},~H_0\Big)=f_{\tilde{\mathbf{Z}}_{N,0}}\Big(\tilde{\mathbf{r}}_{N,0}[n\cdot M-m]-\mathds{B}[n] \cdot \mathbf{a}_{l}\Big)$, where $\tilde{\mathbf{Z}}_{N,0}[n\cdot M-m]$ has a Gaussian PDF, and we also applied $L_{\textsf{tot}}=MN$.
\subsection{Statistics of the Received Signal for Hypothesis \texorpdfstring{$H_1~$}~}
We next characterize the statistics of $\tilde{\mathbf{R}}_{N,0}[n]$ subject to the $H_1$ hypothesis, and prove that $\tilde{\mathbf{R}}_{N,0}[n\cdot M-m],~m=0,1,...,M-1$, subject to $H_1$ hypothesis, are mutually independent (but not i.i.d):
To that aim we use the following lemma, which follows immediately from the independence of the RVs:
\begin{lemma} \label{lemma_indpendent}
Let $X$ and $Y$ be independent random variables, and $a,b \in \mathcal{C}$ be two constants. The random variables: $\tilde{X}\triangleq X+a$ and $\tilde{Y}\triangleq Y+b$ are independent random variables. 
\end{lemma}
Next, we examine $\tilde{\mathbf{R}}_{N,0}[n\cdot M-m]$ subject to $H_1$ hypothesis:
\begin{align*}
\tilde{\mathbf{R}}_{N,0}[n\cdot M-m]
&=\mathds{B}[n] \cdot \tilde{\mathbf{G}}^{(m)}_{L_{\textsf{tot}},0}[n]+\tilde{\mathbf{Z}}_{N,0}[n\cdot M-m]\\
&=\mathds{B}[n] \cdot \left( \begin{array}{l}
\mathbf{t}_{M-1-m}\\
~\\
~\\
\mathbf{D}^{(m)}_{L_{\textsf{tot}},0}[n]
\end{array}\right)+\tilde{\mathbf{Z}}_{N,0}[n\cdot M-m]\\
&=\mathds{B}[n] \cdot \left( \begin{array}{l}
\mathbf{t}_{M-1-m}\\
~\\
~\\
\mathds{O}_{L_{\textsf{ch}} \times 1}
\end{array}\right)+\mathds{B}[n] \cdot \left( \begin{array}{l}
\mathds{O}_{K \times 1}\\
~\\
~\\
\mathbf{D}^{(m)}_{L_{\textsf{tot}},0}[n]
\end{array}\right)+\tilde{\mathbf{Z}}_{N,0}[n\cdot M-m]~~~~~~~~~~~~~~~\\
&=\mathbf{c}[n,m]+\mathbf{X}[n,m]+\tilde{\mathbf{Z}}_{N,0}[n\cdot M-m],~m=0,1,...,M-1~,
\end{align*}
where $\mathbf{c}[n,m]\triangleq \mathds{B}[n] \cdot \left( \begin{array}{l}
\mathbf{t}_{M-1-m}\\
~\\
~\\
\mathds{O}_{L_{\textsf{ch}} \times 1}
\end{array}\right)\in \mathcal{C}^{~N~\times~1}$.
For a given pair of values of $n$ and $m$ respectively, since $\mathds{B}[n]$ is a constant matrix and $\mathbf{t}_{M-1-m}$ is a constant vector, then $\mathbf{c}[n,m]$ is a constant vector.
Let $ \mathbf{X}[n,m]\triangleq \mathds{B}[n] \cdot \left( \begin{array}{l}
\mathds{O}_{K \times 1}\\
~\\
~\\
\mathbf{D}^{(m)}_{L_{\textsf{tot}},0}[n]
\end{array}\right)\in \mathcal{C}^{~N~\times~1}$.
 From the definition of $\mathbf{D}^{(m)}_{L_{\textsf{tot}},0}[n]$ and $\tilde{\mathbf{Z}}_{N,0}[n\cdot M-m]$ it follows that these are mutually independent vectors over all values of $n$ and $m$, and that each vector has independent elements. Accordingly, since $\mathbf{X}[n,m]$  is a function of $\mathbf{D}^{(m)}_{L_{\textsf{tot}},0}[n]$, then it follows that $\mathbf{X}[n,m]$ and $\tilde{\mathbf{Z}}_{N,0}[n\cdot M-m]$ are mutually independent vectors w.r.t. $n$ and $m$. Thus, the random vectors $\mathbf{P}[n,m]\triangleq \mathbf{X}[n,m]+\tilde{\mathbf{Z}}_{N,0}[n\cdot M-m]$ are mutually independent w.r.t. $n$ and $m$, since they are the sum of two vectors mutually independent vectors over $n$ and $m$. Accordingly, from Lemma \ref{lemma_indpendent}, since $\tilde{\mathbf{R}}_{N,0}[n\cdot M-m]= \mathbf{P}[n,m]+\mathbf{c}[n,m]$, it follows that the vectors $\tilde{\mathbf{R}}_{N,0}[n\cdot M-m]$, subject to the $H_1$ hypothesis, are mutually independent w.r.t. $n$ and $m$, for $m=0,1,...,M-1$.\\
Next, we show that the covariance matrix of $\tilde{\mathbf{R}}_{N,0}[n\cdot M-m],~m=0,1,...,M-1$ subject to the $H_1$ hypothesis, and given $\tilde{\mathbf{g}}^{(m)}_{L_{\textsf{tot}},0}[n]$, is equal to $\mathds{C}_{{\textsf{z}}}$ specified in (\ref{Noise_Matrix}), that is:
\begin{align*}
&\mathds{C}_{\tilde{\mathbf{R}}_{N,0}[n\cdot M-m]|\tilde{\mathbf{g}}^{(m)}_{L_{\textsf{tot}},0}[n],H_1}\\
&\triangleq \mathds{E}\Big\{\tilde{\mathbf{R}}_{N,0}[n\cdot M-m]\cdot\big(\tilde{\mathbf{R}}_{N,0}[n\cdot M-m]\big)^{H}\Big|\tilde{\mathbf{g}}^{(m)}_{L_{\textsf{tot}},0}[n],H_1\Big\}\\
&~~~~~~~~~-\mathds{E}\Big\{\tilde{\mathbf{R}}_{N,0}[n\cdot M-m]\Big|\tilde{\mathbf{g}}^{(m)}_{L_{\textsf{tot}},0}[n],H_1\Big\}\cdot\mathds{E}\Big\{\big(\tilde{\mathbf{R}}_{N,0}[n\cdot M-m]\big)^{H}\Big|\tilde{\mathbf{g}}^{(m)}_{L_{\textsf{tot}},0}[n],H_1\Big\}\\
&=\mathds{C}_{{\textsf{z}}}.
\end{align*}
Beginning with $\mathds{E}\Big\{\tilde{\mathbf{R}}_{N,0}[n\cdot M-m]\Big|\tilde{\mathbf{g}}^{(m)}_{L_{\textsf{tot}},0}[n],H_1\Big\}$, then plugging the definition for $\tilde{\mathbf{S}}^{(\textsf{sw})}_{L_{\textsf{tot}},0}[n]$ and applying similar steps to those leading to (\ref{conditional_1}) we obtain that:   
\begin{align} \label{conditional_H_1_1}
&\bigg[\mathds{E}\Big\{\tilde{\mathbf{R}}_{N,0}[n\cdot M-m]\Big|\tilde{\mathbf{g}}^{(m)}_{L_{\textsf{tot}},0}[n],H_1\Big\}\bigg]_{a_1+1,1} \nonumber\\
&=\sum\limits_{l=0}^{L_{\textsf{ch}}-a_1} h[-a_1,l]f_{L_{\textsf{sw}}-1-(l+mK)}+\sum\limits_{l=K-a_1}^{L_{\textsf{ch}}} h\big[-a_1,l\big]\cdot s\big[nL_{\textsf{tot}}-mN-a_1-l\big],\nonumber\\
&~\forall n\in \mathcal{Z},~m=0,1,...,M-1,~a_1=0,1,...,K-1.
\end{align}
Using (\ref{conditional_H_1_1}) we can derive the elements of matrix $\tilde{\mathds{V}}[n.m]\triangleq \mathds{E}\Big\{\tilde{\mathbf{R}}_{N,0}[n\cdot M-m]\Big|\tilde{\mathbf{g}}^{(m)}_{L_{\textsf{tot}},0}[n],H_1\Big\}\cdot\mathds{E}\Big\{\big(\tilde{\mathbf{R}}_{N,0}[n\cdot M-m]\big)^{H}\Big|\tilde{\mathbf{g}}^{(m)}_{L_{\textsf{tot}},0}[n],H_1\Big\}$ as follows:
Defining the sets $\mathcal{I}_1=\{0,1,...,K-L_{\textsf{ch}}-1\}$ and $\mathcal{I}_2=\{K-L_{\textsf{ch}},K-(L_{\textsf{ch}}-1),...,K-1\}$, we write for $a_1,a_2\in \{0,1,...,K-1\}:$
\begin{align*}
&\big[\tilde{\mathds{V}}[n,m]\big]_{a_1+1,a_2+1} \nonumber \\
&~~~~~~=\mathds{E}\bigg\{R\big[(n\cdot M-m)N-a_1\big]\bigg|\tilde{\mathbf{g}}^{(m)}_{L_{\textsf{tot}},0}[n],H_1\bigg\}\cdot \mathds{E}\bigg\{\Big(R\big[(n\cdot M-m)N-a_2\big]\Big)^*\bigg|\tilde{\mathbf{g}}^{(m)}_{L_{\textsf{tot}},0}[n],H_1\bigg\}\nonumber
\end{align*}
and obtain that $\big[\tilde{\mathds{V}}[n,m]\big]_{a_1+1,a_2+1}$ depend on $a_1$ and $a_2$ as follows
\begin{align*} 
&\underline{a_1,a_2\in \mathcal{I}_1}\nonumber\\
&\big[\tilde{\mathds{V}}[n,m]\big]_{a_1+1,a_2+1}\nonumber\\
&=\sum\limits_{l_1=0}^{L_{\textsf{ch}}}\sum\limits_{l_2=0}^{L_{\textsf{ch}}} h\big[-a_1,l_1\big]\cdot\Big(h\big[-a_2,l_2\big]\Big)^*\cdot f_{L_{\textsf{sw}}-1-(l_1+mK+a_1)}\cdot\Big(f_{L_{\textsf{sw}}-1-(l_2+mK+a_2)}\Big)^*\nonumber~~~~~~~~~~~~~~~~~~~~~~\\
& \underline{a_1\in \mathcal{I}_1 ,~a_2\in \mathcal{I}_2}\nonumber\\ 
&\big[\tilde{\mathds{V}}[n,m]\big]_{a_1+1,a_2+1}\nonumber\\
&=\sum\limits_{l_1=0}^{L_{\textsf{ch}}}\sum\limits_{l_2=0}^{K-1-a_2} h\big[-a_1,l_1\big]\cdot\Big(h\big[-a_2,l_2\big]\Big)^* \cdot f_{L_{\textsf{sw}}-1-(l_1+mK+a_1)}\cdot\Big(f_{L_{\textsf{sw}}-1-(l_2+mK+a_2)}\Big)^*\nonumber\\
&~~+\sum\limits_{l_1=0}^{L_{\textsf{ch}}}\sum\limits_{l_2=K-a_2}^{L_{\textsf{ch}}} h\big[-a_1,l_1\big]\cdot\Big(h\big[-a_2,l_2\big]\Big)^*\cdot f_{L_{\textsf{sw}}-1-(l_1+mK+a_1)}\cdot \Big(s\big[nL_{\textsf{tot}}-mN-a_2-l_2\big]\Big)^*\nonumber\\
& \underline{a_1\in \mathcal{I}_2 ,~a_2\in \mathcal{I}_1}\nonumber\\
&\big[\tilde{\mathds{V}}[n,m]\big]_{a_1+1,a_2+1}\nonumber\\
&=\sum\limits_{l_1=0}^{K-1-a_1}\sum\limits_{l_2=0}^{L_{\textsf{ch}}} h\big[-a_1,l_1\big]\cdot\Big(h\big[-a_2,l_2\big]\Big)^*\cdot f_{L_{\textsf{sw}}-1-(l_1+mK+a_1)}\cdot\Big(f_{L_{\textsf{sw}}-1-(l_2+mK+a_2)}\Big)^*\nonumber\\
&~~+\sum\limits_{l_1=K-a_1}^{L_{\textsf{ch}}}\sum\limits_{l_2=0}^{L_{\textsf{ch}}} h\big[-a_1,l_1\big]\cdot\Big(h\big[-a_2,l_2\big]\Big)^*\cdot s\big[nL_{\textsf{tot}}-mN-a_1-l_1\big]\cdot \Big(f_{L_{\textsf{sw}}-1-(l_2+mK+a_2)}\Big)^*
\end{align*} 
\begin{align} \label{V_tilde_Matrix}
&\underline{a_1,a_2\in \mathcal{I}_2}\nonumber\\
&\big[\tilde{\mathds{V}}[n,m]\big]_{a_1+1,a_2+1}\nonumber\\
&=\sum\limits_{l_1=0}^{K-1-a_1}\sum\limits_{l_2=0}^{K-1-a_2} h\big[-a_1,l_1\big]\cdot\Big(h\big[-a_2,l_2\big]\Big)^* \cdot f_{L_{\textsf{sw}}-1-(l_1+mK+a_1)}\cdot\Big(f_{L_{\textsf{sw}}-1-(l_2+mK+a_2)}\Big)^*\nonumber\\
&+\sum\limits_{l_1=0}^{K-1-a_1}\sum\limits_{l_2=K-a_2}^{L_{\textsf{ch}}} h\big[-a_1,l_1\big]\cdot \Big(h\big[-a_2,l_2\big]\Big)^*\cdot f_{L_{\textsf{sw}}-1-(l_1+mK+a_1)}\cdot\Big(s\big[nL_{\textsf{tot}}-mN-a_2-l_2\big]\Big)^*\nonumber\\
&+\sum\limits_{l_1=0}^{K-1-a_1}\sum\limits_{l_2=K-a_2}^{L_{\textsf{ch}}} h\big[-a_1,l_1\big]\cdot\Big(h\big[-a_2,l_2\big]\Big)^*\cdot f_{L_{\textsf{sw}}-1-(l_1+mK+a_1)}\cdot\Big(s\big[nL_{\textsf{tot}}-mN-a_2-l_2\big]\Big)^*\nonumber\\
&+\sum\limits_{l_1=K-a_1}^{L_{\textsf{ch}}}\sum\limits_{l_2=K-a_2}^{L_{\textsf{ch}}} h\big[-a_1,l_1\big]\cdot \Big(h\big[-a_2,l_2\big]\Big)^*\cdot s\big[nL_{\textsf{tot}}-mN-a_1-l_1)\big]\nonumber\\
&~~~~~~~~~~~~~~~~~~~~~~~~~~~~~~~~~~~~~~~~~~~~~~~~~~~~~~~~~~~~~~~~~~~~~~~~~~~~~~~~~\cdot\Big(s\big[nL_{\textsf{tot}}-mN-a_2-l_2\big]\Big)^*
\end{align}
Next, we explicitly express the elements of the matrix $\tilde{\mathds{U}}[n,m]\triangleq \mathds{E}\Big\{\tilde{\mathbf{R}}_{N,0}[n\cdot M-m]\cdot\big(\tilde{\mathbf{R}}_{N,0}[n\cdot M-m]\big)^{H}$ $\Big|\tilde{\mathbf{g}}^{(m)}_{L_{\textsf{tot}},0}[n],H_1\Big\},$ at the $(a_1+1)$'th row, and $(a_2+1)$'th column $a_1,a_2\in \{0,1,...,K-1\}$. Using steps similar to those leading to (\ref{V_tilde_Matrix}) we obtain:
\begin{align} \label{V_Matrix_H_1}
\big[\tilde{\mathds{U}}&[n,m]\big]_{a_1+1,a_2+1}~~~~~~~~~~~~~~~~~~~~~~~~~~~~~~~~~~~~~~~~~~~~~~~~~~~~~~~~~~~~~~~~~~~\nonumber\\
&=\mathds{E}\bigg\{R\big[(n\cdot M-m)N-a_1\big]\cdot\Big(R\big[(n\cdot M-m)N-a_2\big]\Big)^{*}\bigg|\tilde{\mathbf{g}}^{(m)}_{L_{\textsf{tot}},0}[n],H_1\bigg\}\nonumber\\
&=\big[\tilde{\mathds{V}}[n,m]\big]_{a_1+1,a_2+1}+c_{\textsf{z}}\big[-a_2,a_2-a_1\big]
\end{align}
Noting that $\mathds{C}_{\tilde{\mathbf{R}}_{N,0}[n\cdot M-m]|\mathbf{g}^{(m)}_{L_{\textsf{tot}},0}[n],H_0}=\tilde{\mathds{U}}[n,m]-\tilde{\mathds{V}}[n,m]$, it follows that the element at the $(a_1+1)\mbox{'th}$ row and the $(a_2+1)\mbox{'th}$ column, where $~a_1,a_2\in \{0,1,...,K-1\}$ is given by:
\begin{ceqn}
\begin{align} 
&\Big[\mathds{C}_{\tilde{\mathbf{R}}_{N,0}[n\cdot M-m]|\tilde{\mathbf{g}}^{(m)}_{L_{\textsf{tot}},0}[n],H_1}\Big]_{a_1+1,a_2+1}\nonumber~~~~~~~~~~~~~~~~~~~~~~~~~~~~~~~~~~~~~~ \\
&~~~~~~~~~~=\big[\tilde{\mathds{U}}[n,m]\big]_{a_1+1,a_2+1} - \big[\tilde{\mathds{V}}[n,m]\big]_{a_1+1,a_2+1}\nonumber\\
&~~~~~~~~~~\stackrel{(a)}{=}c_{\textsf{z}}\big[-a_2,a_2-a_1\big] \label{C_Z_poof}
\end{align}
\end{ceqn}
where (a) follows from (\ref{V_tilde_Matrix}) and (\ref{V_Matrix_H_1}). 
Hence, $\mathds{C}_{\tilde{\mathbf{R}}_{N,0}[n\cdot M-m]|\tilde{\mathbf{g}}^{(m)}_{L_{\textsf{tot}},0}[n],H_1}=\mathds{C}_{{\textsf{z}}}$.
Next, for representing the PDF of $\mathbf{R}^{(\mathsf{P})}_{L_{\textsf{tot}},0}[n]$ subject to the $H_1$ hypothesis, we define the following vectors:
\begin{ceqn}
\begin{align*}
&\tilde{\mathbf{a}}_{l,k} \triangleq
\left( \begin{array}{l}
\mathbf{t}_{M-1-k} \\
~\\
~\\
s_{(m_0 \bmod N_{\textsf{s}})}\\
s_{(m_1 \bmod N_{\textsf{s}})}\\
\vdots \\
s_{(m_{L_{\textsf{ch}}-1} \bmod N_{\textsf{s}})}
\end{array} \right)\in \mathcal{C}^{N~\times~1},
\end{align*}
\end{ceqn}
where $m_d=\left\lfloor \frac{l}{(N_{\textsf{s}})^d} \right\rfloor,~l=0,1,...,(N_{\textsf{s}})^{L_{\textsf{ch}}}-1,~d=0,1,...,L_{\textsf{ch}}-1,~k=0,1,...,M-1$, where $s_i$, for $i=0,1,...,N_{\textsf{s}}-1$, are the elements of the constellation set $\mathcal{S},~ |\mathcal{S}|= N_{\textsf{s}}$.
The set $\{\tilde{\mathbf{a}}_{l,k}\}_{l=0,k=0}^{(N_{\textsf{s}})^{L_{\textsf{ch}}}-1,M-1}$ represents $M\cdot(N_{\textsf{s}})^{L_{\textsf{ch}}}$ vectors of length $N$, each containing a different combination of $L_{\textsf{ch}}$ symbols from $\mathcal{S}$, which appends a block of $K$ SW symbols out of the $L_{\textsf{sw}}$ complex SW symbols.
The PDF of $\mathbf{R}^{(\mathsf{P})}_{L_{\textsf{tot}},0}[n]$ under the $H_1$ hypothesis is expressed as:
\begin{align*}
&f\Big(\mathbf{R}^{(\mathsf{P})}_{L_{\textsf{tot}},0}[n]\Big|H_1\Big)\\
&\stackrel{(a)}{=}\prod_{m=0}^{M-1} f\Big(\tilde{\mathbf{R}}_{N,0}[n\cdot M-m]\Big|H_1\Big)\\
&\stackrel{(b)}{=} \sum\limits_{u=0}^{(N_{\textsf{s}})^{ML_{\textsf{ch}}}-1}  \prod_{m=0}^{M-1} \frac{1}{(N_{\textsf{s}})^{L_{\textsf{ch}}}}\cdot f\Big(\tilde{\mathbf{R}}_{N,0}[n\cdot M-m]\Big|\tilde{\mathbf{g}}^{(m)}_{L_{\textsf{tot}},0}[n]=\tilde{\mathbf{a}}_{\tilde{q}_{(u,m)},m},~H_1\Big)~~~~~~~~~~~~~~~~~~~~~~~~~~~~~\\
&\stackrel{(c)}{=} \sum\limits_{u=0}^{(N_{\textsf{s}})^{ML_{\textsf{ch}}}-1} \frac{1}{(N_{\textsf{s}})^{ML_{\textsf{ch}}}} \prod_{m=0}^{M-1}\frac{1}{\pi^K \cdot \big|\mathds{C}_{{\textsf{z}}}\big|} \\
&~~~~~\cdot \mbox{exp}\bigg(-\Big(\big(\tilde{\mathbf{r}}_{N,0}[n\cdot M-m]-\mathds{B}[n] \cdot \tilde{\mathbf{a}}_{\tilde{q}_{(u,m)},m} \big)^H\cdot \mathds{C}^{-1}_{{\textsf{z}}}\cdot \big(\tilde{\mathbf{r}}_{N,0}[n\cdot M-m]-\mathds{B}[n] \cdot\tilde{\mathbf{a}}_{\tilde{q}_{(u,m)},m}\big)\Big)\bigg) \\
&=\frac{1}{(N_{\textsf{s}})^{ML_{\textsf{ch}}} \cdot \Big(\pi^{K} \cdot  \big|\mathds{C}_{{\textsf{z}}}\big|\Big)^{M}}\cdot \sum\limits_{u=0}^{(N_{\textsf{s}})^{ML_{\textsf{ch}}}-1} \mbox{exp}\bigg(-\sum\limits_{m=0}^{M-1}\Big(\big(\tilde{\mathbf{r}}_{N,0}[n\cdot M-m]-\mathds{B}[n] \cdot \tilde{\mathbf{a}}_{\tilde{q}_{(u,m)},m} \big)^H \\
&~~~~~~~~~~~~~~~~~~~~~~~~~~~~~~~~~~~~~~~~~~~~~~~~~~~~~~~~~~\cdot \mathds{C}^{-1}_{{\textsf{z}}}\cdot \big(\tilde{\mathbf{r}}_{N,0}[n\cdot M-m]-\mathds{B}[n] \cdot \tilde{\mathbf{a}}_{\tilde{q}_{(u,m)},m}\big)\Big)\bigg)
\end{align*}
where $(a)$ follows since after the post-processing $\mathsf{P}\{\cdot\}$ the vectors $\tilde{\mathbf{R}}_{N,0}[n\cdot M-m]$ are mutually independent (but not i.i.d), in $(b)$ we set
$\tilde{q}_{(u,m)}=\left( \left\lfloor \frac{u}{(N_{\textsf{s}})^{mL_{\textsf{ch}}}}\right\rfloor \mod (N_{\textsf{s}})^{L_{\textsf{ch}}} \right)$, $(c)$ follows since when $\tilde{\mathbf{g}}^{(m)}_{L_{\textsf{tot}},0}[n]$ is given, then for $H_1$, $\tilde{\mathbf{R}}_{N,0}[n\cdot M-m]=\mathds{B}[n] \cdot \tilde{\mathbf{g}}^{(m)}_{L_{\textsf{tot}},0}[n]+\tilde{\mathbf{Z}}_{N,0}[n\cdot M-m]$, therefore
\begin{align*}
&f\Big(\tilde{\mathbf{R}}_{N,0}[n\cdot M-m]\Big|\tilde{\mathbf{g}}^{(m)}_{L_{\textsf{tot}},0}[n]=\tilde{\mathbf{a}}_{u,m},~H_1\Big)=f_{\tilde{\mathbf{Z}}_{N,0}}\Big(\tilde{\mathbf{R}}_{N,0}[n\cdot M-m]-\mathds{B}[n] \cdot \tilde{\mathbf{a}}_{u,m}\Big)   
\end{align*}
 and we use the fact that $\tilde{\mathbf{Z}}_{N,0}[n\cdot M-m]$ is distributed according to a Gaussian PDF.
%
%
%
\subsection{The LRT}
\label{subsec:LRT}
Using the derived expressions for the statistics of the received signal subject to $H_0$ and to $H_1$, the LRT for the case of \emph{known CIR} can be expressed as \cite[Ch. 2.2.1]{van2004detection}:
\begin{align*}
\mbox{LRT}\big(\mathbf{R}^{(\mathsf{P})}_{L_{\textsf{tot}},0}[n]\big)&=\frac{f\Big(\mathbf{R}^{(\mathsf{P})}_{L_{\textsf{tot}},0}[n]\Big|H_0\Big)}{f\Big(\mathbf{R}^{(\mathsf{P})}_{L_{\textsf{tot}},0}[n]\Big|H_1\Big)}  \\
&= \frac{1}{(N_{\textsf{s}})^{MK}}\cdot \frac{\left(\begin{array}{l}
 \sum\limits_{l=0}^{(N_{\textsf{s}})^{L_{\textsf{tot}}}-1} \mbox{exp}\bigg(-\sum\limits_{m=0}^{M-1}\Big(\big(\tilde{\mathbf{r}}_{N,0}[n\cdot M-m]\\
~~~~~~~~~~~~~~~~~~~~~-\mathds{B}[n] \cdot \mathbf{a}_{q_{(l,m)}} \big)^H \cdot \mathds{C}^{-1}_{{\textsf{z}}}\\
~~~~~~~~~~~~~~~~~~~~~~~~~~~~\cdot \big(\tilde{\mathbf{r}}_{N,0}[n\cdot M-m]-\mathds{B}[n] \cdot \mathbf{a}_{q_{(l,m)}}\big)\Big)\bigg)
\end{array}\right)}{\left(\begin{array}{l}
 \sum\limits_{u=0}^{(N_{\textsf{s}})^{ML_{\textsf{ch}}}-1} \mbox{exp}\bigg(-\sum\limits_{m=0}^{M-1}\Big(\big(\tilde{\mathbf{r}}_{N,0}[n\cdot M-m]\\
~~~~~~~~~~~~~~~~~~~~~~~~-\mathds{B}[n] \cdot \tilde{\mathbf{a}}_{\tilde{q}_{(u,m)},m} \big)^H \cdot \mathds{C}^{-1}_{{\textsf{z}}}\\
~~~~~~~~~~~~~~~~~~~~~~~~~~~~~~\cdot \big(\tilde{\mathbf{r}}_{N,0}[n\cdot M-m]-\mathds{B}[n] \cdot \tilde{\mathbf{a}}_{\tilde{q}_{(u,m)},m}\big)\Big)\bigg)
\end{array}\right)} \underset{H_1}{\overset{H_0}{\gtrless}} \lambda.  
\end{align*} 
Finally, taking the natural logarithm of both sides we obtain the following expression for the exact LRT:
\begin{align}
& \log_e \left(\frac{\left(\begin{array}{l}
 \sum\limits_{l=0}^{(N_{\textsf{s}})^{L_{\textsf{tot}}}-1} \mbox{exp}\bigg(-\sum\limits_{m=0}^{M-1}\Big(\big(\tilde{\mathbf{r}}_{N,0}[n\cdot M-m]\\
~~~~~~~~~~~~~~~~~~~~~~-\mathds{B}[n] \cdot \mathbf{a}_{q_{(l,m)}} \big)^H \cdot \mathds{C}^{-1}_{{\textsf{z}}}\\
~~~~~~~~~~~~~~~~~~~~~~~~~\cdot \big(\tilde{\mathbf{r}}_{N,0}[n\cdot M-m]-\mathds{B}[n] \cdot \mathbf{a}_{q_{(l,m)}}\big)\Big)\bigg)
\end{array}\right)}{\left(\begin{array}{l}
 \sum\limits_{u=0}^{(N_{\textsf{s}})^{ML_{\textsf{ch}}}-1} \mbox{exp}\bigg(-\sum\limits_{m=0}^{M-1}\Big(\big(\tilde{\mathbf{r}}_{N,0}[n\cdot M-m]\\
~~~~~~~~~~~~~~~~~~~~~~~~-\mathds{B}[n] \cdot \tilde{\mathbf{a}}_{\tilde{q}_{(u,m)},m} \big)^H \cdot \mathds{C}^{-1}_{{\textsf{z}}}\\
~~~~~~~~~~~~~~~~~~~~~~~~~~~~\cdot \big(\tilde{\mathbf{r}}_{N,0}[n\cdot M-m]-\mathds{B}[n] \cdot \tilde{\mathbf{a}}_{\tilde{q}_{(u,m)},m}\big)\Big)\bigg)
\end{array}\right)} \right) \underset{H_1}{\overset{H_0}{\gtrless}} \log_e\Big( \lambda \cdot( N_{\textsf{s}})^{MK }\Big). 
\label{Test}
\end{align}
\textbf{Computational Complexity:} The  computational  complexity  of  the  LRT was derived in
in terms of complex multiplications (CMs) and complex additions (CAs) in
Appendix \ref{app:Complexity_Analysis}. The total complexity of the LRT detector in Eq. (\ref{Test})~is:
\begin{ceqn}
\begin{align*}
&\mbox{CM}\big(\mbox{LRT}\big)=MK\cdot(N+K+1)\cdot\\
&~~~~~~~~~~~~~~~~~~~~~~~~~~~~~~~~~~ \big((N_{\textsf{s}})^{L_{\textsf{tot}}}+(N_{\textsf{s}})^{ML_{\textsf{ch}}}\big)+1\\
&\mbox{CA}\big(\mbox{LRT}\big)=M\cdot\big(KN+(K-1)\cdot(K+1)\big)\\
&~~~~~~~~~~~~~~~~~~~~~~~~~~~~~~~~~~~~~~~~~~~~~\cdot \big((N_{\textsf{s}})^{L_{\textsf{tot}}}+(N_{\textsf{s}})^{ML_{\textsf{ch}}}\big)~. 
\end{align*}
\end{ceqn}
We observe that the expression (\ref{Test}) is clearly very \emph{intensive computationally} and it also requires \emph{knowledge of the CIR coefficients}. In the next section we apply approximations to obtain a computationally practical expression which does not require apriori knowledge of the channel coefficients.


\section{The RALRT Frame Synchronization Algorithm: Reducing Complexity of the LRT}
\label{chap:SALRT_Algorithm_Derivation}

The main disadvantage of the optimal LRT in $(\ref{Test})$ is its high computational complexity which increases exponentially with the length of the SW. This follows as computation is carried out over $(N_{\textsf{s}})^{L_{\textsf{tot}}}-1$ elements. In the following we reduce the complexity of the LRT in
two steps: First, we replace the summation over the exponents with the difference between two minimizations. We refer to the resulting detector as the approximate LRT (ALRT).
Then, we propose a reduced complexity computation of the two minimizations, by reducing the search space for the minima. We refer to the resulting detection as the reduced approximate LRT (RALRT).
\subsection{The ALRT: Approximating the LRT as the Difference of Two Minima}
We begin the approximation of the difference of the logarithms of sums of exponents by recalling  the following relationship for real numbers $\{x_l\}_{l=1}^{L}$ \cite[Eq. (5)]{nielsen2016guaranteed}:
\begin{ceqn}
\begin{align}
&\max \{x_1,x_2,...,x_L\} \leq \log_e \left(\sum\limits_{l=1}^{L} e^{x_l}\right)\leq \max \{x_1,x_2,...,x_L\} + \log_e (L)  \label{Ineq} .
\end{align}
\end{ceqn} 
Using (\ref{Ineq}) we can write the following inequality for two sequence: $\{x_i\}_{i=1}^{L_0}$ and $\{y_i\}_{i=1}^{L_1}$
\begin{align}
&\max \{x_1,x_2,...,x_{L_0}\} - \max \{y_1,y_2,...,y_{L_1}\} - \log_e (L_1) ~~~~~~~~~~~~~~~~~~~~~~~~~~~~~~~;;\nonumber \\ \label{Sec}
&~~~~~~~~~~~~\leq \log_e \left(\sum\limits_{l_0=1}^{L_0} e^{x_{l_0}}\right) - \log_e \left(\sum\limits_{l_1=1}^{L_1} e^{y_{l_1}}\right) \nonumber\\
&~~~~~~~~~~~~~~~~~~~~~~~~~~~~\leq \max \{x_1,x_2,...,x_{L_0}\} - \max \{y_1,y_2,...,y_{L_1}\} + \log_e (L_0)  
\end{align}
If $L_0\gg  L_1$ then, after multiplying (\ref{Sec}) by $\frac{1}{L_0} $, the following approximation can be used for sufficiently large ${L_0}$:
\begin{align} \label{Aprox} 
&\frac{\log_e \left(\sum\limits_{l_0=1}^{L_0} e^{x_{l_0}}\right) - \log_e \left(\sum\limits_{l_1=1}^{L_1} e^{y_{l_1}}\right)}{L_0}\approx \frac{\max \{x_1,x_2,...,x_{L_0}\} - \max \{y_1,y_2,...,y_{L_1}\}}{L_0} 
\end{align} 
Applying (\ref{Aprox}) to (\ref{Test}) and letting $L_0=(N_{\textsf{s}})^{L_{\textsf{tot}}},~L_1=(N_{\textsf{s}})^{ML_{\textsf{ch}}}$, it follows that the assumption $L_0 \gg L_1$ is satisfied in practical scenarios, and after some manipulations (see  Appendix \ref{app:Derivations_Chapter_SALRT} for details) we arrive at  the approximate LRT (ALRT) detector:
\begin{align}
&\frac{1}{(N_{\textsf{s}})^{L_{\textsf{tot}}}}\cdot \Bigg( \min  \Bigg\{ \Tr \bigg\{ \Big( \big(\mathds{B}[n] \big)^{H}\cdot \mathds{C}^{-1}_{{\textsf{z}}} \cdot \mathds{B}[n] \Big) \cdot  \mathds{D}^{\textsf{(sw)}}_{l_1}\bigg\}   \nonumber \\
& ~~~~~~~~~~ -2\operatorname{Re}\Big( \sum\limits_{m=0}^{M-1}\big(\tilde{\mathbf{r}}_{N,0}[n \cdot M-m] \big)^H\cdot \mathds{C}^{-1}_{{\textsf{z}}}\cdot \mathds{B}[n] \cdot \tilde{\mathbf{a}}_{\tilde{q}_{(l_1,m)}}\Big) \Bigg\}_{l_1=0}^{(N_{\textsf{s}})^{ML_{\textsf{ch}}}-1}  \nonumber  \\
& -  \min  \Bigg\{\Tr \bigg\{ \Big( \big(\mathds{B}[n] \big)^{H}\cdot \mathds{C}^{-1}_{{\textsf{z}}} \cdot \mathds{B}[n] \Big) \cdot  \mathds{D}^{\textsf{(data)}}_{l_0}\bigg\}  \nonumber \\
& ~~~~~~~~~~ -2\operatorname{Re}\Big(\sum\limits_{m=0}^{M-1}\big(\tilde{\mathbf{r}}_{N,0}[n \cdot M-m] \big)^H\cdot \mathds{C}^{-1}_{{\textsf{z}}}\cdot \mathds{B}[n] \cdot \mathbf{a}_{q_{(l_0,m)}}\big) \Bigg\} _{l_0=0}^{(N_{\textsf{s}})^{L_{\textsf{tot}}}-1} \Bigg) \nonumber \\
&~~~~~~~~~~~~~~~~~~~~~~~~~~~~~~~~~~~~~~~~~~~~~~~~~\underset{H_1}{\overset{H_0}{\gtrless}} \bigg(\log_e\Big( \lambda \cdot( N_{\textsf{s}})^{MK }\Big)  \bigg)\cdot \frac{1}{(N_{\textsf{s}})^{L_{\textsf{tot}}}}~, \label{Symplfy_Trace_2}
\end{align}
where 
\begin{ceqn}
\begin{subequations}
\begin{align}
&\mathds{D}^{\textsf{(data)}}_{l_0}\triangleq \sum\limits_{m=0}^{M-1} \mathbf{a}_{q_{(l_0,m)}} \big( \mathbf{a}_{q_{(l_0,m)}}\big)^{H},~l_0 = 0,1,...,(N_{\textsf{s}})^{L_{\textsf{tot}}}-1~, \label{D_data_Matrix}\\
&\mathds{D}^{\textsf{(sw)}}_{l_1}\triangleq \sum\limits_{m=0}^{M-1} \tilde{\mathbf{a}}_{\tilde{q}_{(l_1,m)},m} \big( \tilde{\mathbf{a}}_{\tilde{q}_{(l_1,m)},m}\big)^{H},~l_1 = 0,1,...,(N_{\textsf{s}})^{ML_{\textsf{ch}}}-1~. \label{D_sw_Matrix}
\end{align}
\end{subequations}
\end{ceqn}
We note that for a known channel matrix $\mathds{B}[n]$, the expressions $\Tr \bigg\{ \Big( \big(\mathds{B}[n] \big)^{H}\cdot \mathds{C}^{-1}_{{\textsf{z}}} \cdot \mathds{B}[n] \Big) \cdot  \mathds{D}^{\textsf{(sw)}}_{l_1}\bigg\}_{l_1=0}^{(N_{\textsf{s}})^{ML_{\textsf{ch}}}-1}$ and $\Tr \bigg\{ \Big( \big(\mathds{B}[n] \big)^{H}\cdot \mathds{C}^{-1}_{{\textsf{z}}} \cdot \mathds{B}[n] \Big) \cdot  \mathds{D}^{\textsf{(data)}}_{l_0}\bigg\}_{l_0=0}^{(N_{\textsf{s}})^{L_{\textsf{tot}}}-1}$ can be a-priori calculated before frame synchronization begins, which  considerably reduces the computational complexity of the ALRT. 
\subsection{The RALRT: Reducing the Complexity of the Grid Searches}
\label{Reducing_Grid_Search}
In order to further reduce the complexity of  computing the minimization, we propose that prior to the application of the post-processing $\mathsf{P}\{\mathbf{r}_{L_{\textsf{tot}},0}[n]\}$ at the detector input, a simple hard decision detector be applied to $\mathbf{r}_{L_{\textsf{tot}},0}[n]$, to obtain the estimates:
\begin{align} \label{Hard_Decision_Detector_Data}
&\hat{\mathbf{s}}_{L_{\textsf{tot}},0}^{(\textsf{data})}[n]=\left( \begin{array}{l}
\hat{\mathbf{g}}^{(0)}_{L_{\textsf{tot}},0}[n]\\
\hat{\mathbf{g}}^{(1)}_{L_{\textsf{tot}},0}[n]\\
\vdots\\
\hat{\mathbf{g}}^{(M-1)}_{L_{\textsf{tot}},0}[n]
\end{array}\right)=\argmin_{\mathbf{b}_{l_0}^{(\textsf{data})} }\left\{\|\mathbf{r}_{L_{\textsf{tot}},0}[n]- \mathbf{b}_{l_0}^{(\textsf{data})} \|^2\right\},~l_0 = 0,1,...,(N_{\textsf{s}})^{L_{\textsf{tot}}}-1~,
\end{align}
where
\begin{align*}
&\mathbf{b}_{l_0}^{(\textsf{data})} \triangleq
\left( \begin{array}{l}
\mathbf{a}_{q_{(l_0,0)}}\\
\mathbf{a}_{q_{(l_0,1)}}\\
\vdots \\
~\\
~\\
\mathbf{a}_{q_{(l_0,M-1)}}\\ 
\end{array} \right) \in \mathcal{C}^{L_{\textsf{tot}}~\times~1},~l_0 = 0,1,...,(N_{\textsf{s}})^{L_{\textsf{tot}}}-1~,  \nonumber
\end{align*}
and the estimates
\begin{align} \label{Hard_Decision_Detector_SW}
&\hat{\mathbf{s}}_{L_{\textsf{tot}},0}^{(\textsf{sw})}[n]=\left( \begin{array}{l}
\hat{\tilde{\mathbf{g}}}^{(0)}_{L_{\textsf{tot}},0}[n]\\
\hat{\tilde{\mathbf{g}}}^{(1)}_{L_{\textsf{tot}},0}[n]\\
\vdots\\
\hat{\tilde{\mathbf{g}}}^{(M-1)}_{L_{\textsf{tot}},0}[n]
\end{array}\right)=\argmin_{\mathbf{b}_{l_1}^{(\textsf{sw})} }\left\{\|\mathbf{r}_{L_{\textsf{tot}},0}[n]- \mathbf{b}_{l_1}^{(\textsf{sw})} \|^2\right\},~l_1=0,1,...,(N_{\textsf{s}})^{ML_{\textsf{ch}}}-1~, 
\end{align}
where
\begin{align*}
&\mathbf{b}_{l_1}^{(\textsf{sw})} \triangleq
\left( \begin{array}{l}
\tilde{\mathbf{a}}_{\tilde{q}_{(l_1,0)},0}\\
\tilde{\mathbf{a}}_{\tilde{q}_{(l_1,1)},1}\\
\vdots \\
~\\
~\\
\tilde{\mathbf{a}}_{\tilde{q}_{(l_1,M-1)},M-1}\\ 
\end{array} \right) \in \mathcal{C}^{L_{\textsf{tot}}~\times~1},~l_1=0,1,...,(N_{\textsf{s}})^{ML_{\textsf{ch}}}-1~, \nonumber
\end{align*}
and we note that, by definition of $\mathbf{b}_{l_0}^{(\textsf{data})}$ and $\mathbf{b}_{l_1}^{(\textsf{sw})}$, all the elements corresponding to the data belong to $\mathcal{S}$,  the  constellation set for the transmitted symbols. The vectors $\big\{\mathbf{b}_{l_0}^{(\textsf{data})}\big\}_{l_0=0}^{(N_{\textsf{s}})^{L_{\textsf{tot}}}-1}$ represent all the possible combinations of $L_{\textsf{tot}}$ data symbols in an $L_{\textsf{tot}}$-length vector, and the vectors $\big\{\mathbf{b}_{l_1}^{(\textsf{sw})}\big\}_{l_1=0}^{(N_{\textsf{s}})^{ML_{\textsf{ch}}}-1}$ represent all the possible combinations of $ML_{\textsf{ch}}$ data symbols and the transmitted SW of length $L_{\textsf{sw}}$ in an $L_{\textsf{tot}}$-length vector ordered according to the transmission structure. Let  $\hat{\mathbf{s}}_{L_{\textsf{tot}},0}[n]$ represent the hard-decision at the receiver, where it is understood that subject to $H_0$,  $\hat{\mathbf{s}}_{L_{\textsf{tot}},0}[n]=\hat{\mathbf{s}}_{L_{\textsf{tot}},0}^{(\textsf{data})}[n]$, consists only of data symbols and subject to $H_1$,   $\hat{\mathbf{s}}_{L_{\textsf{tot}},0}[n]=\hat{\mathbf{s}}_{L_{\textsf{tot}},0}^{(\textsf{sw})}[n]$, consists of a combination of data symbols and SW symbols taken from $\mathbf{f}_{\textsf{sw}}$, with the appropriate ordering, while the sub vectors $\hat{\mathbf{g}}^{(m)}_{L_{\textsf{tot}},0}[n]$ and $\hat{\tilde{\mathbf{g}}}^{(m)}_{L_{\textsf{tot}},0}[n],~m=0,1,...,M-1,$ represent hard-decision of the sub vectors $\mathbf{g}^{(m)}_{L_{\textsf{tot}},0}[n]$ and $\tilde{\mathbf{g}}^{(m)}_{L_{\textsf{tot}},0}[n],~m=0,1,...,M-1$, defined in Eqs. (\ref{Data_block_m}) and (\ref{Sync_block_m}) respectively. Next, we define $e_{\textsf{r}_0},~e_{\textsf{r}_1},~0 \leq e_{\textsf{r}_0} \leq N,~0 \leq e_{\textsf{r}_1} \leq L_{\textsf{ch}},~e_{\textsf{r}_0},e_{\textsf{r}_1} \in \mathcal{N}$, with which we construct  the sets $\mathcal{Q}_0$,~$\mathcal{Q}_1$, to contain all the vectors $\mathbf{b}_{l_0}^{(\textsf{data})}$ and $\mathbf{b}_{l_1}^{(\textsf{sw})}$, respectively, that are composed of the sub-vectors $\mathbf{a}_{q_{(l_0,m)}}$ and $\tilde{\mathbf{a}}_{\tilde{q}_{(l_1,0)},m},~m=0,1,...,M-1$, respectively, with a maximum of $e_{\textsf{r}_0}$ or $e_{\textsf{r}_1}$ different coordinates from the hard-decision sub-vectors $\hat{\mathbf{g}}^{(m)}_{L_{\textsf{tot}},0}[n]$ and $\hat{\tilde{\mathbf{g}}}^{(m)}_{L_{\textsf{tot}},0}[n],~m=0,1,...,M-1$, respectively. In order to write mathematically the construction of the sets $\mathcal{Q}_0$, $\mathcal{Q}_1$, we define the indicator function $I(b),~I(b)=1~\mbox{if}~b=0~\mbox{and}~I(b)=0~\mbox{if}~b\neq0,~b\in \mathcal{C}$, and for a vector $\mathbf{v}=\big(v_1,v_2,\ldots,v_n\big)^T,~\mathbf{v}\in \mathcal{C}^{n~\times~1}$, let $I(\mathbf{v})=\big(I(v_1),I(v_2),\ldots,I(v_n)\big)^T$.
Accordingly, the sets $\mathcal{Q}_0$, $\mathcal{Q}_1$ are  mathematically defined as:
\begin{ceqn}
\begin{align}
&\mathbf{b}_{l_0}^{(\textsf{data})} \in \mathcal{Q}_0,~~\mbox{if}~~~\Big(I\big(\hat{\mathbf{g}}^{(m)}_{L_{\textsf{tot}},0}[n]-\mathbf{a}_{q_{(l_0,m)}}\big)\Big)^T \cdot I\big(\hat{\mathbf{g}}^{(m)}_{L_{\textsf{tot}},0}[n]-\mathbf{a}_{q_{(l_0,m)}}\big) \leq e_{\textsf{r}_0},~\forall m=0,1,...M-1,~\nonumber\\
&~~~~~~~~~l_0 = 0,1,...,(N_{\textsf{s}})^{L_{\textsf{tot}}}-1.  \label{Grid_Search_Data} \\
&\mathbf{b}_{l_1}^{(\textsf{sw})} \in \mathcal{Q}_1,~~\mbox{if}~~~\Big(I\big(\hat{\tilde{\mathbf{g}}}^{(m)}_{L_{\textsf{tot}},0}[n]-\tilde{\mathbf{a}}_{\tilde{q}_{(l_1,0)},m}\big)\Big)^T\cdot I\big(\hat{\tilde{\mathbf{g}}}^{(m)}_{L_{\textsf{tot}},0}[n]-\tilde{\mathbf{a}}_{\tilde{q}_{(l_1,0)},m}\big) \leq e_{\textsf{r}_1},~\forall m=0,1,...M-1,~\nonumber\\
&~~~~~~~~~l_1 = 0,1,...,(N_{\textsf{s}})^{ML_{\textsf{ch}}}-1.  \label{Grid_Search_sw}
\end{align}
\end{ceqn}
The two minimizations in $(\ref{Symplfy_Trace_2})$ are now evaluated using only the sequences in the sets $\mathcal{Q}_0$, $\mathcal{Q}_1$, whose sizes are determined by the parameters $e_{\textsf{r}_0}$ and $e_{\textsf{r}_1}$. Therefore, by decreasing $e_{\textsf{r}_0}$, $e_{\textsf{r}_1}$, we can decrease the number of sequences used in the computation of the ALRT, which reduces the computational complexity at the cost of a decrease in optimally of the ALRT detector stated in $(\ref{Symplfy_Trace_2})$. The sets of indexes $\mathcal{L}_0,~\mathcal{L}_1$ for the new grid search follow from the sets $\mathcal{Q}_0,~\mathcal{Q}_1$, respectively, as:
\begin{ceqn}
\begin{align}
&l_0 \in \mathcal{L}_0~ \mbox{iff}~ \mathbf{b}_{l_0}^{(\textsf{data})}  \in \mathcal{Q}_0,~~l_0 = 0,1,...,(N_{\textsf{s}})^{L_{\textsf{tot}}}-1 \label{Grid_Search_indexes_data}\\ 
&l_1 \in \mathcal{L}_1 ~ \mbox{iff}~ \mathbf{b}_{l_1}^{(\textsf{sw})}  \in \mathcal{Q}_1,~~l_1 = 0,1,...,(N_{\textsf{s}})^{ML_{\textsf{ch}}}-1 \label{Grid_Search_indexes_sw}.
\end{align}
\end{ceqn}
Evaluating the ALRT detector within the reduced grid search will be referred in the following to as the reduced approximate LRT (RALRT), which is stated as:
\begin{align}
&\frac{1}{(N_{\textsf{s}})^{L_{\textsf{tot}}}}\cdot \Bigg( \min  \Bigg\{ \Tr \bigg\{ \Big( \big(\mathds{B}[n] \big)^{H}\cdot \mathds{C}^{-1}_{{\textsf{z}}} \cdot \mathds{B}[n] \Big) \cdot  \mathds{D}^{\textsf{(sw)}}_{l_1}\bigg\}   \nonumber \\
& ~~~~~~~~~~~~~~~~~~~~~~~ -2\operatorname{Re}\Big( \sum\limits_{m=0}^{M-1}\big(\tilde{\mathbf{r}}_{N,0}[n \cdot M-m] \big)^H  \cdot \mathds{C}^{-1}_{{\textsf{z}}}\cdot \mathds{B}[n]\cdot \tilde{\mathbf{a}}_{\tilde{q}_{(l_1,m)}}\Big) \Bigg\}_{l_1 \in \mathcal{L}_1}  \nonumber  \\
& ~~~~~~~~~~~~~~~~~-  \min  \Bigg\{\Tr \bigg\{ \Big( \big(\mathds{B}[n] \big)^{H}\cdot \mathds{C}^{-1}_{{\textsf{z}}} \cdot \mathds{B}[n] \Big) \cdot  \mathds{D}^{\textsf{(data)}}_{l_0}\bigg\}  \nonumber \\
&~~~~~~~~~~~~~~~~~~~~~~~~~~~~~~~-2\operatorname{Re}\Big(\sum\limits_{m=0}^{M-1}\big(\tilde{\mathbf{r}}_{N,0}[n \cdot M-m] \big)^H \cdot \mathds{C}^{-1}_{{\textsf{z}}}\cdot \mathds{B}[n] \cdot \mathbf{a}_{q_{(l_0,m)}}\big) \Bigg\}_{l_0 \in \mathcal{L}_0} \Bigg) \nonumber \\
&~~~~~~~~~~~~~~~~~~~~~~~~~~~~~~~~~~~~~~~~~~~~~~~~~~~~~~~~~~~~~~~~~~~~\underset{H_1}{\overset{H_0}{\gtrless}}  \bigg(\log_e\Big( \lambda \cdot( N_{\textsf{s}})^{MK }\Big) \bigg)\cdot  \frac{1}{(N_{\textsf{s}})^{L_{\textsf{tot}}}}~. \label{RALRT}
\end{align}

\subsection{Computational Complexity of the ALRT and the RALRT}
\label{subsec:complexity_alrt_ralrt}
The computational complexities of the  ALRT and the RALRT were also derived in Appendix \ref{app:Complexity_Analysis} for the purpose of comparing the computational complexity of the different detectors. It is concluded that the total complexity of the ALRT detector in Eq. (\ref{Symplfy_Trace_2}) is:
\begin{ceqn}
\begin{align*}
&\mbox{CM}\big(\mbox{ALRT}\big)=\big(MK\cdot (N+1)+1\big)\cdot\big((N_{\textsf{s}})^{L_{\textsf{tot}}}+(N_{\textsf{s}})^{ML_{\textsf{ch}}}\big)+1\\
&\mbox{CA}\big(\mbox{ALRT}\big)=M\cdot \big((N-1)\cdot K+(K-1)\big)\cdot\big((N_{\textsf{s}})^{L_{\textsf{tot}}}+(N_{\textsf{s}})^{ML_{\textsf{ch}}}\big)+1.
\end{align*}
\end{ceqn}
For the RALRT detector in Eq. (\ref{RALRT}) the computational complexity is equal to:
\begin{align*}
&\mbox{CM}\big(\mbox{RALRT}\big)=\big(MK\cdot (N+1)+1\big)\cdot\big(|\mathcal{Q}_1|+|\mathcal{Q}_0|\big)+1\\
&\mbox{CA}\big(\mbox{RALRT}\big)=M\cdot \big((N-1)\cdot K+(K-1)\big)\cdot\big(|\mathcal{Q}_1|
+|\mathcal{Q}_0|\big)+1.
\end{align*}
In Section \ref{chap:Simulation} we show via numerical simulations  that the proposed complexity reduction approach of the RALRT can provide at least an order of magnitude reduction in complexity w.r.t. the 	ALRT without a noticeable loss in detection performance.
\label{pg:RALRT-end}
\section{Novel SALRT Frame Synchronization Algorithm: Lowering Complexity without Channel Knowledge}
\label{Sec:SALRT}
The major disadvantage of the RALRT \eqref{RALRT} as well as of  
the LRT (\ref{Test}) is that it requires knowledge of the CIR. However, as frame synchronization is typically the initial step in reception, it is not likely that the receiver has a reliable estimate of the CIR at the time in which it attempts to synchronize with the beginning of the frame. We note that we assume that the statistical properties of the noise vary much slower than the channel, hence they {\it do not change  throughout the time duration it takes to synchronize}. Consequently, the receiver is assumed to have a reliable estimate of the noise correlation prior to synchronization.  In the following we derive a suboptimal detector based on $(\ref{Test})$ with a significantly reduced computational complexity, which can be applied also to scenarios in which the matrix $\mathds{B}[n]$, which contains the CIR information needed to for comparing the LRT, is unknown. To overcome this difficulty,  we propose a channel estimation procedure which facilitates implementation of the LRT without an  apriori knowledge of the CIR.
\subsection{Blind Estimation of the Channel Matrix}
Next we address the lack of knowledge of the channel matrix $\mathds{B}[n]$ which is needed for computing the RALRT in (\ref{RALRT}). To overcome this issue, we propose to estimate the CIR coefficients. We note that as most of the signal parameters are not synchronized at this point, such an estimation will typically be very noisy, and therefore it is not expected to facilitate sufficient decoding performance. Nevertheless, the simulation study reported in Section \ref{chap:Simulation} demonstrates that this noisy estimation is completely sufficient for obtaining good LRT performance. For the $H_1$ hypothesis we propose to use the method developed in \cite[Eq. (16)]{morelli2000carrier}, for estimating the channel, based on the knowledge of pilot symbols at the receiver.  In order to estimate the CIR for computing the LRT, we  process $L_{\textsf{EQ}}$ received samples as depicted in Fig. \ref{Structure of Transmited Sw}, where $L_{\textsf{tot}}<L_{\textsf{EQ}}<L_{\textsf{d}}-ML_{\textsf{ch}}$, to obtain an estimate of $L_{\textsf{est}}=L_{\textsf{EQ}}-P_{\textsf{h}}\cdot L_{\textsf{ch}}$ information symbols via the blind equalization scheme proposed in \cite[Eq. (28)]{godard1980self}, \cite[Ch. 10.5-2]{proakis2001digital}. To that aim the receiver first collects $L_{\textsf{tot}}$ samples of the received signal $r[m]$, required for generating $\mathbf{r}_{L_{\textsf{tot}},0}[n]$, as depicted in Figure \ref{Structure of Transmited Sw}. Before computing the LRT, additional $L_{\textsf{EQ}}=J \cdot P_{\textsf{h}} ,~J=\xi \cdot ( L_{\textsf{ch}}+1) , ~\xi \geq 3,~\xi \in \mathcal{N} $ samples are collected to facilitate the application of the equalizer proposed in \cite[Eq. (28)]{godard1980self}, \cite[Ch. 10.5-2]{proakis2001digital}. Define next the following vectors:
\begin{ceqn}
\begin{align*}
&\mathbf{r}_{\textsf{eqz}}^{(i,l)}[n]\triangleq\left( \begin{array}{l}
r[n-i] \\
r[n-P_{\textsf{h}}-i] \\
r[n-2 \cdot P_{\textsf{h}}-i] \\
\vdots\\
r[n-l \cdot P_{\textsf{h}}-i]\\
\end{array} \right) \in \mathcal{C}^{(l+1)~\times~1},~i = 0,1,...,P_{\textsf{h}}-1 ,~l~\in \mathcal{N}. \\
&\mathbf{u}^{(i,k)}\triangleq\left( \begin{array}{l}
u_0^{(i,k)} \\
~\\
~\\
u_1^{(i,k)} \\
\\
\vdots\\
\\
\\
u_{L_{\textsf{ch}}}^{(i,k)}\\
\end{array} \right) \in \mathcal{C}^{(L_{\textsf{ch}}+1)~\times~1},~i = 0,1,...,P_{\textsf{h}}-1 ,  ~k = 0,1,...,J-(L_{\textsf{ch}}+1).
\end{align*}
\end{ceqn}
$\mathbf{u}^{(i,k)}$ represents the equalizer taps at iteration number $k$ for the input $\mathbf{r}_{\textsf{eqz}}^{(i,L_{\textsf{ch}})}[n]$, corresponding to the $i$'th time instant within the period of the channel  (which consists of $P_{\textsf{h}}$ time instants). Accordingly, we define the equalizer output after the $k$'th iteration as:
\begin{ceqn}
\begin{align} \label{Equalizer_d_Output}
&\hat{d}^{(i,k)}=\Big(\mathbf{u}^{(i,k)}\Big)^{T} \cdot \mathbf{r}_{\textsf{eqz}}^{(i,L_{\textsf{ch}})}[n-k \cdot P_{\textsf{h}}],~i = 0,1,...,P_{\textsf{h}}-1,  
~k = 0,1,...,J-(L_{\textsf{ch}}+2).
\end{align}
\end{ceqn}
For each time instant within the channel period $P_{\textsf{h}}$, the iterative update equation for the equalizer is given by:
\begin{ceqn}
\begin{align}
\mathbf{u}^{(i,k+1)}&=\mathbf{u}^{(i,k)}+\Delta_p \cdot \big(\mathbf{r}_{\textsf{eqz}}^{(i,L_{\textsf{ch}})}[n-k\cdot P_{\textsf{h}}]\big)^{*} \cdot \hat{d}^{(i,k)}\cdot (\gamma_{\textsf{s}}^2-|\hat{d}^{(i,k)}|^2), \nonumber\\
&i = 0,1,...,P_{\textsf{h}}-1,~k = 0,1,...,J-(L_{\textsf{ch}}+2),~\gamma_{\textsf{s}}^2 \triangleq \frac{\mathds{E}\big[\big|s[m]\big|^4\big]}{\mathds{E}\big[\big|s[m]\big|^2\big]}~,
\label{Equalizer_Finite_Impulse_Response}
\end{align}
\end{ceqn}
where $\Delta_p \in \mathcal{R}$ is the equalizer's step size, which is selected in order to ensure convergence for the relevant scenarios \cite{ungerboeck1972theory},  while attaining a sufficiently small mean squared error (MSE). The initial vector of equalizer taps for the iterative algorithm, $\mathbf{u}^{(i,0)}$, is set to $\mathbf{u}^{(i,0)}=\left( \begin{array}{llll}
u_0^{(i,0)}, & u_1^{(i,0)}, & \ldots, & u_{L_{\textsf{ch}}}^{(i,0)}\\
\end{array} \right)^T
=\left( \begin{array}{llll}
1, & 0, &  \ldots, & 0
\end{array} \right)^T,~i = 0,1,...,P_{\textsf{h}}-1$.
The iterative algorithm output is $\mathbf{u}^{(i,J-(L_{\textsf{ch}}+1))}$, which is then used for estimating the $L_{\textsf{est}}$ information symbols from the received $L_{\textsf{EQ}}$ samples via a hard decision rule:
\begin{ceqn}
\begin{align}
\hat{s}\big[n-k \cdot P_{\textsf{h}}-i\big]&=\argmin_{\rho \in \mathcal{S}}{\Big\| \Big(\mathbf{u}^{\big(i,J-(L_{\textsf{ch}}+1)\big)}\Big)^T\cdot \mathbf{r}_{\textsf{eqz}}^{(i,L_{\textsf{ch}})}[n-k \cdot P_{\textsf{h}}]- \rho \Big\|^2 },\nonumber\\
&i = 0,1,...,P_{\textsf{h}}-1,~k = 0,1,...,J-(L_{\textsf{ch}}+1)~.
\label{Symbol_Estimation}
\end{align}
\end{ceqn}
Next we define $\Psi \in \mathcal{N}$ as the smallest integer satisfying $\Psi\cdot P_{\textsf{h}} > L_{\textsf{ch}}+1$ and the matrices $\hat{\mathds{G}}^{(i)}[n] \in \mathcal{C}^{(\Omega+1)~\times~(L_{\textsf{ch}}+1)}~,~\Omega=\big(J-(L_{\textsf{ch}}+\Psi)\big)~,~i = 0,1,...,P_{\textsf{h}}-1$ as follows:
\begin{ceqn}
\begin{align*}
&\Big[\hat{\mathds{G}}^{(i)}[n]\Big]_{a_1+1,a_2+1}\triangleq \hat{s}\big[n-a_1 \cdot P_{\textsf{h}}-a_2-i\big],~a_1=0,1,...,\Omega,~a_2=0,1,...,L_{\textsf{ch}}.
\end{align*}
\end{ceqn}
With these quantities we estimate a single period of the CIR $\tilde{\mathbf{h}}[n,k]$ defined in (\ref{Vec_Def}), via a least sum of squared errors (LSSE) channel estimator \cite{crozier1991least}, \cite[Eq. (16)]{morelli2000carrier}, that is :
\begin{ceqn}
\begin{align}
&\hat{\tilde{\mathbf{h}}}[n,i]=\big((\hat{\mathds{G}}^{(i)}[n])^H \cdot \hat{\mathds{G}}^{(i)}[n]\big)^{-1} \cdot \big(\hat{\mathds{G}}^{(i)}[n]\big)^H \cdot \mathbf{r}_{\textsf{eqz}}^{(i,\Omega)}[n],~i = 0,1,...,P_{\textsf{h}}-1~. \label{CIR_Estimation}
\end{align}
\end{ceqn}
Lastly, using the periodicity of the CIR, $\hat{\tilde{\mathbf{h}}}[n,i]=\hat{\tilde{\mathbf{h}}}[n,i+P_{\textsf{h}}]$, the entire estimated channel matrix $\mathds{\hat{B}}[n]$ is constructed via Eq. (\ref{eqn:Construct_B}).
Lastly, applying the estimated CIR to the RALRT we obtain the suboptimal approximate LRT, denoted as SALRT, which is expressed as:
\begin{align}
&\frac{1}{(N_{\textsf{s}})^{L_{\textsf{tot}}}}\cdot\Bigg( \min  \Bigg\{ \Tr  \bigg\{ \Big( \big(\mathds{\hat{B}}[n] \big)^{H} \cdot \mathds{C}^{-1}_{{\textsf{z}}} \cdot \mathds{\hat{B}}[n] \Big) \cdot  \mathds{D}^{\textsf{(sw)}}_{l_1}\bigg\}    \nonumber \\
&~~~~~~~~~~~~~~~~~~~~~~~~~  -2\operatorname{Re}\Big( \sum\limits_{m=0}^{M-1}\Big(\tilde{\mathbf{r}}_{N,0}[n \cdot M-m] \Big)^H\cdot \mathds{C}^{-1}_{{\textsf{z}}}\cdot \mathds{\hat{B}}[n] \cdot \tilde{\mathbf{a}}_{\tilde{q}_{(l_1,m)},m}\Big) \Bigg\} _{l_1 \in \mathcal{L}_1}  \nonumber  \\
& ~~~~~~~~~~~~~~~~-  \min  \Bigg\{\Tr \bigg\{ \Big( \big(\mathds{\hat{B}}[n] \big)^{H}\cdot \mathds{C}^{-1}_{{\textsf{z}}} \cdot \mathds{\hat{B}}[n] \Big) \cdot  \mathds{D}^{\textsf{(data)}}_{l_0}\bigg\} 
\nonumber \\
&~~~~~~~~~~~~~~~~~~~~~~~~~~~~~~~~~ -2\operatorname{Re}\Big(\sum\limits_{m=0}^{M-1}\Big(\tilde{\mathbf{r}}_{N,0}[n \cdot M-m] \big)^H \cdot \mathds{C}^{-1}_{{\textsf{z}}}\cdot \mathds{\hat{B}}[n] \cdot \mathbf{a}_{q_{(l_0,m)}}\Big) \Bigg\} _{l_0 \in \mathcal{L}_0} \Bigg) \nonumber\\
&~~~~~~~~~~~~~~~~~~~~~~~~~~~~~~~~~~~~~~~~~~~~~~~~~~~~~~~~~~~~~~~~~~~~~\underset{H_1}{\overset{H_0}{\gtrless}} \bigg(\log_e\Big( \lambda \cdot( N_{\textsf{s}})^{MK }\Big)\bigg)\cdot \frac{1}{(N_{\textsf{s}})^{L_{\textsf{tot}}}} \label{Suboptimal_LRT_detector}
\end{align}
\subsection{Summary:The Steps of the SALRT Frame Synchronization Algorithm}
\textbf{Initialization:} The receiver obtains the following parameters:
\begin{itemize}
  \item $\mathcal{S}~$- Constellation set of transmitted symbols.
  \item $[f_0,f_1,..,f_{L_{\textsf{sw}}-1}]~$- Transmitted SW.
  \item $M~$- Number the SW sub-blocks.
  \item  $P_{\textsf{h}}~$- The period of CIR.
  \item  $P_{\textsf{z}}~$- The period of the ACGN.
  \item $L_{\textsf{ch}}~$- Length of channel memory.
  \item  $N~$- A common multiple of the discrete periods of the CIR and ACGN, $P_{\textsf{h}}$ and $P_{\textsf{z}}$, which satisfies   $k_1P_{\textsf{h}}=k_2P_{\textsf{z}}=N,~k_1,k_2\in \mathcal{N},~\mbox{with} ~N>L_{\textsf{ch}}$. We note that this parameter can be computed independently by the receiver. 
  \item $\mathds{C}_{{\textsf{z}}} \in \mathcal{R}^{K~\times~K} ~$- The covariance matrix of the ACGN, $K=N-L_{\textsf{ch}}$.
  \item $e_{\textsf{r}_0}$, $e_{\textsf{r}_1}~$- Control parameters for the cardinality of the grid search sets, $\mathcal{Q}_0$ and $\mathcal{Q}_1$.
\end{itemize}
\begin{MyRemark}
\label{rem:KnowledgePh}
In general, $N$ and $L_{\textsf{ch}}$ are sufficient for applying the algorithm. 
Knowledge of $P_{\textsf{h}}$ can reduce the complexity of the channel estimation step, as it facilitates the estimation of a single period of the CIR via \eqref{CIR_Estimation}, and then obtaining all required $N$ CIR values for constructing the matrix $\mathds{B}[n]$ by using the periodicity of the CIR. Without such knowledge, the estimator \eqref{CIR_Estimation} has to be applied $N$ times, where $N>P_{\textsf{h}}$.
\end{MyRemark}
\begin{MyRemark}
While $L_{\textsf{ch}}$ may be over estimated, the period of the model can be estimated by dedicated blind algorithms e.g, \cite{houcke2003blind}. If the channel is not periodic, then the algorithm is directly applicable without the need for the period estimation part.
\end{MyRemark}

\textbf{Steps of the SALRT Algorithm:} The steps of the proposed suboptimal approximate LRT frame synchronization detector are detailed below:
\begin{enumerate}
\item Construct the matrices $\mathds{D}^{\textsf{(data)}}_{l_0},~l_0 = 0,1,...,(N_{\textsf{s}})^{L_{\textsf{tot}}}-1$
and $\mathds{D}^{\textsf{(sw)}}_{l_1},~l_1 = 0,1,...,(N_{\textsf{s}})^{ML_{\textsf{ch}}}-1$, via Eqs.  (\ref{D_data_Matrix}) and (\ref{D_sw_Matrix}).
\item Collect $L_{\textsf{tot}}$ channel output samples to assemble
$\mathbf{r}_{L_{\textsf{tot}},0}[n]$, the incoming vector samples.
\item Estimate channel matrix $\mathds{B}[n]$:
  \begin{enumerate}[(a)]
  \item Collect additional $L_{\textsf{EQ}}$ channel output samples as depicted in Fig. \ref{Structure of Transmited Sw}, and compute the finite impulse response (FIR) of the channel equalizer, $\mathbf{u}^{(i,J-(L_{\textsf{ch}}+1))}$, via Eqs. (\ref{Equalizer_d_Output}) and  (\ref{Equalizer_Finite_Impulse_Response}).
  \item Using $\mathbf{u}^{(i,J-(L_{\textsf{ch}}+1))}$ estimate the $L_{\textsf{est}}=L_{\textsf{EQ}}-P_{\textsf{h}}\cdot L_{\textsf{ch}}$ transmitted symbols from the $L_{\textsf{EQ}}$ channel output samples via
  Eq. (\ref{Symbol_Estimation}).
  \item Estimate the CIR  vectors,  $\tilde{\mathbf{h}}[n,i],~i = 0,1,...,P_{\textsf{h}}-1$ via
  Eq. (\ref{CIR_Estimation}), and construct the estimated channel matrix, $\mathds{\hat{B}}[n]$ via Eq. \eqref{eqn:Construct_B}.
  \end{enumerate}
  \item Reducing the complexity of the grid search:
  \begin{enumerate}[(a)]
  \item From $\mathbf{r}_{L_{\textsf{tot}},0}[n]$, evaluate the hard decision estimates $\hat{\mathbf{s}}_{L_{\textsf{tot}},0}^{(\textsf{data})}[n]$ and $\hat{\mathbf{s}}_{L_{\textsf{tot}},0}^{(\textsf{sw})}[n]$, via Eqs. (\ref{Hard_Decision_Detector_Data}) and  (\ref{Hard_Decision_Detector_SW}).
  \item Compute the sets of indexes $\mathcal{L}_0,~\mathcal{L}_1$ for the new grid search via Eqs. (\ref{Grid_Search_Data}), (\ref{Grid_Search_sw}), (\ref{Grid_Search_indexes_data}) and (\ref{Grid_Search_indexes_sw}).
  \end{enumerate}
\item Apply post-processing to $\mathbf{r}_{L_{\textsf{tot}},0}[n]$ according to Eq. (\ref{Post_Processing}), and obtain $\mathsf{P}\{\mathbf{r}_{L_{\textsf{tot}},0}[n]\}=\mathbf{r}^{(\mathsf{P})}_{L_{\textsf{tot}},0}[n]$.
\item Evaluate the SALRT and choose between $H_0$ and $H_1$ via Eq. (\ref{Suboptimal_LRT_detector}).
\end{enumerate}
\subsection{Complexity of the SALRT Detector}
\label{Section:complexity_analysis}
The computational complexity of the proposed SALRT detector is detailed in terms of complex multiplications (CMs) and complex additions (CAs) of each step of the algorithm at Table \ref{table:Computational_Complexity_1} with the details of the derivations at Appendix \ref{app:Complexity_Analysis}.
Accordingly it follows that the total complexity of the SALRT detector in Eq. (\ref{Suboptimal_LRT_detector}) is equal to:
\begin{ceqn}
\begin{align*}
\mbox{CM}\big(\mbox{SALRT}\big)&= 
\big( \begin{array}{l}
MK\cdot (N+1)+1+N^3
\end{array}\big)
\cdot\big(|\mathcal{Q}_1|+|\mathcal{Q}_0|\big)+KN\cdot(K+N)+1\\
&~~~+P_{\textsf{h}} \cdot
 \big(J-(L_{\textsf{ch}}+1)\big)\cdot\big(2\cdot (L_{\textsf{ch}}+1)+3\big)\\
&~~~~+P_{\textsf{h}}\cdot \big(L_{\textsf{ch}}+2\big) \cdot
\big((\Omega+1)\cdot (L_{\textsf{ch}}+1)+(L_{\textsf{ch}}+1)^2\big)\\
&~~~~~+(L_{\textsf{EQ}}-P_{\textsf{h}} \cdot L_{\textsf{ch}})\cdot (N_{\textsf{s}}+L_{\textsf{ch}}+1)
+\big(c_1+c_2\big)\cdot L_{\textsf{tot}}
\end{align*}
\end{ceqn}
\begin{ceqn}
\begin{align*}
\mbox{CA}\big(\mbox{SALRT}\big)&=
\big( \begin{array}{l}
M\cdot \big(
(N-1)\cdot K+(K-1)\big)+(N-1)\cdot N^2+N
\end{array}\big)\cdot\big(|\mathcal{Q}_1|
+|\mathcal{Q}_0|\big)\\
&~~~~~+N\cdot (K-1)\cdot (K+N)+1+P_{\textsf{h}} \cdot
\big(J-(L_{\textsf{ch}}+1)\big)\cdot\big(2\cdot (L_{\textsf{ch}}+1)\big)\\
&~~~~~~~~+P_{\textsf{h}}\cdot
\big(\Omega \cdot (L_{\textsf{ch}}+1)\cdot (L_{\textsf{ch}}+2)+L_{\textsf{ch}}\cdot(L_{\textsf{ch}}+1)+(L_{\textsf{ch}})^3\big)\\
&~~~~~~~~~~+(L_{\textsf{EQ}}-P_{\textsf{h}} \cdot L_{\textsf{ch}})\cdot(N_{\textsf{s}}+L_{\textsf{ch}})+\big(c_1+c_2\big)\cdot(2L_{\textsf{tot}}-1)
\end{align*}
\end{ceqn}
As explained in Appendix \ref{app:Complexity_Analysis}, $c_1\ll (N_{\textsf{s}})^{L_{\textsf{tot}}}$ and $c_2\ll (N_{\textsf{s}})^{ML_{\textsf{ch}}},~c_1,c_2\in \mathcal{N}$, are constants that can be computed from the constellation set $\mathcal{S}$ used for transmission.

A very common detector for frame synchronization in channels with memory is based on the correlation metric \cite{chiani2010noncoherent} :
\begin{ceqn}
\begin{equation}
\mbox{Corr}\Big(\mathbf{r}^{(\mathsf{cor})}_{L_{\textsf{sw}},0}[n]\Big)=\Big\|\Big(\mathbf{r}^{(\mathsf{cor})}_{L_{\textsf{sw}},0}[n]\Big)^H\cdot \mathbf{f}_{\textsf{sw}}\Big\|^2\underset{H_1}{\overset{H_0}{\gtrless}} \lambda~,  \label{Correlation_Test}
\end{equation}
\end{ceqn}
where $\mathbf{r}^{(\mathsf{cor})}_{L_{\textsf{sw}},0}[n] \in \mathcal{C}^{L_{\textsf{sw}}~\times~1}$ is a vector of the received channel output without the pre-processing and the post-processing procedures described in Section \ref{chap:signal_Model_and_Problem_Formulation}.
The total complexity of the  correlator detector of Eq. (\ref{Correlation_Test}) is shown in Appendix \ref{app:Complexity_Analysis} to be: 
\begin{ceqn}
\begin{align*}
\mbox{CM}\big(\mbox{Corr}\big)=&~L_{\textsf{sw}}+1\\
\mbox{CA}\big(\mbox{Corr}\big)=&~L_{\textsf{sw}}-1~. 
\end{align*}
\end{ceqn}
We compare the computational complexity expressions of the different schemes considered focusing on the number of CMs, as a CM takes more systems resources to implement than a CA \cite{karatsuba1995complexity}. Comparing and $\mbox{CM}\big(\mbox{SALRT}\big)$ with $\mbox{CM}\big(\mbox{LRT}\big)$ (see Section \ref{subsec:LRT}),  $\mbox{CM}\big(\mbox{ALRT}\big)$, $\mbox{CM}\big(\mbox{RALRT}\big)$ (see Section \ref{subsec:complexity_alrt_ralrt}), and with $\mbox{CM}\big(\mbox{Corr}\big)$,  we observe that the the computational complexity of ALRT detector is a somewhat smaller than the computational complexity of the LRT detector, but the computational complexity of the RALRT detector is much smaller than that of LRT and of the ALRT detectors since $\big(|\mathcal{Q}_1|
+|\mathcal{Q}_0|\big) \ll \big((N_{\textsf{s}})^{L_{\textsf{tot}}}+(N_{\textsf{s}})^{ML_{\textsf{ch}}}\big)$. The computational complexity of the SALRT detector is higher than that of the RALRT detector since SALRT estimates the CIR prior to applying the RALRT detector. Nevertheless, this complexity is much smaller than that of the ALRT, and of the LRT detectors, since it does not depend on the quantity $\big((N_{\textsf{s}})^{L_{\textsf{tot}}}+(N_{\textsf{s}})^{ML_{\textsf{ch}}}\big)$, which is the dominant term in the computational complexity of the ALRT and LRT detectors.
Lastly, we note that the correlator detector has the lowest computational complexity among all the detectors considered above. 

 Table \ref{table:Computational_Complexity_Order}  summarizes the CM complexity order for the different detectors as a function of the length of the SW,  $L_{\textsf{sw}}$, and the number of blocks into which the SW is partitioned, $M$. We note that while for each detector, the exact number of CMs differs from the exact number of CAs, their asymptotics w.r.t. $L_{\textsf{sw}}$ and $M$ is the same. Accordingly, we refer to this asymptotics in Table \ref{table:Computational_Complexity_Order} as ``CM/CA complexity order".  Examining Table \ref{table:Computational_Complexity_Order}, we first observe that CM/CA complexities for the LRT and for the ALRT are exponentially dependent on $M$ and on $L_{\textsf{sw}}$. This is because these parameters determine the search space for these algorithms. The CM/CA complexities of the RALRT and of the SALRT are inversely proportional to $M$, which follows as when $M$ increases the dimensions of the operations used in these estimators decrease. The CM/CA complexities for the RALRT, the SALRT and the correlator are polynomial in $L_{\textsf{sw}}$, where for the correlator the CM/CA complexity increases linearly in $L_{\textsf{sw}}$, for the RALRT the CM/CA complexity increases as the square of $L_{\textsf{sw}}$ and for the SALRT the CM/CA complexity increase as the cubic power of $L_{\textsf{sw}}$. Thus, the approximation used for the RALRT decreases the CM/CA complexity order from exponential to polynomial, yet it is an order of magnitude larger than that of the correlator -- which evidently has a very low CM/CA complexity. The CM/CA complexity of the SALRT is an order of magnitude larger than that of the RALRT, which is due to the additional complexity required for channel estimation.
\renewcommand{\arraystretch}{1.6}%
\begin{table}[H]
\caption{Detectors computational complexity order }
\centering
\begin{tabular}{ |c|c|c| }
\hline
Detector&\multicolumn{1}{|c|}{CM/CA complexity order} \\
\hline
 LRT  & $\mathcal{O}\Big(\frac{2}{M}\cdot(N_{\textsf{s}})^{ML_{\textsf{ch}}}\cdot\big(L^2_{\textsf{sw}}\cdot(N_{\textsf{s}})^{L_{\textsf{sw}}}\big) \Big)$   \\
\hline
ALRT &$\mathcal{O}\Big(\frac{1}{M}\cdot(N_{\textsf{s}})^{ML_{\textsf{ch}}}\cdot(L^2_{\textsf{sw}}\cdot(N_{\textsf{s}})^{L_{\textsf{sw}}}) \Big)$ \\
\hline
RALRT&$\mathcal{O}\Big(\frac{1}{M}\cdot|\mathcal{Q}_0|\cdot L^2_{\textsf{sw}} \Big)$ \\
\hline
Correlator &$\mathcal{O}\big(L_{\textsf{sw}} \big)$   \\
\hline
SALRT &$\mathcal{O}\Big(\frac{1}{M}\cdot|\mathcal{Q}_0|\cdot(L^2_{\textsf{sw}}+L^3_{\textsf{sw}}) \Big)$\\
\hline
\end{tabular}
\label{table:Computational_Complexity_Order}
\noindent
\end{table}

\section{Numerical Examples and Discussion}
\label{chap:Simulation}
In this section we compare the performance of the three proposed algorithms for frame synchronization, the ALRT stated in Eq. (\ref{Symplfy_Trace_2}), the RALRT stated in Eq. (\ref{RALRT}) and the SALRT stated in Eq. (\ref{Suboptimal_LRT_detector}), together with the performance of the optimal LRT detector derived in Eq. (\ref{Test}), and with the widely adopted correlation metric detector \cite{chiani2010noncoherent}, stated in Eq. (\ref{Correlation_Test}).
The performance of hypothesis testing detectors are evaluated via their receiver operating characteristics (ROC) and the area under the curve (AUC) of the ROC, where a larger AUC is associated with a better detector \cite{fawcett2006introduction}.
The ROC plots depict the probability of successfully detecting a SW when it is present (i.e., given that the $H_1$ hypothesis is true), denoted with $p_{\textsf{d}}$, vs. the probability of declaring SW detection when no SW is present (i.e., when the $H_0$ hypothesis is true), denoted with $p_{\textsf{fa}}$.
We tested the performance of the algorithms for two different scenarios: The first scenario, referred to hereafter as 'Scenario 1', consists of a non-periodic CIR and WSS additive noise, and the second scenario, referred to hereafter as 'Scenario 2', consists of a periodic CIR and ACGN, as described in Section \ref{chap:signal_Model_and_Problem_Formulation}.
The transmitted symbols are selected from a BPSK constellation, $\mathcal{S}=\{-1,1\}$, and accordingly $N_{\textsf{s}}=2$. The parameters of the simulated channels are as follows:\\
\smallskip
\underline{Scenario 1}\\
We consider a slowly time varying frequency selective fading channel, such that the
CIR remains constant throughout the transmission of each frame. The CIR, $h[m,l]$, has complex values taken from \cite[Table 1]{kumar2019novel}:
\begin{ceqn}
\begin{align*}
&h[m,0]=1.05-0.82j,~h[m,1]=0.71+0.45j,~h[m,2]=0.63-0.72j,\\
&h[m,k]=0,~k\neq \{0,1,2\},~\forall ~m \in \mathcal{Z}.  
\end{align*}
\end{ceqn}
Accordingly, the channel period is $P_{\textsf{h}}=1$ with a finite memory $L_{\textsf{h}}=2$. 
The noise is a zero-mean baseband proper complex  additive WSS Gaussian process with a finite memory $L_{\textsf{z}}=2$ and with the following correlation function:
\begin{ceqn}
\begin{align*}
&c_{\textsf{z}}\big[m,l\big]=\sigma_z^2 \cdot \delta[l]+0.5\cdot\sigma_z^2 \cdot \delta[l+1]+0.5\cdot\sigma_z^2 \cdot \delta[l-1]+0.3\cdot\sigma_z^2 \cdot \delta[l+2]\\
&~~~~~~~~~~~~~~+0.3\cdot\sigma_z^2 \cdot \delta[l-2]
,~\sigma_z \in \mathcal{R}^{++}, ~m,l\in \mathcal{Z}, 
\end{align*}
\end{ceqn}
where $\sigma_z^2$ is selected to satisfy the target signal to noise ratio (SNR), which is derived explicitly later. Since the corresponding noise correlation matrix does not depend on $m$, its period is $P_{\textsf{z}}=1$. The memory of the channel is $L_{\textsf{ch}}=\mbox{max}\{L_{\textsf{z}},L_{\textsf{h}}\}=2$ and the period of the channel, denoted with
$N$, is the LCM of the discrete periods $P_{\textsf{h}}$ and $P_{\textsf{z}}$ that satisfies $N>L_{\textsf{ch}}$ that is $N\geq 3$.  We choose $N=8$ to facilitate a reasonable partition (not too fragmented) of the SW as will be described later in this section. The noise correlation matrix, $\mathds{C}_{{\textsf{z}}}$ is defined in Eq. (\ref{Noise_Matrix}), where $K=N-L_{\textsf{ch}}=8-2=6$.\\
\smallskip
\underline{Scenario 2}\\
We consider a time-varying frequency selective fading channel, such that the
channel is LPTV with a complex CIR, $h[m,l]$, whose values are taken from \cite[Table 1]{kumar2019novel}:
\begin{ceqn}
\begin{subequations}
\begin{align}
&h[2\cdot m_1,0]=1.05-0.82j,\nonumber\\
&h[2\cdot m_1,1]=0.71+0.45j,\nonumber\\
&h[2\cdot m_1,2]=0.63-0.72j, \nonumber\\
&h[2\cdot m_1+1,0]=0.53+0.62j,\nonumber\\
&h[2\cdot m_1+1,1]=0.41+0.37j,\nonumber\\
&h[2\cdot m_1+1,2]=0.20-0.34j,~\forall m_1 \in \mathcal{Z},\nonumber\\
&h[m_2,k_2]=0,~k_2\neq \{0,1,2\},~\forall ~m_2 \in \mathcal{Z}.\nonumber
\end{align}
\end{subequations}
\end{ceqn}
Accordingly the channel period is $P_{\textsf{h}}=2$ with a finite memory $L_{\textsf{h}}=2$.
The noise is a zero mean, baseband proper complex ACGN process with a period $P_{\textsf{z}}=8$ and a finite correlation memory of $L_{\textsf{z}}=2$, and is generated by filtering a memoryless proper complex WSCS Gaussian process with zero mean and variance :
\begin{align*}
\sigma_z^2[m]&=\tilde{\sigma}_z^2\cdot\Big(2+\cos\Big(\frac{2 \pi}{P_{\textsf{z}}}\cdot m\Big)\Big),~\tilde{\sigma}_z \in \mathcal{R}^{++},~\forall m \in \mathcal{Z}~, 
\end{align*}
where $\tilde{\sigma}_z^2$ is selected according to the target SNR.
The filter used for noise generation is a casual LTI filter, denoted $h_{\textsf{z}}[m]$, with an exponentially decaying CIR:
\begin{ceqn}
\begin{align*}
&h_{\textsf{z}}[0]=0.6,~h_{\textsf{z}}[1]=0.2,~h_{\textsf{z}}[2]=0.066,~h_{\textsf{z}}[k]=0~ \mbox{for}~k\neq \{0,1,2\},~ k \in \mathcal{Z}~,
\end{align*}
\end{ceqn}
resulting in $L_{\textsf{z}}=2$. The resulting noise correlation function is given by:
\begin{ceqn}
\begin{align*}
c_{\textsf{z}}\big[m,l\big]&=\sum\limits_{k=0}^{L_{\textsf{z}}} h_{\textsf{z}}[k]\cdot \sigma_z[m-k]\cdot h_{\textsf{z}}[l+k]\\
&= h_{\textsf{z}}[0]\cdot \sigma_z[m]\cdot h_{\textsf{z}}[l]+h_{\textsf{z}}[1]\cdot \sigma_z[m-1]\cdot h_{\textsf{z}}[l+1]+h_{\textsf{z}}[2]\cdot \sigma_z[m-2]\cdot h_{\textsf{z}}[l+2]\\
&= 0.6\cdot \sigma_z[m]\cdot h_{\textsf{z}}[l]+0.4\cdot \sigma_z[m-1]\cdot h_{\textsf{z}}[l+1]+0.2\cdot \sigma_z[m-2]\cdot h_{\textsf{z}}[l+2], ~m,l\in \mathcal{Z}.
\end{align*}
\end{ceqn}
The memory of the channel $L_{\textsf{ch}}=\mbox{max}\{L_{\textsf{z}},L_{\textsf{h}}\}=2$ and the period of the channel, denoted with
$N$, is the LCM of the discrete periods $P_{\textsf{h}}$ and $P_{\textsf{z}}$ that satisfies $N>L_{\textsf{ch}}$, and is set to $N=8$.\\
\smallskip
\underline{Synchronization Sequence and Equalizer Parameters}\\
For both scenarios the parameters defining the structure of the transmitted frame, corresponding to Fig. \ref{Structure of Transmited Sw}, are: $N=8,~M=2,~L_{\textsf{tot}}=MN=16,~\mbox{and}~L_{\textsf{sw}}=MK=12$. The parameters for the channel estimation and the channel equalizer described in Section \ref{Sec:SALRT} are: $\gamma_{\textsf{s}}^2=1,~ L_{\textsf{EQ}}=300,~ e_{\textsf{r}_0}=3,~e_{\textsf{r}_1}=2$, accordingly $|Q_0| = 8649~\mbox{and}~|Q_1| = 16$. For Scenario 1 we set $ J=300,~\Omega = 4$ and for Scenario 2 we set $J=150,~\Omega = 2$. \\
\smallskip
\underline{Identifying the Optimal Synchronization Sequence}\\
We first carried out a simulation study for identifying the synchronization sequence that results in the best SALRT algorithm performance, that is, finding the synchronization sequence with the largest AUC for the SALRT ROC. 
The evaluation of the PDF and of the CDF of the AUC was carried out at $SNR=-5~[\mbox{dB}]$, where for each SW, the AUC value was obtained by numerically integrating over the ROC, which was evaluated by averaging over $3000$ Monte-Carlo simulations for each point in the ROC. \label{Pg:AUC-Setup}

For the purpose of the simulation study, the $SNR$ is defined as:
\begin{ceqn}
\begin{align} \label{SNR Eqn}
&SNR=\frac{ \sigma_s^2 \cdot \Tr \Big\{\big( \mathds{A}_{{L_{\textsf{tot}}},0}[n] \big)^H \cdot \mathds{A}_{{L_{\textsf{tot}}},0}[n]\Big\}}{\Tr \bigg\{\mathds{E} \Big\{  \mathbf{z}_{L_{\textsf{tot}},0}[n]\cdot \big(\mathbf{z}_{L_{\textsf{tot}},0}[n]\big)^H \Big\}\bigg\}}~. 
\end{align}
\end{ceqn}
A detailed derivation of the $SNR$ expression in Eq. (\ref{SNR Eqn}) is presented in Appendix  \ref{app:SNR_Expression_Evaluation}. For the first simulation scenario, the $SNR$ expression of Eq. (\ref{SNR Eqn}) specializes to:
\begin{ceqn}
\begin{align}
&SNR=\frac{ \sigma_s^2}{\sigma_z^2 \cdot L_{\textsf{tot}}}\cdot \Tr \Big\{\big( \mathds{A}_{{L_{\textsf{tot}}},0}[n] \big)^H \cdot \mathds{A}_{{L_{\textsf{tot}}},0}[n]\Big\}     
\end{align}
\end{ceqn}
The search for the synchronization sequence with the largest AUC was carried out by evaluating the ROC for each synchronization sequence within the set of $|\mathcal{S}|^{L_{\textsf{sw}}}=2^{12}=4096$ different synchronization sequences. The synchronization sequence with the largest AUC of the SALRT ROC for Scenario 1 was identified as:
\begin{ceqn}
    \begin{align*}
    &\mathbf{f}^T_{\textsf{sw}_1}=[-1,1,-1,1,-1,1,1,-1,1,-1,1,-1]~,
    \end{align*}
\end{ceqn}
while for Scenario 2 the largest AUC is obtained with :
\begin{ceqn}
    \begin{align*}
    &\mathbf{f}^T_{\textsf{sw}_2}=[1,1,1,1,1,1,1,1,1,1,1,1]~.~~~~~~~~~~~~~~~~~~
    \end{align*}
\end{ceqn}
\label{sim-proporty-pg}
We observe that in the presence of CIR memory and noise correlation, the optimal SW does not have the property of low correlation. This indeed demonstrates that when the channel has memory and/or the noise is correlated, new approaches, different from the correlator, need to be applied to achieve high detection performance.

\begin{figure}
~~~~~~~~~~~~~~~~~~~~~~~~~~~~~~\begin{subfigure}[t]{0.5\textwidth}
  \centering
  \includegraphics[width=90mm, height=80mm]{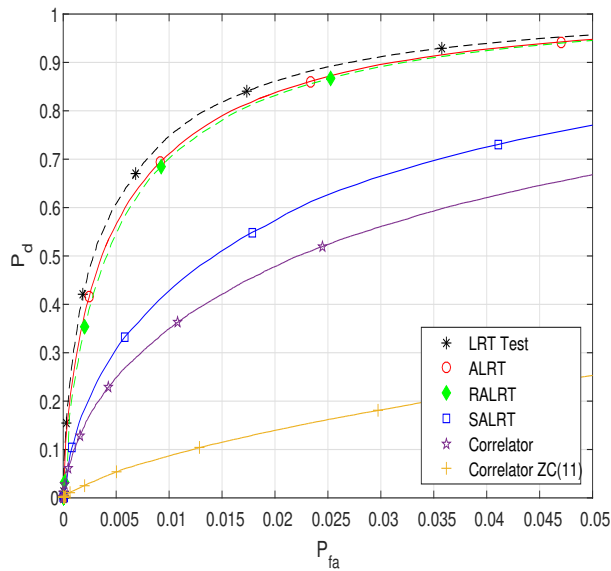}
  \captionsetup{justification=centering}
  \caption{Scenario 1, $SNR=-5$ [dB]}
  \label{fig:sfig1_stationary_minus_5_dB}
\end{subfigure} \\%
\begin{subfigure}[t]{0.5\textwidth}
  \centering
  \includegraphics[width=90mm, height=80mm]{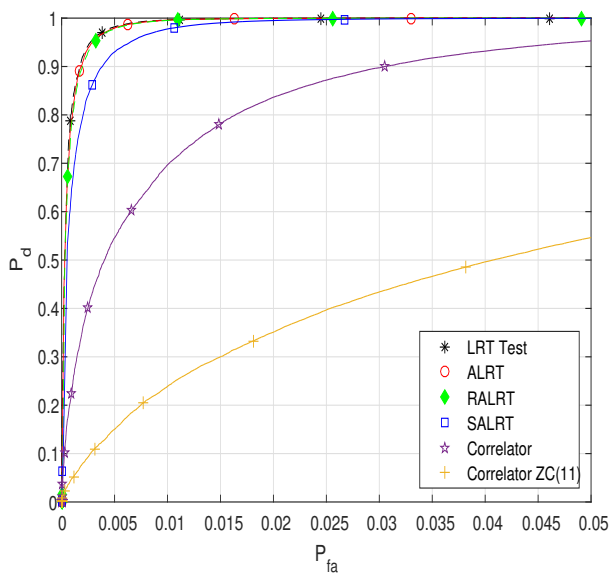}
  \captionsetup{justification=centering}
  \caption{Scenario 1, $SNR=0$ [dB]}
  \label{fig:sfig2_stationary_0_dB}
\end{subfigure}
\begin{subfigure}[t]{0.5\textwidth}
  \centering
  \includegraphics[width=90mm, height=80mm]{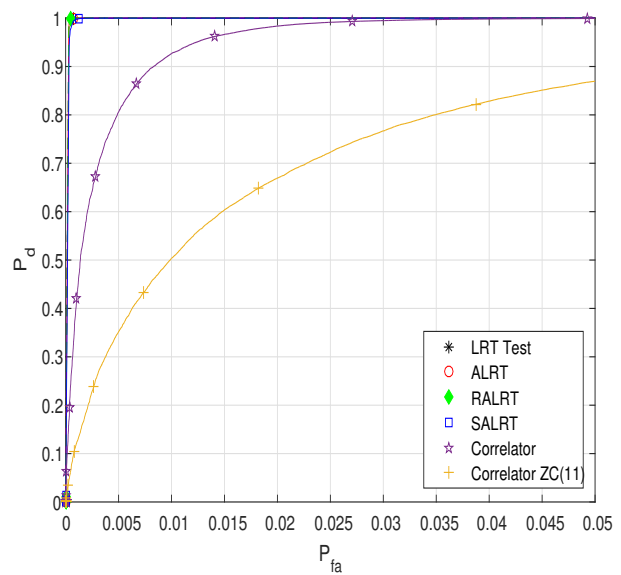}
  \captionsetup{justification=centering}
  \caption{Scenario 1, $SNR=5$ [dB]}
  \label{fig:sfig3_stationary_5_dB}
\end{subfigure}
\captionsetup{justification=raggedright}
\caption[ROC performance for the detectors]{ROC performance for the different detectors: LRT with channel knowledge, ALRT with channel knowledge, RALRT with channel knowledge, SALRT without channel knowledge, and correlator without channel knowledge for Scenario 1 at $SNRs=-5,0,5$ [dB], for the optimal synchronization sequence $\mathbf{f}^T_{\textsf{sw}_1}=[-1,1,-1,1,-1,1,1,-1,1,-1,1,-1]$. The ROC for the correlator with the ZC(11) sequence and without channel knowledge is also included. \label{ROC:stationary}}
\end{figure}

\begin{figure}
~~~~~~~~~~~~~~~~~~~~~~~~~~~~~~\begin{subfigure}[t]{0.5\textwidth}
  \centering
  \includegraphics[width=90mm, height=80mm]{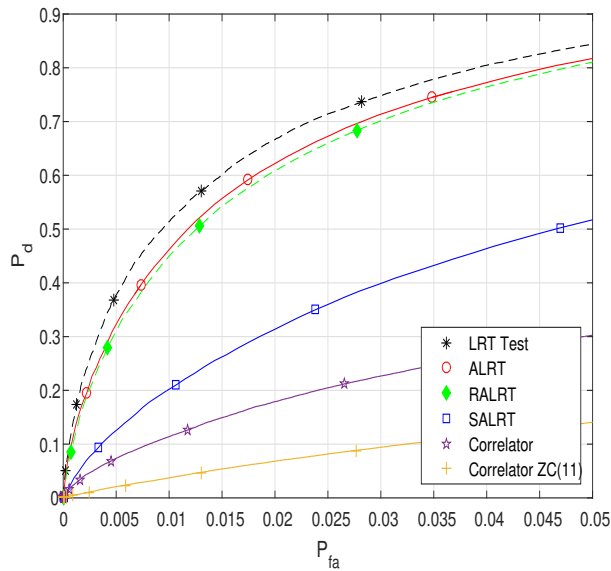}
  \captionsetup{justification=centering}
  \caption{Scenario 2, $SNR=-5$ [dB]}
  \label{fig:sfig1_periodic_minus_5_dB}
\end{subfigure} \\%
\begin{subfigure}[t]{0.5\textwidth}
  \centering
  \includegraphics[width=90mm, height=80mm]{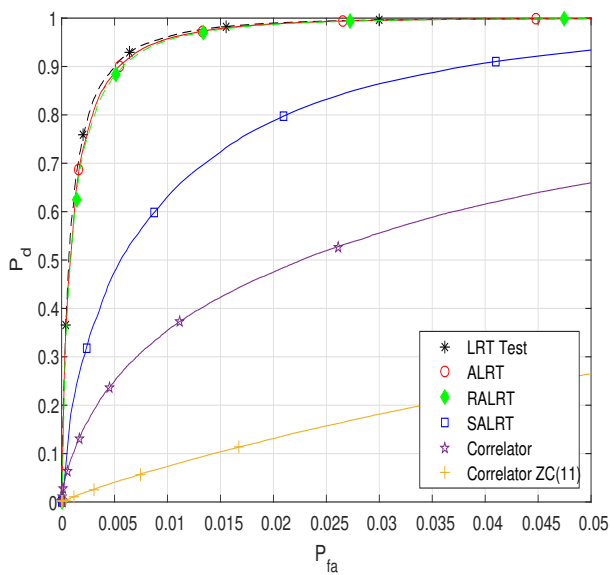}
  \captionsetup{justification=centering}
  \caption{Scenario 2, $SNR=0$ [dB]}
  \label{fig:sfig2_periodic_0_dB}
\end{subfigure}
\begin{subfigure}[t]{0.5\textwidth}
  \centering
  \includegraphics[width=90mm, height=80mm]{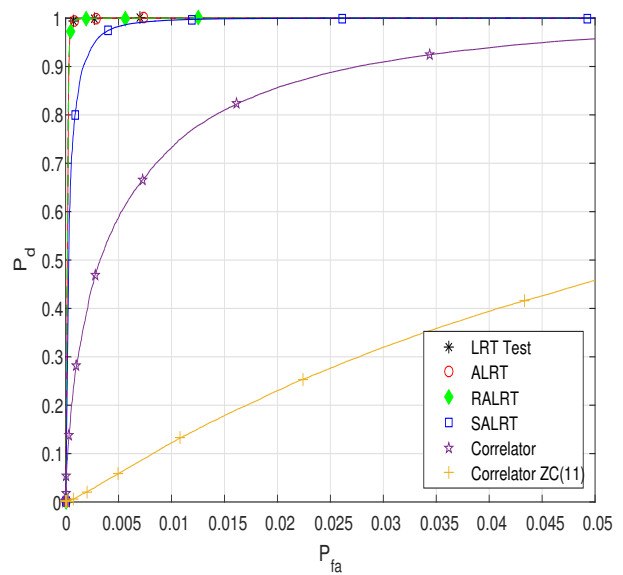}
  \captionsetup{justification=centering}
  \caption{Scenario 2, $SNR=5$ [dB]}
  \label{fig:sfig3_periodic_5_dB}
\end{subfigure}
\captionsetup{justification=raggedright}
\caption[ROC performance for the detectors]{ROC performance for the different detectors: LRT with channel knowledge, ALRT with channel knowledge, RALRT with channel knowledge, SALRT without channel knowledge, and correlator without channel knowledge for Scenario 2 at $SNRs=-5,0,5$ [dB], for the optimal synchronization sequence $\mathbf{f}^T_{\textsf{sw}_2}=[1,1,1,1,1,1,1,1,1,1,1,1]$.  The ROC for the correlator with the ZC(11) sequence and without channel knowledge is also included. \label{ROC:Periodic}}
\end{figure}

The ROC curves for the best synchronization sequence for Scenario 1 is
depicted in Figs. \ref{fig:sfig1_stationary_minus_5_dB}, \ref{fig:sfig2_stationary_0_dB}, \ref{fig:sfig3_stationary_5_dB}, and for Scenario 2 in  Figs. \ref{fig:sfig1_periodic_minus_5_dB}, \ref{fig:sfig2_periodic_0_dB}, \ref{fig:sfig3_periodic_5_dB}. For each scenario, the ROC curves obtained with 200000 Monte Carlo simulations for $SNR=-5~[\mbox{dB}],~SNR=0~[\mbox{dB}],~SNR=5~[\mbox{dB}]$. The PDF and CDF for the AUC of the SALRT ROC, computed over all possible synchronization sequences are presented in Fig. \ref{PDF_CDF} for Scenario 1 and in Fig. \ref{PDF_CDF_complex_2} for the Scenario 2. 

\subsection{Performance Evaluation}
Simulation results for Scenario 1 and 2 are depicted in Figs. \ref{fig:sfig1_stationary_minus_5_dB}, \ref{fig:sfig2_stationary_0_dB}, \ref{fig:sfig3_stationary_5_dB}, and Figs. \ref{fig:sfig1_periodic_minus_5_dB}, \ref{fig:sfig2_periodic_0_dB}, \ref{fig:sfig3_periodic_5_dB} respectively. In each figure we also include that ROC for the correlator evaluated with the Zadoff-Chu sequence with parameters $k=11$ and $L=12$, see \cite[Eq. (2.22)]{FazelMultiCarrierBook}, to demonstrate the impact of the low-correlation property on performance. This Zadoff-Chu sequence is referred to in the following as `ZC(11)'. We observe that the performance of the ALRT is  close to that of the optimal LRT detector
in both scenarios for $SNRs=0$ [dB] and $SNRs=5$ [dB]. However, at the lowest $SNR$, of $-5$ [dB], the suboptimality of the ALRT becomes evident implying that the log-sum approximation if not useful at that values. The performance of the ALRT and of the RALRT detectors (when the channel is known) are practically indistinguishable, although the RALRT has a smaller computational complexity. The SALRT has better performance than the correlator detector, and, as expected, worse performance than the ALRT and RALRT detectors. It is noted that an advantage of the SALRT and the RALRT detectors over the LRT and the ALRT detector is their facilitating a controlled trade-off between computational complexity and performance via selection of the search grid sizes, a property that does not exist in the ALRT and the LRT detectors. Recall also that the ALRT and the RALRT detectors {\em require knowledge of the channel coefficients}, while the SALRT detector does not, which further highlights the advantage of SALRT over the LRT, the ARLT, and the RALRT detectors.
 We also observe that the correlator with the ZC(11) sequence is considerably inferior to the SALRT with the optimal SW and also considerably inferior to the correlator with the optimal SW. This clearly indicates that low-correlation sequences are not good synchronization sequences when the CIR has memory and/or the noise is correlated.\label{pg:Corr_ZC}

The computational complexities for the parameters used in Scenario 1 and in Scenario 2 are summarized in Table \ref{table:Computational_Complexity_comparison_scenario_1} (see Section \ref{Section:complexity_analysis} for the analysis).
\renewcommand{\arraystretch}{1.2}%
\begin{table}[H]
\caption{Detectors computational complexity for Scenario 1 and 2 examples }
\centering
\begin{tabular}{ |c|c|l|l| }
\hline
Scenario&Detector&\multicolumn{1}{|c|}{CM $\big[10^6\big]$} &\multicolumn{1}{c|}{CA $\big[10^6\big]$} \\
\hline
\multirow{5}{*}{1} & LRT  & 11.8 & 10.9  \\
 \cline{2-4}
&ALRT &7.15 & 6.16\\
 \cline{2-4}
&RALRT&0.94 &0.81 \\
 \cline{2-4}
&Correlator   &$1.3\cdot 10^{-5}$ & $1.1\cdot 10^{-5}$  \\
 \cline{2-4}
&SALRT$^*$ &5.38&4.77\\
\hline
\multirow{5}{*}{2} &   LRT  & 11.8 & 10.9  \\
 \cline{2-4}
&ALRT &7.15 & 6.16\\
 \cline{2-4}
&RALRT&0.94 &0.81 \\
 \cline{2-4}
&Correlator   &$1.3\cdot 10^{-5}$ & $1.1\cdot 10^{-5}$  \\
 \cline{2-4}
&SALRT$^*$ &5.38&4.77\\
\hline
\end{tabular}
\label{table:Computational_Complexity_comparison_scenario_1}

\noindent
$(*)$ For the BPSK constellation set, $c_1=1$ and $c_2=1$.  
\end{table}
\smallskip
Observe that for both scenarios the CM and CA are essentially the same (the least significant digits differ between the values for both scenarios - which is not shown in the table). This is because the dimensions of the operations are mostly affected by the memory of the channel which is the same in both scenarios. For the LRT (known CIR), the CM  is approximately $11.8\cdot 10^6$, for the ALRT (known CIR) we achieve a relatively small decrease in CM complexity to approximately $7.15\cdot 10^6$, yet for the RALRT (known CIR) CM complexity is significantly reduced to approximately $10^6$. When the CIR is unknown, the applicable detector is the SALRT, whose CM complexity is approximately $5.4 \cdot 10^6$. 
The correlator has a very small complexity however its performance are considerably inferior to that of the SALRT in both scenarios. Comparing the SALRT and the RALRT we observe that the computational complexity cost for handling an unknown CIR is an increase by approximately $5$ times in the computational complexity, but it is still half the complexity of the LRT.
In our opinion the SALRT constitutes a good approach for achieving a tradeoff between performance and complexity in FS.

\begin{figure}[t]
\begin{subfigure}[t]{0.5\textwidth}
  \centering
  \includegraphics[width=90mm, height=80mm]{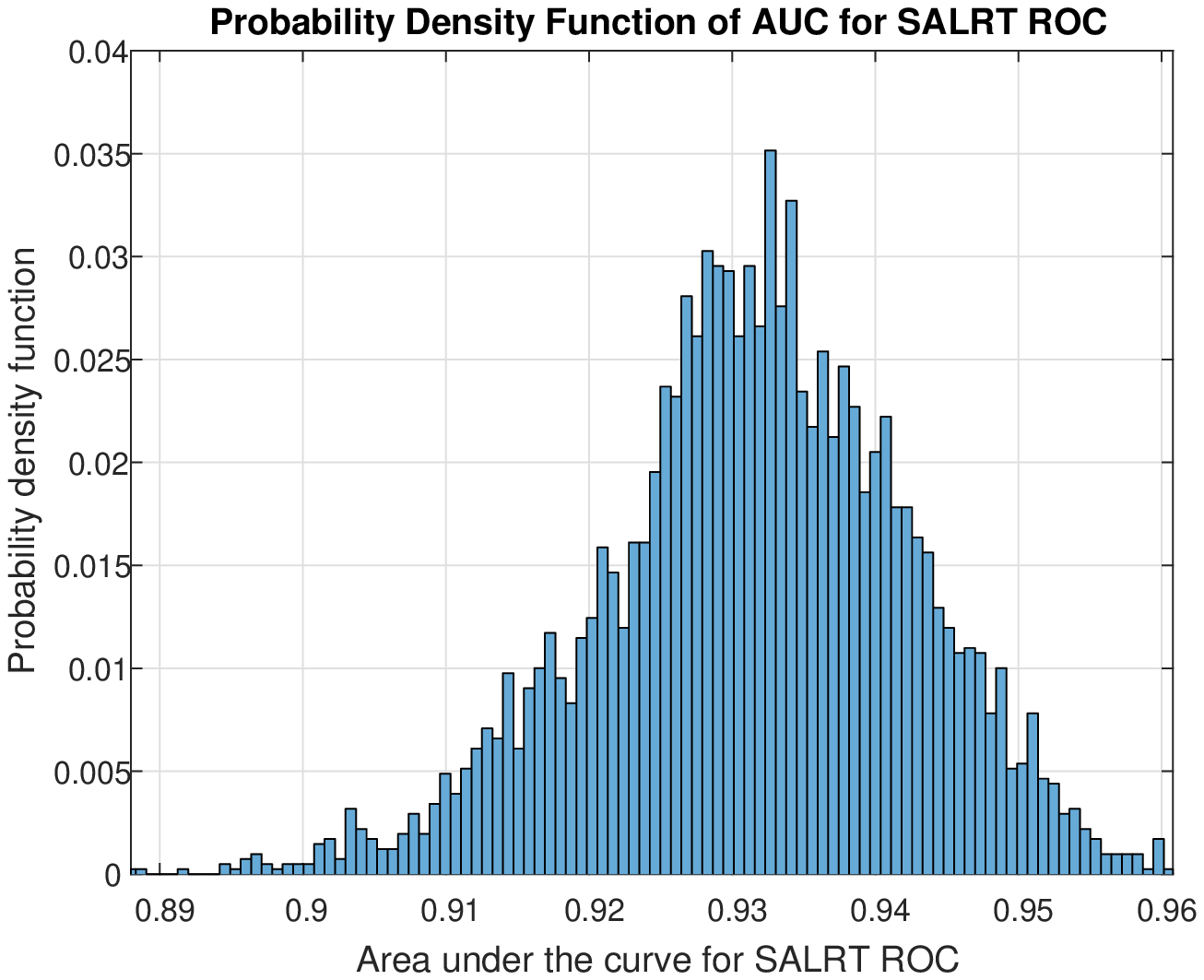}
  \captionsetup{justification=centering}
  \caption{Probability Density Function}
  \label{fig:PDF}
\end{subfigure} %
\begin{subfigure}[t]{0.5\textwidth}
  \centering
  \includegraphics[width=90mm, height=80mm]{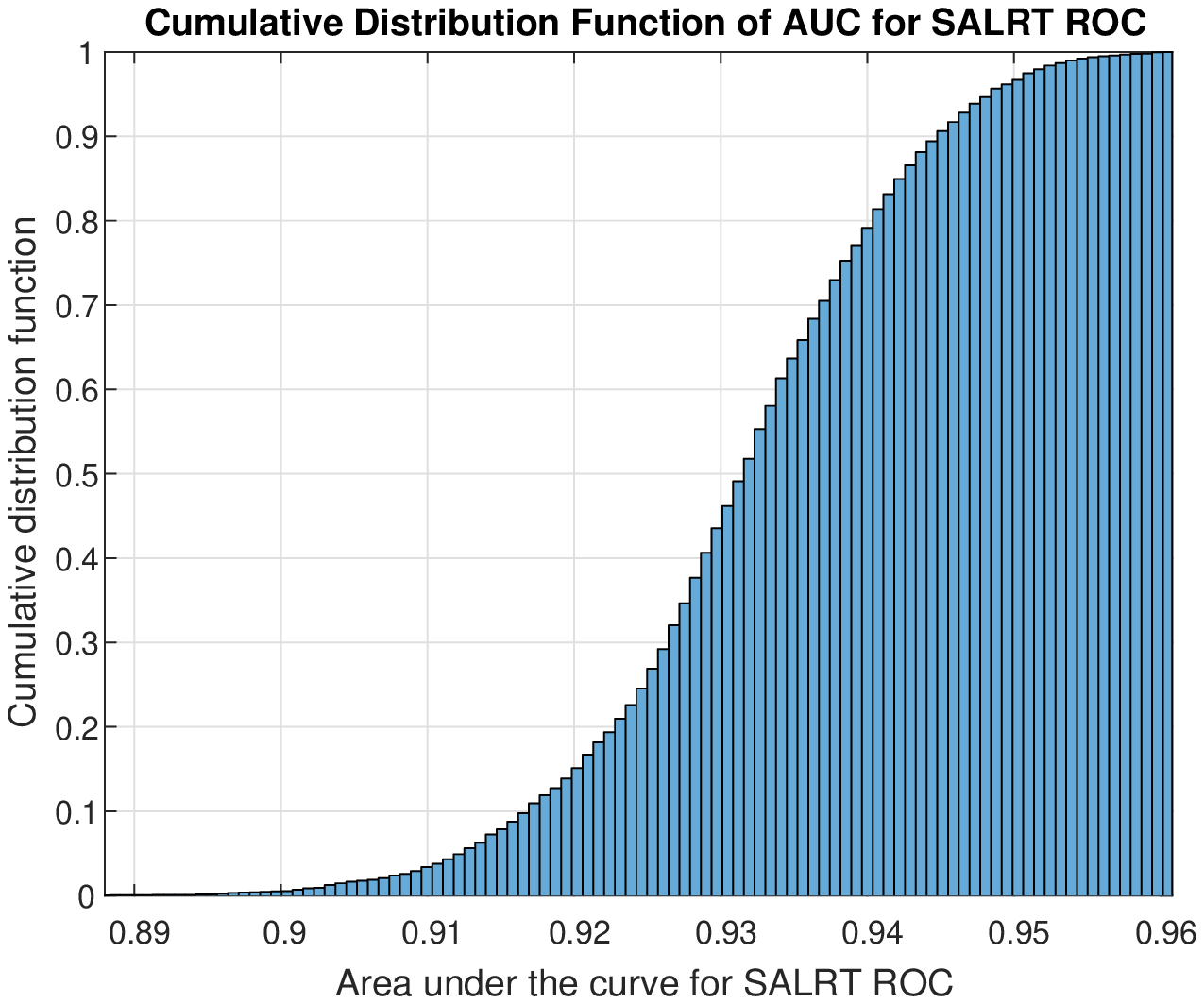}
  \captionsetup{justification=centering}
  \caption{Cumulative Distribution Function}
  \label{fig:CDF}
\end{subfigure}
\captionsetup{justification=raggedright}
\caption[PDF and CDF for AUC of SALRT ROC]{Probability density function (a) and cumulative distribution function (b) of the AUC for the SALRT ROC for Scenario 1 for all possible synchronization sequences at $SNR=-5$ [dB]. \label{PDF_CDF}}
\end{figure}
\begin{figure}[t]
\begin{subfigure}[t]{0.5\textwidth}
  \centering
  \includegraphics[width=90mm, height=80mm]{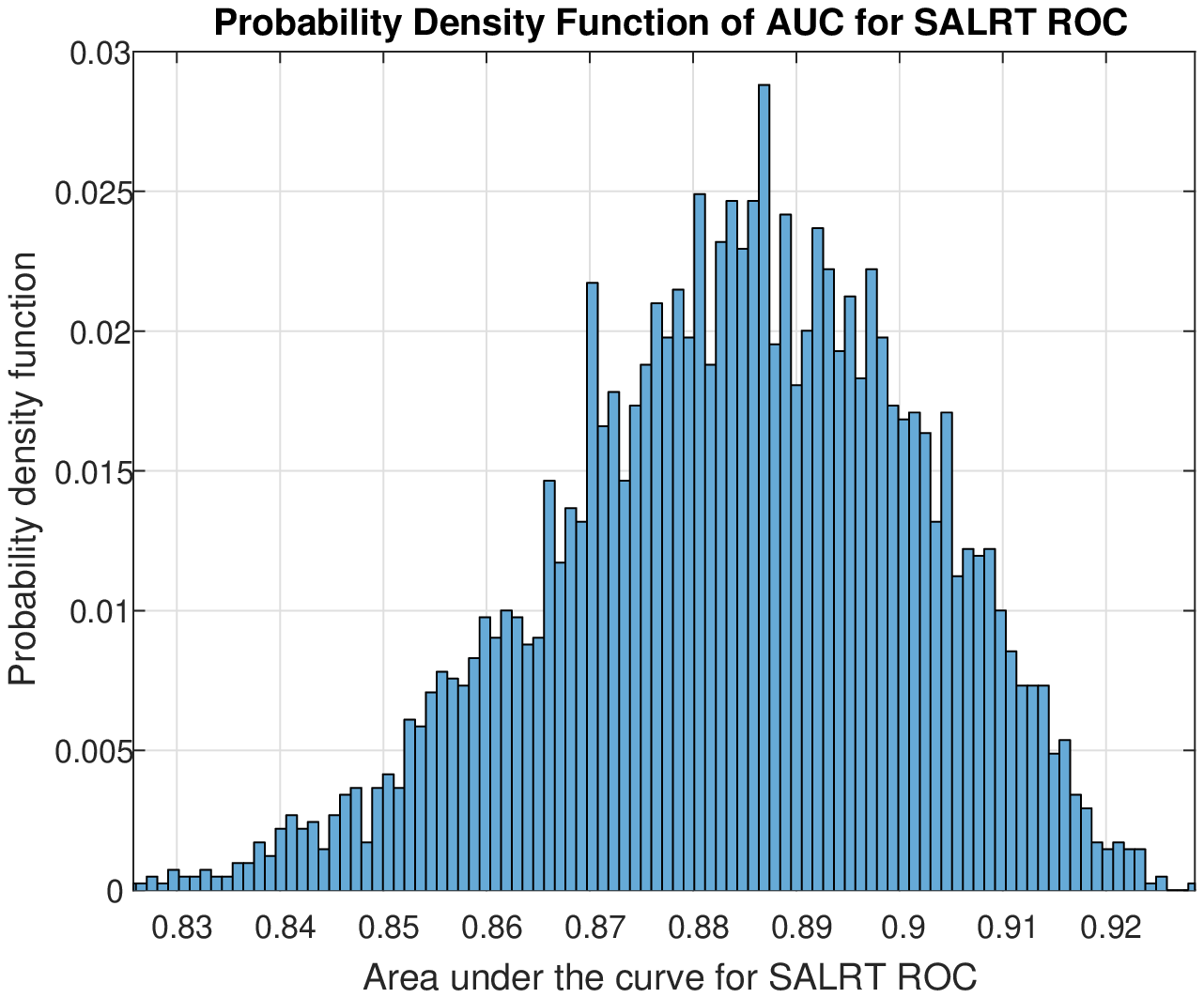}
  \captionsetup{justification=centering}
  \caption{Probability Density Function}
  \label{fig:PDF_complex_2}
\end{subfigure} %
\begin{subfigure}[t]{0.5\textwidth}
  \centering
  \includegraphics[width=90mm, height=80mm]{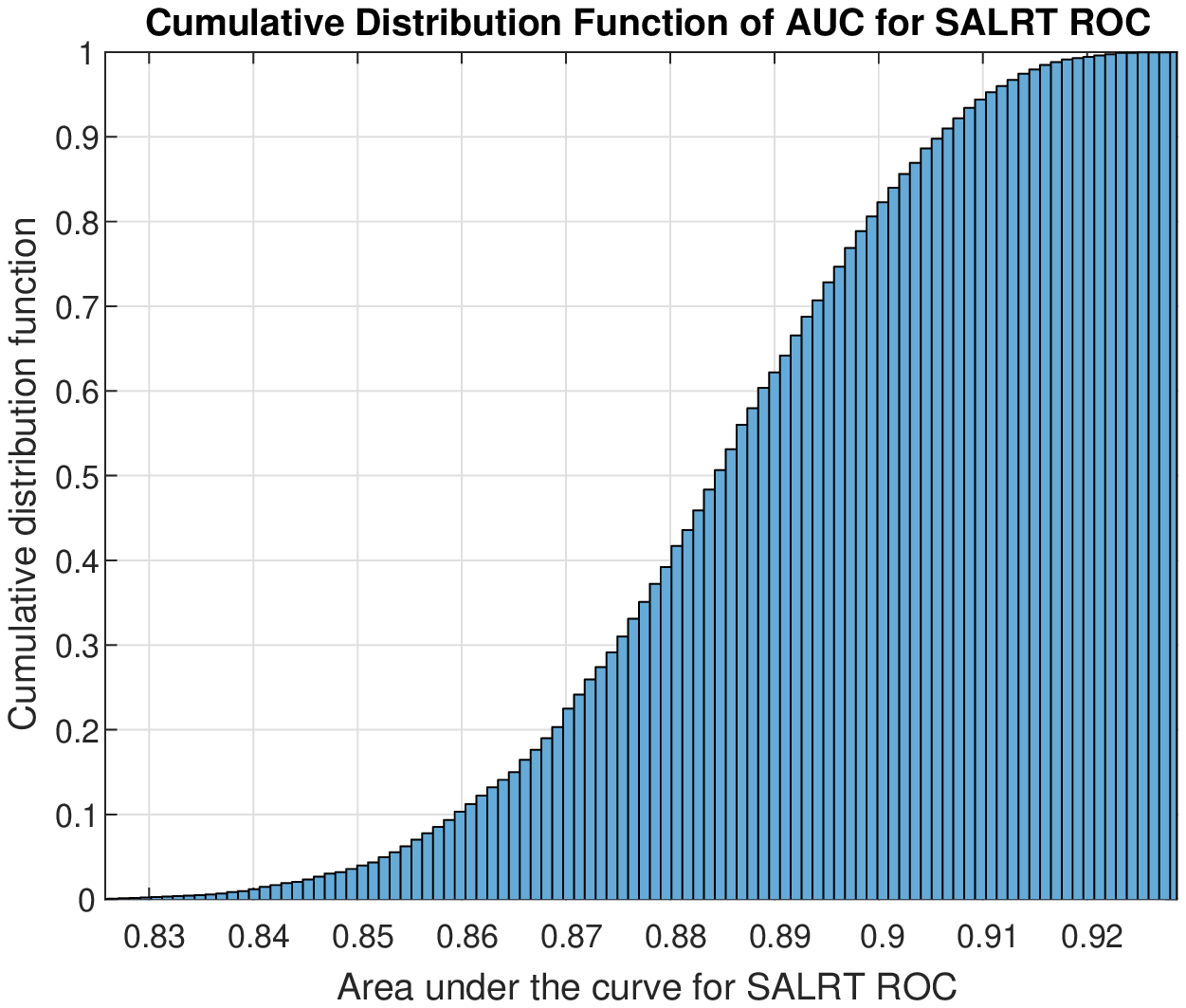}
  \captionsetup{justification=centering}
  \caption{Cumulative Distribution Function}
  \label{fig:CDF_complex_2}
\end{subfigure}
\captionsetup{justification=raggedright}
\caption[PDF and CDF for AUC of SALRT ROC]{Probability density function (a) and cumulative distribution function (b) of the AUC for the SALRT ROC for Scenario 2 for all possible synchronization sequences at $SNR=-5$ [dB]. \label{PDF_CDF_complex_2}}
\end{figure}

The AUC of the SALRT ROC for the best synchronization
sequence in Fig. \ref{fig:sfig1_stationary_minus_5_dB} and Fig. \ref{fig:sfig1_periodic_minus_5_dB} for Scenario 1 and Scenario 2 are $0.9608$ and $0.9286$, respectively, and the corresponding AUCs for the worst synchronization sequence are $0.8880$ and $0.8258$, respectively. Based on the CDFs depicted in Figs. \ref{fig:CDF}, \ref{fig:CDF_complex_2}, we can propose a simple procedure for identifying a nearly optimal synchronization sequence without testing all possible synchronization sequences with different CIRs:
To that aim, first randomly select 100 synchronization sequences, and evaluate the SALRT ROC and the corresponding AUC for each of these sequences. Then, choose the synchronization sequence with the largest AUC as the (nearly) optimal sequence.
From the CDF depicted in Fig. \ref{fig:CDF} and in Fig. \ref{fig:CDF_complex_2} for Scenarios 1 and 2, we note that the range of the AUC whose probability is above 0.9 is $0.945< \mbox{AUC}_1<0.9608$ and $0.9065<\mbox{AUC}_2<0.9286$ respectively. The overall range of the AUC values for scenarios 1 and 2 are $0.880< \mbox{AUC}_1<0.9608$ and $0.8258< \mbox{AUC}_2<0.9286$, respectively. It follows that the percentage of AUC values with probability higher than $90\%$ for Scenario 1 is $\frac{0.9608-0.945}{0.9608-0.8880}=21.7\%$, and for Scenario 2 it is $\frac{0.9286-0.9065}{0.9286-0.8258}=21.5\%$, out of the overall range of the AUC values. Thus, for 100 randomly selected synchronization sequences, approximately 10 sequences will have an AUC higher than $0.945$ and $0.9065$, for Scenario 1 and 2, respectively. 
It is noted that 100 sequences is a relatively small fraction of the set of all the possible synchronization sequences e.g., for the example scenarios considered in the simulations using a BPSK constellation with a synchronization sequence whose length is $L_{\textsf{sw}}=12$, there are $4096$ different possible synchronization sequences.
%
%


\section{Conclusions}
\label{chap:Conclusions}

In this paper we studied frame synchronization for LPTV channels with ACGN, where the CIR coefficients are assumed deterministic and unknown.
We derived a frame synchronization algorithm which accounts for the channel memory, as previously proposed LRT-based detectors did not account for channel memory, due to the associated computational complexity, see \cite{gansman1997optimum,chiani2004optimum,chiani2005practical,chiani2010noncoherent,liang2015sequential}. Prior works which proposed FS algorithms accounting for channel memory, were based either on ad-hoc considerations, such as the correlator, or on ML criterion \cite{moon1991ml}, hence did not facilitate a tradeoff between the probability of successful detection, $p_{\textsf{d}}$, and the probability of false alarm, $p_{\textsf{fa}}$, generally resulting in suboptimal algorithms. 
We first proposed an ALRT detector, which was shown in the simulations to achieve performance which is very close to that of the optimal LRT detector.  
We then proposed a new approximate LRT detector which further reduces the computational complexity by a significant ratio at the cost of an additional decrease in optimally of w.r.t. the LRT referred to as the RALRT. As all above algorithms use knowledge of the channel coefficients in computing the test metric, yet such knowledge is typically not available at the frame synchronization stage, we propose a new blind algorithm, referred to as the SALRT, which handles the lack of CIR knowledge via a non-data-aided iterative CIR estimation.
We note that while the SALRT algorithm does not need a-priori knowledge of the channel coefficients, it achieves has better performance than the correlation detector while in terms of the computational complexity it has a lower computational complexity than the LRT and the ALRT detectors, but a higher computational complexity than the RALRT and the correlator detectors.



\begin{appendix}
\addcontentsline{toc}{chapter}{Appendices}
\renewcommand\thesection{\Alph{section}}
\counterwithin{figure}{section}
\subsection{Derivation of the Baseband Equivalent Noise Model for Real Passband ACGN}
\label{app:Noise_Model}
\begin{figure}[!ht]
\includegraphics[scale=0.5]{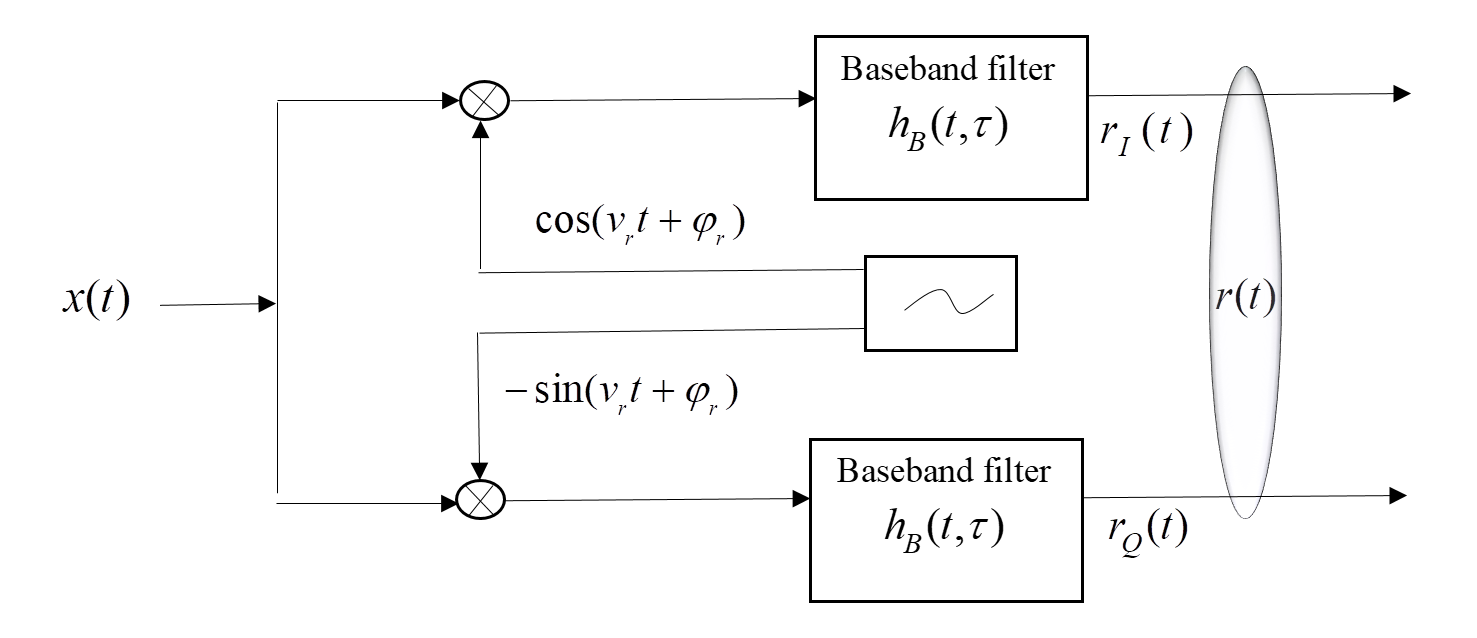}
\centering
\captionsetup{justification=raggedright}
\caption{Schematic description of obtaining the baseband representation of a real passband signal \label{Baseband_System}}
\end{figure}

\numberwithin{equation}{section}
In this appendix we show that the baseband representation of a real bandlimited passband ACGN is a proper complex cyclostationary Gaussian noise and highlight the associated assumptions. Let 
\begin{ceqn}
\begin{align*}
&Z[m]=Z_R[m]+j\cdot Z_I[m],~Z_R[m]\in \mathcal{R},~Z_I[m] \in \mathcal{R},~m \in \mathcal{Z} ~,   
\end{align*}
\end{ceqn}
denote a discrete-time (DT) complex random process, and define the following two quantities:
\begin{itemize}
\item $c_z[m,l]\triangleq \mathds{E}\big\{Z[m+l]Z^*[m]\big\}\Big\},~m,l \in \mathcal{Z}$, is the autocorrelation function of $Z[m]$.
\item $\tilde{c}_z[m,l]\triangleq \mathds{E}\big\{Z[m+l]Z[m]\big\},~m,l \in \mathcal{Z}$, is the pseudo-autocorrelation function of $Z[m]$.
\end{itemize} 
%
\begin{definition} \label{Proper_Complex_definition}
(Proper complex random process \cite{neeser1993proper}) A DT complex random process $Z[m]$, with $\mathds{E}\big\{Z[m]\big\}=0$, is called proper complex if its pseudo-autocorrelation function satisfies $\tilde{c}_z[m,l]=0$.
\end{definition}
Next, we examine $\tilde{c}_z[m,l]$,
\begin{align*}
\tilde{c}_z[m,l]&=\mathds{E}\big\{Z[m+l]Z[m]\big\}\\
&=\mathds{E}\Big\{\big( Z_R[m+l]+j\cdot Z_I[m+l]\big)\cdot\big(Z_R[m]+j\cdot Z_I[m]\big)\Big\}\\
&=\mathds{E}\Big\{\big( Z_R[m+l]Z_R[m]-Z_I[m+l]Z_I[m]\big)\\
&~~~~~~~~~~~~~~~~~+j\cdot\big(Z_R[m+l]Z_I[m]+Z_I[m+l]Z_R[m]  \big)\Big\}
\end{align*}
\begin{align*}
&\stackrel{(a)}{=}\Big( \mathds{E}\big\{ Z_R[m+l]Z_R[m] \big\}-\mathds{E}\big\{ Z_I[m+l]Z_I[m]  \big\} \Big)\\
&~~~~~~~+j\cdot \Big( \mathds{E}\big\{ Z_R[m+l]Z_I[m] \big\}+\mathds{E}\big\{Z_I[m+l]Z_R[m] \big\} \Big)  
\end{align*}
where $(a)$ follows since the stochastic expectation is a linear operator.
It follows that $\tilde{c}_z[m,l]=0$ if and only if it satisfies the following three properties, see also \cite{neeser1993proper}:
\begin{enumerate}
\item[C.1] $\mathds{E}\big\{Z_R[m+l]Z_R[m]\big\}=\mathds{E}\big\{Z_I[m+l]Z_I[m]\big\}$ 
\item[C.2] $\mathds{E}\big\{Z_R[m+l] Z_I[m]\big\}=-\mathds{E}\big\{Z_I[m+l] Z_R[m]\big\}$
\item[C.3] All four correlation functions in conditions C.1 and C.2 above are finite.
\end{enumerate}
Lastly, we recall that from Def. \ref{WSCS_definition}, a zero-mean proper-complex random process $Z[m]$ is cyclostationary if:
\begin{ceqn}
\begin{align*}
&c_z[m,l]=c_z[m+P_{\textsf{z}},l],~P_{\textsf{z}}=\mbox{constant}\in \mathcal{N},l \in \mathcal{Z}. 
\end{align*}
\end{ceqn}
%
Let $Z(t)$ be a real continuous-time (CT) passband ACGN which satisfies $ \mathds{E}\big\{Z(t)\big\}=0,$
with an autocorrelation function $ c_z(t,\tau)=\mathds{E}\big\{Z(t+\tau)Z(t)\big\}$,
$ Z(t) \sim \mathcal{N}\big(0,c_z(t,\tau)\big)$, $t,\tau \in \mathcal{R}$.
Due to cyclostationary, the autocorrelation function $c_z(t,\tau)$ is periodic with some period $T_{\textsf{z}}$:
\begin{ceqn}
\begin{align}
&c_z(t,\tau)=c_z(t+k\cdot T_{\textsf{z}},\tau),~~T_{\textsf{z}}=\mbox{positive constant},~\forall k \in \mathcal{Z}.
\end{align}
\end{ceqn}
 We denote the Fourier coefficients of $c_z(t,\tau)$ w.r.t $\tau$ with $C_{z}(t,v)$:
\begin{ceqn}
\begin{align*}
&C_{z}(t,v)=\int_{\tau=-\infty}^{\infty}c_z(t,\tau) e^{-j\tau v}d\tau 
\end{align*}
\end{ceqn}
\textbf{\textit{AS1}}: It is assumed that $C_{z}(t,v)$ is bandlimited around a center frequency $v_c$, that is: 
\begin{ceqn}
\begin{align}
&C_{z}(t,v)\neq 0, ~~\forall |v-v_c|\leq BW_z,~|v+v_c|\leq BW_z, \nonumber\\
&BW_z~ \mbox{is a positive constant}~,v_c\gg BW_z~. 
\end{align}
\end{ceqn}
We denote the Fourier series elements of $C_z(t,v)$ w.r.t $t$ with $K_z (p,v)$, that is:
\begin{ceqn}
\begin{align*}
&C_{z}(t,v)= \sum_{p=-\infty }^{\infty} K_z (p,v) \cdot e^{j \omega_{\textsf{z}} \cdot p \cdot t} ,~ \omega_{\textsf{z}}=\frac{2 \pi}{T_{\textsf{z}}},~v_c\gg \omega_{\textsf{z}}~.
\end{align*}
\end{ceqn}
\textbf{\textit{AS2}}: It is assumed that the number of Fourier series coefficient for $C_{z}(t,v)$ is finite, that is for a very large index $L_{\textsf{p}}\gg 1,~ L_{\textsf{p}} \in \mathcal{N},~K_z (p_1,v) =0,\mbox{if}~|p_1|>L_{\textsf{p}}$, accordingly:
\begin{ceqn}
\begin{align} \label{Finite_Fourier_series}
&C_{z}(t,v)= \sum_{p=-L_{\textsf{p}} }^{L_{\textsf{p}}} K_z (p,v) \cdot e^{j \omega_{\textsf{z}} \cdot p \cdot t} ,~ \omega_{\textsf{z}}=\frac{2 \pi}{T_{\textsf{z}}}~,v_c\gg \omega_{\textsf{z}} \cdot L_{\textsf{p}}.
\end{align}
\end{ceqn}
%
We consider a linear periodic time varying (LPTV) \emph{baseband} filter $h_{B}(t,\tau)$, whose period is $T_{\textsf{h}}$, that is $h_{B}(t,\tau)=h_{B}(t+T_{\textsf{h}},\tau)$. Let $H_{B}(t,v)$ denote the Fourier transform of $h_{B}(t,\tau)$ w.r.t $\tau ,~H_{B}(t,v)=\int_{\tau=-\infty}^{\infty}h_{B}(t,\tau) e^{-j\tau v}d\tau,$ where
$H_{B}(t,v)$ satisfies:
\begin{align*}
&\mbox{\textbf{\textit{AS3}}: }H_{B}(t,v)=0,~\forall |v|\geq BW_h,~v_c\gg BW_h\gg BW_z~.    
\end{align*}
Lastly, let $\delta(\cdot)$ denote the Dirac impulse function. The generalized linear convolution operator between $h_{B}(t,\tau)$ and $x(t)$ is defined as:
\begin{ceqn}
\begin{align} 
&\big(x(t)*h_{B}(t,\tau)\big)(t)=\int_{\tau=-\infty}^{\infty}x(t-\tau)h_{B}(t,\tau)d\tau ~. \label{Generalized_Convolotion}   
\end{align}
\end{ceqn}
The channel impulse response (CIR) of the LPTV channel model of (\ref{Generalized_Convolotion}), for input $x(t)=\delta(t-t_0)$ is given by $h_{B}(t,t-t_0),~t,t_0 \in \mathcal{R}$. Next, we recall the process of down-converting a passband signal into a baseband signal \cite[Ch. 5, p. 108]{chiueh2012baseband} depicted in Fig. \ref{Baseband_System}.
\\
The passband signal $x(t)$ with a center frequency $v_c$
is first multiplied by $\cos(v_rt+\varphi_r)$ and by $-\sin(v_rt+\varphi_r)$ to extract the in-phase and the quadrature parts of the signal $x(t)$. For analytical tractability, in the following we assume that $v_r=v_c$. Then, the in-phase and the quadrature parts are each filtered with a baseband filter $h_{B}(t,\tau)$, and the results, $r_I(t)$ and $r_Q(t)$, respectively, are collected into a vector $\mathbf{r}(t)$ which constitutes the baseband representation of $x(t)$:
\begin{ceqn}
\begin{align*}
\mathbf{r}(t)=
\begin{bmatrix}
r_I(t)\\
r_Q(t)\\
\end{bmatrix}  \quad
=
\begin{bmatrix}
\Big(\big(\cos(v_rt+\varphi_r)\cdot x(t)\big)*h_{B}(t,\tau)\Big)(t)\\
\Big(\big(-\sin(v_rt+\varphi_r)\cdot x(t)\big)*h_{B}(t,\tau)\Big)(t)\\
\end{bmatrix}
\end{align*}
\end{ceqn}
%
Next, we define the random processes and random vector corresponding to the transformations detailed above, when applied to $Z(t)$:
 \begin{ceqn}
\begin{align*}
Z_{B,R}(t)&\triangleq Z_{I}(t)=\Big(\big(\cos(v_ct+\varphi_c)\cdot Z(t)\big)*h_{B}(t,\tau)\Big)(t)\\
Z_{B,I}(t)&\triangleq Z_{Q}(t)=\Big(\big(-\sin(v_ct+\varphi_c)\cdot Z(t)\big)*h_{B}(t,\tau)\Big)(t)\\
Z_{B}(t)&\triangleq Z_{B,R}(t)+j\cdot Z_{B,I}(t) \Longleftrightarrow 
\mathbf{Z_{B}}(t)\triangleq
\begin{bmatrix}
Z_{B,R}(t)\\
Z_{B,I}(t)
\end{bmatrix}
\equiv
\begin{bmatrix}
Z_{I}(t)\\
Z_{Q}(t)\\
\end{bmatrix}~.
 \end{align*}
  \end{ceqn}
From the theory of complex random variables it follows that that the PDF  
$f_{\mathbf{Z}}(u,s)\equiv f_{Z}(u+j\cdot s),~u,s\in \mathcal{R}$, with the appropriate complex variables operations.
The random processes $Z_{B,R}(t)$ and $Z_{B,I}(t)$ are sampled at a rate $T_{\textsf{s}}$, resulting in the following DT random processes:
 \begin{ceqn}
\begin{align*}
Z_{B,R}[m]&=Z_{B,R}(m\cdot T_{\textsf{s}})=\Big(\big(\cos(v_c\cdot t+\varphi_c)\cdot Z(t)\big)*h_{B}(t,\tau)\Big)(m\cdot T_{\textsf{s}})\\
Z_{B,I}[m]&= Z_{B,I}(m\cdot T_{\textsf{s}})=\Big(\big(-\sin(v_c\cdot t+\varphi_c)\cdot Z(t)\big)*h_{B}(t,\tau)\Big)(m\cdot T_{\textsf{s}})\\
Z_{B}[m]&=Z_{B}(m\cdot T_{\textsf{s}})=Z_{B,R}(m\cdot T_{\textsf{s}})+j\cdot Z_{B,I}(m\cdot T_{\textsf{s}})=Z_{B,R}[m]+j\cdot Z_{B,I}[m],~m\in \mathcal{Z}.
\end{align*}
\end{ceqn}
\textbf{\textit{AS4}}: We assume that $T_{\textsf{s}}$ is a divisor of an integer multiple of $T_{\textsf{h}}$, and of an integer multiple of $T_{\textsf{z}}$, that is:
\begin{ceqn}
\begin{align} \label{Period_constants_1}
\exists \alpha_1,\alpha_2 \in \mathcal{N}~s.t.~~ \frac{\alpha_1 \cdot T_{\textsf{h}}}{T_{\textsf{s}}}=P_{\textsf{h}}\in \mathcal{N},~\frac{\alpha_2 \cdot T_{\textsf{z}}}{T_{\textsf{s}}}=P_{\textsf{tmp}}\in \mathcal{N}
\end{align}
\end{ceqn}
Let $P_{\textsf{z}}$ be the least common multiple (LCM) of $P_{\textsf{h}}$ and $P_{\textsf{tmp}}$, namely $P_{\textsf{z}}$ is the smallest positive integer that is divisible by both $P_{\textsf{h}}$ and $P_{\textsf{tmp}}$:
\begin{ceqn}
\begin{align} \label{Period_constants_2}
&\exists k_1,k_2\in \mathcal{N}~s.t.~P_{\textsf{z}}=k_1\cdot P_{\textsf{h}},~P_{\textsf{z}}=k_2\cdot P_{\textsf{tmp}},.
\end{align}
\end{ceqn}
%
We next prove that $Z_B[m]$ is proper complex noise. We begin by showing that $\mathds{E}\big\{Z_B[m]\big\}=0$:
\begin{ceqn}
\begin{align*}
\mathds{E}\big\{Z_B[m]\big\}&=\mathds{E}\big\{Z_{B,R}[m]+j\cdot Z_{B,I}[m]\big\}\\
&\stackrel{(a)}{=}\mathds{E}\big\{Z_{B,R}[m]\big\}+j\cdot \mathds{E}\big\{Z_{B,I}[m]\big\}\\
&=\mathds{E}\bigg\{\Big(\big(\cos(v_c\cdot t+\varphi_c)\cdot Z(t)\big)*h_{B}(t,\tau)\Big)(m\cdot T_{\textsf{s}})\bigg\}\\
&~~~~~~+j\cdot \mathds{E}\bigg\{\Big(\big(-\sin(v_c\cdot t+\varphi_c)\cdot Z(t)\Big)*h_{B}(t,\tau)\big)(m\cdot T_{\textsf{s}})\bigg\}\\
&\stackrel{(b)}{=}\bigg(\Big(\cos(v_c\cdot t+\varphi_c)\cdot \mathds{E}\big\{Z(t)\big\}\Big)*h_{B}(t,\tau)\bigg)(m\cdot T_{\textsf{s}})\\
&~~~~~~+j\cdot \bigg(\Big(-\sin(v_c\cdot t+\varphi_c)\cdot \mathds{E}\big\{Z(t)\big\}\Big)*h_{B}(t,\tau)\bigg)(m\cdot T_{\textsf{s}})\\
&\stackrel{(c)}{=}\Big(\big(\cos(v_c\cdot t+\varphi_c)\cdot 0)*h_{B}(t,\tau)\big)(m\cdot T_{\textsf{s}})\\
&~~~~~~+j\cdot \Big(\big(-\sin(v_c\cdot t+\varphi_c)\cdot 0\big)*h_{B}(t,\tau)\Big)(m\cdot T_{\textsf{s}})\\
&=0+j\cdot 0=0~,
\end{align*}
\end{ceqn}
where $(a)$ follows since the stochastic expectation is a linear operator, and $(b)$ follows again from the linearity of the expectation since only $Z(t)$ is random and the integral is a linear operator, and in $(c)$ we use $ \mathds{E}\big\{Z(t)\big\}=0$.
Next, we show that $\mathds{E}\big\{Z_{B,R}[m+l]Z_{B,R}[m]\big\}=\mathds{E}\big\{Z_{B,I}[m+l]Z_{B,I}[m]\big\} $, or equivalently $\mathds{E}\big\{Z_{B,R}[m+l]Z_{B,R}[m]-Z_{B,I}[m+l]Z_{B,I}[m]\big\}=0$.\\ Consider first:
\begin{align*}
&\mathds{E}\big\{Z_{B,R}[m+l]Z_{B,R}[m]-Z_{B,I}[m+l]Z_{B,I}[m]\big\}\\
&=\mathds{E}\bigg\{\Big(\big(\cos(v_c\cdot t+\varphi_c)\cdot Z(t)\big)*h_{B}(t,\tau)\Big)\big((m+l)\cdot T_{\textsf{s}}\big)\\
&~~~~~~~~~~~~~~~~~~~~~~~~~~~~~~~~~~~~~~\cdot \Big(\big(\cos(v_c\cdot t+\varphi_c)\cdot Z(t)\big)*h_{B}(t,\tau)\Big)\big(m\cdot T_{\textsf{s}}\big)\\
&~~~~~~~~~~~-\Big(\big(-\sin(v_c\cdot t+\varphi_c)\cdot Z(t)\big)*h_{B}(t,\tau)\Big)((m+l)\cdot T_{\textsf{s}})\\
&~~~~~~~~~~~~~~~~~~~~~~~~~~~~~~~~~~~~~~~~~~~~\cdot \Big(\big(-\sin(v_c\cdot t+\varphi_c)\cdot Z(t)\big)*h_{B}(t,\tau)\Big)(m\cdot T_{\textsf{s}}) \bigg\}
\end{align*}
\begin{align*}
&=\mathds{E}\Bigg\{\bigg(\int_{\tau=-\infty}^{\infty} \cos\Big(v_c \big((m+l)\cdot T_{\textsf{s}}-\tau \big)+\varphi_c \Big)\\
&~~~~~~~~~~~~~\cdot Z\big((m+l)\cdot T_{\textsf{s}}-\tau \big)\cdot h_B\big((m+l)T_{\textsf{s}},\tau\big)d\tau \bigg)\\
&~~~~~~~~~~~~~~~~~~~~~~~~~~~\cdot\bigg(\int_{s=-\infty}^{\infty} \cos\Big(v_c \big(m\cdot T_{\textsf{s}}-s \big)+\varphi_c\Big)\\
&~~~~~~~~~~~~~~~~~~~~~~~~~~~~~~~~~~~~~\cdot Z\big(m\cdot T_{\textsf{s}}-s \big)\cdot h_B\big(mT_{\textsf{s}},s\big)ds \bigg)\\
&~~~~~~~~~~~~~-\bigg(\int_{\tau=-\infty}^{\infty} -\sin\Big(v_c \big((m+l)\cdot T_{\textsf{s}}-\tau \big)+\varphi_c \Big)\\
&~~~~~~~~~~~~~~~~~~~~~~~~~~~~\cdot Z\big((m+l)\cdot T_{\textsf{s}}-\tau \big)\cdot h_B\big((m+l)T_{\textsf{s}},\tau\big)d\tau \bigg)\\
&~~~~~~~~~~~~~~~~~~~~~~~~~~~~~~~~~~~~~~~~~~~~~~~~~\cdot\bigg(\int_{s=-\infty}^{\infty} -\sin\Big(v_c \big(m\cdot T_{\textsf{s}}-s \big)+\varphi_c \Big)\\
&~~~~~~~~~~~~~~~~~~~~~~~~~~~~~~~~~~~~~~~~~~~~~~~~~~~~~~~~~~~~~~\cdot Z\big(m\cdot T_{\textsf{s}}-s \big)\cdot h_B\big(mT_{\textsf{s}},s\big)ds \bigg)\Bigg\}\\
&=\mathds{E}\bigg\{\int_{\tau=-\infty}^{\infty} \int_{s=-\infty}^{\infty} \cos\Big(v_c \big((m+l)\cdot T_{\textsf{s}}-\tau\big)+\varphi_c\Big)\cdot\cos\Big(v_c (m\cdot T_{\textsf{s}}-s)+\varphi_c\Big) \\
&~~~~~~\cdot Z\big((m+l)\cdot T_{\textsf{s}}-\tau\big)\cdot Z\big(m\cdot T_{\textsf{s}}-s\big)\cdot h_B\big((m+l)T_{\textsf{s}},\tau\big)\cdot h_B(mT_{\textsf{s}},s)d\tau ds \\
&~~~~~~~~-\int_{\tau=-\infty}^{\infty} \int_{s=-\infty}^{\infty}\sin\Big(v_c \big((m+l)\cdot T_{\textsf{s}}-\tau \big)+\varphi_c\Big)\cdot \sin\Big(v_c (m\cdot T_{\textsf{s}}-s)+\varphi_c\Big)\\
&~~~~~~~~~~\cdot Z\big((m+l)\cdot T_{\textsf{s}}-\tau\big)\cdot Z\big(m\cdot T_{\textsf{s}}-s\big)\cdot h_B\big((m+l)T_{\textsf{s}},\tau\big)\cdot h_B\big(mT_{\textsf{s}},s\big)d\tau ds\bigg\}\\
&=\mathds{E}\bigg\{\int_{\tau=-\infty}^{\infty} \int_{s=-\infty}^{\infty}  \Big(\cos\Big(v_c \big((m+l)\cdot T_{\textsf{s}}-\tau\big)+\varphi_c\Big)\cdot \cos\big(v_c (m\cdot T_{\textsf{s}}-s)+\varphi_c\big) \nonumber\\
&~~~~~~~~~~~-\sin\Big(v_c \big((m+l)\cdot T_{\textsf{s}}-\tau\big)+\varphi_c\Big)\cdot \sin\big(v_c (m\cdot T_{\textsf{s}}-s)+\varphi_c\big)\Big)\nonumber \\
&~~~~~~~~~\cdot \Big(Z((m+l)\cdot T_{\textsf{s}}-\tau)\cdot Z(m\cdot T_{\textsf{s}}-s)\cdot h_B\big((m+l)T_{\textsf{s}},\tau\big)\cdot h_B\big(mT_{\textsf{s}},s\big)\Big)d\tau ds \bigg\}\nonumber\\
&\stackrel{(a)}{=}\mathds{E}\bigg\{\int_{\tau=-\infty}^{\infty} \int_{s=-\infty}^{\infty}  \cos\Big(v_c \big((2m+l)\cdot T_{\textsf{s}}-\tau-s\big)+2\varphi_c\Big)\nonumber \\
&~~~~~~~~\cdot \Big(Z\big((m+l)\cdot T_{\textsf{s}}-\tau\big)\cdot Z\big(m\cdot T_{\textsf{s}}-s)\cdot h_B\big((m+l)T_{\textsf{s}},\tau\big)\cdot h_B\big(mT_{\textsf{s}},s\big)\Big)d\tau ds \bigg\}\nonumber\\
&\stackrel{(b)}{=}\int_{\tau=-\infty}^{\infty} \int_{s=-\infty}^{\infty} \cos\Big(v_c \big((2m+l)\cdot T_{\textsf{s}}-\tau-s\big)+2\varphi_c \Big) \nonumber\\
&~~\cdot \mathds{E}\bigg\{Z\big((m+l)\cdot T_{\textsf{s}}-\tau\big)\cdot Z\big(m\cdot T_{\textsf{s}}-s)\bigg\}\cdot h_B\big((m+l)T_{\textsf{s}},\tau\big)\cdot h_B\big(mT_{\textsf{s}},s\big)d\tau ds \nonumber\\
&=\int_{\tau=-\infty}^{\infty} \int_{s=-\infty}^{\infty}  \cos\Big(v_c \big((2m+l)\cdot T_{\textsf{s}}-\tau-s\big)+2\varphi_c\Big) \nonumber\nonumber \\
&~~~~~~~~~~~~\cdot c_z\big((m+l)\cdot T_{\textsf{s}}-\tau,~\tau-s-l\cdot T_{\textsf{s}}\big)\cdot h_B\big((m+l)T_{\textsf{s}},\tau\big)\cdot h_B\big(mT_{\textsf{s}},s\big)\Big)d\tau ds \nonumber
\end{align*}
\begin{align}  \label{First_Proof}
&\stackrel{(c)}{=}\int_{\tau=-\infty}^{\infty}\bigg( \int_{u=-\infty}^{\infty} \cos\Big(v_c \big(2(m+l)\cdot T_{\textsf{s}}-2\tau+u\big)+2\varphi_c \Big)\cdot c_z\big((m+l)\cdot T_{\textsf{s}}-\tau,u \big)\nonumber \\
&~~~~~~~~~~~~~~~~~~~~~~~~~~~~~~~~~~~\cdot h_B(m\cdot T_{\textsf{s}},\tau-l\cdot T_{\textsf{s}}-u)du\bigg)\cdot h_B\big((m+l)T_{\textsf{s}},\tau\big)d\tau~, 
\end{align}
where $(a)$ follows from the trigonometric identity $\cos(\alpha+\beta)=\cos(\alpha)\cos(\beta)-\sin(\alpha)\sin(\beta)$, $(b)$ follows as only $Z(t)$ is random and since stochastic expectation is a linear operator, and $(c)$ follows from the change of variables $u=\tau-s-l\cdot T_{\textsf{s}}$.\\
Examine now the inner integral: 
\begin{align*}
&\int_{u=-\infty}^{\infty} \cos\Big(v_c \big(2(m+l)\cdot T_{\textsf{s}}-2\tau+u\big)+2\varphi_c \Big)\cdot c_z\big((m+l)\cdot T_{\textsf{s}}-\tau,u \big)\\
&~~~~~~~~~~~~~~~~~~~~~~~~~~~~~~~~~~~~~~~~~~~~~~~~~~~~~~~~~~~~~~~~~~~~~~\cdot h_B(m\cdot T_{\textsf{s}},\tau-l\cdot T_{\textsf{s}}-u)du
\end{align*}
Since $H_{B}(t,v)$ denotes the Fourier transform of $h_{B}(t,\tau)$ w.r.t $\tau$, we can express $h_{B}(t,\tau)$ via the inverse Fourier transform of $H_{B}(t,v)$ as:
\begin{ceqn}
\begin{align*}
h_{B}(t,\tau)=\frac{1}{2\pi}\int_{v=-\infty}^{\infty}H_{B}(t,v)e^{j\tau v}dv.
\end{align*}
\end{ceqn}
Therefore:
\begin{align}
&\ \int_{u=-\infty}^{\infty}  \cos\Big(v_c \big(2(m+l) \cdot T_{\textsf{s}}-2\tau+u\big)+2\varphi_c \Big)\cdot c_z\big((m+l)\cdot T_{\textsf{s}}-\tau,u \big)\nonumber\\
&~~~~~~~~~~~~~~~~~~~~~~~~~~~~~~~~~~~~~~~~~~~~~~~~~~~~~~~~~~~~~~~~~~~~~\cdot h_B(m \cdot T_{\textsf{s}},\tau-l \cdot T_{\textsf{s}}-u)du \nonumber\\
&= \int_{u=-\infty}^{\infty}  \cos\Big(v_c \big(2(m+l)\cdot T_{\textsf{s}}-2\tau+u\big)+2\varphi_c \Big)\cdot c_z\big((m+l)\cdot T_{\textsf{s}}-\tau,u \big)\nonumber\\
&~~~~~~~~~~~~~~~~~~~~~~~~~~~~~~~~~~~~~~~~~~~~~~\cdot \Big(\frac{1}{2\pi}\int_{v=-\infty}^{\infty}H_{B}(m \cdot T_{\textsf{s}},v)e^{j(\tau-l\cdot T_{\textsf{s}}-u)v}dv\Big)du\nonumber\\
&=\int_{v=-\infty}^{\infty} H_{B}(m \cdot T_{\textsf{s}},v)e^{j(\tau-l \cdot T_{\textsf{s}}) v} \nonumber\\
&~~~~~~~~~~~~~~~~~~~~~~\cdot \bigg( \frac{1}{2\pi}\int_{u=-\infty}^{\infty} \cos\Big(v_c \big(u+2(m+l)\cdot T_{\textsf{s}}-2\tau+\frac{2\varphi_c}{v_c}\big) \Big)\nonumber\\
&~~~~~~~~~~~~~~~~~~~~~~~~~~~~~~~~~~~~~~~~~~~~~~~~~~~~~~~~~~\cdot c_z\big((m+l)\cdot T_{\textsf{s}}-\tau,u \big)\cdot e^{-juv}du \bigg)dv\nonumber
\end{align}
Next consider:
\begin{align}
&\frac{1}{2\pi} \int_{u=-\infty}^{\infty} \cos\bigg(v_c \Big(u+2(m+l)\cdot T_{\textsf{s}}-2\tau+\frac{2\varphi_c}{v_c}\Big)\bigg) \nonumber\\
&~~~~~~~~~~~~~~~~~~~~~~~~~~~~~~~~~~~~~~~~~~~~~~~~~~~~~~~~~~~~~~\cdot c_z\big((m+l)\cdot T_{\textsf{s}}-\tau,u\big)\cdot e^{-juv}du\nonumber
\end{align}
\begin{align}
&\stackrel{(a)}{=} \frac{1}{2\pi} \int_{u=-\infty}^{\infty} \Bigg(\frac{e^{jv_c \big(u+2(m+l)\cdot T_{\textsf{s}}-2\tau+\frac{2\varphi_c}{v_c}\big)}+e^{-jv_c \big(u+2(m+l)\cdot T_{\textsf{s}}-2\tau+\frac{2\varphi_c}{v_c}\big)}}{2} \Bigg)\nonumber\\
&~~~~~~~~~~~~~~~~~~~~~~~~~~~~~~~~~~~~~~~~~~~~~~~~~~~~~~~~~~~~~~\cdot c_z\big((m+l)\cdot T_{\textsf{s}}-\tau,u\big)\cdot e^{-juv}du\nonumber\\
&=\frac{1}{4\pi}\Bigg(e^{jv_c \big(2(m+l)\cdot T_{\textsf{s}}-2\tau+\frac{2\varphi_c}{v_c}\big)}\cdot \int_{u=-\infty}^{\infty} c_z\big((m+l)\cdot T_{\textsf{s}}-\tau,u\big)\cdot e^{-ju(v-v_c)}du\nonumber\\
&~~~~~~~~~~+e^{-jv_c \big(2(m+l)\cdot T_{\textsf{s}}-2\tau+\frac{2\varphi_c}{v_c}\big)}\cdot \int_{u=-\infty}^{\infty} c_z\big((m+l)\cdot T_{\textsf{s}}-\tau,u\big)\cdot e^{-ju(v+v_c)}du \Bigg)\nonumber\\ 
\label{middle}
&\stackrel{(b)}{=}\frac{1}{4\pi}\cdot \Big( e^{jv_c \big(2(m+l)\cdot T_{\textsf{s}}-2\tau+\frac{2\varphi_c}{v_c}\big)}\cdot C_z((m+l)T_{\textsf{s}}-\tau,v-v_c) \nonumber \\ 
&~~~~~~~~~~~~~~~~~~~~~~~~~~~~~~~~~~~~+e^{-jv_c \big(2(m+l)\cdot T_{\textsf{s}}-2\tau+\frac{2\varphi_c}{v_c}\big)}\cdot  C_z((m+l)T_{\textsf{s}}-\tau,v+v_c)\big) \Big)~,
\end{align}
where $(a)$ follows from Euler's formula: $\cos(v_cu)=\frac{e^{jv_cu}+e^{-jv_cu}}{2}$, $(b)$ follows from the definition of the Fourier transform of~$c_z(t,\tau)$ w.r.t $\tau$:
\begin{ceqn}
\begin{align*}
C_{z}(t,v+b)= \int_{\tau=-\infty}^{\infty}c_z(t,\tau) e^{-j\tau( v+b)}d\tau~.
\end{align*}
\end{ceqn}
Using (\ref{middle}) we obtain:
\begin{align} \label{First_Proof_1}
&\int_{v=-\infty}^{\infty} H_{B}(m \cdot T_{\textsf{s}},v)e^{j(\tau-l \cdot T_{\textsf{s}}) v} \nonumber \\
&~~~~~~~~~~~~~~~~~~~~~~\cdot \Bigg( \frac{1}{2\pi}\int_{u=-\infty}^{\infty} \cos\bigg(v_c \Big(u+2(m+l)\cdot T_{\textsf{s}}-2\tau+\frac{2\varphi_c}{v_c}\Big) \bigg)\nonumber\\
&~~~~~~~~~~~~~~~~~~~~~~~~~~~~~~~~~~~~~~~~~~~~~~~~~~~~~~~~~~~\cdot c_z\big((m+l)\cdot T_{\textsf{s}}-\tau,u \big)\cdot e^{-juv}du \Bigg)dv \nonumber \\
&=\int_{v=-\infty}^{\infty} H_{B}(m \cdot T_{\textsf{s}},v)e^{j(\tau-l \cdot T_{\textsf{s}}) v} \nonumber\\
&~~~~~~~~~~~~~~~~~~~~~~~ \cdot \frac{1}{4\pi}\cdot \Big( e^{jv_c \big(2(m+l)\cdot T_{\textsf{s}}-2\tau+\frac{2\varphi_c}{v_c}\big)}\cdot C_z\big((m+l)T_{\textsf{s}}-\tau,v-v_c\big)\nonumber \\
&~~~~~~~~~~~~~~~~~~~~~~~~~~~~~~~~~~~+e^{-jv_c \big(2(m+l)\cdot T_{\textsf{s}}-2\tau+\frac{2\varphi_c}{v_c}\big)}\cdot C_z\big((m+l)T_{\textsf{s}}-\tau,v+v_c\big) \Big)dv\nonumber \\
&=\frac{1}{4\pi}\int_{v=-\infty}^{\infty} \bigg( e^{jv_c \big(2(m+l)\cdot T_{\textsf{s}}-2\tau+\frac{2\varphi_c}{v_c}\big)}\cdot H_{B}(m \cdot T_{\textsf{s}},v) \cdot C_z\big((m+l)T_{\textsf{s}}-\tau,v-v_c\big)\nonumber \\
&~~~+e^{-jv_c \big(2(m+l)\cdot T_{\textsf{s}}-2\tau+\frac{2\varphi_c}{v_c}\big)}\cdot H_{B}(m \cdot T_{\textsf{s}},v) \cdot C_z\big((m+l)T_{\textsf{s}}-\tau,v+v_c\big) \bigg)e^{j(\tau-l \cdot T_{\textsf{s}}) v}dv.
\end{align}
Next, plugging the expression of Eqn (\ref{First_Proof_1}) into Eq. (\ref{First_Proof}) we arrive at:
\begin{align*}
&\int_{\tau=-\infty}^{\infty}\bigg( \frac{1}{4\pi}\int_{v=-\infty}^{\infty} \bigg( e^{jv_c \big(2(m+l)\cdot T_{\textsf{s}}-2\tau+\frac{2\varphi_c}{v_c}\big)}\cdot H_{B}(m \cdot T_{\textsf{s}},v) \cdot C_z\big((m+l)T_{\textsf{s}}-\tau,v-v_c\big)\nonumber \\
&~~~~~~~~+e^{-jv_c \big(2(m+l)\cdot T_{\textsf{s}}-2\tau+\frac{2\varphi_c}{v_c}\big)}\cdot H_{B}(m \cdot T_{\textsf{s}},v) \cdot C_z\big((m+l)T_{\textsf{s}}-\tau,v+v_c\big) \bigg)e^{j(\tau-l \cdot T_{\textsf{s}}) v}dv\bigg)\\
&~~~~~~~~~~~~~~~~~~~~~~~~~~~~~~~~~~~~~~~~~~~~~~~~~~~~~~~~~~~~~~~~~~~~~~~~~~~~~~~~~~~~~\cdot h_B\big((m+l)T_{\textsf{s}},\tau\big)d\tau\\
&\stackrel{(a)}{=}\int_{\beta=-\infty}^{\infty}\bigg( \frac{1}{4\pi}\int_{v=-\infty}^{\infty} \bigg( e^{j2v_c \big(\beta+\frac{\varphi_c}{v_c}\big)}\cdot H_{B}(m \cdot T_{\textsf{s}},v) \cdot C_z\big(\beta,v-v_c\big)\nonumber \\
&~~~~~~~~~~~~~~~~~~~+e^{-2jv_c \big(\beta+\frac{\varphi_c}{v_c}\big)}\cdot H_{B}(m \cdot T_{\textsf{s}},v) \cdot C_z\big(\beta,v+v_c\big) \bigg)\cdot e^{j\big((m+l)\cdot T_{\textsf{s}}-\beta-l \cdot T_{\textsf{s}}\big) v}dv\bigg)\\
&~~~~~~~~~~~~~~~~~~~~~~~~~~~~~~~~~~~~~~~~~~~~~~~~~~~~~~~~~~~~~~~~~~\cdot h_B\big((m+l)T_{\textsf{s}},(m+l)\cdot T_{\textsf{s}}-\beta\big)d\beta\\
&=\frac{1}{4\pi} \cdot \int_{\beta=-\infty}^{\infty} \bigg( \int_{v=-\infty}^{\infty}  e^{j2v_c \big(\beta+\frac{\varphi_c}{v_c}\big)}\cdot H_{B}(m \cdot T_{\textsf{s}},v) \cdot C_z\big(\beta,v-v_c\big) \nonumber \\
&~~~~~~~~~~~~~~~~~~~~~~~~~~~~~\cdot e^{j\big((m+l)\cdot T_{\textsf{s}}-\beta-l \cdot T_{\textsf{s}}\big) v}dv\bigg)\cdot h_B\big((m+l)T_{\textsf{s}},(m+l)\cdot T_{\textsf{s}}-\beta\big)d\beta\\
&~~~~~+\frac{1}{4\pi} \cdot \int_{\beta=-\infty}^{\infty}\bigg( \int_{v=-\infty}^{\infty}  e^{-2jv_c \big(\beta+\frac{\varphi_c}{v_c}\big)}\cdot H_{B}(m \cdot T_{\textsf{s}},v) \cdot C_z\big(\beta,v+v_c\big) \nonumber \\
&~~~~~~~~~~~~~~~~~~~~~~~~~~~~~\cdot e^{j\big((m+l)\cdot T_{\textsf{s}}-\beta-l \cdot T_{\textsf{s}}\big) v}dv\bigg)\cdot h_B\big((m+l)T_{\textsf{s}},(m+l)\cdot T_{\textsf{s}}-\beta\big)d\beta \\
&=\frac{1}{4\pi} \cdot e^{j2\varphi_c} \cdot \int_{v=-\infty}^{\infty}  H_{B}(m \cdot T_{\textsf{s}},v)  \cdot e^{-j l \cdot T_{\textsf{s}}\cdot v}\cdot \bigg( \int_{\beta=-\infty}^{\infty}  e^{j2v_c \beta} \cdot C_z\big(\beta,v-v_c\big) \nonumber \\
&~~~~~~~~~~~~~~~~~~~~~~~~~~~~~~~~~~~\cdot e^{j\big((m+l)\cdot T_{\textsf{s}}-\beta\big) v}\cdot h_B\big((m+l)T_{\textsf{s}},(m+l)\cdot T_{\textsf{s}}-\beta\big)d\beta \bigg)dv\\
&~~~~~+\frac{1}{4\pi} \cdot  e^{-2j \varphi_c} \cdot \int_{v=-\infty}^{\infty}  H_{B}(m \cdot T_{\textsf{s}},v) \cdot  e^{-j l \cdot T_{\textsf{s}}\cdot  v} \cdot \bigg( \int_{\beta=-\infty}^{\infty}  e^{-2jv_c\beta} \cdot C_z\big(\beta,v+v_c\big) \nonumber \\
&~~~~~~~~~~~~~~~~~~~~~~~~~~~~~~~~~~~~~\cdot e^{j\big((m+l)\cdot T_{\textsf{s}}-\beta \big) v}\cdot h_B\big((m+l)T_{\textsf{s}},(m+l)\cdot T_{\textsf{s}}-\beta\big)d\beta \bigg)dv\\
&\stackrel{(b)}{=}\frac{1}{4\pi} \cdot e^{j2\varphi_c} \cdot \int_{v=-\infty}^{\infty}  H_{B}(m \cdot T_{\textsf{s}},v)  \cdot e^{-j l \cdot T_{\textsf{s}}\cdot v} \nonumber \\
&~~~~~~~~~~~~~~~~~~~~~~~~\cdot \bigg( \int_{\beta=-\infty}^{\infty}  e^{j2v_c \beta} \cdot \Big(\sum_{p=-L_{\textsf{p}} }^{L_{\textsf{p}}} K_z (p,v-v_c)\cdot e^{j \omega_{\textsf{z}} \cdot p \beta}   \Big)\\
&~~~~~~~~~~~~~~~~~~~~~~~~~~~~~~~~~~~\cdot e^{j\big((m+l)\cdot T_{\textsf{s}}-\beta\big) v}\cdot h_B\big((m+l)T_{\textsf{s}},(m+l)\cdot T_{\textsf{s}}-\beta\big)d\beta \bigg)dv\\
&~~~~~+\frac{1}{4\pi} \cdot  e^{-2j \varphi_c} \cdot \int_{v=-\infty}^{\infty}  H_{B}(m \cdot T_{\textsf{s}},v) \cdot  e^{-j l \cdot T_{\textsf{s}}\cdot  v} \nonumber \\
&~~~~~~~~~~~~~~~~~~~~~~~~~~~~~~~~~\cdot \bigg( \int_{\beta=-\infty}^{\infty}  e^{-2jv_c\beta} \cdot \Big( \sum_{p=-L_{\textsf{p}} }^{L_{\textsf{p}}} K_z (p,v+v_c)\cdot e^{j \omega_{\textsf{z}} \cdot p \beta}  \Big )\\
&~~~~~~~~~~~~~~~~~~~~~~~~~~~~~~~~~~~~~\cdot e^{j\big((m+l)\cdot T_{\textsf{s}}-\beta \big) v}\cdot h_B\big((m+l)T_{\textsf{s}},(m+l)\cdot T_{\textsf{s}}-\beta\big)d\beta \bigg)dv
\end{align*}
\begin{align*} 
&\stackrel{(c)}{=}\frac{1}{4\pi} \cdot e^{j2\varphi_c} \cdot \int_{v=-\infty}^{\infty}  H_{B}(m \cdot T_{\textsf{s}},v)  \cdot e^{-j l \cdot T_{\textsf{s}}\cdot v} \nonumber \\
&~~~~~~~~~~~~~~~~~~~~~~~~\cdot \bigg( \sum_{p=-L_{\textsf{p}} }^{L_{\textsf{p}}} K_z (p,v-v_c) \cdot \int_{\beta=-\infty}^{\infty}  e^{j2v_c \beta} \cdot  e^{j \omega_{\textsf{z}} \cdot p \beta}  \nonumber \\
&~~~~~~~~~~~~~~~~~~~~~~~~~~~~~~~~~~~\cdot e^{j\big((m+l)\cdot T_{\textsf{s}}-\beta\big) v}\cdot h_B\big((m+l)T_{\textsf{s}},(m+l)\cdot T_{\textsf{s}}-\beta\big)d\beta \bigg)dv\nonumber\\
&~~~~~+\frac{1}{4\pi} \cdot  e^{-2j \varphi_c} \cdot \int_{v=-\infty}^{\infty}  H_{B}(m \cdot T_{\textsf{s}},v) \cdot  e^{-j l \cdot T_{\textsf{s}}\cdot  v} \nonumber \\
&~~~~~~~~~~~~~~~~~~~~~~~~~~~~~~~~~\cdot \bigg( \sum_{p=-L_{\textsf{p}} }^{L_{\textsf{p}}} K_z (p,v+v_c) \cdot \int_{\beta=-\infty}^{\infty}  e^{-2jv_c\beta} \cdot  e^{j \omega_{\textsf{z}} \cdot p \beta}  \nonumber\\
&~~~~~~~~~~~~~~~~~~~~~~~~~~~~~~~~~~~~~\cdot e^{j\big((m+l)\cdot T_{\textsf{s}}-\beta \big) v}\cdot h_B\big((m+l)T_{\textsf{s}},(m+l)\cdot T_{\textsf{s}}-\beta\big)d\beta \bigg)dv\nonumber\\
&\stackrel{(d)}{=}\frac{1}{4\pi} \cdot e^{j2\varphi_c} \cdot \int_{v=-\infty}^{\infty}  H_{B}(m \cdot T_{\textsf{s}},v)  \cdot e^{-j l \cdot T_{\textsf{s}}\cdot v} \nonumber \\
&~~~~~~~~~~~~~~~~\cdot \bigg( \sum_{p=-L_{\textsf{p}} }^{L_{\textsf{p}}} K_z (p,v-v_c) \cdot \int_{\mu=-\infty}^{\infty}  e^{j2v_c \cdot \big((m+l)T_{\textsf{s}}-\mu\big)} \cdot  e^{j \omega_{\textsf{z}} \cdot p \cdot \big((m+l)T_{\textsf{s}}-\mu\big)} \nonumber \\
&~~~~~~~~~~~~~~~~~~~~~~~~~~~~~~~~~~~~~~~~~~~~~~~~~~~~~~~~~~~~~~~~~~~~~~~\cdot e^{j \mu  v}\cdot h_B\big((m+l)T_{\textsf{s}},\mu \big)d\mu \bigg)dv \nonumber\\
&~~~~~+\frac{1}{4\pi} \cdot  e^{-2j \varphi_c} \cdot \int_{v=-\infty}^{\infty}  H_{B}(m \cdot T_{\textsf{s}},v) \cdot  e^{-j l \cdot T_{\textsf{s}}\cdot  v} \nonumber \\
&~~~~~~~~~~~~~~~~~~~~~~~~\cdot \bigg( \sum_{p=-L_{\textsf{p}} }^{L_{\textsf{p}}} K_z (p,v+v_c) \cdot \int_{\mu=-\infty}^{\infty}  e^{-2jv_c\cdot \big((m+l)T_{\textsf{s}}-\mu\big)} \cdot  e^{j \omega_{\textsf{z}} \cdot p \cdot \big((m+l)T_{\textsf{s}}-\mu\big)}  \nonumber \\
&~~~~~~~~~~~~~~~~~~~~~~~~~~~~~~~~~~~~~~~~~~~~~~~~~~~~~~~~~~~~~~~~~~~~~~~~~~\cdot e^{j \mu  v}\cdot h_B\big((m+l)T_{\textsf{s}},\mu \big)d\mu \bigg)dv \nonumber \\
&=\frac{1}{4\pi} \cdot e^{j\big(2\varphi_c+2v_c \cdot (m+l)T_{\textsf{s}}\big)} \cdot \int_{v=-\infty}^{\infty}  H_{B}(m \cdot T_{\textsf{s}},v)  \cdot e^{-j l \cdot T_{\textsf{s}}\cdot v} \nonumber \\
&~~~~~~\cdot \bigg( \sum_{p=-L_{\textsf{p}} }^{L_{\textsf{p}}} e^{j p \cdot  \omega_{\textsf{z}} \cdot (m+l)T_{\textsf{s}}}\cdot K_z (p,v-v_c) \nonumber \\
&~~~~~~~~~~~~~~~~~~~~~~~~~~~~~~~~\cdot \int_{\mu=-\infty}^{\infty}  h_B\big((m+l)T_{\textsf{s}},\mu \big) \cdot e^{j(v-2v_c-\omega_{\textsf{z}} \cdot p )\cdot \mu)}d\mu \bigg)dv \nonumber\\
&~~~~~+\frac{1}{4\pi} \cdot  e^{-j\big(2 \varphi_c+2v_c\cdot (m+l)T_{\textsf{s}}\big)} \cdot \int_{v=-\infty}^{\infty}  H_{B}(m \cdot T_{\textsf{s}},v) \cdot  e^{-j l \cdot T_{\textsf{s}}\cdot  v} \nonumber \\
&~~~~~~~~~\cdot \bigg( \sum_{p=-L_{\textsf{p}} }^{L_{\textsf{p}}}  e^{j p \cdot \omega_{\textsf{z}} \cdot(m+l)T_{\textsf{s}}}\cdot K_z (p,v+v_c)   \nonumber \\
&~~~~~~~~~~~~~~~~~~~~~~~~~~~~~~~~~~~~~~~~~~\cdot \int_{\mu=-\infty}^{\infty}  h_B\big((m+l)T_{\textsf{s}},\mu \big)\cdot e^{j( v+2v_c- \omega_{\textsf{z}} \cdot p)\cdot \mu} d\mu \bigg)dv \nonumber\\
\end{align*}
\begin{align} \label{First_proof_b}
&\stackrel{(e)}{=}\frac{1}{4\pi} \cdot e^{j\big(2\varphi_c+2v_c \cdot (m+l)T_{\textsf{s}}\big)} \cdot \int_{v=-\infty}^{\infty}  e^{-j l \cdot T_{\textsf{s}}\cdot v}\cdot  \sum_{p=-L_{\textsf{p}} }^{L_{\textsf{p}}} e^{j p \cdot  \omega_{\textsf{z}} \cdot (m+l)T_{\textsf{s}}} \nonumber \\
&~~~~~~~~~~~~~~~~\cdot K_z (p,v-v_c)\cdot  H_B\big((m+l)T_{\textsf{s}},v-2v_c-\omega_{\textsf{z}} \cdot p  \big)\cdot  H_{B}(m \cdot T_{\textsf{s}},v) dv  \nonumber\\
&~~~~~+\frac{1}{4\pi} \cdot  e^{-j\big(2 \varphi_c+2v_c\cdot (m+l)T_{\textsf{s}}\big)} \cdot \int_{v=-\infty}^{\infty}  e^{-j l \cdot T_{\textsf{s}}\cdot  v} \cdot  \sum_{p=-L_{\textsf{p}} }^{L_{\textsf{p}}}  e^{j p \cdot \omega_{\textsf{z}} \cdot(m+l)T_{\textsf{s}}} \nonumber \\
&~~~~~~~~~~~~~~~~~~\cdot K_z (p,v+v_c)\cdot  H_B\big((m+l)T_{\textsf{s}},v+2v_c- \omega_{\textsf{z}} \cdot p \big) \cdot H_{B}(m \cdot T_{\textsf{s}},v)  dv~,
\end{align}
where $(a)$ follows from the change of variables $\beta=(m+l)T_{\textsf{s}}-\tau$, $(b)$ follows from the Fourier series expansion of $C_z\big(\beta,v+v_c\big)$ and $C_z\big(\beta,v+v_c\big)$ detailed in Eq. (\ref{Finite_Fourier_series}), and $(c)$ follows since the integral is a linear operator and the sum is finite, $(d)$ follows from the change of variables $\mu=(m+l)T_{\textsf{s}}-\beta$, and $(e)$ follows from Fourier transform definition for $h_B\big((m+l)T_{\textsf{s}},\mu \big)$.
Next, we make the following observations:
\begin{enumerate}
\item 
The frequency response $H_{B}(m \cdot T_{\textsf{s}},v)$, is non zero only for
$-BW_h \leq v \leq BW_h$.
\item 
The frequency response $H_{B}(m \cdot T_{\textsf{s}},v+2v_c- \omega_{\textsf{z}} \cdot p)$, is non zero for\\ $-BW_h-2v_c + \omega_{\textsf{z}} \cdot p \leq v \leq BW_h-2v_c + \omega_{\textsf{z}}\cdot p$.
\item 
The frequency response $H_{B}(m \cdot T_{\textsf{s}},v-2v_c-\omega_{\textsf{z}} \cdot p)$, is non zero for\\ $-BW_h+2v_c + \omega_{\textsf{z}} \cdot p \leq v \leq BW_h+2v_c + \omega_{\textsf{z}}\cdot p$.
\end{enumerate}
From the facts 1, 2, and 3, we conclude that $H_B\big((m+l)T_{\textsf{s}},v+2v_c- \omega_{\textsf{z}} \cdot p \big) \cdot H_{B}(m \cdot T_{\textsf{s}},v)= 0$ if:
\begin{align} \label{condition_1}
&BW_h-2v_c + \omega_{\textsf{z}} \cdot p < -BW_h~ \Longleftrightarrow~ 2v_c >2BW_h+\omega_{\textsf{z}} \cdot p~,~|p|\leq L_{\textsf{P}},~ p \in \mathcal{Z}~,   
\end{align}
and $H_B\big((m+l)T_{\textsf{s}},v-2v_c-\omega_{\textsf{z}} \cdot p  \big)\cdot  H_{B}(m \cdot T_{\textsf{s}},v)= 0$ if:
\begin{align} \label{condition_2}
&-BW_h+2v_c + \omega_{\textsf{z}} \cdot p > BW_h~ \Longleftrightarrow~ 2v_c >2BW_h-\omega_{\textsf{z}} \cdot p~,~|p|\leq L_{\textsf{P}},~ p \in \mathcal{Z}~,  
\end{align}
since $v_c \gg BW_h$ and $v_c \gg \omega_{\textsf{z}} \cdot L_{\textsf{P}} $, the conditions in Eq. (\ref{condition_1}) and (\ref{condition_2}) are satisfied, accordingly $H_B\big((m+l)T_{\textsf{s}},v+2v_c- \omega_{\textsf{z}} \cdot p \big) \cdot H_{B}(m \cdot T_{\textsf{s}},v)= 0$ and $H_B\big((m+l)T_{\textsf{s}},v-2v_c-\omega_{\textsf{z}} \cdot p  \big)\cdot  H_{B}(m \cdot T_{\textsf{s}},v)= 0$, and consequently we have that
\begin{align} 
&\mathds{E}\big\{Z_R[m+l]Z_R[m]-Z_I[m+l]Z_I[m]\big\} \nonumber \\
&=\int_{\tau=-\infty}^{\infty}\bigg( \int_{u=-\infty}^{\infty}  \cos\Big(v_c \big(2(m+l)\cdot T_{\textsf{s}}-2\tau+u\big)+2\varphi_c \Big)\cdot c_z\big((m+l)\cdot T_{\textsf{s}}-\tau,u \big) \nonumber \\
&~~~~~~~~~~~~~~~~~~~~~~~~~~~~~~~~~~~~~~\cdot h_B(m\cdot T_{\textsf{s}},\tau-l\cdot T_{\textsf{s}}-u)du\bigg)\cdot h_B\big((m+l)\cdot T_{\textsf{s}},\tau\big)d\tau \nonumber
\end{align}
\begin{align} \label{DOT_1}
&\stackrel{(a)}{=}\frac{1}{4\pi} \cdot e^{j(2\varphi_c+2v_c \cdot \big((m+l)T_{\textsf{s}})} \cdot \int_{v=-\infty}^{\infty}\bigg(  e^{-j l \cdot T_{\textsf{s}}\cdot v}\cdot  \sum_{p=-L_{\textsf{p}} }^{L_{\textsf{p}}} e^{j p \cdot  \omega_{\textsf{z}} \cdot (m+l)T_{\textsf{s}}}\cdot K_z (p,v-v_c)\cdot  0 \bigg)dv \nonumber \\
&~~~~~+\frac{1}{4\pi} \cdot  e^{-j(2 \varphi_c+2v_c\cdot \big((m+l)T_{\textsf{s}})} \cdot \int_{v=-\infty}^{\infty} \bigg( e^{-j l \cdot T_{\textsf{s}}\cdot  v} \cdot  \sum_{p=-L_{\textsf{p}} }^{L_{\textsf{p}}}  e^{j p \cdot \omega_{\textsf{z}} \cdot(m+l)T_{\textsf{s}}}\cdot K_z (p,v+v_c)\cdot  0 \bigg) dv  \nonumber \\
&=0 ,~~~~~~~\mbox{Q.E.D}
\end{align}
where (a) follows from Eq. (\ref{First_proof_b}) and the discussion above.\\
Next, we consider the second condition for proper complexity, namely $\mathds{E}\big\{Z_{B,R}[m+l] Z_{B,I}[m]\big\}=-\mathds{E}\big\{Z_{B,I}[m+l] Z_{B,R}[m]\big\}$, or equivalently $\mathds{E}\big\{Z_{B,R}[m+l] Z_{B,I}[m]+Z_{B,I}[m+l] Z_{B,R}[m]\big\}=0$.\\
\begin{align*}
&\mathds{E}\big\{Z_{B,R}[m+l] Z_{B,I}[m]+Z_{B,I}[m+l] Z_{B,R}[m]\big\}\\
&=\mathds{E}\bigg\{\Big(\big(\cos(v_c\cdot t+\varphi_c)\cdot Z(t)\big)*h_{B}(t,\tau)\Big)\big((m+l)\cdot T_{\textsf{s}}\big)\\
&~~~~~~~~~~~~~~~~~~~~~~~~~~~~~~~~~~~~\cdot \Big(\big(-\sin(v_c\cdot t+\varphi_c)\cdot Z(t)\big)*h_{B}(t,\tau)\Big)(m\cdot T_{\textsf{s}})\\
&~~~~~~~~~~~~~~~~~~~~+\Big(\big(-\sin(v_c\cdot t+\varphi_c)\cdot Z(t)\big)*h_{B}(t,\tau)\Big)\big((m+l)\cdot T_{\textsf{s}}\big)\\
&~~~~~~~~~~~~~~~~~~~~~~~~~~~~~~~~~~~~~~~~~~~~~\cdot \Big(\big(\cos(v_c\cdot t+\varphi_c)\cdot Z(t)\big)*h_{B}(t,\tau)\Big)(m\cdot T_{\textsf{s}})\bigg\}\\
&=\mathds{E}\Bigg\{\bigg(\int_{\tau=-\infty}^{\infty} \cos\Big(v_c \big((m+l)\cdot T_{\textsf{s}}-\tau\big)+\varphi_c\Big)\\
&~~~~~~~~~~~~~~~~~~~~~\cdot Z\big((m+l)\cdot T_{\textsf{s}}-\tau\big)\cdot h_B\big((m+l)T_{\textsf{s}},\tau\big)d\tau \bigg)\\
&~~~~~~~~~~~~~~~~~\cdot \bigg(\int_{s=-\infty}^{\infty}- \sin\big(v_c (m\cdot T_{\textsf{s}}-s)+\varphi_c\big)\\
&~~~~~~~~~~~~~~~~~~~~~~~~~~~~~~~~~~~~~~~~~~~~\cdot Z(m\cdot T_{\textsf{s}}-s)\cdot h_B(m\cdot T_{\textsf{s}},s)ds \bigg)\\
&~~~~~~~~~~~~~~+\bigg(\int_{\tau=-\infty}^{\infty} -\sin\Big(v_c \big((m+l)\cdot T_{\textsf{s}}-\tau\big)+\varphi_c\Big)\\
&~~~~~~~~~~~~~~~~~~~~~~~~~~~~~~~~~~~~~~~~~~~\cdot Z\big((m+l)\cdot T_{\textsf{s}}-\tau\big)\cdot h_B\big((m+l)T_{\textsf{s}},\tau\big)d\tau \bigg)\\
&~~~~~~~~~~~~~~~~~~~~~~~~~~~~~~~~~~~\cdot \bigg(\int_{s=-\infty}^{\infty} \cos\big(v_c (m\cdot T_{\textsf{s}}-s)+\varphi_c\big)\\
&~~~~~~~~~~~~~~~~~~~~~~~~~~~~~~~~~~~~~~~~~~~~~~~~~~~~~~~~~~~~~~\cdot Z(m\cdot T_{\textsf{s}}-s)\cdot h_B(m\cdot T_{\textsf{s}},s)ds \bigg)\Bigg\}
\end{align*}
\begin{align*}
&=\mathds{E}\Bigg\{-\int_{\tau=-\infty}^{\infty} \int_{s=-\infty}^{\infty} \bigg( \cos\Big(v_c \big((m+l)\cdot T_{\textsf{s}}-\tau\big)+\varphi_c\Big)\cdot \sin\big(v_c (m\cdot T_{\textsf{s}}-s)+\varphi_c\big)\\
&~~~~~~~~~~~~~~~~~~~~~~~+\sin\Big(v_c \big((m+l)\cdot T_{\textsf{s}}-\tau\big)+\varphi_c\Big)\cdot \cos\big(v_c (m\cdot T_{\textsf{s}}-s)+\varphi_c\big)\bigg) \\
&~~~~~~~~\cdot Z\big((m+l)\cdot T_{\textsf{s}}-\tau\big)\cdot Z(m\cdot T_{\textsf{s}}-s)\cdot h_B(m\cdot T_{\textsf{s}},s)\cdot  h_{B}\big((m+l)\cdot T_{\textsf{s}},\tau\big)ds d\tau \Bigg\}\\
&\stackrel{(a)}{=}\mathds{E}\bigg\{-\int_{\tau=-\infty}^{\infty} \int_{s=-\infty}^{\infty}  \sin\Big(v_c \big((2m+l)\cdot T_{\textsf{s}}-\tau-s\big)+2\varphi_c\Big)\\
&~~~~~~~~\cdot Z\big((m+l\big)\cdot T_{\textsf{s}}-\tau)\cdot Z(m\cdot T_{\textsf{s}}-s)\cdot h_{B}(m\cdot T_{\textsf{s}},s)\cdot h_{B}\big((m+l)\cdot T_{\textsf{s}},\tau\big)ds d\tau \bigg\}\\
&\stackrel{(b)}{=}-\int_{\tau=-\infty}^{\infty} \int_{s=-\infty}^{\infty} \sin\Big(v_c \big((2m+l)\cdot T_{\textsf{s}}-\tau-s\big)+2\varphi_c\Big)\\
&~~~\cdot \mathds{E}\Big\{Z\big((m+l)\cdot T_{\textsf{s}}-\tau\big)\cdot Z(m\cdot T_{\textsf{s}}-s)\Big\}\cdot h_{B}(m\cdot T_{\textsf{s}},s)\cdot h_{B}\big((m+l)\cdot T_{\textsf{s}},\tau\big)ds d\tau \\
&=-\int_{\tau=-\infty}^{\infty} \int_{s=-\infty}^{\infty} \sin\Big(v_c \big((2m+l)\cdot T_{\textsf{s}}-\tau-s\big)+2\varphi_c\Big)\\
&~~~~~~~~~\cdot c_z\big((m+l)\cdot T_{\textsf{s}}-\tau,~\tau-s-l\cdot T_{\textsf{s}}\big)\cdot h_{B}(m\cdot T_{\textsf{s}},s)\cdot h_{B}\big((m+l)\cdot T_{\textsf{s}},\tau\big)ds d\tau\\
&\stackrel{(c)}{=}-\int_{\tau=-\infty}^{\infty} \bigg(\int_{u=-\infty}^{\infty} \sin\Big(v_c \big(u+2(m+l)\cdot T_{\textsf{s}}-2\tau\big)+2\varphi_c\Big)\\
&~~~~~~\cdot c_z\big((m+l)\cdot T_{\textsf{s}}-\tau,u\big)\cdot h_{B}(m\cdot T_{\textsf{s}},\tau-l\cdot T_{\textsf{s}} -u)du\bigg)\cdot h_{B}\big((m+l)\cdot T_{\textsf{s}},\tau\big)d\tau
\end{align*}
where $(a)$ follows from the trigonometric identity $\sin(\alpha+\beta)=\cos(\alpha)\sin(\beta)+\sin(\alpha)\cos(\beta)$ , $(b)$ follows since $Z(t)$ is a random process and stochastic expectation is a linear operator, and $(c)$ follows from change of variables $u=\tau-s-l\cdot T_{\textsf{s}}$. The rest of the proof that $\mathds{E}\big\{Z_{B,R}[m+l]\cdot Z_{B,I}[m]+Z_{B,I}[m+l]\cdot Z_{B,R}[m]\big\}=0$ is similar to the proof of $\mathds{E}\big\{Z_{B,R}[m+l]\cdot Z_{B,R}[m]-Z_{B,I}[m+l]\cdot Z_{B,I}[m]\big\}=0$, with the only difference being the use of the identity $\sin(v_cu)=\frac{e^{jv_cu}-e^{-jv_cu}}{2j}$ instead of the identity $\cos(v_cu)=\frac{e^{jv_cu}+e^{-jv_cu}}{2}$. We not that changing the proof of the first property by introducing the common factor of the imaginary unit, $j=\sqrt{-1}$, into the denominator and by replacing of the plus sign between the exponents with a minus sign, does not change the overall proof, therefore we obtain that:
\begin{ceqn}
\begin{align}
&\mathds{E}\big\{Z_{B,R}[m+l]\cdot Z_{B,I}[m]+Z_{B,I}[m+l]\cdot Z_{B,R}[m]\big\}=0~~ Q.E.D. \label{DOT_DOT_1} 
\end{align}
\end{ceqn}
It follows from (\ref{DOT_1}) and (\ref{DOT_DOT_1}) that $Z_{B}[m]$ is proper complex, next we will prove that $Z_{B}[m]$ is also cyclostationary. First recall that $\mathds{E}\big\{Z_B[m]\big\}=0$, hence the mean is periodic with any arbitrary period. Next we examine the periodicity of the correlation function:
\begin{ceqn}
\begin{align*}
c_{z_B}[m,l]&=\mathds{E}\big\{Z_{B}[m+l]Z_{B}^*[m]\big\}\\
&=\mathds{E}\Big\{\big( Z_{B,R}[m+l]+j\cdot Z_{B,I}[m+l]\big)\big(Z_{B,R}[m]+j\cdot Z_{B,I}[m]\big)^{*}\Big\}\\
&=\mathds{E}\Big\{\big( Z_{B,R}[m+l]+j\cdot Z_{B,I}[m+l]\big)\big(Z_{B,R}[m]-j\cdot Z_{B,I}[m]\big)\Big\}\\
&=\mathds{E}\Big\{\big( Z_{B,R}[m+l]Z_{B,R}[m]+Z_{B,I}[m+l]Z_{B,I}[m]\big)\\
&~~~~~~~~~~~~~~~+j\cdot\big(Z_{B,I}[m+l]Z_{B,R}[m]-Z_{B,R}[m+l]Z_{B,I}[m]\big)\Big\}\\
&\stackrel{(a)}{=} \mathds{E}\Big\{ Z_{B,R}[m+l]Z_{B,R}[m]+ Z_{B,I}[m+l]Z_{B,I}[m] \Big\} \\
&~~~~~~~~~~~~~~+j\cdot \mathds{E}\Big\{Z_{B,I}[m+l]Z_{B,R}[m]- Z_{B,R}[m+l]Z_{B,I}[m] \Big\}~, 
\end{align*}
\end{ceqn}
where $(a)$ follows since stochastic expectation is a linear operator. This implies that $c_{z_B}[m,l]=c_{z_B}[m+P_{\textsf{z}},l],~P_{\textsf{z}}=\mbox{constant},~P_{\textsf{z}}\in \mathcal{N},~l \in \mathcal{Z}$ if the following 2 conditions are satisfied:
\begin{enumerate}
\item $\mathds{E}\big\{ Z_{B,R}[m+l]Z_{B,R}[m]+ Z_{B,I}[m+l]Z_{B,I}[m]\big\}=\mathds{E}\big\{ Z_{B,R}[m+P_{\textsf{z}}+l]Z_{B,R}[m+P_{\textsf{z}}]+ Z_{B,I}[m+P_{\textsf{z}}+l]Z_{B,I}[m+P_{\textsf{z}}] \big\}$  
\item $\mathds{E}\big\{Z_{B,I}[m+l]Z_{B,R}[m]- Z_{B,R}[m+l]Z_{B,I}[m] \big\}=\mathds{E}\big\{Z_{B,I}[m+P_{\textsf{z}}+l]Z_{B,R}[m+P_{\textsf{z}}]- Z_{B,R}[m+P_{\textsf{z}}+l]Z_{B,I}[m+P_{\textsf{z}}] \big\}$
\end{enumerate} 
We first consider Condition 1. :
\begin{align*}
\mathds{E}\big\{&Z_{B,R}[m+l]Z_{B,R}[m]+Z_{B,I}[m+l]Z_{B,I}[m]\big\}\\
&=\mathds{E}\bigg\{\Big(\big(\cos(v_c\cdot t+\varphi_c)\cdot Z(t)\big)*h_{B}(t,\tau)\Big)\big((m+l)\cdot T_{\textsf{s}}\big)\\
&~~~~~~~~~~~~~~~~~~~~~~~~~~~~~~~~~~~~~~~~\cdot \Big(\big(\cos(v_c\cdot t+\varphi_c)\cdot Z(t)\big)*h_{B}(t,\tau)\Big)\big(m\cdot T_{\textsf{s}}\big)\\
&~~~~~~~~~+\Big(\big(-\sin(v_c\cdot t+\varphi_c)\cdot Z(t)\big)*h_{B}(t,\tau)\Big)((m+l)\cdot T_{\textsf{s}})\\
&~~~~~~~~~~~~~~~~~~~~~~~~~~~~~~~~~~~~~~~~\cdot \Big(\big(-\sin(v_c\cdot t+\varphi_c)\cdot Z(t)\big)*h_{B}(t,\tau)\Big)(m\cdot T_{\textsf{s}}) \bigg\}
\end{align*}
\begin{align*}
&=\mathds{E}\Bigg\{\bigg(\int_{\tau=-\infty}^{\infty} \cos\Big(v_c \big((m+l)\cdot T_{\textsf{s}}-\tau \big)+\varphi_c \Big)\cdot Z\big((m+l)\cdot T_{\textsf{s}}-\tau \big)\\
&~~~~~~~~~~~~~~~~~~~~~~~~~~~~~~~~~~~~~~~~~~~~~~~~~~~~~~~~~~~\cdot h_{B}\big((m+l)\cdot T_{\textsf{s}},\tau\big)d\tau \bigg)\\
&~~~~~~~~~~~~~~~~~~~~~~~~~~\cdot \bigg(\int_{s=-\infty}^{\infty} \cos\Big(v_c \big(m\cdot T_{\textsf{s}}-s \big)+\varphi_c\Big)\cdot Z\big(m\cdot T_{\textsf{s}}-s \big)\\
&~~~~~~~~~~~~~~~~~~~~~~~~~~~~~~~~~~~~~~~~~~~~~~~~~~~~~~~~~~~~~~~~~~~~~~~\cdot h_{B}\big(m\cdot T_{\textsf{s}},s\big)ds \bigg)\\
&~~~~~~~~~~~+\bigg(\int_{\tau=-\infty}^{\infty} -\sin\Big(v_c \big((m+l)\cdot T_{\textsf{s}}-\tau \big)+\varphi_c \Big)\cdot Z\big((m+l)\cdot T_{\textsf{s}}-\tau \big)\\
&~~~~~~~~~~~~~~~~~~~~~~~~~~~~~~~~~~~~~~~~~~~~~~~~~~~~~~~~~~~~~~~~~~~~\cdot h_{B}\big((m+l)\cdot T_{\textsf{s}},\tau\big)d\tau \bigg)\\
&~~~~~~~~~~~~~~~~~~~~~~~~~~~~~~~\cdot \bigg(\int_{s=-\infty}^{\infty} -\sin\Big(v_c \big(m\cdot T_{\textsf{s}}-s \big)+\varphi_c \Big)\cdot Z\big(m\cdot T_{\textsf{s}}-s \big)\\
&~~~~~~~~~~~~~~~~~~~~~~~~~~~~~~~~~~~~~~~~~~~~~~~~~~~~~~~~~~~~~~~~~~~~~~~~~~~~~~\cdot h_{B}\big(m\cdot T_{\textsf{s}},s\big)ds \bigg)\Bigg\}\\
&=\mathds{E}\bigg\{\int_{\tau=-\infty}^{\infty} \int_{s=-\infty}^{\infty} \cos\Big(v_c \big((m+l)\cdot T_{\textsf{s}}-\tau\big)+\varphi_c\Big)\cdot\cos\Big(v_c (m\cdot T_{\textsf{s}}-s)+\varphi_c\Big) \\
&~~~~~\cdot Z\big((m+l)\cdot T_{\textsf{s}}-\tau\big)\cdot Z\big(m\cdot T_{\textsf{s}}-s\big)\cdot h_{B}\big((m+l)\cdot T_{\textsf{s}},\tau\big)\cdot h_{B}(m\cdot T_{\textsf{s}},s)d\tau ds \\
&~~~~~~~+\int_{\tau=-\infty}^{\infty} \int_{s=-\infty}^{\infty}\sin\Big(v_c \big((m+l)\cdot T_{\textsf{s}}-\tau \big)+\varphi_c\Big)\cdot \sin\Big(v_c (m\cdot T_{\textsf{s}}-s)+\varphi_c\Big)\\
&~~~~\cdot Z\big((m+l)\cdot T_{\textsf{s}}-\tau\big)\cdot Z\big(m\cdot T_{\textsf{s}}-s\big)\cdot h_{B}\big((m+l)\cdot T_{\textsf{s}},\tau\big)\cdot h_{B}\big(m\cdot T_{\textsf{s}},s\big)d\tau ds\bigg\}\\
&=\mathds{E}\Bigg\{\int_{\tau=-\infty}^{\infty} \int_{s=-\infty}^{\infty}  \bigg(\cos\Big(v_c \big((m+l)\cdot T_{\textsf{s}}-\tau\big)+\varphi_c\Big)\cdot \cos\big(v_c (m\cdot T_{\textsf{s}}-s)+\varphi_c\big)\\
&~~~~~~~~~~~~~~~~~~~~~~~~~~~+\sin\Big(v_c \big((m+l)\cdot T_{\textsf{s}}-\tau\big)+\varphi_c\Big)\cdot \sin\big(v_c (m\cdot T_{\textsf{s}}-s)+\varphi_c\big)\bigg) \\
&~~~~~~~~~~~~~~~~~~~~~~~~~~~~~~~~~~~~~~~~~~~\cdot \Big(Z\big((m+l)\cdot T_{\textsf{s}}-\tau\big)\cdot Z(m\cdot T_{\textsf{s}}-s)\\
&~~~~~~~~~~~~~~~~~~~~~~~~~~~~~~~~~~~~~~~~~~~~~~~~~~~~~\cdot h_{B}\big((m+l)\cdot T_{\textsf{s}},\tau\big)\cdot h_{B}\big(m\cdot T_{\textsf{s}},s\big)\Big)d\tau ds \Bigg\}\\
&\stackrel{(a)}{=}\mathds{E}\bigg\{\int_{\tau=-\infty}^{\infty} \int_{s=-\infty}^{\infty}  \cos\Big(v_c \big(l\cdot T_{\textsf{s}}-\tau+s\big)\Big) \\
&~~~~~~~~~~~~~~~~~~~~~~~~~~~~~~~~\cdot \Big(Z\big((m+l)\cdot T_{\textsf{s}}-\tau\big)\cdot Z\big(m\cdot T_{\textsf{s}}-s)\\
&~~~~~~~~~~~~~~~~~~~~~~~~~~~~~~~~~~~~~~~~~~~~~~~~~~~~~\cdot h_{B}\big((m+l)\cdot T_{\textsf{s}},\tau\big)\cdot h_{B}\big(m\cdot T_{\textsf{s}},s\big)\Big)d\tau ds \bigg\}\\
&\stackrel{(b)}{=}\int_{\tau=-\infty}^{\infty} \int_{s=-\infty}^{\infty} \cos\Big(v_c \big(l\cdot T_{\textsf{s}}-\tau+s\big)\Big) \\
&~~~~~~~~~~~~~~~~~~~~~~~~~~~~~~~\cdot \mathds{E}\bigg\{\Big(Z\big((m+l)\cdot T_{\textsf{s}}-\tau\big)\cdot Z\big(m\cdot T_{\textsf{s}}-s)\bigg\} \\
&~~~~~~~~~~~~~~~~~~~~~~~~~~~~~~~~~~~~~~~~~~~~~~~~~~~~~~\cdot h_{B}\big((m+l)\cdot T_{\textsf{s}},\tau\big)\cdot h_{B}\big(m\cdot T_{\textsf{s}},s\big)\Big)d\tau ds+;p
\end{align*}
\begin{align*}
&=\int_{\tau=-\infty}^{\infty} \int_{s=-\infty}^{\infty} \cos\Big(v_c \big(l\cdot T_{\textsf{s}}-\tau+s\big)\Big) \\
&~~~~~~~~\cdot c_z\big((m+l)\cdot T_{\textsf{s}}-\tau,~\tau-s-l\cdot T_{\textsf{s}}\big)\cdot h_{B}\big((m+l)\cdot T_{\textsf{s}},\tau\big)\cdot h_{B}\big(m\cdot T_{\textsf{s}},s\big)\Big)d\tau ds \\
&\stackrel{(c)}{=}\int_{\tau=-\infty}^{\infty} \int_{s=-\infty}^{\infty} \cos\Big(v_c \big(l\cdot T_{\textsf{s}}-\tau+s\big)\Big)\\
&~~~~~~~~~~~~~~~~~~~~\cdot c_z\big((m+l)\cdot T_{\textsf{s}}-\tau+k_2\cdot\alpha_2 \cdot T_{\textsf{z}},~\tau-s-l\cdot T_{\textsf{s}}\big) \\
&~~~~~~~~~~~~~~~~~~~~ \cdot h_{B}\big((m+l)\cdot T_{\textsf{s}}+k_1\cdot \alpha_1 \cdot T_{\textsf{h}},\tau\big)\cdot h_{B}\big(m\cdot T_{\textsf{s}}+k_1\cdot \alpha_1 \cdot T_{\textsf{h}},s\big)\Big)d\tau ds \\
&\stackrel{(d)}{=}\int_{\tau=-\infty}^{\infty} \int_{s=-\infty}^{\infty} \cos\Big(v_c \big(l\cdot T_{\textsf{s}}-\tau+s\big)\Big)\\
&~~~~~~~~~~~~~~~~~~~~~~~~~~~~~~\cdot c_z\big((m+l)\cdot T_{\textsf{s}}-\tau+k_2\cdot P_{\textsf{tmp}}\cdot T_{\textsf{s}},~\tau-s-l\cdot T_{\textsf{s}}\big) \\
&~~~~~~~~~~~~~~~~~~\cdot h_{B}\big((m+l)\cdot T_{\textsf{s}}+k_1\cdot P_{\textsf{h}}\cdot T_{\textsf{s}},\tau\big)\cdot h_{B}\big(m\cdot T_{\textsf{s}}+k_1\cdot P_{\textsf{h}}\cdot T_{\textsf{s}},s\big)\Big)d\tau ds \\
&\stackrel{(e)}{=}\int_{\tau=-\infty}^{\infty} \int_{s=-\infty}^{\infty} \cos\Big(v_c \big(l\cdot T_{\textsf{s}}-\tau+s\big)\Big)\cdot c_z\big((m+l+P_{\textsf{z}})\cdot T_{\textsf{s}}-\tau,~\tau-s-l\cdot T_{\textsf{s}}\big) \\
&~~~~~~~~~~~~~~~~~~~~~~~~~~~~~~~~~~~~~\cdot h_{B}\big((m+l+P_{\textsf{z}})\cdot T_{\textsf{s}},\tau\big)\cdot h_{B}\big((m+P_{\textsf{z}})\cdot T_{\textsf{s}},s\big)\Big)d\tau ds \\
&=\mathds{E}\Big\{ Z_{B,R}[m+P_{\textsf{z}}+l]Z_{B,R}[m+P_{\textsf{z}}]+ Z_{B,I}[m+P_{\textsf{z}}+l]Z_{B,I}[m+P_{\textsf{z}}] \Big\}~~~~~~~\mbox{Q.E.D}
\end{align*}
where $(a)$ follows from the trigonometric identity: $\cos(\alpha-\beta)=\cos(\alpha)\cos(\beta)+\sin(\alpha)\sin(\beta)$, $(b)$ follows as only $Z(t)$ is random and the stochastic expectation is a linear operator,$~(c)$ follows since $c_z(t,\tau)=c_z(t+k\cdot T_{\textsf{z}},\tau)$ and $h_{B}(t,\tau)=h_{B}(t+k\cdot T_{\textsf{h}},\tau),~\forall k \in \mathcal{Z}$, and we note that $\alpha_1,~k_1,~\alpha_2,~k_2 \in \mathcal{Z}$, $(d)$ follows since by (\ref{Period_constants_1}),$~\alpha_2 \cdot T_{\textsf{z}}=P_{\textsf{tmp}}\cdot T_{\textsf{s}},~\alpha_1 \cdot T_{\textsf{h}}=P_{\textsf{h}}\cdot T_{\textsf{s}}$, and lastly, $(e)$ follows since by (\ref{Period_constants_2}), $~P_{\textsf{z}}=k_2\cdot P_{\textsf{tmp}}=k_1\cdot P_{\textsf{h}}$.
Next, we prove Condition 2. :
\begin{align*}
\mathds{E}\Big\{&Z_{B,I}[m+l]Z_{B,R}[m]- Z_{B,R}[m+l]Z_{B,I}[m] \Big\}\\
&=-\mathds{E}\Big\{Z_{B,R}[m+l]\cdot Z_{B,I}[m]-Z_{B,I}[m+l]\cdot Z_{B,R}[m]\Big\}\\
&=-\mathds{E}\bigg\{\Big(\big(\cos(v_c\cdot t+\varphi_c)\cdot Z(t)\big)*h_{B}(t,\tau)\Big)\big((m+l)\cdot T_{\textsf{s}}\big) \\
&~~~~~~~~~~~~~~~~~~~~~~~~~~~~~~~~~~~~~~\cdot \Big(\big(-\sin(v_c\cdot t+\varphi_c)\cdot Z(t)\big)*h_{B}(t,\tau)\Big)(m\cdot T_{\textsf{s}})\\
&~~~~~~~~~~-\Big(\big(-\sin(v_c\cdot t+\varphi_c)\cdot Z(t)\big)*h_{B}(t,\tau)\Big)\big((m+l)\cdot T_{\textsf{s}}\big)\\
&~~~~~~~~~~~~~~~~~~~~~~~~~~~~~~~~~~~~~~~~~~\cdot \Big(\big(\cos(v_c\cdot t+\varphi_c)\cdot Z(t)\big)*h_{B}(t,\tau)\Big)(m\cdot T_{\textsf{s}})\bigg\}\\
\end{align*}
\begin{align*}
&=-\mathds{E}\Bigg\{\bigg(\int_{\tau=-\infty}^{\infty} \cos\Big(v_c \big((m+l)\cdot T_{\textsf{s}}-\tau\big)+\varphi_c\Big)\\
&~~~~~~~~~~~~~~~~~~~~~~~~~~~~~~~~~~~\cdot Z\big((m+l)\cdot T_{\textsf{s}}-\tau\big)\cdot h_{B}\big((m+l)\cdot T_{\textsf{s}},\tau\big)d\tau \bigg)\\
&~~~~~~~~~~~~~~~~~~~\cdot \bigg(\int_{s=-\infty}^{\infty}- \sin\big(v_c (m\cdot T_{\textsf{s}}-s)+\varphi_c\big)\\
&~~~~~~~~~~~~~~~~~~~~~~~~~~~~~~~~~~~~~~~~~~~~~~~~~~~~~\cdot Z(m\cdot T_{\textsf{s}}-s)\cdot h_{B}(m\cdot T_{\textsf{s}},s)ds \bigg)\\
&~~~~~~~~~~~~-\bigg(\int_{\tau=-\infty}^{\infty} -\sin\Big(v_c \big((m+l)\cdot T_{\textsf{s}}-\tau\big)+\varphi_c\Big)\\
&~~~~~~~~~~~~~~~~~~~~~~~~~~~~~~~~~~~~~~~\cdot Z\big((m+l)\cdot T_{\textsf{s}}-\tau\big)\cdot h_{B}\big((m+l)\cdot T_{\textsf{s}},\tau\big)d\tau \bigg)\\
&~~~~~~~~~~~~~~~~~~~~\cdot \bigg(\int_{s=-\infty}^{\infty} \cos\big(v_c (m\cdot T_{\textsf{s}}-s)+\varphi_c\big)\\\
&~~~~~~~~~~~~~~~~~~~~~~~~~~~~~~~~~~~~~~~~~~~~~~~~~~~~~~~~\cdot Z(m\cdot T_{\textsf{s}}-s)\cdot h_{B}(m\cdot T_{\textsf{s}},s)ds \bigg)\Bigg\}\\
&=\mathds{E}\Bigg\{\int_{\tau=-\infty}^{\infty} \int_{s=-\infty}^{\infty} \bigg( \cos\Big(v_c \big((m+l)\cdot T_{\textsf{s}}-\tau\big)+\varphi_c\Big)\cdot \sin\big(v_c (m\cdot T_{\textsf{s}}-s)+\varphi_c\big)\\
&~~~~~~~~~~~~~~~~~-\sin\Big(v_c \big((m+l)\cdot T_{\textsf{s}}-\tau\big)+\varphi_c\Big)\cdot \cos\big(v_c (m\cdot T_{\textsf{s}}-s)+\varphi_c\big)\bigg) \\
&~~~~~~~~\cdot Z\big((m+l)\cdot T_{\textsf{s}}-\tau\big)\cdot Z(m\cdot T_{\textsf{s}}-s)\cdot h_{B}\big((m+l)\cdot T_{\textsf{s}},\tau\big)\cdot h_{B}(m\cdot T_{\textsf{s}},s)ds d\tau \Bigg\}\\
&\stackrel{(a)}{=}\mathds{E}\Bigg\{\int_{\tau=-\infty}^{\infty} \int_{s=-\infty}^{\infty} - \sin\Big(v_c \big(l\cdot T_{\textsf{s}}-\tau+s\big)\Big)\\
&~~~~~~~~\cdot Z\big((m+l\big)\cdot T_{\textsf{s}}-\tau)\cdot Z(m\cdot T_{\textsf{s}}-s)\cdot h_{B}\big((m+l)\cdot T_{\textsf{s}},\tau\big)\cdot h_{B}(m\cdot T_{\textsf{s}},s)ds d\tau \Bigg\}\\
&\stackrel{(b)}{=}\int_{\tau=-\infty}^{\infty} \int_{s=-\infty}^{\infty}  -\sin\Big(v_c \big(l\cdot T_{\textsf{s}}-\tau+s\big)\Big)\\
&~~~\cdot \mathds{E}\Big\{Z\big((m+l\big)\cdot T_{\textsf{s}}-\tau)\cdot Z(m\cdot T_{\textsf{s}}-s)\Big\}\cdot h_{B}\big((m+l)\cdot T_{\textsf{s}},\tau\big)\cdot h_{B}(m\cdot T_{\textsf{s}},s)ds d\tau \\
&=\int_{\tau=-\infty}^{\infty} \int_{s=-\infty}^{\infty}  -\sin\Big(v_c \big(l\cdot T_{\textsf{s}}-\tau+s\big)\Big)\\
&~~~~~~~~~~\cdot c_z\big((m+l)\cdot T_{\textsf{s}}-\tau,~\tau-s-l\cdot T_{\textsf{s}}\big)\cdot h_{B}\big((m+l)\cdot T_{\textsf{s}},\tau\big)\cdot h_{B}(m\cdot T_{\textsf{s}},s)ds d\tau \\
&\stackrel{(c)}{=}\int_{\tau=-\infty}^{\infty} \int_{s=-\infty}^{\infty}  -\sin\Big(v_c \big(l\cdot T_{\textsf{s}}-\tau+s\big)\Big)\\
&~~~~~~~~~~~~\cdot c_z\big((m+l)\cdot T_{\textsf{s}}-\tau+k_2\cdot \alpha_2 \cdot T_{\textsf{z}},~\tau-s-l\cdot T_{\textsf{s}}\big)\\
&~~~~~~~~~~~~~~~~~~~~~~ \cdot h_{B}\big((m+l)\cdot T_{\textsf{s}}+k_1\cdot \alpha_1 \cdot T_{\textsf{h}},\tau\big)\cdot h_{B}(m\cdot T_{\textsf{s}}+k_1\cdot \alpha_1 \cdot T_{\textsf{h}},s) ds d\tau 
\end{align*}
\begin{align*}
&\stackrel{(d)}{=}\int_{\tau=-\infty}^{\infty} \int_{s=-\infty}^{\infty}  -\sin\Big(v_c \big(l\cdot T_{\textsf{s}}-\tau+s\big)\Big)\\
&~~~~~~~~~~~~~~\cdot c_z\big((m+l)\cdot T_{\textsf{s}}-\tau+k_2 \cdot P_{\textsf{tmp}}\cdot T_{\textsf{s}},~\tau-s-l\cdot T_{\textsf{s}}\big)\\
&~~~~~~~~~~~~~~~~~~~~~ \cdot h_{B}\big((m+l)\cdot T_{\textsf{s}}+k_1\cdot P_{\textsf{h}}\cdot T_{\textsf{s}},\tau\big)\cdot h_{B}(m\cdot T_{\textsf{s}}+k_1\cdot P_{\textsf{h}}\cdot T_{\textsf{s}},s) ds d\tau \\
&\stackrel{(e)}{=}\int_{\tau=-\infty}^{\infty} \int_{s=-\infty}^{\infty}  -\sin\Big(v_c \big(l\cdot T_{\textsf{s}}-\tau+s\big)\Big)\cdot c_z\big((m+l+P_{\textsf{z}})\cdot T_{\textsf{s}}-\tau,~\tau-s-l\cdot T_{\textsf{s}}\big)\\
&~~~~~~~~~~~~~~~~~~~~~~~~~~~~~~~~~~~~~~~~ \cdot h_{B}\big((m+l+P_{\textsf{z}})\cdot T_{\textsf{s}},\tau\big)\cdot h_{B}\big((m+P_{\textsf{z}})\cdot T_{\textsf{s}},s\big) ds d\tau \\
&=\mathds{E}\Big\{Z_{B,I}[m+P_{\textsf{z}}+l]Z_{B,R}[m+P_{\textsf{z}}]- Z_{B,R}[m+P_{\textsf{z}}+l]Z_{B,I}[m+P_{\textsf{z}}] \Big\},~~~~~~~\mbox{Q.E.D}
\end{align*}
where $(a)$ follows from the trigonometric identity: $-\sin(\alpha-\beta)=\cos(\alpha)\sin(\beta)-\sin(\alpha)\cos(\beta)$, $(b)$ follows since $Z(t)$ is a random process and the stochastic expectation is a linear operator,$~(c)$ follows since $c_z(t,\tau)=c_z(t+k\cdot T_{\textsf{z}},\tau)$ and $h_{B}(t,\tau)=h_{B}(t+k\cdot T_{\textsf{h}},\tau),~\forall k \in \mathbb{Z}$,and $h_{B}(t,\tau)=h_{B}(t+k\cdot T_{\textsf{h}},\tau),~\forall k \in \mathcal{Z}$, and we note that $\alpha_1,~k_1,~\alpha_2,~k_2 \in \mathcal{Z}$, $(d)$ follows since by (\ref{Period_constants_1}), $~\alpha_2 \cdot T_{\textsf{z}}=P_{\textsf{tmp}}\cdot T_{\textsf{s}},~\alpha_1 \cdot T_{\textsf{h}}=P_{\textsf{h}}\cdot T_{\textsf{s}}$, and lastly $(e)$ follows since by (\ref{Period_constants_2})  $P_{\textsf{z}}=k_2\cdot P_{\textsf{tmp}}=k_1\cdot P_{\textsf{h}}$. Therefore, $c_{z_B}[m,l]=c_{z_B}[m+P_{\textsf{z}},l]$ and recalling that $\mathds{E}\big\{Z_B[m]\big\}=0$, we conclude that $Z_B[m]$ is proper complex and cyclostationary.
The last step in proving that $Z_B[m]$ is proper complex cyclostationary \emph{Gaussian} process, is to show that $Z_B[m]$ has complex Gaussian distribution. By definition \cite{lapidoth2017foundation}, $Z_{B}[m]=Z_{B,R}[m]+j\cdot Z_{B,I}[m]$ is a complex Gaussian process if 
$ \mathbf{Z_B}[m]=
\begin{bmatrix}
Z_{B,R}[m]\\
Z_{B,I}[m]\\
\end{bmatrix}$ 
is a jointly Gaussian vector process. We thus examine
$ \mathbf{Z_B}(t)=
\begin{bmatrix}
Z_{B,R}(t)\\
Z_{B,I}(t)\\
\end{bmatrix}$:
\begin{ceqn}
\begin{align*}
\mathbf{Z_B}(t)&=\begin{bmatrix}
Z_{B,R}(t)\\
Z_{B,I}(t)\\
\end{bmatrix}\\
&=\begin{bmatrix}
\Big(\big(\cos(v_c\cdot t+\varphi_c)\cdot Z(t)\big)*h_{B}(t,\tau)\Big)(t)\\
\Big(\big(-\sin(v_c\cdot t+\varphi_c)\cdot Z(t)\Big)*h_{B}(t,\tau)\big)(t)\\
\end{bmatrix}\\
&=\begin{bmatrix}
\bigg(\int_{s=-\infty}^{\infty} \cos\Big(v_c \big(t-s \big)+\varphi_c\Big)\cdot Z\big(t-s \big)\cdot h\big(t,s\big)ds \bigg)\\
\bigg(\int_{\tau=-\infty}^{\infty}- \sin\Big(v_c \big(t-\tau \big)+\varphi_c \Big)\cdot Z\big(t-\tau \big)\cdot h\big(t,\tau\big)d\tau \bigg)\\
\end{bmatrix}\\
&\stackrel{(a)}{=}\begin{bmatrix}
\bigg(\int_{s=-\infty}^{\infty} \tilde{h}_1\big(t,s\big) \cdot Z\big(t-s \big) ds \bigg)\\
\bigg(\int_{\tau=-\infty}^{\infty} \tilde{h}_2\big(t,\tau\big) \cdot Z\big(t-\tau \big) d\tau \bigg)\\
\end{bmatrix}
\end{align*}
\end{ceqn}
where $(a)$ follows by defining $\tilde{h}_1\big(t,s\big)= \cos\big(v_c (t-s )+\varphi_c\big)\cdot h\big(t,s\big)$ and $\tilde{h}_2\big(t,s\big)= -\sin\big(v_c (t-s )+\varphi_c \big)\cdot h\big(t,s\big)$. $Z_{B,R}(t)$ and $Z_{B,I}(t)$ are linear transformations of a Gaussian process $Z(t)$, that are applied simultaneously to $Z(t)$. Therefore, $Z_{B,R}(t)$ and $Z_{B,I}(t)$ are jointly Gaussian real processes, and since by definition $\mathbf{Z_B}[m]=\mathbf{Z_B}(mT_{\textsf{s}}),~m \in \mathcal{Z}$, then $\mathbf{Z_B}[m]$ is also a jointly Gaussian random vector process. We therefore conclude that $Z_B[m]$ is a complex Gaussian process Q.E.D.
\subsubsection{Summary of Assumption for Baseband Representation}
In this subsection we present the assumptions used for the derivation of the baseband equivalent noise model for passband real ACGN derived in this appendix:
\begin{enumerate}
\item [\textbf{\textit{AS1}}]
It is assumed that $C_{z}(t,v)$ the Fourier transform of $c_{z}(t,\tau)$ w.r.t $\tau$ is bandlimited around a center frequency $v_c$, that is: 
\begin{align*}
&C_{z}(t,v)\neq 0, ~~\forall |v-v_c|\leq BW_z,~|v+v_c|\leq BW_z,
~BW_z~ \mbox{is a positive constant,}\\
&v_c\gg BW_z~. 
\end{align*}
\item [\textbf{\textit{AS2}}]
It is assumed that $H_{B}(t,v)$ denote the Fourier transform of $h_{B}(t,\tau)$ w.r.t $\tau $ satisfies:
\begin{ceqn}
\begin{align*}
& H_{B}(t,v)=0,~\forall |v|\geq BW_h,~v_c\gg BW_h\gg BW_z~.    
\end{align*}
\end{ceqn}
\item [\textbf{\textit{AS3}}]
It is assumed that the number of Fourier series coefficient for $C_{z}(t,v)$ is finite, that is for a very large index $L_{\textsf{p}}\gg 1,~ L_{\textsf{p}} \in \mathcal{N},~K_z (p_1,v) =0,\mbox{if}~|p_1|>L_{\textsf{p}}$, accordingly:
\begin{ceqn}
\begin{align*} 
&C_{z}(t,v)= \sum_{p=-L_{\textsf{p}} }^{L_{\textsf{p}}} K_z (p,v) \cdot e^{j \omega_{\textsf{z}} \cdot p \cdot t} ,~ \omega_{\textsf{z}}=\frac{2 \pi}{T_{\textsf{z}}}~,v_c\gg \omega_{\textsf{z}} \cdot L_{\textsf{p}}.
\end{align*}
\end{ceqn}
\item [\textbf{\textit{AS4}}]
It is assumed that $T_{\textsf{s}}$ is a divisor of an integer multiple of $T_{\textsf{h}}$, and of an integer multiple of $T_{\textsf{z}}$, that is:
\begin{ceqn}
\begin{align*} 
\exists \alpha_1,\alpha_2 \in \mathcal{N}~s.t.~~ \frac{\alpha_1 \cdot T_{\textsf{h}}}{T_{\textsf{s}}}=P_{\textsf{h}}\in \mathcal{N},~\frac{\alpha_2 \cdot T_{\textsf{z}}}{T_{\textsf{s}}}=P_{\textsf{tmp}}\in \mathcal{N}
\end{align*}
\end{ceqn}
\end{enumerate}
\subsection{Derivations for Section \ref{chap:LRT_Derivation}}
\label{app:Derivations_Chapter_LRT}
\subsubsection{Proof that \texorpdfstring{$\mathds{E}\big\{\tilde{\mathbf{R}}_{N,0}[n]\big|H_0\big\}=\mathbf{0}$}~}
\label{app:Proof_H_0_R}
\arraycolsep=1.0pt\def\arraystretch{0.1}
\begin{align}
&\mathds{E}\big\{\tilde{\mathbf{R}}_{N,0}[n]\big|H_0\big\} \nonumber\\ 
&=\mathds{E}\left\{\left. \left( 
\begin{array}{l}
R[nN] \\
R[nN-1] \\
\vdots\\
R[nN-(K-1)]\\
\end{array}\right)\right| H_0\right\} \nonumber \\ 
&\stackrel{(a)}{=}\mathds{E}\left\{\left. \left( 
\begin{array}{l}
\sum\limits_{l=0}^{L_{\textsf{ch}}} h[nN,l]S[nN-l]+Z[nN]\\
\sum\limits_{l=0}^{L_{\textsf{ch}}} h[nN-1,l]S[nN-1-l]+Z[nN-1]\\
\vdots\\
\sum\limits_{l=0}^{L_{\textsf{ch}}} h[nN-(K-1),l]S[nN-(K-1)-l]+Z[nN-(K-1)]\\
\end{array}\right)\right| H_0 \right\} \nonumber\\
&\stackrel{(b)}{=}
\left(\begin{array}{l}
\sum\limits_{l=0}^{L_{\textsf{ch}}} h[nN,l]\cdot \mathds{E}\big\{S[nN-l]\big|H_0\big\}+\mathds{E}\big\{Z[nN]\big|H_0\big\}\\
\sum\limits_{l=0}^{L_{\textsf{ch}}} h[nN-1,l]\cdot\mathds{E}\big\{S[nN-1-l]\big|H_0\big\}+\mathds{E}\big\{Z[nN-1]\big|H_0\big\}\\
\vdots\\
\left(\begin{array}{l}
\sum\limits_{l=0}^{L_{\textsf{ch}}} h[nN-(K-1),l]\cdot\mathds{E}\big\{S[nN-(K-1)-l]\big|H_0\big\}\\
~~~~~~~~~~~~~~~~~~~~~~~~~~~~~~+\mathds{E}\big\{Z[nN-(K-1)]\big|H_0\big\}\end{array}\right)\\
\end{array}\right) \nonumber\\ 
&\stackrel{(c)}{=}
\left(\begin{array}{l}
\sum\limits_{l=0}^{L_{\textsf{ch}}} h[nN,l] \cdot 0+0\\
\sum\limits_{l=0}^{L_{\textsf{ch}}} h[nN-1,l] \cdot 0+0\\
\vdots\\
\sum\limits_{l=0}^{L_{\textsf{ch}}} h[nN-(K-1),l] \cdot 0+0\\
\end{array}\right)\ 
=
\left(\begin{array}{l}
~0~~\\
~\\
~\\
~0~~\\
~\\
~\\
~\vdots~~\\
~\\
~\\
~0~~\\
\end{array}\right), \qquad \forall n \in \mathbb{Z},\label{Fisrt_wss_1}
\end{align}
where $(a)$ follows from the relationship in (\ref{Rreceived_signal_model}), $(b)$ follows since stochastic expectation is a linear operator and only $Z[n]$ and $S[n]$ are random, and $(c)$ follows since by (\ref{Zero_Noise_Expectation}) and (\ref{Zero_Symbols_Expectation}), $\mathds{E}\big\{Z[n]\big|H_0\big\}=0,~\mathds{E}\big\{S[n]\big|H_0\big\}=0,~\forall n \in \mathcal{Z}$ .
\subsubsection{the General Element of \texorpdfstring{$\mathds{E}\big\{\tilde{\mathbf{R}}_{N,0}[n]\cdot(\tilde{\mathbf{R}}_{N,0}[n+l])^{H} |H_0\big\}$}~ Matrix}
\label{app:General_element_H_0_Matrix}
\begin{align*}
&\mathds{E}\big\{\tilde{\mathbf{R}}_{N,0}[n]\cdot(\tilde{\mathbf{R}}_{N,0}[n+l])^{H} |H_0\big\}\\
&=\mathds{E}\left\{ \left(\begin{array}{ll}
R[nN]\cdot\Big(R\big[(n+l)N\big]\Big)^{*}&R[nN]\cdot\Big(R\big[(n+l)N-1\big]\Big)^{*} \\
R[nN-1]\cdot\Big(R\big[(n+l)N\big]\Big)^{*}&R[nN-1]\cdot\Big(R\big[(n+l)N-1\big]\Big)^{*} \\
R[nN-2]\cdot\Big(R\big[(n+l)N\big]\Big)^{*}&R[nN-2]\cdot\Big(R\big[(n+l)N-1\big]\Big)^{*} \\
\vdots&\vdots \\
R\big[nN-(K-1)\big]\cdot\Big(R\big[(n+l)N\big]\Big)^{*}&R\big[nN-(K-1)\big]\cdot\Big(R\big[(n+l)N-1\big]\Big)^{*} \\
\end{array}\right. \right.\\
&~~~~~~~~~~~~~~~~~~~~~~~~~~~~~~~~~\left. \left. \left.\begin{array}{ll}
\ldots&R[nN]\cdot\Big(R\big[(n+l)N-(K-1)\big]\Big)^{*} \\
\ldots&R[nN-1]\cdot\Big(R\big[(n+l)N-(K-1)\big]\Big)^{*} \\
\ldots&R[nN-2]\cdot\Big(R\big[(n+l)N-(K-1)\big]\Big)^{*} \\
\ddots&\vdots \\
\ldots&R\big[nN-(K-1)\big]\cdot\Big(R\big[(n+l)N-(K-1)\big]\Big)^{*} \\
\end{array}\right)\right| H_0 \right\}
\end{align*}
\begin{align} \label{Autocovariance_Matrix}
&=\left( \begin{array}{l}
\mathds{E}\bigg\{R[nN]\cdot\Big(R\big[(n+l)N\big]\Big)^{*}\bigg|H_0\bigg\} \\
~\\
\mathds{E}\bigg\{R[nN-1]\cdot\Big(R\big[(n+l)N\big]\Big)^{*}\bigg|H_0\bigg\} \\
~\\
\mathds{E}\bigg\{R[nN-2]\cdot\Big(R\big[(n+l)N\big]\Big)^{*}\bigg|H_0\bigg\} \\
~\\
\vdots \\
~\\
\mathds{E}\bigg\{R\big[nN-(K-1)\big]\cdot\Big(R\big[(n+l)N\big]\Big)^{*}\bigg|H_0\bigg\} \\
~\\
\end{array}\right. \nonumber\\
&~~~~~~~~~~~~~~~~~\left. \begin{array}{ll}
\mathds{E}\bigg\{R[nN]\cdot\Big(R\big[(n+l)N-1\big]\Big)^{*}\bigg|H_0\bigg\}&\ldots \\
~\\
\mathds{E}\bigg\{R[nN-1]\cdot\Big(R\big[(n+l)N-1\big]\Big)^{*}\bigg|H_0\bigg\}&\ldots \\
~\\
\mathds{E}\bigg\{R[nN-2]\cdot\Big(R\big[(n+l)N-1\big]\Big)^{*}\bigg|H_0\bigg\}&\ldots \\
~\\
\vdots&\ddots \\
~\\
\mathds{E}\bigg\{R\big[nN-(K-1)\big]\cdot\Big(R\big[(n+l)N-1\big]\Big)^{*}\bigg|H_0\bigg\}&\ldots \\
~\\
\end{array}\right. \nonumber\\
&~~~~~~~~~~~~~~~~~~~\left. \begin{array}{lll}
\ldots&\mathds{E}\bigg\{R[nN]\cdot\Big(R\Big[(n+l)N-(K-1)\Big]\Big)^{*}\bigg|H_0\bigg\} \\
~\\
\ldots&\mathds{E}\bigg\{R[nN-1]\cdot\Big(R\Big[(n+l)N-(K-1)\Big]\Big)^{*}\bigg|H_0\bigg\} \\
~\\
\ldots&\mathds{E}\bigg\{R[nN-2]\cdot\Big(R\Big[(n+l)N-(K-1)\Big]\Big)^{*}\bigg|H_0\bigg\} \\
~\\
\ddots&\vdots \\
\ldots&\mathds{E}\bigg\{R\Big[nN-(K-1)\Big]\cdot\Big(R\Big[(n+l)N-(K-1)\Big]\Big)^{*}\bigg|H_0\bigg\} \\
\end{array}\right)
\end{align}

\subsubsection{Derivations of the General Element of \texorpdfstring{$\mathds{R}_{\tilde{\mathbf{R}}_{N,0}|H_0}[n,l]$}~ Matrix}
\label{app:Derivations_H_0_Matrix}
We now examine a general element of the matrix $\mathds{R}_{\tilde{\mathbf{R}}_{N,0}|H_0}[n,l]$ presented in Eq. (\ref{Autocovariance_Matrix}), $\mathds{E}\bigg\{R[nN-a_1]\cdot\Big(R\big[(n+l)N-a_2\big]\Big)^{*}\bigg|H_0\bigg\},~a_1,a_2\in \{0,1,...,K-1\}$:
\begin{align*}
&\mathds{E}\bigg\{R[nN-a_1]\cdot\Big(R\big[(n+l)N-a_2\big]\Big)^{*}\bigg|H_0\bigg\}\\
&=\mathds{E}\bigg\{\Big(\sum\limits_{l_1=0}^{L_{\textsf{ch}}} h[nN-a_1,l_1]S[nN-a_1-l_1]+Z[nN-a_1]\Big)\\
&~~~~~~~~~~~~~~~~~\cdot \Big(\sum\limits_{l_2=0}^{L_{\textsf{ch}}} h[(n+l)N-a_2,l_2]S[(n+l)N-a_2-l_2]+Z[(n+l)N-a_2]\Big)^*\bigg|H_0\bigg\}\\
&=\mathds{E}\bigg\{\Big(\sum\limits_{l_1=0}^{L_{\textsf{ch}}} h[nN-a_1,l_1]S[nN-a_1-l_1]\Big)\\
&~~~~~~~~~~\cdot\Big(\sum\limits_{l_2=0}^{L_{\textsf{ch}}} h\big[(n+l)N-a_2,l_2\big]S\big[(n+l)N-a_2-l_2\big]\Big)^*  \nonumber \\
&~~~~~~~~~~~~~~+\Big(\sum\limits_{l_1=0}^{L_{\textsf{ch}}} h[nN-a_1,l_1]S[nN-a_1-l_1]\Big)\cdot \Big(Z\big[(n+l)N-a_2\big]\Big)^*  \nonumber \\
&~~~~~~~~~~~~~~~~~~~+\Big(\sum\limits_{l_2=0}^{L_{\textsf{ch}}} h\big[(n+l)N-a_2,l_2\big]S\big[(n+l)N-a_2-l_2\big]\Big)^*\cdot Z\big[nN-a_1\big]  \nonumber \\
&~~~~~~~~~~~~~~~~~~~~~~~~~~~~~~~~~~~~~~~~~~~~~~~~~~~~~~~~~~~+Z\big[nN-a_1\big]\cdot \Big(Z\big[(n+l)N-a_2\big]\Big)^*\bigg|H_0 \bigg\}  \nonumber \\
&\stackrel{(a)}{=}\sum\limits_{l_1=0}^{L_{\textsf{ch}}}\sum\limits_{l_2=0}^{L_{\textsf{ch}}} \bigg(h\big[nN-a_1,l_1\big]\cdot\Big(h\big[(n+l)N-a_2,l_2\big]\Big)^*\bigg)\\
&~~~~~~~~~~~~~~~~~~~\cdot \mathds{E}\bigg\{S[nN-a_1-l_1]\cdot\Big(S\big[(n+l)N-a_2-l_2\big]\Big)^*\bigg|H_0\bigg\}  \nonumber \\
&~~~~~+\sum\limits_{l_1=0}^{L_{\textsf{ch}}} h[nN-a_1,l_1]\cdot \mathds{E}\bigg\{S\big[nN-a_1-l_1\big]\cdot\Big(Z\big[(n+l)N-a_2\big]\Big)^*\bigg|H_0\bigg\}  \nonumber \\
&+\sum\limits_{l_2=0}^{L_{\textsf{ch}}} \Big(h\big[(n+l)N-a_2,l_2\big]\Big)^*\cdot \mathds{E}\bigg\{\Big(S\big[(n+l)N-a_2-l_2\big]\Big)^*\cdot Z\big[nN-a_1\big]\bigg|H_0\bigg\}  \nonumber \\
&~~~~~~~~~~~~~~~~~~~~~~~~~~~~~~~~~~~~~~~~~~~~~~~~~~~~~~+\mathds{E}\bigg\{Z\big[nN-a_1\big]\cdot\Big(Z\big[(n+l)N-a_2\big]\Big)^*\bigg|H_0\bigg\}  \nonumber \\
&\stackrel{(b)}{=}\sum\limits_{l_1=0}^{L_{\textsf{ch}}}\sum\limits_{l_2=0}^{L_{\textsf{ch}}} h[nN-a_1,l_1]\cdot\Big(h\big[(n+l)N-a_2,l_2\big]\Big)^* \nonumber\\
&~~~~~~~~~~~~~~~~~~\cdot \mathds{E}\bigg\{S\big[nN-a_1-l_1\big]\cdot\Big(S\big[(n+l)N-a_2-l_2\big]\Big)^*\bigg|H_0\bigg\}  \nonumber \\
&~~~~~~~~~~~~~~~~~~~~~~~~~~~~~~~~~~~~~~~~~~~~~~~~~~+\mathds{E}\bigg\{Z\big[nN-a_1\big]\cdot\Big(Z\big[(n+l)N-a_2\big]\Big)^*\bigg|H_0\bigg\}  \nonumber 
\end{align*}
\begin{align*}
&\stackrel{(c)}{=} \sum\limits_{l_1=0}^{L_{\textsf{ch}}}\sum\limits_{l_2=0}^{L_{\textsf{ch}}} h[-a_1,l_1]\cdot\big(h[-a_2,l_2]\big)^*\nonumber \\
&~~~~~~~~~~~~~~~~~\cdot \mathds{E} \bigg\{S \big[nN-a_1-l_1\big]\cdot\Big(S\big[(n+l)N-a_2-l_2\big]\Big)^*\bigg|H_0\bigg\} \nonumber \\
&~~~~~~~~~~~~~~~~~~~~~~~~~~~~~~~~~~~~~~~~~~~~~~~~~~~~~~~~~~~~~~~~~~~~~~~~+c_{\textsf{z}}\big[nN-a_1,lN+a_1-a_2\big] \nonumber \\ 
&\stackrel{(d)}{=}\sum\limits_{l_1=0}^{L_{\textsf{ch}}}\sum\limits_{l_2=0}^{L_{\textsf{ch}}} h[-a_1,l_1]\cdot\big(h[-a_2,l_2]\big)^* \cdot \sigma_s^2\cdot\delta[lN-a_2-l_2+a_1+l_1] \nonumber\\
&~~~~~~~~~~~~~~~~~~~~~~~~~~~~~~~~~~~~~~~~~~~~~~~~~~~~~~~~~~~~~~~~~~~~~~~~~~~~~+c_{\textsf{z}}[-a_1,lN+a_1-a_2]~,
\end{align*}
where $(a)$ follows since stochastic expectation is a linear operator and only $Z[n]$ and $S[n]$ are random, $(b)$ follows since $Z[n]$ and $S[n]$ are mutually independent and each has a zero stochastic expectation, $(c)$ follows since $N=k_1P_{\textsf{h}}$, thus $h[n,l_1]=h[n+k_1P_{\textsf{h}},l_1]=h[n+N,l_1]$ , $(d)$ follows since 
$\mathds{E}\bigg\{S[nN-a_1-l_1]\cdot\Big(S\big[(n+l)N-a_2-l_2\big]\Big)^*\bigg|H_0\bigg\}=
\left\{ \begin{array}{l}
0~~~~~~,-a_1-l_1\neq lN-a_2-l_2\\
\sigma_s^2~~~~~~,-a_1-l_1 = lN-a_2-l_2\\
\end{array}\right.$ 
and since $N=k_2P_{\textsf{z}}$, we have that $c_z[n,l]=c_z[n+k_2P_{\textsf{z}},l]=c_z[n+N,l]$.
\subsubsection{Derivations of \texorpdfstring{$\mathds{E}\Big\{\tilde{\mathbf{R}}_{N,0}[nM-m]\Big|\tilde{\mathbf{g}}^{(m)}_{L_{\textsf{tot}},0}[n],H_1\Big\}$}~}
\label{app:Derivations_of_H_1_Expectation}
Writing $\mathds{E}\Big\{\tilde{\mathbf{R}}_{N,0}[nM-m]\Big|\tilde{\mathbf{g}}^{(m)}_{L_{\textsf{tot}},0}[n],H_1\Big\}$ explicitly we obtain:
\arraycolsep=1.0pt\def\arraystretch{0.1}
\begin{align*}
&\mathds{E}\Big\{\tilde{\mathbf{R}}_{N,0}[nM-m]\Big|\tilde{\mathbf{g}}^{(m)}_{L_{\textsf{tot}},0}[n],H_1\Big\}\nonumber \\ 
&=\mathds{E}\left\{\left. \left( 
\begin{array}{l}
R\big[(nM-m)N\big] \\
R\big[(nM-m)N-1\big] \\
\vdots\\
R\big[(nM-m)N-(K-L_{\textsf{ch}}-1)\big] \\
R\big[(nM-m)N-(K-L_{\textsf{ch}})\big] \\
R\big[(nM-m)N-(K-L_{\textsf{ch}}+1)\big] \\
\vdots\\
R\big[(nM-m)N-(K-1)\big]
\end{array}\right)\right|\tilde{\mathbf{g}}^{(m)}_{L_{\textsf{tot}},0}[n],H_1\right\} \nonumber \\ 
&=\mathds{E}\left\{\left. \left( 
\begin{array}{l}
\left(\begin{array}{l}
\sum\limits_{l=0}^{L_{\textsf{ch}}} h\big[(nM-m)N,l\big]S\big[nMN-mN-l\big]\\
~~~~~~~~~~~~~~~~~~~~~~~~~~~~~~~~~~~~~~~~+Z\big[nMN-mN\big]
\end{array}\right)\\
\left(\begin{array}{l}
\sum\limits_{l=0}^{L_{\textsf{ch}}} h\big[(nM-m)N-1,l\big]S\big[nMN-mN-1-l\big]\\
~~~~~~~~~~~~~~~~~~~~~~~~~~~~~~~~~~~~~~~~+Z\big[nMN-mN-1\big]
\end{array}\right)\\
\vdots\\
\left(\begin{array}{l}
\sum\limits_{l=0}^{L_{\textsf{ch}}} h\big[(nM-m)N-(K-L_{\textsf{ch}}-1),l\big]\\
~~~~~~~~~~\cdot S\big[nMN-mN-(K-L_{\textsf{ch}}-1)-l\big]\\
~~~~~~~~~~~~~~~~~~~~~~~+Z\big[nMN-mN-(K-L_{\textsf{ch}}-1)\big]
\end{array}\right)\\
\left(\begin{array}{l}
\sum\limits_{l=0}^{L_{\textsf{ch}}} h\big[(nM-m)N-(K-L_{\textsf{ch}}),l\big]\\
~~~~~~~~~~\cdot S\big[nMN-mN-(K-L_{\textsf{ch}})-l\big]\\
~~~~~~~~~~~~~~~~~~~~~~~+Z\big[nMN-mN-(K-L_{\textsf{ch}})\big]
\end{array}\right)\\
\left(\begin{array}{l}
\sum\limits_{l=0}^{L_{\textsf{ch}}} h\big[(nM-m)N-(K-L_{\textsf{ch}}+1),l\big]\\
~~~~~~~~~~\cdot S\big[nMN-mN-(K-L_{\textsf{ch}}+1)-l\big]\\
~~~~~~~~~~~~~~~~~~~~~~~+Z\big[nMN-mN-(K-L_{\textsf{ch}}+1)\big]
\end{array}\right)\\
\vdots\\
~\\
~\\
\left(\begin{array}{l}
\sum\limits_{l=0}^{L_{\textsf{ch}}} h\big[(nM-m)N-(K-1),l\big]\\
~~~~~~~~~~\cdot S\big[nMN-mN-(K-1)-l\big]\\
~~~~~~~~~~~~~~~~~~~~~~~+Z\big[nMN-mN-(K-1)\big]
\end{array}\right)\\
\end{array}\right)\right|\tilde{\mathbf{g}}^{(m)}_{L_{\textsf{tot}},0}[n],H_1\right\}  \nonumber\\ 
\end{align*}
\begin{align*}
&\stackrel{(a)}{=}\mathds{E}\left\{\left. \left( 
\begin{array}{l}
\sum\limits_{l=0}^{L_{\textsf{ch}}} h\big[(nM-m)N,l\big]f_{L_{\textsf{sw}}-1-(l+mK)} +Z[nMN-mN]\\
\sum\limits_{l=0}^{L_{\textsf{ch}}} h\big[(nM-m)N-1,l\big]f_{L_{\textsf{sw}}-1-(l+mK+1)}+Z[nMN-mN-1]\\
\vdots\\
\left(\begin{array}{l}
\sum\limits_{l=0}^{L_{\textsf{ch}}} h\big[(nM-m)N-(K-L_{\textsf{ch}}-1),l\big]f_{L_{\textsf{sw}}-1-(l+mK+K-L_{\textsf{ch}}-1)}\\
~~~~~~~~~~~~~~~~~~~~~~~~~~~~~~~~~~+Z\big[nMN-mN-(K-L_{\textsf{ch}}-1)\big]
\end{array}\right)\\
\left(\begin{array}{l}
\sum\limits_{l=0}^{L_{\textsf{ch}}-1} h\big[(nM-m)N-(K-L_{\textsf{ch}}),l\big]f_{L_{\textsf{sw}}-1-(l+mK+K-L_{\textsf{ch}})}\\
~~~~~~~~~~~~~~~~~~~~~~~~~~~~~~~~~~~~~~~~+Z\big[nMN-mN-(K-L_{\textsf{ch}})\big]\end{array}\right.\\
\left(\begin{array}{l}
\sum\limits_{l=0}^{L_{\textsf{ch}}-2} h\big[(nM-m)N-(K-L_{\textsf{ch}}+1),l\big]f_{L_{\textsf{sw}}-1-(l+mK+K-L_{\textsf{ch}}+1)}\\
~~~~~~~~~~~~~~~~~~~~~~~~~~~~~~~~~~~+Z\big[nMN-mN-(K-L_{\textsf{ch}}+1)\big]\end{array}\right. \\
\vdots\\
~\\
~\\
\left(\begin{array}{l}
h\big[(nM-m)N-(K-1),0\big]f_{L_{\textsf{sw}}-1-(mK+K-1)}\\
~~~~~~~~~~~~~~~~~~~~~~~~~~~~~~~~~~~~~~~~~~~~+Z\big[nMN-mN-(K-1)\big]\end{array}\right.\\
\end{array}\right.\right. \right. \nonumber\\
&~~~~~~~~~~\left.\left. \left. 
\begin{array}{l}
+0\\
~\\
~\\
~\\
~\\
+0\\
~\\
~\\
~\\
~\\
\vdots\\
~\\
~\\
~\\
~\\
+0\\
~\\
~\\
~\\
~\\
\left.\begin{array}{l}
+ h\big[(nM-m)N-(K-L_{\textsf{ch}}),L_{\textsf{ch}}\big]\\
~~~~~~~~~~~~~~~~~~~\cdot s\big[nMN-mN-(K-L_{\textsf{ch}})-L_{\textsf{ch}}\big]\end{array}\right)\\
~\\
~\\
~\\
~\\
\left.\begin{array}{l}
+\sum\limits_{l=L_{\textsf{ch}}-1}^{L_{\textsf{ch}}} h\big[(nM-m)N-(K-L_{\textsf{ch}}+1),l\big]\\
~~~~~~~~~~~~~~~~~\cdot s\big[nMN-mN-(K-L_{\textsf{ch}}+1)-l\big]\end{array}\right)\\
~\\
~\\
~\\
~\\
\vdots\\
~\\
~\\
~\\
~\\
\left.\begin{array}{l}
+\sum\limits_{l=1}^{L_{\textsf{ch}}} h\big[(nM-m)N-(K-1),l\big]\\
~~~~~~~~~~~~~~~~~~~~~~~\cdot s\big[nMN-mN-(K-1)-l\big]
\end{array}\right)
\end{array}\right)\right| \tilde{\mathbf{g}}^{(m)}_{L_{\textsf{tot}},0}[n],H_1\right\} \nonumber\\ 
\end{align*}
\begin{align*}
&\stackrel{(b)}{=}
\left(\begin{array}{l}
\sum\limits_{l=0}^{L_{\textsf{ch}}} h\big[(nM-m)N,l\big]f_{L_{\textsf{sw}}-1-(l+mK)} \\
\sum\limits_{l=0}^{L_{\textsf{ch}}} h\big[(nM-m)N-1,l\big]f_{L_{\textsf{sw}}-1-(l+mK+1)}\\
~~~\vdots\\
\sum\limits_{l=0}^{L_{\textsf{ch}}} h\big[(nM-m)N-(K-L_{\textsf{ch}}-1),l\big]f_{L_{\textsf{sw}}-1-(l+mK+K-L_{\textsf{ch}}-1)}\\
\sum\limits_{l=0}^{L_{\textsf{ch}}-1} h\big[(nM-m)N-(K-L_{\textsf{ch}}),l\big]f_{L_{\textsf{sw}}-1-(l+mK+K-L_{\textsf{ch}})}\\
\sum\limits_{l=0}^{L_{\textsf{ch}}-2} h\big[(nM-m)N-(K-L_{\textsf{ch}}+1),l\big]f_{L_{\textsf{sw}}-1-(l+mK+K-L_{\textsf{ch}}+1)}\\
~~~\vdots\\
~\\
~\\
h\big[(nM-m)N-(K-1),0\big]f_{L_{\textsf{sw}}-1-(mK+K-1)}
\end{array}\right. \nonumber \\
&~~~~~~~~~~~~~~~\left.\begin{array}{l}
+0\\
+0\\
\vdots\\
+0\\
+\left(\begin{array}{l}
h\big[(nM-m)N-(K-L_{\textsf{ch}}),L_{\textsf{ch}}\big]\\
~~~~~~~~~~~~~~~~~~\cdot s\big[nL_{\textsf{tot}}-mN-(K-L_{\textsf{ch}})-L_{\textsf{ch}}\big]\\ \end{array}\right)\\
+\left(\begin{array}{l}
\sum\limits_{l=L_{\textsf{ch}}-1}^{L_{\textsf{ch}}} h\big[(nM-m)N-(K-L_{\textsf{ch}}+1),l\big]\\
~~~~~~~~~~~~~~~~\cdot s\big[nL_{\textsf{tot}}-mN-(K-L_{\textsf{ch}}+1)-l\big]
\end{array}\right)\\
\vdots\\
+\left(\begin{array}{l}\sum\limits_{l=1}^{L_{\textsf{ch}}} h\big[(nM-m)N-(K-1),l\big]\\
~~~~~~~~~~~~~~~~~~~~~~~~\cdot s\big[nL_{\textsf{tot}}-mN-(K-1)-l\big]\end{array}\right)
\end{array}\right. \nonumber \\
&~~~~~~~~~~~~~~~~~~~~~~~~~~~~~~~~~~\left.\begin{array}{l}
+\mathds{E}\Big\{Z[nL_{\textsf{tot}}-mN]|\tilde{\mathbf{g}}^{(m)}_{L_{\textsf{tot}},0}[n],H_1\Big\}\\
+\mathds{E}\Big\{Z\big[nL_{\textsf{tot}}-mN-1]|\tilde{\mathbf{g}}^{(m)}_{L_{\textsf{tot}},0}[n],H_1\Big\}\\
\vdots\\
+\mathds{E}\Big\{Z\big[nL_{\textsf{tot}}-mN-(K-L_{\textsf{ch}}-1)\big]\Big|\tilde{\mathbf{g}}^{(m)}_{L_{\textsf{tot}},0}[n],H_1\Big\}\\
+\mathds{E}\Big\{Z\big[nL_{\textsf{tot}}-mN-(K-L_{\textsf{ch}})\big]\Big|\tilde{\mathbf{g}}^{(m)}_{L_{\textsf{tot}},0}[n],H_1\Big\}\\
+\mathds{E}\Big\{Z\big[nL_{\textsf{tot}}-mN-(K-L_{\textsf{ch}}+1)\big]\big|\tilde{\mathbf{g}}^{(m)}_{L_{\textsf{tot}},0}[n],H_1\Big\} \\
\vdots\\
+\mathds{E}\Big\{Z\big[nL_{\textsf{tot}}-mN-(K-1)\big]\Big|\tilde{\mathbf{g}}^{(m)}_{L_{\textsf{tot}},0}[n],H_1\Big\}
\end{array}\right) \nonumber\\
\end{align*}
\begin{align*}
&\stackrel{(c)}{=}
\left(\begin{array}{l}
\sum\limits_{l=0}^{L_{\textsf{ch}}} h[0,l]f_{L_{\textsf{sw}}-1-(l+mK)} \\
\sum\limits_{l=0}^{L_{\textsf{ch}}} h[-1,l]f_{L_{\textsf{sw}}-1-(l+mK+1)}\\
\vdots\\
\sum\limits_{l=0}^{L_{\textsf{ch}}} h[-(K-L_{\textsf{ch}}-1),l]f_{L_{\textsf{sw}}-1-(l+mK+K-L_{\textsf{ch}}-1)}\\
\sum\limits_{l=0}^{L_{\textsf{ch}}-1} h\big[-(K-L_{\textsf{ch}}),l\big]f_{L_{\textsf{sw}}-1-(l+mK+K-L_{\textsf{ch}})}\\
\sum\limits_{l=0}^{L_{\textsf{ch}}-2} h\big[-(K-L_{\textsf{ch}}+1),l\big]f_{L_{\textsf{sw}}-1-(l+mK+K-L_{\textsf{ch}}+1)} \\
\vdots\\
~\\
~\\
 h[-(K-1),0]f_{L_{\textsf{sw}}-1-(mK+K-1)}
\end{array}\right. \nonumber \\
&~~~~~~\left.\begin{array}{l}
+0\\
+0\\
\vdots\\
+0\\
+\left(\begin{array}{l}
h\big[-(K-L_{\textsf{ch}}),L_{\textsf{ch}}\big]\\
~~~~~~~~~~~~~\cdot s\big[nL_{\textsf{tot}}-mN-(K-L_{\textsf{ch}})-L_{\textsf{ch}}\big]
\end{array} \right)\\
+\left(\begin{array}{l}
\sum\limits_{l=L_{\textsf{ch}}-1}^{L_{\textsf{ch}}} h\big[-(K-L_{\textsf{ch}}+1),l\big]\\
~~~~~~~~~~~\cdot s\big[nL_{\textsf{tot}}-mN-(K-L_{\textsf{ch}}+1)-l\big]
\end{array} \right)\\
\vdots\\
+\sum\limits_{l=1}^{L_{\textsf{ch}}} h\big[-(K-1),l\big]s\big[nL_{\textsf{tot}}-mN-(K-1)-l\big]
\end{array}\right),~ \begin{array}{l} \forall n\in \mathcal{Z}\\ ~\\
~m=0,1,...,M-1~ \end{array} 
\end{align*}
where $(a)$ follows from the structure of $\tilde{\mathbf{g}}^{(m)}_{L_{\textsf{tot}},0}[n]$ , $(b)$ follows since stochastic expectation is a linear operator and given $\tilde{\mathbf{g}}^{(m)}_{L_{\textsf{tot}},0}[n]$ only $Z[n]$ is random, and $NM=L_{\textsf{tot}}$, $(c)$ follows since $h[n,l]=h[n+k_1P_{\textsf{h}},l]=h[n+N,l]$ and $\mathds{E}\big\{Z[n]\big|H_1\big\}=0,~\forall n \in \mathcal{Z}$.
\subsubsection{Derivation of the General Element of the Matrix \texorpdfstring{$\tilde{\mathds{U}}[n,m]$}~}
\label{app:Derivations_H_1_Matrix}
In the following we calculate the elements of the matrix $\tilde{\mathds{U}}[n,m]\triangleq \mathds{E}\Big\{\tilde{\mathbf{R}}_{N,0}[nM-m]\cdot\big(\tilde{\mathbf{R}}_{N,0}[nM-m]\big)^{H}\Big|\tilde{\mathbf{g}}^{(m)}_{L_{\textsf{tot}},0}[n],H_1\Big\},$ at the $(a_1+1)$'th row, and $(a_2+1)$'th column, $a_1,a_2\in \{0,1,...,K-1\}$:
\begin{align*}
\big[&\tilde{\mathds{U}}[n,m]\big]_{a_1+1,a_2+1}=\mathds{E}\bigg\{R\big[(nM-m)N-a_1\big]\Big(R\big[(nM-m)N-a_2\big]\Big)^{*}\bigg|\tilde{\mathbf{g}}^{(m)}_{L_{\textsf{tot}},0}[n],H_1\bigg\}\\
&=\mathds{E}\bigg\{\Big(\sum\limits_{l_1=0}^{L_{\textsf{ch}}} h[(nM-m)N-a_1,l_1]S[(nM-m)N-a_1-l_1]+Z[(nM-m)N-a_1]\Big)\\
&~~~~~~~~~~~~\cdot \Big(\sum\limits_{l_2=0}^{L_{\textsf{ch}}} h[(nM-m)N-a_2,l_2]S[(nM-m)N-a_2-l_2]\\
&~~~~~~~~~~~~~~~~~~~~~~~~~~~~~~~~~~~~~~~~~~~~~~~~~~~~~~~~~~~~+Z[(nM-m)N-a_2]\Big)^*\bigg|\tilde{\mathbf{g}}^{(m)}_{L_{\textsf{tot}},0}[n],H_1\bigg\}
\end{align*}
\begin{align*}
&=\mathds{E}\bigg\{\Big(\sum\limits_{l_1=0}^{L_{\textsf{ch}}} h[(nM-m)N-a_1,l_1]S[(nM-m)N-a_1-l_1]\Big)\\
&~~~~~~~~~~\cdot\Big(\sum\limits_{l_2=0}^{L_{\textsf{ch}}} h\big[(nM-m)N-a_2,l_2\big]S\big[(nM-m)N-a_2-l_2\big]\Big)^*  \nonumber \\
&~~~~~~~~~~~~~~+\Big(\sum\limits_{l_1=0}^{L_{\textsf{ch}}} h[(nM-m)N-a_1,l_1]S[(nM-m)N-a_1-l_1]\Big) \nonumber\\
&~~~~~~~~~~~~~~~~~~~~~~~~~~~~~~~~~~~~~~~~~~~~~~~~~~~~~~~~~~~~~~~\cdot \Big(Z\big[(nM-m)N-a_2\big]\Big)^*  \nonumber \\
&~~~~~~~~~~~~~~~~~~~~~~~~+\Big(\sum\limits_{l_2=0}^{L_{\textsf{ch}}} h\big[(nM-m)N-a_2,l_2\big]S\big[(nM-m)N-a_2-l_2\big]\Big)^*  \nonumber\\
&~~~~~~~~~~~~~~~~~~~~~~~~~~~~~~~~~~~~~~~~~~~~~~~~~~~~~~~~~~~~~~~~~~~~~~~~~~\cdot Z\big[(nM-m)N-a_1\big]  \nonumber \\
&~~~~~~~~~~~~~~~~~~~~~~~~~~+Z\big[(nM-m)N-a_1\big]\cdot \Big(Z\big[(nM-m)N-a_2\big]\Big)^*\bigg|\tilde{\mathbf{g}}^{(m)}_{L_{\textsf{tot}},0}[n],H_1 \bigg\}  \nonumber \\
&\stackrel{(a)}{=}\sum\limits_{l_1=0}^{L_{\textsf{ch}}}\sum\limits_{l_2=0}^{L_{\textsf{ch}}} \bigg(h\big[(nM-m)N-a_1,l_1\big]\Big(h\big[(nM-m)N-a_2,l_2\big]\Big)^*\bigg)\\
&~~~~~~~~~~~~~~~\cdot \mathds{E}\bigg\{S[(nM-m)N-a_1-l_1]\Big(S\big[(nM-m)N-a_2-l_2\big]\Big)^*\bigg|\tilde{\mathbf{g}}^{(m)}_{L_{\textsf{tot}},0}[n],H_1\bigg\}  \nonumber \\
&~~~~~+\sum\limits_{l_1=0}^{L_{\textsf{ch}}} h[(nM-m)N-a_1,l_1]  \nonumber\\
&~~~~~~~~~~~~~~~~~~~~\cdot \mathds{E}\bigg\{S\big[(nM-m)N-a_1-l_1\big]\Big(Z\big[(nM-m)N-a_2\big]\Big)^*\bigg|\tilde{\mathbf{g}}^{(m)}_{L_{\textsf{tot}},0}[n],H_1\bigg\}  \nonumber \\
&~~~~~~~+\sum\limits_{l_2=0}^{L_{\textsf{ch}}} \Big(h\big[(nM-m)N-a_2,l_2\big]\Big)^*\nonumber\\
&~~~~~~~~~~~~~~~~~~~~\cdot \mathds{E}\bigg\{\Big(S\big[(nM-m)N-a_2-l_2\big]\Big)^*Z\big[(nM-m)N-a_1\big]\bigg|\tilde{\mathbf{g}}^{(m)}_{L_{\textsf{tot}},0}[n],H_1\bigg\}  \nonumber \\
&~~~~~~~~~~~~~~~~~~~~~~~~+\mathds{E}\bigg\{Z\big[(nM-m)N-a_1\big]\Big(Z\big[(nM-m)N-a_2\big]\Big)^*\bigg|\tilde{\mathbf{g}}^{(m)}_{L_{\textsf{tot}},0}[n],H_1\bigg\}  \nonumber \\
&\stackrel{(b)}{=}\sum\limits_{l_1=0}^{L_{\textsf{ch}}}\sum\limits_{l_2=0}^{L_{\textsf{ch}}} h[(nM-m)N-a_1,l_1]\Big(h\big[(nM-m)N-a_2,l_2\big]\Big)^* \nonumber\\
&~~~~~~~~\cdot \mathds{E}\bigg\{S\big[(nM-m)N-a_1-l_1\big]\Big(S\big[(nM-m)N-a_2-l_2\big]\Big)^*\bigg|\tilde{\mathbf{g}}^{(m)}_{L_{\textsf{tot}},0}[n],H_1\bigg\}  \nonumber \\
&~~~~~~~~~~~~~~~~~~~~~+\mathds{E}\bigg\{Z\big[(nM-m)N-a_1\big]\Big(Z\big[(nM-m)N-a_2\big]\Big)^*\bigg|\tilde{\mathbf{g}}^{(m)}_{L_{\textsf{tot}},0}[n],H_1\bigg\}  \nonumber\\ 
&\stackrel{(c)}{=} \sum\limits_{l_1=0}^{L_{\textsf{ch}}}\sum\limits_{l_2=0}^{L_{\textsf{ch}}} h[-a_1,l_1]\big(h[-a_2,l_2]\big)^*\nonumber\\
&~~~~~~~~~~~\cdot \mathds{E} \bigg\{S \big[(nM-m)N-a_1-l_1\big]\Big(S\big[(nM-m)N-a_2-l_2\big]\Big)^*\bigg|\tilde{\mathbf{g}}^{(m)}_{L_{\textsf{tot}},0}[n],H_1\bigg\} \nonumber \\
&~~~~~~~~~~~~~~~~~~~~~~~~~~~~~~~~~~~~~~~~~~~~~~~~~~~~~~~~~~~~~~~~~~+c_{\textsf{z}}\big[(nM-m)N-a_1,a_1-a_2\big] \nonumber 
\end{align*}
\begin{align}
&\stackrel{(d)}{=} \sum\limits_{l_1=0}^{L_{\textsf{ch}}}\sum\limits_{l_2=0}^{L_{\textsf{ch}}} h[-a_1,l_1]\big(h[-a_2,l_2]\big)^*\nonumber\\
&~~~~~~~~~~~\cdot \mathds{E} \bigg\{S \big[nL_{\textsf{tot}}-mN-a_1-l_1\big]\Big(S\big[nL_{\textsf{tot}}-mN-a_2-l_2\big]\Big)^*\bigg|\tilde{\mathbf{g}}^{(m)}_{L_{\textsf{tot}},0}[n],H_1\bigg\} \nonumber \\
&~~~~~~~~~~~~~~~~~~~~~~~~~~~~~~~~~~~~~~~~~~~~~~~~~~~~~~~~~~~~~~~~~~~~~~~~~~~~~~~~~+c_{\textsf{z}}\big[-a_1,a_1-a_2\big]\nonumber\\ 
&\stackrel{(e)}{=}\left\{\begin{array}{l}
\sum\limits_{l_1=0}^{L_{\textsf{ch}}}\sum\limits_{l_2=0}^{L_{\textsf{ch}}} h\big[-a_1,l_1\big]\cdot\Big(h\big[-a_2,l_2\big]\Big)^*\cdot f_{L_{\textsf{sw}}-1-(l_1+mK+a_1)}\cdot\Big(f_{L_{\textsf{sw}}-1-(l_2+mK+a_2)}\Big)^*  \\
\left(\begin{array}{l}
\sum\limits_{l_1=0}^{L_{\textsf{ch}}}\sum\limits_{l_2=0}^{K-1-a_2} h\big[-a_1,l_1\big]\cdot\Big(h\big[-a_2,l_2\big]\Big)^*\cdot f_{L_{\textsf{sw}}-1-(l_1+mK+a_1)}\\
~~~~~~~~~~~~~~~~~~~~~~~~~~~~~~~~~~~~~~~~~~~~~~~~~~~~~~~~~~\cdot \Big(f_{L_{\textsf{sw}}-1-(l_2+mK+a_2)}\Big)^*
\end{array} \right.\\
\left(\begin{array}{l}
\sum\limits_{l_1=0}^{K-1-a_1}\sum\limits_{l_2=0}^{L_{\textsf{ch}}} h\big[-a_1,l_1\big]\cdot\Big(h\big[-a_2,l_2\big]\Big)^*\cdot f_{L_{\textsf{sw}}-1-(l_1+mK+a_1)}\\
~~~~~~~~~~~~~~~~~~~~~~~~~~~~~~~~~~~~~~~~~~~~~~~~~~~~~~~~~~~~~~~~~~~~\cdot \Big(f_{L_{\textsf{sw}}-1-(l_2+mK+a_2)}\Big)^*
\end{array} \right.\\
\left(\begin{array}{l}
\sum\limits_{l_1=0}^{K-1-a_1}\sum\limits_{l_2=0}^{K-1-a_2} h\big[-a_1,l_1\big]\cdot\Big(h\big[-a_2,l_2\big]\Big)^*\cdot f_{L_{\textsf{sw}}-1-(l_1+mK+a_1)} \\
~~~~~~~~~~~~~~~~~~~~~~~~~~~~~~~~~~~~~~~~~~~~~~~~~~~~~~~~~~~~ \cdot\Big(f_{L_{\textsf{sw}}-1-(l_2+mK+a_2)}\Big)^*\\
+\sum\limits_{l_1=0}^{K-1-a_1}\sum\limits_{l_2=K-a_2}^{L_{\textsf{ch}}} h\big[-a_1,l_1\big]\cdot\Big(h\big[-a_2,l_2\big]\Big)^*\cdot f_{L_{\textsf{sw}}-1-(l_1+mK+a_1)} \\
~~~~~~~~~~~~~~~~~~~~~~~~~~~~~~~~~~~~~~~~~~~~~~~~~~~~~~~~~~~~ \cdot\Big(s\big[nL_{\textsf{tot}}-mN-a_2-l_2\big]\Big)^*
\end{array} \right. \nonumber
\end{array}\right.\\ 
&~~~\left. 
\begin{array}{l}
~~~~~+0~~~~~~~~~~~~~~~~~~~~~~~~~~~~~~~~~~~~~~~~~~~~~~~~~~~~~~~~~~~~~~~~~~~~~~~~~,~\{a_1,a_2\in \mathcal{I}_1\}\\
\left.\begin{array}{l}
~~~~~+\sum\limits\limits_{l_1=0}^{L_{\textsf{ch}}}\sum\limits_{l_2=K-a_2}^{L_{\textsf{ch}}} h\big[-a_1,l_1\big]\cdot \Big(h\big[-a_2,l_2\big]\Big)^*\\
~~~~~~~~~~~\cdot f_{L_{\textsf{sw}}-1-(l_1+mK+a_1)}\cdot\Big(s\big[nL_{\textsf{tot}}-mN-a_2-l_2)\big]\Big)^*\end{array}\right)~~,\{a_1\in \mathcal{I}_1 ~,~a_2\in \mathcal{I}_2\}\\
\left.\begin{array}{l}
~~~~~+\sum\limits_{l_1=K-a_1}^{L_{\textsf{ch}}}\sum\limits_{l_2=0}^{L_{\textsf{ch}}} h\big[-a_1,l_1\big]\cdot \Big(h\big[-a_2,l_2\big]\Big)^*\\
~~~~~~~~~~~\cdot s\big[nL_{\textsf{tot}}-mN-a_1-l_1)\big]\cdot\Big(f_{L_{\textsf{sw}}-1-(l_2+mK+a_2)}\Big)^{*}
\end{array}\right)~~~,~\{a_1\in \mathcal{I}_2 ,~a_2\in \mathcal{I}_1\}  \\
\left.\begin{array}{l}
~~~~~+\sum\limits_{l_1=0}^{K-1-a_1}\sum\limits_{l_2=K-a_2}^{L_{\textsf{ch}}} h\big[-a_1,l_1\big]\cdot \Big(h\big[-a_2,l_2\big]\Big)^*\\
~~~~~~~~~~~~~~~~~~~\cdot f_{L_{\textsf{sw}}-1-(l_1+mK+a_1)}\cdot\Big(s\big[nL_{\textsf{tot}}-mN-a_2-l_2\big]\Big)^*\\
~~~~~+\sum\limits_{l_1=K-a_1}^{L_{\textsf{ch}}}\sum\limits_{l_2=K-a_2}^{L_{\textsf{ch}}} h\big[-a_1,l_1\big]\cdot \Big(h\big[-a_2,l_2\big]\Big)^*\\
~~~~~~~~\cdot s\big[nL_{\textsf{tot}}-mN-a_1-l_1)\big]\cdot\Big(s\big[nL_{\textsf{tot}}-mN-a_2-l_2\big]\Big)^*
\end{array}\right)~~,~\{a_1,a_2\in \mathcal{I}_2\}
\end{array}\right\} \nonumber \\
&~~~~~~~~~~~~~~~~~~~~~~~~~~~~~~~~~~~~~~~~~~~~~~~~~~~~~~~~~~~~~~~~~~~~~~~~~~~~~~~~~~~~~~~~+c_{\textsf{z}}\big[-a_1,a_1-a_2\big]
\end{align}
where $(a)$ follows since stochastic expectation is a linear operator and only $Z[n]$ and $S[n]$ are random, $(b)$ follows since $Z[n]$ and $S[n]$ are mutually independent and the mean of $Z[n]$ is a zero, $(c)$ follows since $N=k_1P_{\textsf{h}}$, thus $h[n,l_1]=h[n+k_1P_{\textsf{h}},l_1]=h[n+N,l_1]$, $(d)$ follows since $L_{\textsf{tot}}=MN$, and $c_z[n,l]=c_z[n+k_2P_{\textsf{z}},l]=c_z[n+N,l]$, (e) follows since $S \big[(nM-m)N-a_1-l_1\big]\cdot\Big(S\big[(nM-m)N-a_2-l_2\big]\Big)^*$ given $\tilde{\mathbf{g}}^{(m)}_{L_{\textsf{tot}},0}[n]$ is deterministic, and was calculated in (\ref{conditional_H_1_1}) and (\ref{V_tilde_Matrix}), for $a_1,a_2\in \{0,1,...,K-1\}$ according to subsets $\mathcal{I}_1=\{0,1,...,K-L_{\textsf{ch}}-1\}$, and $\mathcal{I}_2=\{K-L_{\textsf{ch}},K-(L_{\textsf{ch}}-1),...,K-1\}$.
\subsection{Derivations for Section \ref{chap:SALRT_Algorithm_Derivation}}
\label{app:Derivations_Chapter_SALRT}
\subsubsection{Derivations of approximate LRT in (\ref{Symplfy_Trace_2})}
Applying (\ref{Aprox}) to (\ref{Test}) and letting $L_0=(N_{\textsf{s}})^{L_{\textsf{tot}}},~L_1=(N_{\textsf{s}})^{ML_{\textsf{ch}}}$, it follows that the assumption $L_0 \gg L_1$ is satisfied in practical scenarios, and eventually we arrive at the following detector, see [full version, Appendix A] for the derivation of the approximate LRT (ALRT):
\begin{align}
&\frac{1}{(N_{\textsf{s}})^{L_{\textsf{tot}}}} \cdot \left( \max  \left\{ -\sum\limits_{m=0}^{M-1}\Big(\big(\tilde{\mathbf{r}}_{N,0}[nM-m]-\mathds{B}[n] \cdot \mathbf{a}_{q_{(l_0,m)}} \big)^H  \right.  \right.\nonumber  \\
&~~~~~~~~~~~~~~~~~~~~~~~~\cdot \mathds{C}^{-1}_{{\textsf{z}}}  \cdot \big(\tilde{\mathbf{r}}_{N,0}[nM-m]-\mathbf{B}[n] \cdot \mathbf{a}_{q_{(l_0,m)}}\big)\Big)  \Bigg\} _{l_0=0}^{(N_{\textsf{s}})^{L_{\textsf{tot}}}-1}  \nonumber \\
&~~~~~~~~~~~~~~~~~~~~- \max \left\{ -\sum\limits_{m=0}^{M-1}\Big(\big(\tilde{\mathbf{r}}_{N,0}[nM-m]-\mathds{B}[n] \cdot \tilde{\mathbf{a}}_{\tilde{q}_{(l_1,m)},m} \big)^H  \right. \nonumber \\
&~~~~~~~~~~~~~~~~~~~~~~~~~~~~~ \cdot \mathds{C}^{-1}_{{\textsf{z}}} \cdot \big(\tilde{\mathbf{r}}_{N,0}[nM-m]-\mathds{B}[n] \cdot \tilde{\mathbf{a}}_{\tilde{q}_{(l_1,m)},m}\big)\Big)  \Bigg\} _{l_1=0}^{(N_{\textsf{s}})^{ML_{\textsf{ch}}}-1}\Bigg)\nonumber\\
&~~~~~~~~~~~~~~~~~~~~~~~~~~~~~~~~~~~~~~~~~~~~~~~~~~~~~~~~~~~~~~~~\underset{H_1}{\overset{H_0}{\gtrless}} \bigg( \log_e\Big( \lambda \cdot( N_{\textsf{s}})^{MK }\Big) \bigg)\cdot \frac{1}{(N_{\textsf{s}})^{L_{\textsf{tot}}}}~ . \nonumber
\end{align}
Next, we simplify the test using the identity
$\max(-x_1,-x_2,\ldots,-x_Q)=-\min(-x_1,-x_2,\ldots,-x_Q),
Q \in \mathcal{N}$, and obtain the following approximate LRT:
\begin{align}
&\frac{1}{(N_{\textsf{s}})^{L_{\textsf{tot}}}}\cdot\left( \min  \left\{ \sum\limits_{m=0}^{M-1}\Big(\big(\tilde{\mathbf{r}}_{N,0}[nM-m]-\mathds{B}[n] \cdot \tilde{\mathbf{a}}_{\tilde{q}_{(l_1,m)},m} \big)^H  \right. \right. \nonumber \\
&~~~~~~~~~~~~~~ \cdot \mathds{C}^{-1}_{{\textsf{z}}} \cdot \big(\tilde{\mathbf{r}}_{N,0}[nM-m]-\mathds{B}[n] \cdot \tilde{\mathbf{a}}_{\tilde{q}_{(l_1,m)},m}\big)\Big)  \Bigg\} _{l_1=0}^{(N_{\textsf{s}})^{ML_{\textsf{ch}}}-1} \nonumber  \\
& ~~~~~~~~~~~~~~~~~~~ -  \min  \left\{ \sum\limits_{m=0}^{M-1}\Big(\big(\tilde{\mathbf{r}}_{N,0}[nM-m]-\mathds{B}[n] \cdot \mathbf{a}_{q_{(l_0,m)}} \big)^H  \right. \nonumber \\
&~~~~~~~~~~~~~~~~~~~~~~~~~~~~~~~~~ \cdot \mathds{C}^{-1}_{{\textsf{z}}} \cdot \big(\tilde{\mathbf{r}}_{N,0}[nM-m]-\mathds{B}[n] \cdot \mathbf{a}_{q_{(l_0,m)}}\big)\Big)  \Bigg\} _{l_0=0}^{(N_{\textsf{s}})^{L_{\textsf{tot}}}-1} \Bigg)\nonumber\\  &~~~~~~~~~~~~~~~~~~~~~~~~~~~~~~~~~~~~~~~~~~~~~~~~~~~~~~~~~~~~~\underset{H_1}{\overset{H_0}{\gtrless}} \bigg(\log_e\Big( \lambda \cdot( N_{\textsf{s}})^{MK }\Big)  \bigg)\cdot \frac{1}{(N_{\textsf{s}})^{L_{\textsf{tot}}}}~ . \label{Max_Test_Stage_1}
\end{align}
In order to further simplify the computational complexity of the LRT, we expand the weighted quadratic expressions in (\ref{Max_Test_Stage_1}):
\begin{align*}
& \min  \left\{ \sum\limits_{m=0}^{M-1}\big(\tilde{\mathbf{r}}_{N,0}[nM-m] \big)^H \cdot \mathds{C}^{-1}_{{\textsf{z}}}\cdot \tilde{\mathbf{r}}_{N,0}[nM-m]\right. \nonumber \\
&~~~~~~~~~-2\sum\limits_{m=0}^{M-1}\operatorname{Re}\Big(\big(\tilde{\mathbf{r}}_{N,0}[nM-m] \big)^H \cdot \mathds{C}^{-1}_{{\textsf{z}}}\cdot \mathds{B}[n] \cdot \tilde{\mathbf{a}}_{\tilde{q}_{(l_1,m)},m}\Big)   \nonumber \\
&~~~~~~~~~~~~~~~~~~ +\bigg( \sum\limits_{m=0}^{M-1}\big(\mathds{B}[n] \cdot \tilde{\mathbf{a}}_{\tilde{q}_{(l_1,m)},m} \big)^H \cdot \mathds{C}^{-1}_{{\textsf{z}}} \cdot \mathds{B}[n] \cdot \tilde{\mathbf{a}}_{\tilde{q}_{(l_1,m)},m}\bigg) \Bigg\} _{l_1=0}^{(N_{\textsf{s}})^{ML_{\textsf{ch}}}-1}  \nonumber  \\
&~~~~~~~~~ -  \min \left\{ \sum\limits_{m=0}^{M-1}\big(\tilde{\mathbf{r}}_{N,0}[nM-m]\big)^H \cdot \mathds{C}^{-1}_{{\textsf{z}}}\cdot \tilde{\mathbf{r}}_{N,0}[nM-m]\right.\nonumber \\
&~~~~~~~~~~~~~~~~~~-2\sum\limits_{m=0}^{M-1}\operatorname{Re}\Big(\big(\tilde{\mathbf{r}}_{N,0}[nM-m] \big)^H \cdot \mathds{C}^{-1}_{{\textsf{z}}}\cdot \mathds{B}[n] \cdot \mathbf{a}_{q_{(l_0,m)}}\Big)   \nonumber \\
&~~~~~~~~~~~~~~~~~~~~~~~~~~~~~~~~~~~~+ \left. \sum\limits_{m=0}^{M-1}\big(\mathds{B}[n] \cdot \mathbf{a}_{q_{(l_0,m)}}\big)^{H}\cdot \mathds{C}^{-1}_{{\textsf{z}}} \cdot \mathds{B}[n] \cdot \mathbf{a}_{q_{(l_0,m)}} \right\} _{l_0=0}^{(N_{\textsf{s}})^{L_{\textsf{tot}}}-1}  \nonumber 
\end{align*}
\begin{align}
&\stackrel{(a)}{=}  \min  \left\{ -2\operatorname{Re}\bigg( \sum\limits_{m=0}^{M-1}\big(\tilde{\mathbf{r}}_{N,0}[nM-m] \big)^H  \right.\cdot \mathds{C}^{-1}_{{\textsf{z}}}\cdot \mathds{B}[n] \cdot \tilde{\mathbf{a}}_{\tilde{q}_{(l_1,m)},m}\bigg)  \nonumber  \\
&~~~~~~~~~~~~  +\bigg( \sum\limits_{m=0}^{M-1}\big( \tilde{\mathbf{a}}_{\tilde{q}_{(l_1,m)},m} \big)^H \Big( \big(\mathds{B}[n] \big)^H \cdot \mathds{C}^{-1}_{{\textsf{z}}} \cdot \mathds{B}[n]  \Big) \cdot \tilde{\mathbf{a}}_{\tilde{q}_{(l_1,m)},m}\bigg)  \Bigg\} _{l_1=0}^{(N_{\textsf{s}})^{ML_{\textsf{ch}}}-1}  \nonumber \\
& ~~~~~~~~~~~~~~~~~~-  \min  \left\{ -2\operatorname{Re}\bigg(\sum\limits_{m=0}^{M-1}\big(\tilde{\mathbf{r}}_{N,0}[nM-m] \big)^H \right. \cdot \mathds{C}^{-1}_{{\textsf{z}}}\cdot \mathds{B}[n] \cdot \mathbf{a}_{q_{(l_0,m)}}\bigg)  \nonumber \\
&~~~~~~~~~~~~~~~~~~~~~~~~~~~~~~~+\bigg( \sum\limits_{m=0}^{M-1}\big( \mathbf{a}_{q_{(l_0,m)}}\big)^{H} \Big( \big(\mathds{B}[n] \big)^{H}\cdot \mathds{C}^{-1}_{{\textsf{z}}} \cdot \mathds{B}[n] \Big) \cdot \mathbf{a}_{q_{(l_0,m)}} \bigg) \Bigg\}_{l_0=0}^{(N_{\textsf{s}})^{L_{\textsf{tot}}}-1} \label{Symplfy_1}
\end{align}
where $(a)$ follows since $\sum\limits_{m=0}^{M-1}\big(\tilde{\mathbf{r}}_{N,0}[nM-m] \big)^H \cdot \mathds{C}^{-1}_{{\textsf{z}}}\cdot \big(\tilde{\mathbf{r}}_{N,0}[nM-m]\big)$ is an additive constant for both minimizations which appears in all values of $l_0$ and $l_1$, and from the identity:  $\min(x_1+a,x_2+a,...,x_N+a) = \min(x_1,x_2,...,x_N)+a,~\mbox{for} ~a=\mbox{constant}$.\\
Next, examine the summation $\sum\limits_{m=0}^{M-1}\big( \mathbf{a}_{q_{(l,m)}}\big)^{H} \Big( \big(\mathds{B}[n] \big)^{H}\cdot \mathds{C}^{-1}_{{\textsf{z}}} \cdot \mathds{B}[n] \Big) \cdot \mathbf{a}_{q_{(l,m)}}$ :\\
\begin{ceqn}
\begin{align}
&\sum\limits_{m=0}^{M-1}\big( \mathbf{a}_{q_{(l,m)}}\big)^{H} \Big( \big(\mathds{B}[n] \big)^{H}\cdot \mathds{C}^{-1}_{{\textsf{z}}} \cdot \mathds{B}[n] \Big) \cdot \mathbf{a}_{q_{(l,m)}} \nonumber \\
&\stackrel{(a)}{=} \sum\limits_{m=0}^{M-1}\Tr \Big\{ \big( \mathbf{a}_{q_{(l,m)}}\big)^{H} \Big( \big(\mathds{B}[n] \big)^{H}\cdot \mathds{C}^{-1}_{{\textsf{z}}} \cdot \mathds{B}[n] \Big) \cdot \mathbf{a}_{q_{(l,m)}} \Big\} \nonumber \\
&\stackrel{(b)}{=}  \sum\limits_{m=0}^{M-1}\Tr \Big\{\Big( \big(\mathds{B}[n] \big)^{H}\cdot \mathds{C}^{-1}_{{\textsf{z}}} \cdot \mathds{B}[n] \Big) \cdot \mathbf{a}_{q_{(l,m)}} \big( \mathbf{a}_{q_{(l,m)}}\big)^{H} \Big\} \nonumber \\
&\stackrel{(c)}{=} \Tr \Big\{ \Big( \big(\mathds{B}[n] \big)^{H}\cdot \mathds{C}^{-1}_{{\textsf{z}}} \cdot \mathds{B}[n] \Big) \cdot  \sum\limits_{m=0}^{M-1} \mathbf{a}_{q_{(l,m)}} \big( \mathbf{a}_{q_{(l,m)}}\big)^{H}\Big\} \label{Symplfy_2}
\end{align}
\end{ceqn}
where $(a)$ follows since for a scalar $g$, $\Tr\{g\}=g$, $(b)$ follows from the cyclic property of the $\Tr\{\cdot\}$ operator \cite[Ch. 3, p. 110]{meyer2000matrix} : $\Tr\{\mathds{A}\mathds{B}\mathds{C}\}=\Tr\{\mathds{C}\mathds{A}\mathds{B}\}=\Tr\{\mathds{B}\mathds{C}\mathds{A}\}$, and $(c)$ follows since $\Tr\{\cdot\}$ is a linear operator. We define the following matrices:
\begin{ceqn}
\begin{subequations}
\begin{align}
&\mathds{D}^{\textsf{(data)}}_{l_0}\triangleq \sum\limits_{m=0}^{M-1} \mathbf{a}_{q_{(l_0,m)}} \big( \mathbf{a}_{q_{(l_0,m)}}\big)^{H},~l_0 = 0,1,...,(N_{\textsf{s}})^{L_{\textsf{tot}}}-1~, \\
&\mathds{D}^{\textsf{(sw)}}_{l_1}\triangleq \sum\limits_{m=0}^{M-1} \tilde{\mathbf{a}}_{\tilde{q}_{(l_1,m)},m} \big( \tilde{\mathbf{a}}_{\tilde{q}_{(l_1,m)},m}\big)^{H},~l_1 = 0,1,...,(N_{\textsf{s}})^{ML_{\textsf{ch}}}-1~, 
\end{align}
\end{subequations}
\end{ceqn}
we note that $\mathds{D}^{\textsf{(data)}}_{l_0},~\mathds{D}^{\textsf{(sw)}}_{l_1}$ can be calculated apriori, before frame synchronization is applied.
Using $(\ref{Symplfy_1})$ and $(\ref{Symplfy_2})$ we arrive at the following approximate LRT (ALRT):
\begin{align*}
&\frac{1}{(N_{\textsf{s}})^{L_{\textsf{tot}}}}\cdot \Bigg( \min  \Bigg\{ \Tr \bigg\{ \Big( \big(\mathds{B}[n] \big)^{H}\cdot \mathds{C}^{-1}_{{\textsf{z}}} \cdot \mathds{B}[n] \Big) \cdot  \mathds{D}^{\textsf{(sw)}}_{l_1}\bigg\}   \nonumber \\
& ~~~~~~~~~~ -2\operatorname{Re}\Big( \sum\limits_{m=0}^{M-1}\big(\tilde{\mathbf{r}}_{N,0}[nM-m] \big)^H\cdot \mathds{C}^{-1}_{{\textsf{z}}}\cdot \mathds{B}[n] \cdot \tilde{\mathbf{a}}_{\tilde{q}_{(l_1,m)}}\Big) \Bigg\}_{l_1=0}^{(N_{\textsf{s}})^{ML_{\textsf{ch}}}-1}  \nonumber  \\
& -  \min  \Bigg\{\Tr \bigg\{ \Big( \big(\mathds{B}[n] \big)^{H}\cdot \mathds{C}^{-1}_{{\textsf{z}}} \cdot \mathds{B}[n] \Big) \cdot  \mathds{D}^{\textsf{(data)}}_{l_0}\bigg\}  \nonumber \\
& ~~~~~~~~~~ -2\operatorname{Re}\Big(\sum\limits_{m=0}^{M-1}\big(\tilde{\mathbf{r}}_{N,0}[nM-m] \big)^H\cdot \mathds{C}^{-1}_{{\textsf{z}}}\cdot \mathds{B}[n] \cdot \mathbf{a}_{q_{(l_0,m)}}\big) \Bigg\} _{l_0=0}^{(N_{\textsf{s}})^{L_{\textsf{tot}}}-1} \Bigg) \nonumber \\
&~~~~~~~~~~~~~~~~~~~~~~~~~~~~~~~~~~~~~~~~~~~~~~~~~\underset{H_1}{\overset{H_0}{\gtrless}} \bigg(\log_e\Big( \lambda \cdot( N_{\textsf{s}})^{MK }\Big)  \bigg)\cdot \frac{1}{(N_{\textsf{s}})^{L_{\textsf{tot}}}}~. 
\end{align*}
\subsection{Detailed Complexity Analysis}
\label{app:Complexity_Analysis}
\subsubsection{Complexity Analysis for the SALRT}
\label{app:SALRT_Complexity_Analysis}
In the following, we detail the derivation of the computational complexity expressions stated in Table \ref{table:Computational_Complexity_2}, in CMs and CAs, that adds up to the total computational complexity of the SALRT detector in Eq.(\ref{Suboptimal_LRT_detector}):
\renewcommand{\arraystretch}{1.5}%
\begin{table}[H] 
\caption{Computational complexity of the SALRT detector part \RNum{1} }
\begin{tabular}{ |c|l|l|c| }
\hline
Step & \multicolumn{2}{c|}{Description}   &\multicolumn{1}{c|}{~~~~~~~~~~~~~~~~~~~Term~~~~~~~~~~~~~~~~~~~~}\\
\hline
\multirow{2}{*}{1} & \multicolumn{2}{c|}{\multirow{2}{14em}{Construct symbols matrices: Eqns. (\ref{D_data_Matrix}), (\ref{D_sw_Matrix})}} & $\left\{\mathds{D}^{\textsf{(data)}}_{l_0}\right\}_{l_0=0}^{l_0=(N_{\textsf{s}})^{L_{\textsf{tot}}}-1}$  \\
\cline{4-0}
~~ &\multicolumn{2}{c|}{} & $\left\{\mathds{D}^{\textsf{(sw)}}_{l_1}\right\}_{l_1=0}^{l_1=(N_{\textsf{s}})^{L_{\textsf{ch}}}-1}$   \\
\cline{3-4} 
\hline
2& \multicolumn{2}{c|}{Collect $L_{\textsf{tot}}$ channel samples: Eqn. (\ref{Received_Vector})}& $\mathbf{r}_{L_{\textsf{tot}},0}[n]$   \\ 
\hline
\multirow{8}{*}{3} & \multirow{7}{5em}{Channel matrix estimation}  & (3a) Evaluate & $\left\{\mathbf{u}^{(i,J-(L_{\textsf{ch}}+1))}\right\}_{i=0}^{i=P_{\textsf{h}}-1}$  \\
& &~~~~~~~equalizer  FIR:&\\
& ~&~~~~~~~Eqns. (\ref{Equalizer_d_Output}),  (\ref{Equalizer_Finite_Impulse_Response})& \\
 \cline{3-4}
&   &(3b) Estimate& $\hat{s}\big[n-k \cdot P_{\textsf{h}}-i\big]$ ,  \\
& &~~~~~~~transmitted symbols:&$i=0,1,...,P_{\textsf{h}}-1$ ,\\
& &~~~~~~~Eqn. (\ref{Symbol_Estimation})  &$k=0,1,...,J-(L_{\textsf{ch}}+1)$\\
 \cline{3-4}
&   &(3c) Estimate CIR:&$\left\{\hat{\tilde{\mathbf{h}}}[n,i]\right\}_{i=0}^{i=P_{\textsf{h}}-1}$ \\
& &~~~~~~~Eqn. (\ref{CIR_Estimation}) & $\mathds{\hat{B}}[n]$\\
\hline
\multirow{6}{*}{4}& \multirow{4}{5em}{Reducing grid search complexity} &(4a) hard decision& $\hat{\mathbf{s}}_{L_{\textsf{tot}},0}^{(\textsf{data})}[n]~ \Big|~ \hat{\mathbf{s}}_{L_{\textsf{tot}},0}^{(\textsf{sw})}[n]$  \\
&  &~~~~~~detector:& \\
&  &~~~~~~Eqns. (\ref{Hard_Decision_Detector_Data}), (\ref{Hard_Decision_Detector_SW}) & \\
\cline{3-4}
&  &(4b) Evaluate new&$\mathcal{Q}_0~|~\mathcal{Q}_1$  \\
&  &~~~~~~grid search: &\\
&  &~~~~~~ Eqns. (\ref{Grid_Search_Data}), (\ref{Grid_Search_sw}) & \\
\hline
5& \multicolumn{2}{c|}{Post-processing: Eqn. (\ref{Post_Processing})}& $\mathbf{r}^{(\mathsf{P})}_{L_{\textsf{tot}},0}[n]$   \\
\hline
6 & \multicolumn{2}{c|}{Suboptimal LRT: Eqn. (\ref{Suboptimal_LRT_detector})}& SALRT$\Big(\mathbf{r}^{(\mathsf{P})}_{L_{\textsf{tot}},0}[n]\Big)$    \\
\hline
\end{tabular}
\label{table:Computational_Complexity_1}
\end{table}
\addtocounter{table}{-1}
\renewcommand{\arraystretch}{1.5}%
\begin{table}[H]
\caption{Computational complexity of the SALRT detector part \RNum{2} }
\begin{tabular}{ |c|l|l| }
\hline
Step&\multicolumn{1}{|c|}{CM} &\multicolumn{1}{c|}{CA} \\
\hline
\multirow{2}{*}{1}&$(N_{\textsf{s}})^{L_{\textsf{tot}}}\cdot(MN^2)$ &$(N_{\textsf{s}})^{L_{\textsf{tot}}}\cdot\big((M-1)\cdot N^2\big)$\\
\cline{2-3}
~&$(N_{\textsf{s}})^{ML_{\textsf{ch}}}\cdot(MN^2)$&$(N_{\textsf{s}})^{ML_{\textsf{ch}}}\cdot\big((M-1)\cdot N^2\big)$\\
\cline{1-3} 
\hline
2&\hrulefill& \hrulefill\\ 
\hline
3a&
$P_{\textsf{h}} \cdot
\left( \begin{array}{l}
\big(J-(L_{\textsf{ch}}+1)\big)\\
~~~~\cdot\big(2\cdot (L_{\textsf{ch}}+1)+3\big)
\end{array}\right)$& $ P_{\textsf{h}} \cdot
\left( \begin{array}{l}
\big(J-(L_{\textsf{ch}}+1)\big)\\
~~~~\cdot\big(2\cdot (L_{\textsf{ch}}+1)\big)
\end{array}\right)$ \\
 \cline{1-3}
3b&$(L_{\textsf{EQ}}-P_{\textsf{h}} \cdot L_{\textsf{ch}})\cdot (N_{\textsf{s}}+L_{\textsf{ch}}+1)$& $(L_{\textsf{EQ}}-P_{\textsf{h}} \cdot L_{\textsf{ch}})\cdot(N_{\textsf{s}}+L_{\textsf{ch}})$\\
 \cline{1-3}
3c&$P_{\textsf{h}}\cdot \big(L_{\textsf{ch}}+2\big) \cdot
\left(\begin{array}{l}
(\Omega+1)\cdot (L_{\textsf{ch}}+1)\\
~~~~~~~+(L_{\textsf{ch}}+1)^2 
\end{array}\right)$
& 
$P_{\textsf{h}}\cdot
\left(\begin{array}{l}
\Omega \cdot (L_{\textsf{ch}}+1)\cdot (L_{\textsf{ch}}+2)\\
+ L_{\textsf{ch}}\cdot(L_{\textsf{ch}}+1)+(L_{\textsf{ch}})^3
\end{array}\right)$\\
\hline
4a&$(*)~c_1\cdot L_{\textsf{tot}}~ \Big|~ c_2\cdot L_{\textsf{tot}}$&$(*)~c_1\cdot(2L_{\textsf{tot}}-1)~ \Big|~ c_2\cdot(2L_{\textsf{tot}}-1)$\\
\cline{1-3}
4b&$~~~~~~~~~~~~~~~~~~~~~(**)$&$~~~~~~~~~~~~~~~~~~~~~~~~(**)$\\
\hline
5&\hrulefill& \hrulefill \\
\hline
6&$ \begin{array}{l}
\left( \begin{array}{l}
MK\cdot (N+1)\\
~~~+1+N^3
\end{array}\right)
\cdot\left(\begin{array}{l}|\mathcal{Q}_1|\\
+|\mathcal{Q}_0|\end{array}\right)\\
~~~~~+KN\cdot(K+N)+1\\
\end{array}$ &$ \begin{array}{l}
\left( \begin{array}{l}
M\cdot\left( \begin{array}{l}
(N-1)\cdot K\\
~+(K-1)
\end{array}\right)\\
~+(N-1)\cdot N^2+N\end{array}\right)\cdot\left(\begin{array}{l}|\mathcal{Q}_1|\\
+|\mathcal{Q}_0|\end{array}\right)\\
~~~~~+N\cdot (K-1)\cdot (K+N)+1\end{array}$ \\
\hline
\end{tabular}
\label{table:Computational_Complexity_2}
\end{table}
\noindent
$(*)$ As will be explained in the following $c_1\ll (N_{\textsf{s}})^{L_{\textsf{tot}}}$ and $c_2\ll (N_{\textsf{s}})^{ML_{\textsf{ch}}},~c_1,c_2\in \mathcal{N}$, are constants that depends on the chosen constellation set $\mathcal{S}$. \\
\noindent
$(**)$ As will be explained in the following step (4b) do not increase the computational complexity in CMs and CAs terms, since an equivalent process that builds the vectors in the sets $\mathcal{Q}_1$ and $\mathcal{Q}_2$ whose sizes is $\left(\sum\limits_{l=0}^{e_{\textsf{r}_0}}{N\choose l}\cdot (|\mathcal{S}|-1)^l\right)^M$ and $\left(\sum\limits_{l=0}^{e_{\textsf{r}_1}}{L_{\textsf{ch}}\choose l}\cdot (|\mathcal{S}|-1)^l\right)^M$ respectively is done, that increases the algorithm running time without implementing CMs and CAs.
\begin{enumerate}
    \item Step 1: Constructing $\left\{\mathds{D}^{\textsf{(data)}}_{l_0}\right\}_{l_0=0}^{(N_{\textsf{s}})^{L_{\textsf{tot}}}-1}$  and $\left\{\mathds{D}^{\textsf{(sw)}}_{l_1}\right\}_{l_1=0}^{(N_{\textsf{s}})^{L_{\textsf{ch}}}-1}$ symbols matrices  according to Eq. (\ref{D_data_Matrix}), (\ref{D_sw_Matrix}):
    \begin{enumerate}
        \item
        $\mathbf{a}_{q_{(l_0,k)}} \big( \mathbf{a}_{q_{(l_0,k)}}\big)^{H}$: A vector multiplication of $\mathbf{a}_{q_{(l_0,k)}} \in \mathcal{C}^{N~\times~1}$ with its transpose, results in an $\mathcal{C}^{N~\times~N}$ matrix at the cost of:  \mybox{CM:$~N^2$~~$|$~~CA:$~0~~$}. 
        \item
        $\mathds{D}^{\textsf{(data)}}_{l_0}\triangleq \sum\limits_{k=0}^{M-1} \mathbf{a}_{q_{(l_0,k)}} \big( \mathbf{a}_{q_{(l_0,k)}}\big)^{H}$: M vector multiplication of step (a) and M-1 matrix additions of $\mathcal{C}^{N~\times~N}$ matrices that results in 1 matrix $\mathds{D}^{\textsf{(data)}}_{l_0}$ at the cost of:
        \mybox{CM:$~MN^2$~~$|$~~CA:$~(M-1)\cdot N^2~~$}.
        \item
        $\left\{\mathds{D}^{\textsf{(data)}}_{l_0}\right\}_{l_0=0}^{(N_{\textsf{s}})^{L_{\textsf{tot}}}-1}$: $(N_{\textsf{s}})^{L_{\textsf{tot}}}$ matrix calculations at the cost of:\\
        \mybox{CM:$~(N_{\textsf{s}})^{L_{\textsf{tot}}}\cdot(MN^2)$~~$|$~~CA:$~(N_{\textsf{s}})^{L_{\textsf{tot}}}\cdot((M-1)\cdot N^2)~~$}.   
    \end{enumerate}
    Repeating the complexity analysis of $\mathds{D}^{\textsf{(data)}}_{l_0}$ for $\left\{\mathds{D}^{\textsf{(sw)}}_{l_1}\right\}_{l_1=0}^{(N_{\textsf{s}})^{L_{\textsf{ch}}}-1}$  we obtain a complexity of:\\
    \mybox{CM:$~(N_{\textsf{s}})^{ML_{\textsf{ch}}}\cdot(MN^2)$~~$|$~~CA:$~(N_{\textsf{s}})^{ML_{\textsf{ch}}}\cdot((M-1)\cdot N^2)~~$} for constructing the set\\ $\left\{\mathds{D}^{\textsf{(sw)}}_{l_1}\right\}_{l_1=0}^{(N_{\textsf{s}})^{L_{\textsf{ch}}}-1}$.
     \item Step 2: Collecting $L_{\textsf{tot}}$ channel samples: Eq. (\ref{Received_Vector}) does not incur a computational complexity.
    \item Step 3: Constructing the channel matrix estimate $\mathds{\hat{B}}[n]$ according to Eqs. (\ref{Equalizer_Finite_Impulse_Response}),  (\ref{Symbol_Estimation}), and (\ref{CIR_Estimation}):
    \begin{enumerate}
    \item Step (3a): Collect additional $L_{\textsf{EQ}}$ channel samples and evaluate equalizer FIR,\\ $\left\{\mathbf{u}^{(i,J-(L_{\textsf{ch}}+1))}\right\}_{i=0}^{i=P_{\textsf{h}}-1}$ via Eqs. (\ref{Equalizer_d_Output}), (\ref{Equalizer_Finite_Impulse_Response}):
        \begin{enumerate}
            \item
            $\hat{d}^{(i,k)}=\Big(\mathbf{u}^{(i,k)}\Big)^{T} \cdot \mathbf{r}_{\textsf{eqz}}^{(i,L_{\textsf{ch}})}[n-k \cdot P_{\textsf{h}}]$ : A vector multiplication of $\mathbf{u}^{(i,k)}\in \mathcal{C}^{(L_{\textsf{ch}}+1)~\times~1}$ and $\mathbf{r}_{\textsf{eqz}}^{(i,L_{\textsf{ch}})}[n-k \cdot P_{\textsf{h}}]\in \mathcal{C}^{(L_{\textsf{ch}}+1)~\times~1}$, has a complexity of:\\ 
            \mybox{CM:$~L_{\textsf{ch}}+1$~~$|$~~CA:$~L_{\textsf{ch}}~~$}
            \item
            $|\hat{d}^{(i,k)}|^2=\Big(\hat{d}^{(i,k)}\Big)^*\cdot \hat{d}^{(i,k)}$: A cost of 1 complex multiplication where $\hat{d}^{(i,k)}\in \mathcal{C}^{1~\times~1}$ is added to the cost of step (i.) resulting in a cumulative complexity of: \mybox{CM:$~L_{\textsf{ch}}+1+1$~~$|$~~CA:$~L_{\textsf{ch}}~~$} 
            \item 
            $\Delta_p \cdot \hat{d}^{(i,k)} \cdot (R_2-|\hat{d}^{(i,k)}|^2)$ : 2 complex multiplications and 1 complex addition are added to the complexity of step (ii.). The cumulative complexity is:
            \mybox{CM:$~L_{\textsf{ch}}+2+2$~~$|$~~CA:$~L_{\textsf{ch}}+1~~$}
            \item
            $\mathbf{u}^{(i,k+1)}=\mathbf{u}^{(i,k)}+\Delta_p \cdot (\mathbf{r}_{\textsf{eqz}}^{(i,L_{\textsf{ch}})}[n-k])^{*} \cdot \hat{d}^{(i,k)} \cdot (R_2-|\hat{d}^{(i,k)}|^2)$ : A cost of 1 vector multiplication of $\mathbf{r}_{\textsf{eqz}}^{(i,L_{\textsf{ch}})}[n-k] \in \mathcal{C}^{(L_{\textsf{ch}}+1)~\times~1}$ with a complex element $\mathcal{C}^{1~\times~1}$, and a cost of 1 complex vector addition of the product with $\mathbf{u}^{(i,k)} \in \mathcal{C}^{(L_{\textsf{ch}}+1)~\times~1}$, are added to the complexity of step (iii.), resulting in a cumulative complexity of:
            \mybox{CM:$~L_{\textsf{ch}}+4+L_{\textsf{ch}}+1$~~$|$~~CA:$~L_{\textsf{ch}}+1+L_{\textsf{ch}}+1~~$}
            \item 
            $\mathbf{u}^{(i,k+1)}=\mathbf{u}^{(i,k)}+\Delta_p \cdot (\mathbf{r}_{\textsf{eqz}}^{(i,L_{\textsf{ch}})}[n-k])^{*} \cdot \hat{d}^{(i,k)} \cdot (R_2-|\hat{d}^{(i,k)}|^2),$\\
            $i = 0,1,...,P_{\textsf{h}}-1,~ k = 0,1,...,J-(L_{\textsf{ch}}+2)$
            , the iterative process is repeated $J-(L_{\textsf{ch}}+1)$ for $P_{\textsf{h}}$ different time indexes within the period, resulting in a total complexity of step (3a) of:\\
            \mybox{CM:$~P_{\textsf{h}} \cdot \big(J-(L_{\textsf{ch}}+1)\big)\cdot\big(2\cdot(L_{\textsf{ch}}+1)+3\big)$}\\
            \mybox{CA:$~P_{\textsf{h}} \cdot\big(J-(L_{\textsf{ch}}+1)\big)\cdot\big(2\cdot(L_{\textsf{ch}}+1)\big))$}
        \end{enumerate}
        \item Step (3b), estimate additional $\left\{\hat{s}\big[n-k \cdot P_{\textsf{h}}-i\big]\right\}_{(i=0,~k=0)}^{\big(P_{\textsf{h}}-1,~J-(L_{\textsf{ch}}+1)\big)}$ transmitted symbols: Eq. (\ref{Symbol_Estimation}):
        \begin{enumerate}
            \item 
            $\bigg(\mathbf{u}^{\big(i,J-(L_{\textsf{ch}}+1)\big)}\bigg)^T\cdot \mathbf{r}_{\textsf{eqz}}^{(i,L_{\textsf{ch}})}[n-k \cdot P_{\textsf{h}}]$: vector multiplication of $\mathbf{u}^{\big(i,J-(L_{\textsf{ch}}+1)\big)}\in \mathcal{C}^{(L_{\textsf{ch}}+1)~\times~1}$ and $\mathbf{r}_{\textsf{eqz}}^{(i,L_{\textsf{ch}})}[n-k \cdot P_{\textsf{h}}]\in \mathcal{C}^{(L_{\textsf{ch}}+1)~\times~1}$, results in a complex number at the cost of:\\ 
            \mybox{CM:$~L_{\textsf{ch}}+1$~~$|$~~CA:$~L_{\textsf{ch}}~~$}
            \item
            $\Bigg\| \bigg(\mathbf{u}^{\big(i,J-(L_{\textsf{ch}}+1)\big)}\bigg)^T\cdot \mathbf{r}_{\textsf{eqz}}^{(i,L_{\textsf{ch}})}[n-k \cdot P_{\textsf{h}}]- \rho \Bigg\|^2$: For each $\rho \in \mathcal{S}$, where $|\mathcal{S}|=N_{\textsf{s}}$, 1 complex addition and 1 complex multiplication are applied in order to choose the minimum value to estimate $\hat{s}\big[n-k \cdot P_{\textsf{h}}-i\big]$. Adding this cost to step (i.) results in a cumulative cost of:\\
            \mybox{CM:$~L_{\textsf{ch}}+1+N_{\textsf{s}}$~~$|$~~CA:$~L_{\textsf{ch}}+N_{\textsf{s}}~~$}
            \item
            A total of $P_{\textsf{h}}\cdot (J-L_{\textsf{ch}})=P_{\textsf{h}}\cdot J-P_{\textsf{h}}\cdot L_{\textsf{ch}}= L_{\textsf{EQ}}-P_{\textsf{h}} \cdot L_{\textsf{ch}}$ symbols are estimated: $\hat{s}\big[n-k \cdot P_{\textsf{h}}-i\big],~$
            $i = 0,1,...,P_{\textsf{h}}-1,~k = 0,1,...,J-(L_{\textsf{ch}}+1)$
            bringing the total cost of step (3b) to:\\          \mybox{CM:$~(L_{\textsf{EQ}}-P_{\textsf{h}}L_{\textsf{ch}})\cdot(N_{\textsf{s}}+L_{\textsf{ch}}+1)$~~$|$~~CA:$~(L_{\textsf{EQ}}-P_{\textsf{h}}L_{\textsf{ch}})\cdot(N_{\textsf{s}}+L_{\textsf{ch}})~~$}
        \end{enumerate}
        \item 
        Step (3c) estimate CIR, $\hat{\tilde{\mathbf{h}}}[n,i]$ via Eq. (\ref{CIR_Estimation}):
        \begin{enumerate}
            \item 
            $(\hat{\mathds{G}}^{(i)}[n])^H \cdot \hat{\mathds{G}}^{(i)}[n]$: Matrix multiplication, where $\hat{\mathds{G}}^{(i)}[n]\in \mathcal{C}^{(\Omega+1)~\times~(L_{\textsf{ch}}+1)}$. This results in a square matrix whose dimensions are $\mathcal{C}^{(L_{\textsf{ch}}+1)~\times~(L_{\textsf{ch}}+1)}$, at the cost of:\\
            \mybox{CM:$~(\Omega+1)\cdot(L_{\textsf{ch}}+1)^2$~~$|$~~CA:$~\Omega\cdot(L_{\textsf{ch}}+1)^2~~$}
            \item
            $\big((\hat{\mathds{G}}^{(i)}[n])^H \cdot \hat{\mathds{G}}^{(i)}[n]\big)^{-1}$: Inverting a square matrix, has a cubic complexity with respect to the matrix size \cite[Chapter~28]{cormen2009introduction}, resulting in an overall complexity of steps (i.) and (ii.) to:\\
            \mybox{CM:$~(\Omega+1)\cdot(L_{\textsf{ch}}+1)^2+(L_{\textsf{ch}}+1)^3$~~$|$~~CA:$~\Omega\cdot(L_{\textsf{ch}}+1)^2+(L_{\textsf{ch}})^3~~$}
            \item
            $(\hat{\mathds{G}}^{(i)}[n])^H \cdot \mathbf{r}_{\textsf{eqz}}^{(i,\Omega)}[n]$: Matrix multiplication, where $\hat{\mathds{G}}^{(i)}[n]\in \mathcal{C}^{(\Omega+1)~\times~(L_{\textsf{ch}}+1)}$ and
            $\mathbf{r}_{\textsf{eqz}}^{(i,\Omega)}[n]\in \mathcal{C}^{(\Omega+1)~\times~1}$,
            results in a vector with dimensions $\mathcal{C}^{(L_{\textsf{ch}}+1)~\times~1}$. The overall complexity of steps i.--iii. is:\\
           \mybox{CM:$~(\Omega+1)\cdot(L_{\textsf{ch}}+1)$~~$|$~~CA:$~\Omega\cdot(L_{\textsf{ch}}+1)~~$}
            \item
            $\hat{\tilde{\mathbf{h}}}[n,i]=\big((\hat{\mathds{G}}^{(i)}[n])^H \cdot \hat{\mathds{G}}^{(i)}[n]\big)^{-1} \cdot (\hat{\mathds{G}}^{(i)}[n])^H \cdot \mathbf{r}_{\textsf{eqz}}^{(i,\Omega)}[n]$: Matrix multiplication, of a square matrix of dimensions $\mathcal{C}^{(L_{\textsf{ch}}+1)~\times~(L_{\textsf{ch}}+1)}$ and a vector of dimensions to $\mathcal{C}^{(L_{\textsf{ch}}+1)~\times~1}$. The overall complexity of steps i.--iv. is:\\
            \mybox{CM:$~(L_{\textsf{ch}}+1)^2+(\Omega+1)\cdot(L_{\textsf{ch}}+1)^2+(L_{\textsf{ch}}+1)^3+(\Omega+1)\cdot(L_{\textsf{ch}}+1)$~~}\\
            \mybox{CA:$~L_{\textsf{ch}}\cdot(L_{\textsf{ch}}+1)+\Omega\cdot(L_{\textsf{ch}}+1)^2+(L_{\textsf{ch}})^3 +\Omega\cdot(L_{\textsf{ch}}+1)~~$}
            \item 
            We repeat the calculation of $\hat{\tilde{\mathbf{h}}}[n,i]$ $P_{\textsf{h}}$ times, one time for each $i$, resulting at the total cost for CIR estimation of:\\
            \mybox{CM:$~P_{\textsf{h}}\cdot\Big(\big(L_{\textsf{ch}}+2\big)\cdot \big((\Omega+1)\cdot (L_{\textsf{ch}}+1)+(L_{\textsf{ch}}+1)^2\big)\Big)$~~}\\
            \mybox{CA:$~P_{\textsf{h}}\cdot\Big(\Omega \cdot (L_{\textsf{ch}}+1)\cdot (L_{\textsf{ch}}+2)+ L_{\textsf{ch}}\cdot(L_{\textsf{ch}}+1)+(L_{\textsf{ch}})^3 \Big)$~~}  
        \end{enumerate}
    \end{enumerate}
    \item Step 4: Reducing grid search complexity
    \begin{enumerate}
        \item In Step (4a) the hard decision estimates,~$\hat{\mathbf{s}}_{L_{\textsf{tot}},0}^{(\textsf{data})}[n]$,~$\hat{\mathbf{s}}_{L_{\textsf{tot}},0}^{(\textsf{sw})}[n]$, are generated via Eqs. (\ref{Hard_Decision_Detector_Data}), (\ref{Hard_Decision_Detector_SW}):
        \begin{enumerate}
            \item
            $\|\mathbf{r}_{L_{\textsf{tot}},0}[n]- \mathbf{b}_{l_0}^{(\textsf{data})} \|^2$: First 1 vector addition is applied. Then, for the squared norm operator, 1 vector complex addition and 1 vector complex multiplication are applied for a vector which belongs to $\in \mathcal{C}^{L_{\textsf{tot}}~\times~1}$. The overall cost is:
            \mybox{CM:$~L_{\textsf{tot}}$~~$|$~~CA:$~2L_{\textsf{tot}}-1~~$}
            \item 
            In a simple implementation of Eq. (\ref{Hard_Decision_Detector_Data}), the squared norm $\|\mathbf{r}_{L_{\textsf{tot}},0}[n]- \mathbf{b}_{l_0}^{(\textsf{data})} \|^2$ is repeated $(N_{\textsf{s}})^{L_{\textsf{tot}}}$ times, and $\hat{\mathbf{s}}_{L_{\textsf{tot}},0}^{(\textsf{data})}[n]$ is selected as the vector $\mathbf{b}_{l_0}$ which yields the minimum value:\\
            $\hat{\mathbf{s}}_{L_{\textsf{tot}},0}^{(\textsf{data})}[n]= \underset{{\mathbf{b}_{l_0}^{(\textsf{data})}}}{\argmin} \left\{\|\mathbf{r}_{L_{\textsf{tot}},0}[n]- \mathbf{b}_{l_0}^{(\textsf{data})} \|^2\right\}, ~l_0 = 0,1,...,(N_{\textsf{s}})^{L_{\textsf{tot}}}-1$. An alternative efficient method of evaluating step ii. would be extracting the phase and amplitude of each coordinate in the vector $\mathbf{r}_{L_{\textsf{tot}},0}[n]$ and through a look up table minimize the number of needed squared norm calculation from $(N_{\textsf{s}})^{L_{\textsf{tot}}}$ to a constant $c_1\ll (N_{\textsf{s}})^{L_{\textsf{tot}}}$ that depends on the transmitted constellation set $\mathcal{S}$. As an example, in a BPSK constellation set, $\mathcal{S}=\{-1,1\}$, instead of repeating the squared norm calculation $2^{L_{\textsf{tot}}}$ times, just checking each coordinate phase will give the resulted symbol hard decision estimation as 1 or -1, that is $c_1=1$. The total cost of steps i. and ii. is:\\
            \mybox{CM:$~c_1\cdot L_{\textsf{tot}}$~~$|$~~CA:$~c_1\cdot(2L_{\textsf{tot}}-1)~~$}
        \end{enumerate}
        Following the same steps of complexity analysis of computing $\hat{\mathbf{s}}_{L_{\textsf{tot}},0}^{(\textsf{data})}[n]$ to the computation of $\hat{\mathbf{s}}_{L_{\textsf{tot}},0}^{(\textsf{sw})}[n]$,  we obtain the total cost for computing $\hat{\mathbf{s}}_{L_{\textsf{tot}},0}^{(\textsf{sw})}[n]$, where $c_2\ll (N_{\textsf{s}})^{ML_{\textsf{ch}}}$ as:\\ \mybox{CM:$~c_2\cdot L_{\textsf{tot}}$~~$|$~~CA:$~c_2\cdot(2L_{\textsf{tot}}-1)~~$}.
        \item 
        Step (4b): Evaluate new grid search $\mathcal{Q}_0$,~,$\mathcal{Q}_1$ via Eqs. (\ref{Grid_Search_Data}), (\ref{Grid_Search_sw}). Since after step (4a), $\hat{\mathbf{s}}_{L_{\textsf{tot}},0}^{(\textsf{data})}[n]$ and $ \hat{\mathbf{s}}_{L_{\textsf{tot}},0}^{(\textsf{sw})}[n]$ are known, we do not have to explicitly calculate Eqs. (\ref{Grid_Search_Data})~\mbox{and}~ (\ref{Grid_Search_sw}) in order to evaluate the new grid sets $\mathcal{Q}_1$ and $\mathcal{Q}_2$. An equivalent process would be to chose up to $e_r$ coordinates from the estimated $L_{\textsf{tot}}$ length vectors and change each coordinate with a different symbol from $\mathcal{S}$, the symbol constellation set, that is building $\left(\sum\limits_{l=0}^{e_{\textsf{r}_0}}{N\choose l}\cdot (|\mathcal{S}|-1)^l\right)^M$ and $\left(\sum\limits_{l=0}^{e_{\textsf{r}_1}}{L_{\textsf{ch}}\choose l}\cdot (|\mathcal{S}|-1)^l\right)^M$  vectors for the sets $\mathcal{Q}_1$ and $\mathcal{Q}_2$ respectively, this process do not increase the computational complexity in CMs and CAs terms, but does increase the algorithm running time.
    \end{enumerate}
    \item Step 5: Post-processing via Eq. (\ref{Post_Processing}) does not incur a computational complexity.
    \item Step 6: SALRT computation via Eq. (\ref{Suboptimal_LRT_detector}):
    \begin{enumerate}
        \item 
        $\Big( \big(\mathds{\hat{B}}[n] \big)^{H}\cdot \mathds{C}^{-1}_{{\textsf{z}}} \cdot \mathds{\hat{B}}[n] \Big):$
         First consider $\mathds{C}^{-1}_{{\textsf{z}}} \cdot \mathds{\hat{B}}[n]$ matrix multiplication, where $\mathds{\hat{B}}[n]\in \mathcal{C}^{K~\times~N}$ and $\mathds{C}^{-1}_{{\textsf{z}}}\in \mathcal{R}^{K~\times~K}$ that results in a matrix of dimensions $\mathcal{C}^{K~\times~N}$ at the cost of:\\
        \mybox{CM:$~K^2N$~~$|$~~CA:$~ KN\cdot(K-1)$}. Next, to complete the calculation, we apply a second matrix multiplication, where $\mathds{C}^{-1}_{{\textsf{z}}} \cdot \mathds{\hat{B}}[n] \in \mathcal{C}^{K~\times~N}$ and $\mathds{\hat{B}}[n] \in \mathcal{C}^{K~\times~N}$ that results in a square matrix belongs to $\mathcal{C}^{N~\times~N}$. The overall complexity of step (a) is thus:\\
        \mybox{CM:$~KN^2+K^2 N$~~$|$~~CA:$~ KN \cdot (K-1)+N^2\cdot (K-1)~~$} 
        \item
        $ \Tr \bigg\{ \Big( \big(\mathds{\hat{B}}[n] \big)^{H}\cdot \mathds{C}^{-1}_{{\textsf{z}}} \cdot \mathds{\hat{B}}[n] \Big) \cdot      \mathds{D}^{\textsf{(sw)}}_{l_1}\bigg\}:$ matrix multiplication, where $\mathds{D}^{\textsf{(sw)}}_{l_1}\in \mathcal{C}^{N~\times~N}$, that results in a square matrix belongs to $\mathcal{C}^{N~\times~N}$, followed by the trace operator. The complexity of step (b) is thus:\\
        \mybox{CM:$~N^3$~~$|$~~CA:$~(N-1)\cdot N^2+N-1~$}.
        \item
        $\Big(\tilde{\mathbf{r}}_{N,0}[nM-m] \big)^H \cdot \mathds{C}^{-1}_{{\textsf{z}}}\cdot \mathds{\hat{B}}[n] \cdot \tilde{\mathbf{a}}_{\tilde{q}_{(l_1,m)},m}\Big)$: First consider $\mathds{C}^{-1}_{{\textsf{z}}} \cdot \mathds{\hat{B}}[n]$ matrix multiplication that results in a matrix belongs to $\mathcal{C}^{K~\times~N}$ that was computed in the first matrix multiplication in step (6a), thus, this quantity does not incur computional complexity. 
        Next, to complete the calculation, we apply a second matrix multiplication, where $\tilde{\mathbf{r}}_{N,0}[nM-m]\in \mathcal{C}^{K~\times~1}$,  $\mathds{C}^{-1}_{{\textsf{z}}} \cdot \mathds{\hat{B}}[n] \in \mathcal{R}^{K~\times~N}$ and $\tilde{\mathbf{a}}_{\tilde{q}_{(l_1,m)},m}\in \mathcal{C}^{N~\times~1}$ that results in a complex number. The complexity of step (c) is thus:
        \mybox{CM:$~NK+K$~~$|$~~CA:$~(N-1)\cdot K+(K-1)~$}. 
        \item
        $-2\operatorname{Re}\Big(\sum\limits_{m=0}^{M-1}\Big(\tilde{\mathbf{r}}_{N,0}[nM-m] \big)^H \cdot \mathds{C}^{-1}_{{\textsf{z}}}\cdot \mathds{\hat{B}}[n] \cdot \tilde{\mathbf{a}}_{\tilde{q}_{(l_1,m)},m}\Big)\Big)$: $M$ calculation of step (6c) plus 1 complex multiplication. Thus the combined complexities of steps (c), and (d) is:\\
        \mybox{CM:$~MK\cdot (N+1)+1$~~$|$~~CA:$~M \cdot \big((N-1)\cdot K+(K-1)\big)~$}.
        \item 
        The overall complexity for the minimization over $\mathcal{Q}_1$ in Eq. (\ref{Suboptimal_LRT_detector}) is obtained by noting that step (d) is done once, while the evaluations (6b) and (6d) are repeated for each $l_1 \in \mathcal{L}_1$. Accordingly the overall complexity of the minimization over $\mathcal{L}_1$ is:\\
        \mybox{CM:$~\big(MK\cdot (N+1)+1+N^3\big)\cdot|\mathcal{Q}_1|+KN^2+K^2N$}\\
        \mybox{CA:$~ \begin{array}{l}\Big(M\big((N-1)K+(K-1)\big)+(N-1)\cdot N^2+N-1+1\Big)\cdot|\mathcal{Q}_1|\\
        ~~~~~~~+N\cdot (K-1)\cdot(K+N)\end{array}$}
        \item
        Repeating calculations (6b)-(6e) for all $l_0 \in \mathcal{L}_0$ to implement the minimization over $\mathcal{Q}_0$. The overall complexity of Eq. (\ref{Suboptimal_LRT_detector}), accounting for the multiplication by  $\frac{1}{(N_{\textsf{s}})^{L_{\textsf{tot}}}}$ is obtained as:\\
        \mybox{CM:~~$\begin{array}{l}
        \big(MK\cdot (N+1)+1+N^3\big)\cdot\big(|\mathcal{Q}_1|+|\mathcal{Q}_0|\big)\\
        ~~~~+KN\cdot(K+N)+1\\
        \end{array}$}\\
        \mybox{CA:$~ \begin{array}{l}
        \Big( M\cdot \big((N-1)\cdot K+(K-1)\big)+(N-1)\cdot N^2+N\Big)\cdot\big(|\mathcal{Q}_1|
        +|\mathcal{Q}_0|\big)\\
        ~~~~~~~+N\cdot (K-1)\cdot (K+N)+1\end{array}$}
    \end{enumerate}
\end{enumerate}
The total computational complexity following steps (1)-(6) is:\\
\mybox{$ \mbox{CM}:\begin{array}{l}
~\ \begin{array}{l}
\left( \begin{array}{l}
MK\cdot (N+1)+1+N^3\\
\end{array}\right)
\cdot\big(|\mathcal{Q}_1|+|\mathcal{Q}_0|\big)+KN\cdot(K+N)+1\\
\end{array}\\
~~~~+P_{\textsf{h}} \cdot
 \begin{array}{l}
\big(J-(L_{\textsf{ch}}+1)\big)\cdot\big(2\cdot (L_{\textsf{ch}}+1)+3\big)\\
\end{array}\\
~~~~~~~+P_{\textsf{h}}\cdot \big(L_{\textsf{ch}}+2\big) \cdot
\Big(\begin{array}{l}
(\Omega+1)\cdot (L_{\textsf{ch}}+1)+(L_{\textsf{ch}}+1)^2\\
\end{array}\Big)\\
~~~~~~~~~~+(L_{\textsf{EQ}}-P_{\textsf{h}} \cdot L_{\textsf{ch}})\cdot (N_{\textsf{s}}+L_{\textsf{ch}}+1)+\big( (N_{\textsf{s}})^{L_{\textsf{tot}}}+(N_{\textsf{s}})^{ML_{\textsf{ch}}}\big)\cdot L_{\textsf{tot}}
\end{array}$}\\
\mybox{$\mbox{CA}:~\begin{array}{l} \begin{array}{l}
\Big( M\cdot \big((N-1)\cdot K+(K-1)\big)+(N-1)\cdot N^2+N\Big)\cdot\big(|\mathcal{Q}_1|
        +|\mathcal{Q}_0|\big)\\
~~~~+N\cdot (K-1)\cdot (K+N)+1+P_{\textsf{h}} \cdot
\begin{array}{l}
\big(J-(L_{\textsf{ch}}+1)\big)\cdot\big(2\cdot (L_{\textsf{ch}}+1)\big)
\end{array}\end{array}\\
~~~~~~+P_{\textsf{h}}\cdot
\left(\begin{array}{l}
\Omega \cdot (L_{\textsf{ch}}+1)\cdot (L_{\textsf{ch}}+2) +L_{\textsf{ch}}\cdot(L_{\textsf{ch}}+1)+(L_{\textsf{ch}})^3\\
\end{array}\right)\\
~~~~~~~~~+(L_{\textsf{EQ}}-P_{\textsf{h}} \cdot L_{\textsf{ch}})\cdot(N_{\textsf{s}}+L_{\textsf{ch}})+\big((N_{\textsf{s}})^{L_{\textsf{tot}}}+(N_{\textsf{s}})^{ML_{\textsf{ch}}}\big)\cdot(2L_{\textsf{tot}}-1)
\end{array}$}~,\\
in the total complexity calculation, the complexity of Step 1. calculation is omitted, since it can be carried out offline prior to the transmission, therefore it does not affect the real-time complexity of the frame synchronization algorithm.
\subsubsection{Complexity Analysis for the ALRT and the RALRT }
\label{app:ALRT_Complexity_Analysis}
In the following, we overview the considerations used for obtaining the total computational complexity of the ALRT and of the RALRT detector in Eqs. (\ref{Symplfy_Trace_2}) and (\ref{RALRT}), respectively, in CMs and CAs:
The computational complexity of the ALRT and of the RALRT is equivalent to the computational complexity derived in step (6) for SALRT detector in Appendix \ref{app:SALRT_Complexity_Analysis} except the following changes:
\begin{enumerate}
    \item 
    Since in ALRT and RALRT detectors, we assume a known channel matrix $\mathds{B}[n]$, the expressions: $\Tr \bigg\{ \Big( \big(\mathds{B}[n] \big)^{H}\cdot \mathds{C}^{-1}_{{\textsf{z}}} \cdot \mathds{B}[n] \Big) \cdot  \mathds{D}^{\textsf{(sw)}}_{l_1}\bigg\}_{l_1=0}^{(N_{\textsf{s}})^{ML_{\textsf{ch}}}-1}$ and $\Tr \bigg\{ \Big( \big(\mathds{B}[n] \big)^{H}\cdot \mathds{C}^{-1}_{{\textsf{z}}} \cdot \mathds{B}[n] \Big) \cdot  \mathds{D}^{\textsf{(data)}}_{l_0}\bigg\}_{l_0=0}^{(N_{\textsf{s}})^{L_{\textsf{tot}}}-1}$ can be apriori calculated before frame synchronization begins, therefore we omit step (6a) and step (6b) computational complexity calculation in Appendix \ref{app:SALRT_Complexity_Analysis} from ALRT and RALRT total computational complexity.
    \item
    Since the ALRT does not include the reducing of the grid search described in Section \ref{Reducing_Grid_Search}, the grid search is done on all possible symbols combination, where RALRT is done on the reduced search grid.   
\end{enumerate}
Accordingly the total computational complexity of ALRT detector consists of steps 2,5 and modified 6, is:\\
        \mybox{CM:~~$
        \big(MK\cdot (N+1)+1\big)\cdot\big((N_{\textsf{s}})^{L_{\textsf{tot}}}+(N_{\textsf{s}})^{ML_{\textsf{ch}}}\big)+1$}\\
         \mybox{CA:$~M\cdot \big((N-1)\cdot K+(K-1)\big)
        \cdot\big((N_{\textsf{s}})^{L_{\textsf{tot}}}+(N_{\textsf{s}})^{ML_{\textsf{ch}}}\big)+1$} .\\
        and the total computational complexity of RALRT detector consists of steps 2, 4, 5 and modified 6 is:\\
        \mybox{CM:~~$
        \big(MK\cdot (N+1)+1\big)\cdot\big(|\mathcal{Q}_1|+|\mathcal{Q}_0|\big)+1+\big( (N_{\textsf{s}})^{L_{\textsf{tot}}}+(N_{\textsf{s}})^{ML_{\textsf{ch}}}\big)\cdot L_{\textsf{tot}}$}\\
        \mybox{CA:$\left(\begin{array}{l}~M\cdot \big((N-1)\cdot K+(K-1)\big)
        \cdot\big(|\mathcal{Q}_1|
        +|\mathcal{Q}_0|\big)+1\\
        ~~~~~~~~~+2\cdot\big((N_{\textsf{s}})^{L_{\textsf{tot}}}+(N_{\textsf{s}})^{ML_{\textsf{ch}}}\big)\cdot(2L_{\textsf{tot}}-1)\end{array}\right)$}~.
\subsubsection{Complexity Analysis for the LRT}
\label{app:LRT_Complexity_Analysis}
In the following, we detail the derivations of the computational complexity the total computational complexity of the ALRT detector in Eq. (\ref{Test}) in terms of CMs and CAs:
\begin{enumerate}
    \item 
   Step 1: Evaluate the exponent argument,\\ $\sum\limits_{k=0}^{M-1}\Big(\big(\tilde{\mathbf{r}}_{N,0}[nM-k]-\mathds{B}[n] \cdot \mathbf{a}_{q_{(l_0,k)}} \big)^H \cdot \mathds{C}^{-1}_{{\textsf{z}}}\cdot \big(\tilde{\mathbf{r}}_{N,0}[nM-k]-\mathds{B}[n] \cdot \mathbf{a}_{q_{(l_0,k)}}\big)\Big)$ :
    \begin{enumerate}
        \item
        $\big(\tilde{\mathbf{r}}_{N,0}[nM-k]-\mathds{B}[n] \cdot \mathbf{a}_{q_{(l_0,k)}} \big)$ : First consider $\mathds{B}[n] \cdot \mathbf{a}_{q_{(l_0,k)}}$ matrix multiplication, where $\mathds{B}[n]\in \mathcal{C}^{K~\times~N}$ and $\mathbf{a}_{q_{(l_0,k)}}\in \mathcal{R}^{N~\times~1}$ that results in a vector belongs to $\mathcal{C}^{K~\times~1}$ at the cost of:
        \mybox{CM:$~NK$~~$|$~~CA:$~ (N-1)\cdot K$}. Next, to complete the calculation, we apply a vector addition, where $\mathds{B}[n] \cdot \mathbf{a}_{q_{(l_0,k)}} \in \mathcal{R}^{K~\times~1}$ and $\tilde{\mathbf{r}}_{N,0}[nM-k] \in \mathcal{C}^{K~\times~1}$ that results in a vector belongs to $\mathcal{C}^{K~\times~1}$. Thus, the overall complexity of step (a) is:\\
        \mybox{CM:$~NK$~~$|$~~CA:$~(N-1)\cdot K+K~~$} 
        \item
        $\mathds{C}^{-1}_{{\textsf{z}}}\cdot \big(\tilde{\mathbf{r}}_{N,0}[nM-k]-\mathds{B}[n] \cdot \mathbf{a}_{q_{(l_0,k)}}\big)$ : The matrix multiplication, where $\big(\tilde{\mathbf{r}}_{N,0}[nM-k]-\mathds{B}[n] \cdot \mathbf{a}_{q_{(l_0,k)}}\big) \in \mathcal{C}^{K~\times~1}$ and $\mathds{C}^{-1}_{{\textsf{z}}}\in \mathcal{R}^{K~\times~K}$ results in a vector whose dimensions are $\mathcal{C}^{K~\times~1}$. The overall complexity of steps (a) and (b) is thus:\\
        \mybox{CM:$~NK+K^2$~~$|$~~CA:$~(N-1)\cdot K+K+(K-1)\cdot K$}.
        \item
        $\Big(\big(\tilde{\mathbf{r}}_{N,0}[nM-k]-\mathds{B}[n] \cdot \mathbf{a}_{q_{(l_0,k)}} \big)^H \cdot \mathds{C}^{-1}_{{\textsf{z}}}\cdot \big(\tilde{\mathbf{r}}_{N,0}[nM-k]-\mathds{B}[n] \cdot \mathbf{a}_{q_{(l_0,k)}}\big)\Big)$ :
         A vector multiplication where $\mathds{C}^{-1}_{{\textsf{z}}}\cdot \big(\tilde{\mathbf{r}}_{N,0}[nM-k]-\mathds{B}[n] \cdot \mathbf{a}_{q_{(l_0,k)}}\big) \in \mathcal{C}^{K~\times~1}$ and $\big(\tilde{\mathbf{r}}_{N,0}[nM-k]-\mathds{B}[n] \cdot \mathbf{a}_{q_{(l_0,k)}}\big)\in \mathcal{R}^{K~\times~1}$ results in a complex number. The overall complexity of steps (a), (b) and (c) is:\\
        \mybox{CM:$~NK+K^2+K$~~$|$~~CA:$~(N-1)\cdot K+K+(K-1)\cdot K+K-1$}, which can be simplified as:
        \mybox{CM:$~K\cdot(N+K+1)$~~$|$~~CA:$~KN+(K-1)\cdot(K+1)$}.
        \item
        Finally, evaluating step (1c) $M$ times, one for each element in the summation,  brings the computational complexity of evaluations of the exponential term to:\\
        \mybox{CM:$~MK\cdot(N+K+1)$~~$|$~~CA:$~M\cdot\big(KN+(K-1)\cdot(K+1)\big)$}.
    \end{enumerate}
    \item 
    Step 2: Computing the numerator of Eq. (\ref{Test}):
    \begin{ceqn}
    \begin{align*}
    &\sum\limits_{l_0=0}^{(N_{\textsf{s}})^{L_{\textsf{tot}}}-1} e^{-\sum\limits_{k=0}^{M-1}\Big(\big(\tilde{\mathbf{r}}_{N,0}[nM-k]-\mathds{B}[n] \cdot \mathbf{a}_{q_{(l_0,k)}} \big)^H \cdot \mathds{C}^{-1}_{{\textsf{z}}}\cdot \big(\tilde{\mathbf{r}}_{N,0}[nM-k]-\mathds{B}[n] \cdot \mathbf{a}_{q_{(l_0,k)}}\big)\Big)}~,
    \end{align*}
    \end{ceqn}
    requires $(N_{\textsf{s}})^{L_{\textsf{tot}}}$ evaluations of the exponential term of Step 1. The computational complexity is:\\
    \mybox{CM:$~ MK\cdot(N+K+1)\cdot (N_{\textsf{s}})^{L_{\textsf{tot}}}$}\\
    \mybox{CA:$~ M\cdot\big(KN+(K-1)\cdot(K+1)\big)\cdot (N_{\textsf{s}})^{L_{\textsf{tot}}}$}
    \item 
    Step 3: Following the same computations done for Step 1 and Step 2, we obtain that the computational complexity of evaluating the exponential at the denominator
    \begin{ceqn}
    \begin{align*}
    &\sum\limits_{l_1=0}^{(N_{\textsf{s}})^{ML_{\textsf{ch}}}-1} e^{-\sum\limits_{k=0}^{M-1}\Big(\big(\tilde{\mathbf{r}}_{N,0}[nM-k]-\mathds{B}[n] \cdot \tilde{\mathbf{a}}_{\tilde{q}_{(l_1,k)},k} \big)^H \cdot \mathds{C}^{-1}_{{\textsf{z}}} \cdot \big(\tilde{\mathbf{r}}_{N,0}[nM-k]-\mathds{B}[n] \cdot \tilde{\mathbf{a}}_{\tilde{q}_{(l_1,k)},k}\big)\Big)}
    \end{align*}
    \end{ceqn}
    , to be:\\
    \mybox{CM:$~ MK\cdot(N+K+1)\cdot (N_{\textsf{s}})^{ML_{\textsf{ch}}}$}\\
    \mybox{CA:$~M\cdot\big(KN+(K-1)\cdot(K+1)\big) \cdot (N_{\textsf{s}})^{ML_{\textsf{ch}}}$}
\end{enumerate}
From Steps 1-3 it follows that the total computational complexity of the LRT detector is:\\ 
\mybox{CM:$~MK\cdot(N+K+1)\cdot \big((N_{\textsf{s}})^{L_{\textsf{tot}}}+N_{\textsf{s}})^{ML_{\textsf{ch}}}\big)+1$}\\
\mybox{CA:$~M\cdot\big(KN+(K-1)\cdot(K+1)\big)\cdot \big((N_{\textsf{s}})^{L_{\textsf{tot}}}+N_{\textsf{s}})^{ML_{\textsf{ch}}}\big)$}\\
, when the added complex multiplication is due to taking the division of the two exponential terms.
\subsubsection{Complexity Analysis for the Correlator}
\label{app:Correlator_Complexity_Analysis}
In the following, we detail the derivations of the computational complexity of the correlator detector in Eq. (\ref{Correlation_Test}) in terms of CMs and CAs:
\begin{enumerate}
   \item 
        $\Big(\mathbf{r}^{(\mathsf{cor})}_{L_{\textsf{sw}},0}[n]\Big)^H\cdot \mathbf{f}_{\textsf{sw}}$: vector multiplication of $\mathbf{r}^{(\mathsf{cor})}_{L_{\textsf{sw}},0}[n]\in \mathcal{C}^{L_{\textsf{sw}}~\times~1}$ and $\mathbf{f}_{\textsf{sw}}\in \mathcal{C}^{L_{\textsf{sw}}~\times~1}$, results in a complex number. The computational complexity is:\\ 
        \mybox{CM:$~L_{\textsf{sw}}$~~$|$~~CA:$~L_{\textsf{sw}}-1~~$} 
        \item
        $\Big\|\Big(\mathbf{r}^{(\mathsf{cor})}_{L_{\textsf{sw}},0}[n]\Big)^H\cdot \mathbf{f}_{\textsf{sw}}\Big\|^2$: we execute additional 1 complex multiplication. Thus, the total computional complexity of the correlator is:\\
        \mybox{CM:$~L_{\textsf{sw}}+1$~~$|$~~CA:$~L_{\textsf{sw}}-1~~$}
\end{enumerate}
\subsection{SNR Expression Evaluation}
\label{app:SNR_Expression_Evaluation}
In the following, we evaluate the signal to noise ratio ($SNR$) expression, corresponding to the received signal model in (\ref{Pre_Process}), the SNR is as follows:
\begin{ceqn}
\begin{align*}
SNR&=\frac{\mathds{E}\Big\{\big(\mathds{A}_{{L_{\textsf{tot}}},0}[n] \cdot \mathbf{s}_{L_{\textsf{tot}},0}[n]\big)^H\cdot \mathds{A}_{{L_{\textsf{tot}}},0}[n] \cdot \mathbf{s}_{L_{\textsf{tot}},0}[n]\Big\}}{\mathds{E}\Big\{\big(\mathbf{z}_{L_{\textsf{tot}},0}[n]\big)^H \cdot \mathbf{z}_{L_{\textsf{tot}},0}[n]\Big\}}\\
&\stackrel{(a)}{=}\frac{\mathds{E}\bigg\{\Tr \Big\{\big( \mathbf{s}_{L_{\textsf{tot}},0}[n]\big)^H\cdot\big(\mathds{A}_{{L_{\textsf{tot}}},0}[n] \big)^H \cdot \mathds{A}_{{L_{\textsf{tot}}},0}[n] \cdot \mathbf{s}_{L_{\textsf{tot}},0}[n]\Big\}\bigg\}}{\mathds{E}\bigg\{\Tr \Big\{\big(\mathbf{z}_{L_{\textsf{tot}},0}[n]\big)^H \cdot \mathbf{z}_{L_{\textsf{tot}},0}[n]\Big\}\bigg\}}\\
&\stackrel{(b)}{=}\frac{\mathds{E}\bigg\{\Tr \Big\{\big( \mathds{A}_{{L_{\textsf{tot}}},0}[n] \big)^H \cdot \mathds{A}_{{L_{\textsf{tot}}},0}[n] \cdot \mathbf{s}_{L_{\textsf{tot}},0}[n]\cdot \big(\mathbf{s}_{L_{\textsf{tot}},0}[n]\big)^H\Big\}\bigg\}}{\mathds{E}\bigg\{\Tr \Big\{  \mathbf{z}_{L_{\textsf{tot}},0}[n]\cdot \big(\mathbf{z}_{L_{\textsf{tot}},0}[n]\big)^H \Big\}\bigg\}}\\
&\stackrel{(c)}{=}\frac{\Tr \bigg\{\mathds{E} \Big\{\big( \mathds{A}_{{L_{\textsf{tot}}},0}[n] \big)^H \cdot \mathds{A}_{{L_{\textsf{tot}}},0}[n] \cdot \mathbf{s}_{L_{\textsf{tot}},0}[n]\cdot \big(\mathbf{s}_{L_{\textsf{tot}},0}[n]\big)^H\Big\}\bigg\}}{\Tr \bigg\{\mathds{E} \Big\{  \mathbf{z}_{L_{\textsf{tot}},0}[n]\cdot \big(\mathbf{z}_{L_{\textsf{tot}},0}[n]\big)^H \Big\}\bigg\}}\\
&\stackrel{(d)}{=}\frac{\Tr \bigg\{\big( \mathds{A}_{{L_{\textsf{tot}}},0}[n] \big)^H \cdot \mathds{A}_{{L_{\textsf{tot}}},0}[n] \cdot \mathds{E} \Big\{ \mathbf{s}_{L_{\textsf{tot}},0}[n]\cdot \big(\mathbf{s}_{L_{\textsf{tot}},0}[n]\big)^H\Big\}\bigg\}}{\Tr \bigg\{\mathds{E} \Big\{  \mathbf{z}_{L_{\textsf{tot}},0}[n]\cdot \big(\mathbf{z}_{L_{\textsf{tot}},0}[n]\big)^H \Big\}\bigg\}}\\
&\stackrel{(e)}{=}\frac{\Tr \bigg\{\big( \mathds{A}_{{L_{\textsf{tot}}},0}[n] \big)^H \cdot \mathds{A}_{{L_{\textsf{tot}}},0}[n] \cdot \sigma_s^2 \cdot \mathds{I}_{(L_{\textsf{tot}}+L_{\textsf{ch}}) \times (L_{\textsf{tot}}+L_{\textsf{ch}})}\Big\}\bigg\}}{\Tr \bigg\{\mathds{E} \Big\{  \mathbf{z}_{L_{\textsf{tot}},0}[n]\cdot \big(\mathbf{z}_{L_{\textsf{tot}},0}[n]\big)^H \Big\}\bigg\}}\\
&\stackrel{(f)}{=}\frac{ \sigma_s^2}{\Tr \bigg\{\mathds{E} \Big\{  \mathbf{z}_{L_{\textsf{tot}},0}[n]\cdot \big(\mathbf{z}_{L_{\textsf{tot}},0}[n]\big)^H \Big\}\bigg\}}\cdot \Tr \Big\{\big( \mathds{A}_{{L_{\textsf{tot}}},0}[n] \big)^H \cdot \mathds{A}_{{L_{\textsf{tot}}},0}[n]\Big\}
\end{align*}
\end{ceqn}
where $(a)$ follows since for a scalar $g$, $\Tr\{g\}=g$, $(b)$ follows from the cyclic property of the $\Tr\{\cdot\}$ operator \cite{lipschutz2009linear} : $\Tr\{\mathds{A}\mathds{B}\mathds{C}\}=\Tr\{\mathds{C}\mathds{A}\mathds{B}\}=\Tr\{\mathds{B}\mathds{C}\mathds{A}\}$, and $(c)$ follows since $\Tr\{\cdot\}$ is a linear operator, $(d)$ follows since only $\mathbf{s}_{L_{\textsf{tot}},0}[n]$ and $\mathbf{z}_{L_{\textsf{tot}},0}[n]$ are random. $(e)$ follows since from section \ref{Received Signal Model}, the symbols $S[n]$ are i.i.d with variance $\sigma_s^2$, and $(f)$ follows since trace is a linear operator. We note, that in the spacial case when the noise is a stationary Gaussian process with variance $\sigma_z^2$, the trace of the noise auto-covariance matrix is its size times its variance, accordingly
\begin{ceqn}
\begin{align*}
&SNR=\frac{ \sigma_s^2}{\sigma_z^2 \cdot L_{\textsf{tot}}}\cdot \Tr \Big\{\big( \mathds{A}_{{L_{\textsf{tot}}},0}[n] \big)^H \cdot \mathds{A}_{{L_{\textsf{tot}}},0}[n]\Big\}  
\end{align*}
\end{ceqn}
\end{appendix}

\bibliographystyle{IEEEtran}
\bibliography{Ref}

\begin{thebibliography}{10}
\providecommand{\url}[1]{#1}
\csname url@samestyle\endcsname
\providecommand{\newblock}{\relax}
\providecommand{\bibinfo}[2]{#2}
\providecommand{\BIBentrySTDinterwordspacing}{\spaceskip=0pt\relax}
\providecommand{\BIBentryALTinterwordstretchfactor}{4}
\providecommand{\BIBentryALTinterwordspacing}{\spaceskip=\fontdimen2\font plus
\BIBentryALTinterwordstretchfactor\fontdimen3\font minus
  \fontdimen4\font\relax}
\providecommand{\BIBforeignlanguage}[2]{{%
\expandafter\ifx\csname l@#1\endcsname\relax
\typeout{** WARNING: IEEEtran.bst: No hyphenation pattern has been}%
\typeout{** loaded for the language `#1'. Using the pattern for}%
\typeout{** the default language instead.}%
\else
\language=\csname l@#1\endcsname
\fi
#2}}
\providecommand{\BIBdecl}{\relax}
\BIBdecl

\bibitem{scholtz1980frame}
R.~Scholtz, ``Frame synchronization techniques,'' \emph{IEEE Transactions on
  Communications}, vol.~28, no.~8, pp. 1204--1213, Aug. 1980.

\bibitem{robertson1995optimal}
P.~Robertson, ``Optimal frame synchronization for continuous and packet data
  transmission,'' Ph.D. dissertation, Bundeswehr University Munich, 1995.

\bibitem{mengali2013synchronization}
U.~Mengali, \emph{Synchronization Techniques for Digital Receivers}.\hskip 1em
  plus 0.5em minus 0.4em\relax Springer Science \& Business Media, 2013.

\bibitem{gardner2006cyclostationarity}
W.~A. Gardner, A.~Napolitano, and L.~Paura, ``Cyclostationarity: Half a century
  of research,'' \emph{Signal Processing}, vol.~86, no.~4, pp. 639--697, Apr.
  2006.

\bibitem{nassar2012local}
M.~Nassar, J.~Lin, Y.~Mortazavi, A.~Dabak, I.~H. Kim, and B.~L. Evans, ``Local
  utility power line communications in the 3--500 khz band: channel
  impairments, noise, and standards,'' \emph{IEEE Signal Processing Magazine},
  vol.~29, no.~5, pp. 116--127, Sept. 2012.

\bibitem{shaked2017joint}
R.~Shaked, N.~Shlezinger, and R.~Dabora, ``Joint estimation of carrier
  frequency offset and channel impulse response for linear periodic channels,''
  \emph{IEEE Transactions on Communications}, vol.~66, no.~1, pp. 302--319,
  Aug. 2017.

\bibitem{Andrews:2014}
J.~G. Andrews, S.~Buzzi, W.~Choi, S.~V. Hanly, A.~Lozano, A.~C.~K. Soong, and
  J.~C. Zhang, ``What will 5g be?'' \emph{IEEE journal on selected areas in
  communications}, vol.~32, no.~6, pp. 1065--1082, Jun. 2014.

\bibitem{Dai:2015}
L.~Dai, B.~Wang, Y.~Yuan, S.~Han, C.-L. I, and Z.~Wang, ``Non-orthogonal
  multiple access for 5g: solutions, challenges, opportunities, and future
  research trends,'' \emph{IEEE communications magazine}, vol.~53, no.~9, pp.
  74--81, Sep. 2015.

\bibitem{Hong:2009}
X.~Hong, Z.~Chen, C.-X. Wang, S.~A. Vorobyov, and J.~S. Thompson, ``Cognitive
  radio networks: Interference cancelation and management techniques,''
  \emph{IEEE Vehicular Technology Magazine}, vol.~4, no.~4, pp. 76--84, Dec.
  2009.

\bibitem{Campbell:1983}
J.~Campbell, A.~J. Gibbs, and B.~M. Smith, ``The cyclostationary nature of
  crosstalk interference from digital signals in multipair cable - part i:
  Fundamentals,'' \emph{IEEE Transactions on communications}, vol.~31, no.~5,
  pp. 629--637, May 1983.

\bibitem{McHenry:2015}
M.~A. McHenry, D.~Roberson, and R.~J. Matheson, ``Phone to fridge: Shut up,''
  \emph{IEEE Spectrum}, vol.~52, no.~9, pp. 50--56, Aug. 2015.

\bibitem{massey1972optimum}
J.~L. Massey, ``Optimum frame synchronization,'' \emph{IEEE Transactions on
  Communications}, vol.~20, no.~2, pp. 115--119, Apr. 1972.

\bibitem{ramakrishnan2010frame}
B.~Ramakrishnan, ``Frame synchronization with large carrier frequency offsets:
  Point estimation versus hypothesis testing,'' in \emph{Proceedings of
  International Symposium on Communication Systems, Networks \& Digital Signal
  Processing (CSNDSP)}, Newcastle upon Tyne, UK, 2010, pp. 45--50.

\bibitem{liang2015sequential}
Y.~Liang, D.~Rajan, and O.~E. Eliezer, ``Sequential frame synchronization based
  on hypothesis testing with unknown channel state information,'' \emph{IEEE
  Transactions on Communications}, vol.~63, no.~8, pp. 2972--2984, Aug. 2015.

\bibitem{nielsen1973some}
P.~Nielsen, ``Some optimum and suboptimum frame synchronizers for binary data
  in {G}aussian noise,'' \emph{IEEE Transactions on Communications}, vol.~21,
  no.~6, pp. 770--772, Jun. 1973.

\bibitem{lui1987frame}
G.~Lui and H.~Tan, ``Frame synchronization for {G}aussian channels,''
  \emph{IEEE Transactions on Communications}, vol.~35, no.~8, pp. 818--829,
  Aug. 1987.

\bibitem{choi2002frame}
Z.~Y. Choi and Y.~H. Lee, ``Frame synchronization in the presence of frequency
  offset,'' \emph{IEEE Transactions on Communications}, vol.~50, no.~7, pp.
  1062--1065, Nov. 2002.

\bibitem{moon1991ml}
B.~H. Moon and S.~Soliman, ``M{L} frame synchronization for the {G}aussian
  channel with {ISI},'' in \emph{Proceedings of the {IEEE} International
  Conference on Communications {(ICC)}}, Denver, CO, 1991, pp. 1698--1702.

\bibitem{wang2004continuous}
Y.~Wang, K.~Shi, and E.~Serpedin, ``Continuous-mode frame synchronization for
  frequency-selective channels,'' \emph{IEEE Transactions on Vehicular
  Technology}, vol.~53, no.~3, pp. 865--871, May 2004.

\bibitem{gansman1997optimum}
J.~A. Gansman, M.~P. Fitz, and J.~V. Krogmeier, ``Optimum and suboptimum frame
  synchronization for pilot-symbol-assisted modulation,'' \emph{IEEE
  Transactions on Communications}, vol.~45, no.~10, pp. 1327--1337, Oct. 1997.

\bibitem{chiani2004optimum}
M.~Chiani and M.~G. Martini, ``Optimum synchronization of frames with unknown
  variable lengths on {G}aussian channels,'' in \emph{Proceedings of the {IEEE}
  Global Telecommunications Conference {(GLOBECOM)}}, vol.~6, Dallas, TX, 2004,
  pp. 4087--4091.

\bibitem{chiani2005practical}
------, ``Practical frame synchronization for data with unknown distribution on
  {AWGN} channels,'' \emph{IEEE Communications Letters}, vol.~9, no.~5, pp.
  456--458, May 2005.

\bibitem{chiani2010noncoherent}
M.~Chiani, ``Noncoherent frame synchronization,'' \emph{IEEE Transactions on
  Communications}, vol.~58, no.~5, pp. 1536--1545, May 2010.

\bibitem{hasselmann1981techniques}
K.~Hasselmann and T.~Barnett, ``Techniques of linear prediction for systems
  with periodic statistics,'' \emph{Journal of the Atmospheric Sciences},
  vol.~38, no.~10, pp. 2275--2283, Nov. 1981.

\bibitem{cavers1991analysis}
J.~K. Cavers, ``An analysis of pilot symbol assisted modulation for {R}ayleigh
  fading channels,'' \emph{IEEE Transactions on vehicular technology}, vol.~40,
  no.~4, pp. 686--693, Nov. 1991.

\bibitem{giannakis1998cyclostationary}
G.~B. Giannakis, ``Cyclostationary signal analysis,'' in \emph{Digital Signal
  Processing Handbook}, V.~K. Madisetti and D.~B. Williams, Eds.\hskip 1em plus
  0.5em minus 0.4em\relax CRC Press, 1999, ch.~17.

\bibitem{parzen1979approach}
E.~Parzen and M.~Pagano, ``An approach to modeling seasonally stationary time
  series,'' \emph{Journal of Econometrics}, vol.~9, no. 1-2, pp. 137--153, Jan.
  1979.

\bibitem{newton1982using}
H.~J. Newton, ``Using periodic autoregressions for multiple spectral
  estimation,'' \emph{Technometrics}, vol.~24, no.~2, pp. 109--116, May 1982.

\bibitem{chiani2006sequential}
M.~Chiani and M.~G. Martini, ``On sequential frame synchronization in {AWGN}
  channels,'' \emph{IEEE Transactions on Communications}, vol.~54, no.~2, pp.
  339--348, Feb. 2006.

\bibitem{van2004detection}
H.~L. Van~Trees, \emph{Detection, Estimation, and Modulation Theory, Part I:
  Detection, Estimation, and Linear Modulation Theory}.\hskip 1em plus 0.5em
  minus 0.4em\relax John Wiley \& Sons, 2004.

\bibitem{mendenhall2012introduction}
W.~Mendenhall, R.~J. Beaver, and B.~M. Beaver, \emph{Introduction to
  Probability and Statistics}.\hskip 1em plus 0.5em minus 0.4em\relax Cengage
  Learning, 2012.

\bibitem{nielsen2016guaranteed}
F.~Nielsen and K.~Sun, ``Guaranteed bounds on the {K}ullback-{L}eibler
  divergence of univariate mixtures using piecewise log-sum-exp inequalities,''
  \emph{arXiv preprint arXiv:1606.05850}, 2016.

\bibitem{morelli2000carrier}
M.~Morelli and U.~Mengali, ``Carrier-frequency estimation for transmissions
  over selective channels,'' \emph{IEEE Transactions on Communications},
  vol.~48, no.~9, pp. 1580--1589, Sept. 2000.

\bibitem{godard1980self}
D.~Godard, ``Self-recovering equalization and carrier tracking in
  two-dimensional data communication systems,'' \emph{IEEE Transactions on
  Communications}, vol.~28, no.~11, pp. 1867--1875, Nov. 1980.

\bibitem{proakis2001digital}
J.~G. Proakis and M.~Salehi, \emph{Digital Communications}.\hskip 1em plus
  0.5em minus 0.4em\relax McGraw-Hill, New York, 2001, vol.~4.

\bibitem{ungerboeck1972theory}
G.~Ungerboeck, ``Theory on the speed of convergence in adaptive equalizers for
  digital communication,'' \emph{IBM Journal of Research and Development},
  vol.~16, no.~6, pp. 546--555, Nov. 1972.

\bibitem{crozier1991least}
S.~Crozier, D.~D. Falconer, and S.~Mahmoud, ``Least sum of squared errors
  ({LSSE}) channel estimation,'' in \emph{IEE Proceedings F-Radar and Signal
  Processing}, vol. 138, no.~4, 1991, pp. 371--378.

\bibitem{houcke2003blind}
S.~Houcke, A.~Chevreuil, and P.~Loubaton, ``Blind equalization-case of an
  unknown symbol period,'' \emph{IEEE transactions on signal processing},
  vol.~51, no.~3, pp. 781--793, Mar. 2003.

\bibitem{karatsuba1995complexity}
A.~A. Karatsuba, ``The complexity of computations,'' \emph{Proceedings of the
  Steklov Institute of Mathematics-Interperiodica Translation}, vol. 211, pp.
  169--183, Jan. 1995.

\bibitem{fawcett2006introduction}
T.~Fawcett, ``An introduction to {ROC} analysis,'' \emph{Pattern Recognition
  Letters}, vol.~27, no.~8, pp. 861--874, Jun. 2006.

\bibitem{kumar2019novel}
M.~Kumar and R.~Dabora, ``A novel sampling frequency offset estimation
  algorithm for {OFDM} systems based on cyclostationary properties,''
  \emph{IEEE Access}, vol.~7, pp. 100\,692--100\,705, 2019.

\bibitem{FazelMultiCarrierBook}
K.~Fazel and S.~Kaiser, \emph{Multi-Carrier and Spread Spectrum Systems: From
  OFDM and MC-CDMA to LTE and WiMAX}.\hskip 1em plus 0.5em minus 0.4em\relax
  John Wiley \& Sons, Ltd., 2nd ed., 2008.

\bibitem{neeser1993proper}
F.~D. Neeser and J.~L. Massey, ``Proper complex random processes with
  applications to information theory,'' \emph{IEEE Transactions on Information
  Theory}, vol.~39, no.~4, pp. 1293--1302, Jul. 1993.

\bibitem{chiueh2012baseband}
T.-D. Chiueh, P.-Y. Tsai, L.~I-Wei, and T.-D. Chiueh, \emph{Baseband receiver
  design for wireless MIMO-OFDM communications}.\hskip 1em plus 0.5em minus
  0.4em\relax Wiley Online Library, 2012.

\bibitem{lapidoth2017foundation}
A.~Lapidoth, \emph{A Foundation in Digital Communication}.\hskip 1em plus 0.5em
  minus 0.4em\relax Cambridge University Press, 2017.

\bibitem{meyer2000matrix}
C.~D. Meyer, \emph{Matrix analysis and applied linear algebra}.\hskip 1em plus
  0.5em minus 0.4em\relax Siam, 2000, vol.~71.

\bibitem{cormen2009introduction}
T.~H. Cormen, C.~E. Leiserson, R.~L. Rivest, and C.~Stein, \emph{Introduction
  to Algorithms}.\hskip 1em plus 0.5em minus 0.4em\relax MIT press, 2009.

\bibitem{lipschutz2009linear}
S.~Lipschutz and M.~Lipson, \emph{Linear Algebra: Schaum's Outlines}.\hskip 1em
  plus 0.5em minus 0.4em\relax McGraw-Hill, 2009.

\end{thebibliography}


\end{document}